%% file: main.tex
\newcommand{\aj}{Astronomical Journal}

\newcommand{\araa}{Annual Review of Astron and Astrophys}
\newcommand{\apj}{Astrophysical Journal}
\newcommand{\apjl}{Astrophysical Journal, Letters}
\newcommand{\apjs}{Astrophysical Journal, Supplement}

\newcommand{\apss}{Astrophysics and Space Science}
\newcommand{\aap}{Astronomy and Astrophysics}
\newcommand{\aapr}{Astronomy and Astrophysics Reviews}

\newcommand{\jcap}{Journal of Cosmology and Astroparticle Physics}

\newcommand{\mnras}{Monthly Notices of the RAS}

\newcommand{\prd}{Physical Review D}

\newcommand{\prl}{Physical Review Letters}
\newcommand{\pasa}{Publications of the Astron. Soc. of Australia}

\newcommand{\pasj}{Publications of the ASJ}

\newcommand{\ssr}{Space Science Reviews}

\newcommand{\nat}{Nature}

\newcommand{\physrep}{Physics Reports}

\documentclass[
11pt, 
english, 
doublespacing,
nolistspacing, 
liststotoc, 
toctotoc, 
headsepline, 
]{MastersDoctoralThesis} 
\usepackage{sectsty}
\usepackage[utf8]{inputenc} 
\usepackage[LGR,T1]{fontenc} 
\usepackage{mathpazo} 
\usepackage{lscape}
\usepackage{enumitem}
\usepackage{amsmath}
\usepackage{amssymb}
\usepackage[table]{xcolor}
\usepackage{listings}
\usepackage{textcomp}
\usepackage{rotating}
\usepackage{subcaption}
\usepackage{hyphenat}
\usepackage{tikz}
\usetikzlibrary{arrows.meta,automata,positioning}

\usepackage{multirow}

\definecolor{listinggray}{rgb}{225,224, 224}
\definecolor{lbcolor}{rgb}{255, 255, 255}
\usepackage[authoryear]{natbib}

\definecolor{frontmattercolor}{HTML}{660099}
\definecolor{chaptercolor}{HTML}{660099}
\definecolor{sectioncolor}{HTML}{335C67}
\definecolor{subsectioncolor}{HTML}{335C67}
\definecolor{subsubsectioncolor}{HTML}{335C67}

\definecolor{boldcolor}{HTML}{2A9D8F}
\newcommand{\bcf}{\bf \color{boldcolor}}

\definecolor{urlcolor}{HTML}{BE9827}
\definecolor{hyperlinkcolor}{HTML}{BE9827}
\definecolor{citationcolor}{HTML}{BE9827}
\definecolor{simulationinsightcolor}{HTML}{AB260A}

\chapterfont{\color{chaptercolor}}
\sectionfont{\color{sectioncolor}}
\subsectionfont{\color{subsectioncolor}}
\subsubsectionfont{\color{subsubsectioncolor}}

\usepackage[autostyle=true]{csquotes} 

\newcommand{\annoterel}[3][]{%
  \overset{%
    \substack{\hidewidth\text{#2}\hidewidth\\#1\downarrow}%
  }{#3}%
}

\newcommand{\textgreek}[1]{\begingroup\fontencoding{LGR}\selectfont#1\endgroup}

\newcommand{\simulationinsight}[1]{{\color{simulationinsightcolor}{#1}}}
\newcommand{\velociraptor}{\textsc{VELOCIraptor}\,}
\newcommand{\swift}{\textsc{SWIFT}\,}

\defcitealias{CC02}{CC02}
\defcitealias{CM02}{CM02}

\thesistitle{Building models of the Universe with hydrodynamic simulations} 
\supervisor{Prof. Scott T. Kay} 
\examiner{Prof. Anna M.M. Scaife} 
\degree{Doctor of Philosophy} 
\author{Dr. Edoardo Altamura} 
\addresses{} 

\subject{Computational cosmology} 
\university{\href{https://www.manchester.ac.uk/}{The University of Manchester}} 
\department{\href{https://www.physics.manchester.ac.uk/}{Department of Physics and Astronomy in the School of Natural Sciences}} 
\group{\href{http://www.jodrellbank.manchester.ac.uk/}{Jodrell Bank Centre for Astrophysics}} 
\faculty{\href{http://www.se.manchester.ac.uk/}{Faculty of Science and Engineering}} 

\AtBeginDocument{
\hypersetup{pdftitle=\ttitle} 
\hypersetup{pdfauthor=\authorname} 
\hypersetup{pdfkeywords=\keywordnames} 
}

\newcommand{\changeurlcolor}[1]{\hypersetup{urlcolor=#1}}

\begin{document}


\pagestyle{plain} 


\begin{titlepage}
\begin{center}
\changeurlcolor{frontmattercolor}

\begin{figure}
    \centering
    \includegraphics[width=0.5\textwidth]{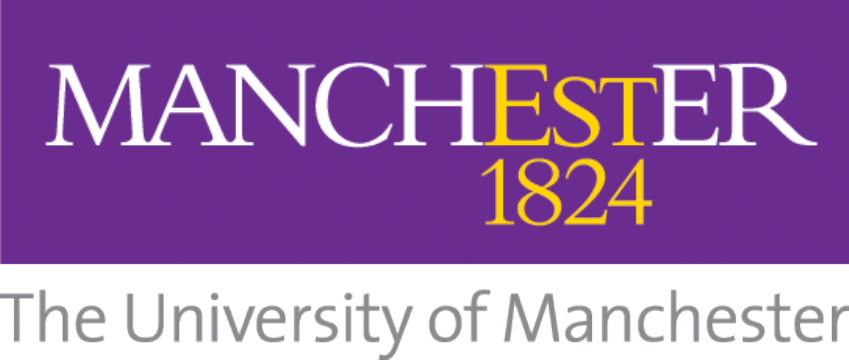}
\end{figure}

 \vspace*{.01\textheight}
\textsc{\Large Doctoral Thesis}\\[0.5cm] 

\HRule \\[0.4cm] 
{\huge \bfseries \textcolor{sectioncolor}{\ttitle}\par}\vspace{0.4cm} 
\HRule \\[1.cm] 
 
\begin{minipage}[t]{0.4\textwidth}
\begin{flushleft} \large
\emph{Author:}\\
\href{https://research.manchester.ac.uk/en/persons/edoardo.altamura}{\authorname} 
\end{flushleft}
\end{minipage}
\begin{minipage}[t]{0.4\textwidth}
\begin{flushright} \large
\emph{Supervisors:} \\
\href{https://research.manchester.ac.uk/en/persons/scott.kay}{\supname}\\ 
\href{https://research.manchester.ac.uk/en/persons/jens.chluba}{Prof. Jens Chluba}\\ 
\end{flushright}
\end{minipage}
\begin{minipage}[c]{0.4\textwidth}
\begin{center}\large
\emph{Examiners:}\\
\href{https://www.turing.ac.uk/people/researchers/anna-scaife}{\examname}\\
\href{https://www.ljmu.ac.uk/about-us/staff-profiles/faculty-of-engineering-and-technology/astrophysics-research-institute/ian-mccarthy}{Prof. Ian G. McCarthy}
\end{center}
\end{minipage}\\[1cm] 
\vfill
\large \textit{A thesis submitted to the University of Manchester\\ for the degree of \degreename}\\[0.3cm] 
\textit{in the}\\[0.4cm]
\deptname\\\facname\\[1cm] 
 
\vfill

{\large {\the\year}}\\[2cm] 

\vfill

\end{center}
\end{titlepage}

\begin{flushright}
    Blank page
\end{flushright}

\pagebreak

\tableofcontents 

\listoffigures 

\listoftables 


\begin{abbreviations}{ll} 
\textcolor{white}{Blank line to prevent indent}\\
{\bcf AGN} & {\bcf A}ctive {\bcf G}alactic {\bcf N}ucleus\\
{\bcf BH} & {\bcf B}lack {\bcf H}ole\\
{\bcf SMBH} & {\bcf S}uper {\bcf M}assive {\bcf B}lack {\bcf H}ole\\
{\bcf SN} & {\bcf S}uper{\bcf n}ova\\
{\bcf SNe} & {\bcf S}uper{\bcf n}ova{\bcf e}\\
{\bcf BCG} & {\bcf B}rightest {\bcf C}entral {\bcf G}alaxy\\
{\bcf GR} & {\bcf G}eneral {\bcf R}elativity\\
{\bcf SPH} & {\bcf S}moothed {\bcf P}article {\bcf H}ydrodynamics\\
{\bcf CC} & {\bcf C}ool-{\bcf C}ore (clusters)\\
{\bcf NCC} & {\bcf N}on-{\bcf C}ool-{\bcf C}ore (clusters)\\
{\bcf HPC} & {\bcf H}igh-{\bcf P}erformance {\bcf C}omputing\\
{\bcf DM} & {\bcf D}ark {\bcf M}atter\\
{\bcf CDM} & {\bcf C}old {\bcf D}ark {\bcf M}atter\\
{\bcf DMO} & {\bcf D}ark-{\bcf M}atter-{\bcf O}nly (simulation)\\
{\bcf HMF} & {\bcf H}alo {\bcf M}ass {\bcf F}unction\\
{\bcf PH} & {\bcf P}eano-{\bcf H}ilbert (curve)\\
{\bcf KH} & {\bcf K}elvin-{\bcf H}elmholtz (instability)\\
{\bcf MPI} & {\bcf M}essage-{\bcf P}assing {\bcf I}nterface\\
{\bcf SZ} & {\bcf S}unyaev-{\bcf Z}eldovich (effects)\\
{\bcf tSZ} & {\bcf t}hermal {\bcf S}unyaev-{\bcf Z}eldovich (effect)\\
{\bcf kSZ} & {\bcf k}inetic {\bcf S}unyaev-{\bcf Z}eldovich (effect)\\
{\bcf rkSZ} & {\bcf r}otational {\bcf k}inetic {\bcf S}unyaev-{\bcf Z}eldovich (effect)\\
{\bcf CMB} & {\bcf C}osmic {\bcf M}icrowave {\bcf B}ackground\\
{\bcf ICM} & {\bcf I}ntra-{\bcf C}luster {\bcf M}edium\\
{\bcf IGM} & {\bcf I}nter-{\bcf G}alactic {\bcf M}edium\\
{\bcf LoS} & {\bcf L}ine {\bcf o}f {\bcf S}ight\\
{\bcf NFW} & {\bcf N}avarro-{\bcf F}renk-{\bcf W}hite (profile)\\
{\bcf ZA} & {\bcf Z}eldovich {\bcf A}pproximation\\
{\bcf TTT} & {\bcf T}idal {\bcf T}orque {\bcf T}heory\\
{\bcf EdS} & {\bcf E}instein-{\bcf d}e {\bcf S}itter (universe)\\
{\bcf CoP} & {\bcf C}entre {\bcf o}f {\bcf P}otential\\
{\bcf CoM} & {\bcf C}entre {\bcf o}f {\bcf M}ass\\
{\bcf FoF} & {\bcf F}riends-{\bcf o}f-{\bcf F}riends (algorithm)\\
{\bcf STHC} & {\bcf S}pherical {\bcf T}op-{\bcf H}at {\bcf C}ollapse (model)\\
{\bcf ICL} & {\bcf I}ntra-{\bcf C}luster {\bcf L}ight\\
{\bcf LPT} & {\bcf L}agrangian {\bcf P}erturbation {\bcf T}heory\\
{\bcf DES} & {\bcf D}ark {\bcf E}nergy {\bcf S}urvey\\

\end{abbreviations}










\begin{spacing}{1.}
\begin{abstract}
\addchaptertocentry{\abstractname} 
Hydrodynamic simulations have become irreplaceable in modern cosmology for exploring complex systems and making predictions to guide future observations. In this doctoral thesis, we address these challenges and explore the role of numerical simulations in advancing our understanding of galaxy formation. In Chapter \ref{chapter:1}, we begin our discussion by describing a philosophical framework for understanding the role of simulations in science. We argue that simulations can bridge the gap between empirical knowledge and fundamental, universal knowledge. The validation of simulation outcomes is crucial, and stresses the importance of achieving a balance between trustworthiness and scepticism in the scientific community. Next, Chapter \ref{chapter:2} introduces relevant cosmological concepts, outlines the processes leading to the formation of structures at early times and how numerical simulations can probe the growth of perturbations in the non-linear regime at late times. We discuss aspects of non-linear structure formation with hydrodynamics and baryonic physics, allowing for direct comparisons between synthetic and observational data. Chapter \ref{chapter:3} provides technical details of numerical simulations, including the production pipeline of zoom-in simulations used to model individual objects in detail. We discuss the development of novel methods to mitigate known shortcomings. Then, we assess the weak-scaling performance of the \swift code in solving a purely hydrodynamic problem and found it to be one of the hydrodynamic codes with the highest parallel efficiency. In Chapter \ref{chapter:4}, we study the rotational kinetic Sunyaev-Zeldovich (rkSZ) effect for high-mass galaxy clusters from the MACSIS simulations. We find a maximum signal of $\gtrsim$100 $\mu$K, approximately 30 times stronger than early predictions based on self-similar models, opening prospects for future observations. In Chapter \ref{chapter:5}, we address a tension between the distribution of entropy measured from observations and predicted by simulations of groups and clusters of galaxies. We find that most recent hydrodynamic simulations systematically over-predict the entropy profiles by up to one order of magnitude, leading to profiles that are shallower and higher than the power-law-like entropy profiles that have been observed. We discuss the dependence of the entropy distribution on different hydrodynamic and sub-grid parameters using variations of the EAGLE model, setting the directives for future work. Chapter \ref{chapter:6} explores the evolution of global properties of groups and clusters of galaxies, together with the entropy profiles as a function of cosmic time. We identify four key phases in the evolution of these systems and report power-law-like entropy profiles at high redshift for both objects. However, at late times, an entropy plateau develops and alters the shape of the profile. 
\end{abstract}


\begin{popularabstract}
\addchaptertocentry{\popularabstractname} 
Numerical simulations are tools for solving complex problems and achieving breakthroughs in the field of cosmology and extra-galactic astrophysics. In this work, we discuss the role of simulations in bridging the gap between the fundamental laws of physics and the complex astronomical objects we can observe. As a next step, in Chapter \ref{chapter:2}, we describe how galaxies and clusters of galaxies are thought to have formed in the early Universe, and the role of supercomputer simulations in probing the more complex evolution of structures in the late-time Universe until the present day. Chapter \ref{chapter:3} illustrates the technical details of numerical simulations and the production pipeline of zoom-in simulations. Chapter \ref{chapter:4} presents the prediction of the Sunyaev-Zeldovich effect arising from the rotation of the hot gas in simulated galaxy clusters, and Chapter \ref{chapter:5} probes a tension between the temperature and density of the gas measured from observations and predicted by simulations of groups and clusters of galaxies. Finally, Chapter \ref{chapter:6} briefly illustrates the evolution of global properties of the group and cluster of galaxies, together with the entropy profiles as a function of cosmic time. This work contributes to the advancement of computational cosmology by providing insights into the advantages and disadvantages of simulations, outlining the technical details of numerical simulations, and exploring new pathways to study properties of the Universe using simulations.
\end{popularabstract}


\begin{declaration}
\addchaptertocentry{\authorshipname} 
\noindent I, \authorname, declare that this thesis titled, \enquote{\ttitle} and the work presented in it are my own. \\[0.3cm]

\noindent I declare that:

\begin{itemize} 
\item This work was done wholly while in candidature for a research degree at this University.
\item No portion of the work referred to in the thesis has been submitted in support of an application for another degree or qualification of this or any other University or other Institute of learning.
\item Where I have consulted the published work of other researchers, this is always clearly attributed.
\item Where I have quoted from the work of others, the source is always given. With the exception of such quotations, this thesis is entirely the author's own work.
\item I have acknowledged all main sources of support.
\item Where the thesis is based on work done by the author jointly with others, it is clearly specified what was done by others and what the author has contributed.\\
\end{itemize}
 
 
\end{declaration}

\chapter*{Publications}
This doctoral thesis is accompanied by the following supporting research papers:
\begin{itemize}
    \item \textbf{Edoardo Altamura}, Scott T. Kay, Jens Chluba, Imogen Towler, \textit{Galaxy cluster rotation revealed in the MACSIS simulations with the kinetic Sunyaev-Zeldovich effect}; Monthly Notices of the Royal Astronomical Society, Volume 524, Issue 2, September 2023, Pages 2262–2289, \href{https://doi.org/10.1093/mnras/stad1841}{https://doi.org/10.1093/mnras/stad1841}\\
    This paper forms the basis of Chapter \ref{chapter:4}.
    
    \item \textbf{Edoardo Altamura}, Scott T. Kay, Richard G Bower, Matthieu Schaller, Yannick M. Bah{\'e}, Joop Schaye, Josh Borrow, Imogen Towler, \textit{EAGLE-like simulation models do not solve the entropy core problem in groups and clusters of galaxies}; Monthly Notices of the Royal Astronomical Society, Volume 520, Issue 2, April 2023, Pages 3164–3186; \href{https://doi.org/10.1093/mnras/stad342}{https://doi.org/10.1093/mnras/stad342}\\
    This paper forms the basis of Chapter \ref{chapter:5}.
    
    \item \textbf{Edoardo Altamura} and Scott T. Kay, \textit{Merging activity can disrupt low entropy cores in EAGLE-like simulations models}, in preparation.\\
    This working paper forms the basis of Chapter \ref{chapter:6}.
\end{itemize}

In addition, the author contributed to the following research papers:
\begin{itemize}
    \item Josh Borrow, Matthieu Schaller, Yannick M. Bah{\'e}, Joop Schaye, Aaron D. Ludlow, Sylvia Ploeckinger, Folkert S.J. Nobels, \textbf{Edoardo Altamura}, \textit{The impact of stochastic modeling on the predictive power of galaxy formation simulations}; Pre-print, November 2022; \href{https://ui.adsabs.harvard.edu/abs/2022arXiv221108442B}{https://ui.adsabs.harvard.edu/abs/2022arXiv221108442B}. Submitted to \mnras~and under review.
    
    \item Imogen Towler, Scott T. Kay, \textbf{Edoardo Altamura}, \textit{Gas clumping and its effect on hydrostatic bias in the MACSIS simulations}; Monthly Notices of the Royal Astronomical Society, Volume 520, Issue 4, April 2023, Pages 5845–5857;\\  \href{https://doi.org/10.1093/mnras/stad453}{https://doi.org/10.1093/mnras/stad453}
\end{itemize}

\addcontentsline{toc}{chapter}{Publications}

\changeurlcolor{urlcolor}

\cleardoublepage


\begin{copyrightstatement}

\addchaptertocentry{\copyrightname} 

\begin{enumerate}[label=(\roman*)]
\item The author of this thesis (including any appendices and/or
schedules to this thesis) owns certain copyright or related rights
in it (the ``Copyright'') and s/he has given The University of
Manchester certain rights to use such Copyright, including for
administrative purposes.\\[0.3cm]
\item Copies of this thesis, either in full or in extracts and whether in
hard or electronic copy, may be made {\bcf only} in accordance with the
Copyright, Designs and Patents Act 1988 (as amended) and
regulations issued under it or, where appropriate, in accordance
with licensing agreements which the University has from time to
time. This page must form part of any such copies made.\\[0.3cm]
\item The ownership of certain Copyright, patents, designs, trademarks
and other intellectual property (the ``Intellectual Property'') and
any reproductions of copyright works in the thesis, for example
graphs and tables (``Reproductions''), which may be described in
this thesis, may not be owned by the author and may be owned
by third parties. Such Intellectual Property and Reproductions
cannot and must not be made available for use without the prior
written permission of the owner(s) of the relevant Intellectual
Property and/or Reproductions.\\[0.3cm]
\item Further information on the conditions under which disclosure,
publication and commercialisation of this thesis, the Copyright
and any Intellectual Property and/or Reproductions described in
it may take place is available in the University IP Policy (see
\href{http://documents.manchester.ac.uk/DocuInfo.aspx?DocID=2442
0}{documents.manchester.ac.uk}), in any relevant Thesis restriction declarations deposited in the
University Library, The University Library’s regulations (see
\href{http://www.library.manchester.ac.uk/about/regulations/}{www.library.manchester.ac.uk/about/regulations/}) and in
The University’s policy on Presentation of Theses.
\end{enumerate}
\end{copyrightstatement}
\end{spacing}
\cleardoublepage






\begin{acknowledgements}
\addchaptertocentry{\acknowledgementname} 
Just like gravitational forces, the experience of a doctoral program is influenced by short- and long-range interactions. Professionally, no interaction was closer and more valuable to me than that with my supervisor Scott Kay. He has been an outstanding teacher and mentor, effectively since my first year of undergraduate studies nearly 8 years ago, then through my Masters project and PhD. 

I was honoured and fortunate to collaborate with Richard Bower, whose enthusiasm and positive energy were unstoppable. The time spent working together on EAGLE-XL and ExCALIBUR was one of the highlights of my PhD, and a turning point in terms of the approach to science that he helped me develop.

I would like to thank Matthieu Schaller for being such a great host during my countless visits in Leiden, for his support and advice while writing my first paper and navigating the job market.

Working in the Manchester office, rather than from home, would have never been so memorable without office mates like Thomas Sweetnam, Danielius Banys, Nialh McCallum, Sankarshana Srinivasan, Valerio Gilles, Manish Patel, Eunseong Lee, Imogen Towler and Luke Hart. I am grateful to be able to call you colleagues and friends, and very soon also \textit{philosophi\ae~doctores} (PhD)!

I would like to thank my parents for supporting me during highs and lows in a myriad of ways; and those I was blessed to have crossed paths with during my past 8 years in Manchester, and who I consider my ''acquired`` family. To all those special individuals: I hope reading this paragraph will put a huge smile on your face, because I surely had one on mine while writing this thesis and thinking about you. 

I want to conclude with an alphabetically-ordered and possibly incomplete summary of the talented people I crossed paths with during my studies.

\newpage
\thispagestyle{plain}
\begin{spacing}{1.}
\begin{small}
\topskip0pt

\noindent Aaron Liang, Adriana Rossi, Agatino Rifatto, Alamghir Miah, Alamghir Miah, Aleksandra Lesniewska, Alessio Travasi, Alex Godeanu, Alex Jenkins, Alex Lisboa-Wright, Alexandra Bonta, Alice Eleanor Matthews, Amy Suddards, Andrea Sante, Andy Blackett-May, Angela Esposito, Angelo Angeletti, Anna Wolter, Anthony Wallace Cross, Antonio Josue Espin, Aodhan Burke, Arsim Ahmedi, Asger Jung Laursen, Bannawit Pimpanuwat, Bart Wlodarczyk-Sroka, Beatrice Blanes, Becky Brown, Ben Carter, Ben Symons, Ben Wilson, Beth Cropper, Brandon Wang, Brijesh Patel, Bruce Basset, Bryan Alexander Polski, Camille Lorfing, Charis Pooni, Charlotte Collier, Chen Zhaoting, Chenfu Shi, Chuamping Xi, Claudia Jano, Clive Dickinson, Collette Pakuza, Costanza Improta, Dan Thomas, Dani da Costa-Kendall, Daniel Cookman, Daniel Dupkala, Daniel Seal, Daniele Teresi, Danielius Banys, Danny Valentine, Dario Lorenzoni, Dave Beynon, David Whitworth, Dean Anthony Daggett, Dean Arksey, Du Zibin, Duncan Austin, Eddie West, Edmondo Mazzoncini, Elia Pizzati, Elizabeth Lee, Ellen Leahy, Ellie Callcut, Elliy Ocock, Emily Cuffin-Munday, Emily Nevinson, Emily Woodroofe, Emma Alexander, Erasmo Taglioni, Eu-Bin Kim, Eunseong Lee, Evgenii Chaikin, Fabrizio Simeoni, Faye Yan Zhang, Ferdinando Potentini, Fernando Tinaut, Filip Husko, Flavio Ryu, Folkert Nobels, Francesca Pearce, Francesco Barabucci, Francesco Pace, Francesco Trono, Francois de Tournemire, Gaargi Jain, Gabriele Panebianco, Gaetano Valentini, George Pap, Georgie Stroud, Gianclaudio Ciampechini, Gibran Yael Andrade, Giovanni Altamura, Giulia Natalicchi, Gruff Jones, Gualtiero Campodonico, Guanchen Peng, Gwen Bovill Hale, Hannu Parviainen, Harry Bevins, Harry Howell, Harry Waring, Hawys Williams, Heidi Korhonen, Henry Philip, Holly McGrath, Hongming Tang, Hugo Morel, Imogen Towler, Iuliana Nitu, Jade Ajagun-Brauns, Jaime Garcia Iglesias, Jakob Wenninger, Jakub Lusnak, James Simpson, James Stringer, Jamie Bailey, Jamie Gooding, Janni Harju, Jaroslav Merc, Jaspal Singh, Jelle Houtman, Jenifer Hanki, Jenni Shackleton, Jens Chluba, Joe Tomlinson, Joe Walsh, Joel Williams, Joey Braspenning, Johan Peter Uldall Fynbo, John Proctor, Johnny Escobar, Jonathan Wong, Joop Schaye, Josh Borrow, Josh Hayes, Josh Lindsay, Joshua Winter, Julia Sisk Reynes, Junaid Ali Bokhari, Karolis Markauskas, Katie Margate Price, Kavishen Ramsamy, Kevin Golan, Kevin Kilburn, Kiana Salimipour, Krishan Jethwa, Lars Lindberg Christensen, Laura Driessen, Leonardo Villalba Rocha, Leroy Huang, Levi Evans, Liandro D'Ascenzo, Lizz Martindale, Lloyd Cawthorne, Lora Weingerl, Luan Cassal, Lucas Holik, Lucas Nogueira, Lucia Maltoni, Lucy Grayson, Luismi San Martin Fernandez, Luke Dyks, Luke Hart, Luke Jones, Mahmoud Elhawati, Malcolm David Gray, Manish Patel, Manlio Bellesi, Marc Sumner, Marcus Allan, Marek Murin, Margaux Zaffran, Maria Pia Morresi, Maria Poncela, Maria Potentini, Marina Guzzo, Mario Mangiapia, Martin Murin, Martin Page, Martina Jorgensen, Massimiliano Canzi, Mateusz Wasilewski, Matt Wheeler, Matteo Albano, Matthew Robinson, Matthieu Schaller, Mauro Dolci, Max Potter, Mayang Kinanti Puteri, Meirin Oan Evans, Melanie Wendl, Mercedes Anjelica Thompson, Michael Garrett, Michael Schulz, Michael Wright, Miroslav Fedurco, Mithran Radha Krishnan, Muirine D'Alexei, Nada Ihanec, Nial Perry, Nialh McCallum, Niamh Fearon, Oana Bazavan, Olga Khabuktanova, Oskar Nummedal, Pal Csoke, Paolo Arru, Patrick Odagiu, Paul John Hanafin, Pep Rey Altimir, Periklis Petropoulos, Philip Macdonald, Piera Battista, Pietro Mazzaglia, Pip van den Elzen, Raghav Vashishtha, Raluca Chitic, Raoul Scribbans, Rebecca Barosi, Rene Breton, Riccardo Iena, Riccardo Trono, Richard G Bower, Rio Tamargo, Robert Naylor, Robert White, Robin Upham, Roi Kugel, Roke Cepeda, Rory Varley, Rudolfs Grobins, Rudolfs Treilis, Ruifan Tang, Ryun O'Neil, Sabina Ascenzi, Sabrina Cheung, Sam Bradley, Sam Markovic, Sam Parkinson, Sam Poland, Sandra Nogueira Dos Reis, Sankarshana Srinivasan, Scott T Kay, Sean Gunness, Sean O'Braonain, Selwyn Lilley, Sergio Belmonte, Shahbaz Chaudhry, Shaojin Liu, Shresht Jain, Shuchong Ding, Simon Finney, Simon Hellewell, Simon Song, Simona Todorova, Simonetta Regoli, Sofia Perelli-Rocco, Sol Cotton, Sotirios Karamitsos, Soumya Shreeram, Stefano Coretta, Steph England, Stephanie Buttigieg, Sue Pritchard, Susanna Azzoni, Swapnil Singh, Syed Ali-Akbar Alamdar Hassan, Symeon Ndirittu, Syrian Fis, Tadas Indrele, Takaaki Vincent Joya, Thomas Wynne Condliffe, Tianyue Chen, Timothy Chong, Toby Overton, Tom Armitage, Tom Kite, Tom Parry, Tom Sweetnam, Toni Mary Newell, Tuvshinzaya Erdenehuu, Tymoteusz Chatys, Tymoteusz Miara, Valentina Campodonico, Valentina Pivetta, Valerio Gilles, Valters Jekabs Zakrevskis, Vlad-Marius Griguta, Wincenty Szymura, Xiangrui Zheng, Yasuhiro Yamamoto, Ynyr Harris, Yuankang Liu, Yunho Jeong, Yvonne Chow, Zainab Ahmed, Zak Lawrence, Zixin Wang, and Zongyi Li.

\end{small}
\end{spacing}

\end{acknowledgements}


\dedicatory{To those who have been my sources of inspiration throughout my astrophysical journey.} 

\changeurlcolor{urlcolor}


\begin{prefacestatement}
\addchaptertocentry{\prefacename} 
\section*{The author}
\authorname~is an Italian-British astrophysicist and data scientist, who published research on galaxy formation, the core-entropy problem and the kinetic Sunyaev\hyp{}Zeldovich effect. He is a specialist in massively parallel hydrodynamic simulations of galaxy formation, big data pipelines and high-performance computing. 

Originally from Macerata, a town in central Italy, \authorname~moved to the UK in 2015 to read for a Master in Physics and Astrophysics (MPhys) at the University of Manchester. In 2019, he graduated with first-class honours and a thesis titled ``\textit{Substructure in hydrodynamic simulations of galaxy clusters}``, based on research done with Martin Murin and supervised by \supname~on the kinetic Sunyaev-Zeldovich effect.

In the third quarter of 2019, \authorname~was awarded a UKRI-STFC scholarship and the President’s Doctoral Scholarship (PDS) to fund his research as a PhD candidate at Jodrell Bank Centre for Astrophysics under the guidance of \supname. In 2021, his research proposal on the development of ``\textit{AGN feedback schemes for the \swift code and software optimisation for Exascale high-performance computers}`` was shortlisted by the HPC-Europa3 Commission. The support included funding for a 3-month visit at Leiden University (the Netherlands) to collaborate with Dr. Matthieu Schaller and the computational cosmology group at Leiden Observatory led by Prof. Joop Schaye. The prize included computing time for 50k CPU-hours on the Surf-SARA Dutch supercomputing facility.

After the submission of this thesis, \authorname~continued working as a Postdoctoral Research Associate in the cosmology group at the University of Manchester, in collaboration with the Virgo Consortium for computational cosmology and the SPH-Exascale group in the ExCALIBUR Collaboration.

\end{prefacestatement}



\pagestyle{thesis} 


\include{Chapters/Chapter1}
\include{Chapters/Chapter2}
\include{Chapters/Chapter3}
\include{Chapters/Chapter4} 
\include{Chapters/Chapter5}

\include{Chapters/Chapter6} 
\include{Chapters/Chapter7} 


\appendix 


\include{Appendices/AppendixA}


\label{Bibliography}  
\bibliographystyle{mnras}  
\bibliography{Bibliography}


\end{document}

%% file: Chapters/Chapter1.tex
\chapter{Introduction}
\label{chapter:1}

\section{Philosophy of science}
\label{sec:intro-philosophy}

\begin{figure}
    \centering
    \includegraphics[width=\textwidth]{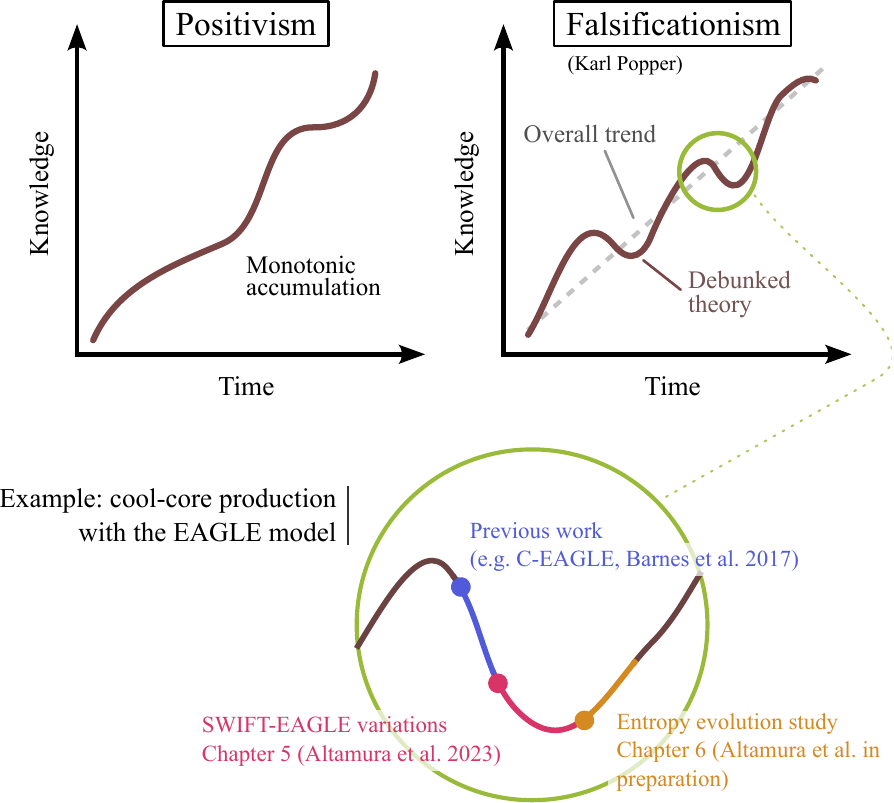}    
    \smallskip
    \caption[Philosophy of science diagram]{In the top row, graphical illustration of the positivist thought process (left) compared to the Popper's falsificationism (right). From the top right panel, we highlight the overall trend of the scientific progress over time with a dashed line, and indicate an example of debunked theory. In the inset (bottom), we illustrate the logic of Popper's idea of scientific falsifiability in the context of the production of cool-cores galaxy groups and clusters in the EAGLE model, analysed in previous studies \citep[e.g.][]{ceagle.barnes.2017} and further investigated in Chapters \ref{chapter:5} \citep{2023MNRAS.520.3164A} and \ref{chapter:6} (Altamura et al. in preparation).}
    \label{fig:philosophy}
\end{figure}

Modern science advances in a non-linear way. Theories and ideas can be confuted and debunked, before new ones are developed and, once again, put to the test. This logic is a powerful tool for interpreting the evolution of astrophysics research, and particularly the study of clusters of galaxies, on a high level.

Numerous areas of research follow these dynamics. As an example, the measurement of the baryon content of the Universe from the Cosmic Microwave Background (CMB) at high redshift was found to be at odds the baryon census in low-redshift clusters \citep{1999ApJ...514....1C, 2012ApJ...759...23S}. This discrepancy, known as the \textit{missing baryons problem} in the literature, was recently resolved by attributing the unaccounted gas to the warm–hot intergalactic medium \citep[WHIM,][]{2018Natur.558..406N, 2021MNRAS.503.1798C}, which is subject of active research \citep[e.g.][]{2023arXiv230107771T, 2023MNRAS.519.2251B}. Testing the validity of existing results from hydrodynamic simulations of galaxy clusters and exposing their limitations is the philosophical foundation of this thesis.

Utilising the principles of scientific progress introduced by Karl Popper, a 20$^{\rm th}$\hyp{}century philosopher, we aim to establish a connection between the work outlined in later chapters and the broader, current state of cluster astrophysics research as the focus of this thesis. 
For Popper, the ability to prove a theory wrong is the fundamental engine that drives science forward \citep{popper1934logic}. According to this line of thought, the set of the human knowledge can experience pull-backs before advancing again with more accurate and general scientific theories, as illustrated in the \textit{Logik der Forschung} \citep[translated as \textit{The Logic of Scientific Discovery},][]{popper1966logik}. Scientific progress would, in fact, be centered on the idea that theories can never be proven to be (generally) true, but can only be falsified through observations and experimentation. In particular, a theory is defined as scientific if it can be tested and logically contradicted through empirical evidence. This means that a scientific theory must be able to allow predictions that can be probed and potentially confuted. If a prediction is shown to be incorrect, then the theory must be revised or discarded \citep{kuhn1962historical, ladyman2007every}. This criterion is often referred to as \textit{principle of demarcation}, as it discriminates science from non-science, and encapsulates the philosophy of \textit{falsificationism}.

In contrast to Popper's view, the modern-positivist approach conceived the human knowledge as a progressive collection of data and empirical evidence stacking up over time \citep{comte1865general}. Positivists believed that scientific theories can be proven to be true through the \textit{accumulation} of evidence and that the empirical experience (not necessarily combined with logical abstractions) is the most reliable way to gain knowledge about the world \citep[e.g.][Chapter 4]{hempel1965aspects}. The key difference between Popper's view and positivism is that Popper emphasises the role of falsifiability and logic in the scientific process, while positivists tend to focus on the accumulation of evidence as a way of proving the validity of a theory.
In Fig.~\ref{fig:philosophy}, we illustrate schematically the main difference between the positivist (left) and Popperian (right) view of scientific progress. Assuming that the knowledge can be quantified and changes over time, we show a monotonically increasing relationship in the positivist case and a curve with local minima in the Popperian case. The work of the \textit{entropy core problem} presented in Chapters \ref{chapter:4} \citep{2023MNRAS.520.3164A} and \ref{chapter:5} (Altamura et al. in preparation) can be used as a case-study of the falsificationism (bottom panel). Despite the tremendous success of the EAGLE model \citep{2015MNRAS.446..521S} in describing the formation galaxies and groups in the Universe, the work by \cite{ceagle.barnes.2017} exposed its limitations in reproducing the observed populations of cool-core and non-cool-core clusters (blue). Following up on this result, \cite{2023MNRAS.520.3164A} noted that this shortcoming is widespread among numerous projects involving hydrodynamic simulations and probed variations of current EAGLE-class models to identify the possible causes. Conceptually, this work aims at deepening the understanding of the point of failure in the EAGLE model, i.e. reaching the local minimum of the curve in Fig.~\ref{fig:philosophy} (pink), before suggesting possible solutions and develop future models with the strengths of EAGLE, and also capable of reproducing the correct entropy in cluster cores (orange).

Interestingly, Popper's philosophy was inspired by Einstein's General theory of Relativity (GR), which had significant impact not only among scientists, but also in philosophy and literature in the first half of the 1900s \citep{popper2014two}. In fact, the publication by \cite{einstein1916gr} was followed by other works attempting to make predictions from GR and design experiments to test the theory \citep[e.g.][]{dyson1920ix}. Popper saw in this succession of events a logical pattern which also other theories of science at the time had in common. Incidentally, the theory of GR has passed all the tests performed in over a century. Even today numerous research groups continue to invest resources in an attempt to measure deviations from GR predictions and potentially confute it. This result would be the key to uncovering new physics \citep[see][for a review]{2015CQGra..32x3001B}. Crucially, GR lead to the formulation of what is known as the \textit{standard model of cosmology}, which describes the evolution of baryons, dark energy ($\Lambda$) and cold dark matter (CDM) over cosmic time. We will outline the principles of $\Lambda$CDM cosmology in the next chapter.

\section{The epistemic core of simulations}
\label{sec:role-of-simulations}

When asked the question ``\textit{why should we create simulations of the Universe?}``, most computational cosmologists and astrophysicists might argue that simulations are the key to capturing the complexity of stars, gas, black holes, and even dark matter in a \textbf{controlled setting}. Computers are able to machine these complex interactions, following a relatively small set of rules and physical laws which we define \textit{ab initio}.

The freedom to tune and tweak the physical model at the foundation of a cosmological simulation can allow us to answer ``\textit{what if [...]?}`` questions. Consciously or subconsciously, simulation scientists construct their models to recreate a hypothetical scenario that answers questions of this kind.\footnote{``\textit{What if [...]?}`` questions make up two non-fiction books from the popular web-comic \href{https://xkcd.com/}{\textit{xkcd}}: \textit{What If?: Serious Scientific Answers to Absurd Hypothetical Questions} \citep{xkcd_whatif_2014} and \textit{What If? 2} \citep{xkcd_whatif_2022}. In \textit{What if?}, author Randall Munroe considers hypothetical scenarios submitted by \textit{xkcd} readers on various scientific topics, and uses back-of-the-envelope calculations with physically-motivated logic to argument what would happen in that improbable situation. Although further from reality, on a fundamental and epistemic level Munroe's approach is aligned with that adopted by cosmology simulation scientists.} For instance, we may ask ``\textit{what if there were no supernova explosions in the Universe?}`` or ``\textit{what if heavy elements in the inter-galactic medium did not radiate energy away?}``. These hypothetical questions will be discussed in detail in Chapter \ref{chapter:4} and are designed to advance our understanding of more fundamental questions about the real Universe, such as ``\textit{how can cool-core galaxy clusters maintain their thermodynamic state?}``, ``\textit{what pathways of the baryon cycle are crucial for the evolution of galaxies?}`` and ultimately ``\textit{why do galaxies evolve the way they do in the Universe?}``.

Clearly, we would not be able to prevent supernovae from exploding or metals from radiating in the real Universe, or restart the Universe from its initial state to study the long-term effects of those alterations. Despite being \textit{approximations} of the real Universe, simulations allow these changes and can grant access to valuable information about hypothetical universes which we would not otherwise be able to study with the same level of precision or complexity.

Over the past five decades, scientists viewed simulations as an accelerant and catalyst for new astrophysics and cosmology research. They devoted large amounts of resources into developing simulation codes as soon as the right technology, both software and hardware, became available. Today, the field of cosmological and astrophysical simulations is so diverse that each team could motivate their projects in vastly different ways. Nonetheless, most simulations-based investigation methods, on a high level, share a common pattern: they leverage a long history of observations to gain deeper knowledge into \textit{how} a set of physical processes works and interacts. This logical process is the archetype of the branch of philosophy known as epistemology.

From the Greek \textgreek{ἐπιστήμη} (knowledge, science) and \textgreek{λόγος} (explanation, reason), epistemology attempts to rationalise the acquisition of knowledge by distinguishing between the \textit{knowing that} [e.g., galaxy clusters form and maintain cool cores] and the \textit{knowing how} [is that so] \citep{ryle1949concept}. The former is often referred to as \textit{knowledge by acquaintance}, while the latter as \textit{knowledge by description} \citep{russell1912problems}. 

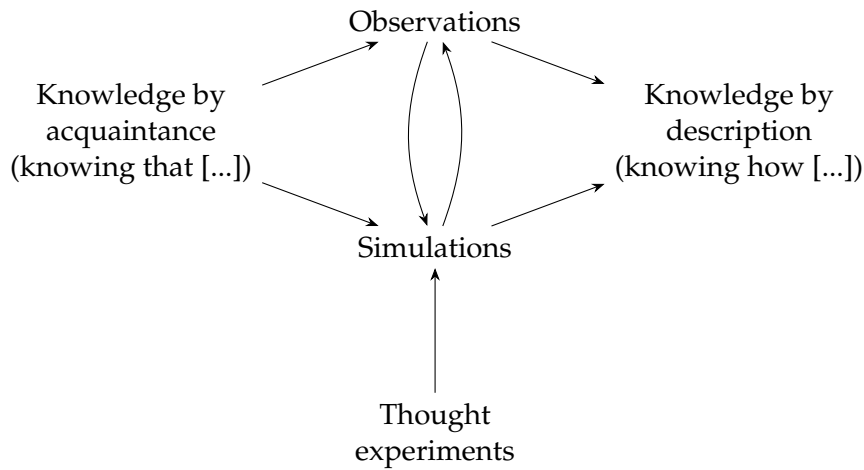
\begin{figure}
\centering
\begin{minipage}{0.85\textwidth}   
\begin{tikzpicture}
\begin{scope}[every node/.style={align=center}]
    \node (A) at (0,0) {Knowledge by\\acquaintance\\(knowing that [...])};
    \node (B) at (8,0) {Knowledge by\\description\\(knowing how [...])};
    \node (C) at (4,1.5) {Observations};
    \node (D) at (4,-1.5) {Simulations};
    \node (E) at (4,-4) {Thought\\experiments};
\end{scope}

\begin{scope}[>={Stealth[black]},
              every node/.style={}
              every edge/.style={draw=black,very thick}]
    \path [->] (A) edge node {} (C);
    \path [->] (C) edge node {} (B);
    \path [->] (A) edge node {} (D);
    \path [->] (D) edge node {} (B);
    \path [->] (C) edge[bend right=20] node {} (D); 
    \path [->] (D) edge[bend left=-20] node {} (C); 
    \path [->] (E) edge node {} (D); 
\end{scope}
\end{tikzpicture}
\end{minipage}
    \caption{Diagram of the epistemic process from knowledge by acquaintance to knowledge by description via empirical observations and via simulations or thought experiments.}
    \label{fig:epistemic-process}
\end{figure}

\vspace{5mm}
\begin{center}
\begin{minipage}{0.85\textwidth}   
The chief importance of knowledge by description is that it enables us to pass beyond the limits of our experience. In spite of the fact that we can only know truths which are wholly composed of terms which we have experienced in acquaintance, we can yet have knowledge by description of things which we have never experienced. In view of the very narrow range of our immediate experience, this result is vital, and until it is understood, much of our knowledge must remain mysterious and therefore doubtful.
\vspace{-2pt}
\par\noindent\rule{\textwidth}{0.5pt} 
\vspace{-20pt}
\begin{flushright}
Bertrand Russell, The problems of philosophy (\citeyear{russell1912problems})\\
Chapter 5: \textit{Knowledge by acquaintance and knowledge by description}
\end{flushright}
\end{minipage}
\end{center}
\vspace{5mm}
The transition from knowledge by acquaintance to knowledge by description is historically executed by conducting more observations and measurements, i.e. empirical experience. This logical process can be represented using the graph in Fig.~\ref{fig:epistemic-process} (top branch). However, the use of \textit{thought experiments} proved to be effective for analysing hypothetical scenarios, transcending empirical measurements and deriving a more general understanding of physics. The special theory of relativity, for instance, was born of thought experiments \citep{norton1991thought}.\footnote{Physicist Albert Einstein acknowledged explicitly the use of thought experiments (in German, \textit{Gedankenexperiment}) to develop the conceptual foundations of his theory of relativity.} The advent of computer simulations took the concept of thought experiments a step further. Both thought experiments and simulations provide a way to derive the a more general knowledge of physics from empirical facts (bottom branch), but computer simulations can often harness complexity beyond the limits of thought experiments because of the larger capacity to perform numerical calculations.

\subsubsection*{Are simulations replacing thought experiments?}
Galaxy formation is an excellent example of a complex problem that requires computer simulations beyond the limit of thought experiments. Thought experiments can form the basis for the gravitational spherical top-hat collapse model (Section \ref{sec:late-perturbations}) and the Zeldovich ``pancake`` collapse model (Section \ref{sec:zeldovich-approximation}), but computers are required to capture (i) the non-linear evolution of $\sim 10^{10}$ particles (or resolution elements) in detail and (ii) the interaction between different astrophysical components. 

This argument may suggest that thought experiments have been demoted from the role they had a century ago, because of their limit in scaling with complexity. Nonetheless, they can still contribute greatly in \textit{designing} modern simulation models. In fact, the ``\textit{what if [...]?}`` question often sparks from a thought experiment, before being translated and implemented into a simulation model. In many cases, thought experiments could still assist the development of cosmological simulations from their inception through the first stages of implementation \citep[see also][]{chandrasekharan2012computational}. This relation is represented by the bottom link in Fig. \ref{fig:epistemic-process}.
Finally, computing technology is pivotal for analysing and interpreting simulations, which often appear very complex because of the many processes involves, but can be understood at a macroscopic level with simple analytical reasoning, akin to thought experiments.

\subsubsection*{Calibration and predictions}
While simulated realisations of the Universe are hypothetical, they must adhere to the physical principles to be of any use to the scientific community. In addition to the basic laws of physics, e.g. gravity and thermodynamics, simulations often adopt \textit{fiducial} metrics derived from observations. For example, the cosmological parameters are often kept fixed to values measures from e.g. \cite{2020A&A...641A...6P} (unless one wishes to explore galaxy formation in different cosmologies), or the fact that galaxies in massive dark matter halos host super-massive black holes. Such pieces on information are derived from observations and indicate how observations can inform simulation models (observation-simulation link in Fig. \ref{fig:epistemic-process}).

When simulations are designed to match the observed Universe, this connection tightens: more metrics are inspired by observations, such as the stellar mass function and galaxy sizes. Then, the model parameters are tuned to find the best match between simulated and observed Universe, and the simulation model is now \textit{calibrated}. The calibration aims to provide the best possible alignment between simulations and observations, so that new insights can be derived from the artificial data set to make \textit{predictions}, to be probed by future observations (simulation\hyp{}observation link in Fig. \ref{fig:epistemic-process}; see \citealt{grune2010epistemology}, for a review).

\subsubsection*{Trustworthiness and skepticism}
Simulation models are based on a finite set of rules, and assumptions. Both the choice of rules and the validity of the assumptions influence the robustness of the results from complex runs, and the trustworthiness of the predictions that follow \citep{grune2010epistemology}. Over the past two decades, success stories documented the synergy between the Planck satellite observational team and the simulations team for the modelling of foregrounds, or the study of X-rays from galaxy clusters to determine cluster masses \citep[][]{2021A&A...650A.104C}. However, a wide-spread skepticism towards simulation results is still present within the community, sometimes for a good reason.

Using Large-Hadron Collider (LHC) simulations as a case-study, \cite{mattig2021trustworthy} investigates the criteria that a simulation should meet to be considered trustworthy. Some of these criteria can also become points of failure which can impact the resonance of a computational finding among the community. \textit{Validating} the simulations is one of the key steps to establish trustworthiness (and often physical significance) in a data set. In the context of galaxy formation simulations, the validation is often achieved by calibrating the model, as explained above. However, one cannot derive absolute certainty from simulations, simply because some assumptions on, e.g., a free parameter allow for some flexibility, or the observational \textit{priors} carry an uncertainty. Simulations can only be as precise as these priors allow, and cannot be considered trustworthy outside these bounds.
\vspace{5mm}
\begin{center}
\begin{minipage}{0.85\textwidth}   
A simulation can only be trustworthy in property regions that have been probed by material measurements. Outside of the probed region, a simulation cannot be considered trustworthy. [...]

\qquad [which follows]

Within its uncertainty range, the simulation cannot provide any trustworthy statement. A trustworthy simulation can only be meaningfully applied in the property region outside of the simulation’s uncertainty.
\vspace{-2pt}
\par\noindent\rule{\textwidth}{0.5pt} 
\vspace{-20pt}
\begin{flushright}
Peter M{\"a}ttig, Trustworthy simulations\\ and their epistemic hierarchy (\citeyear{mattig2021trustworthy})\\
\end{flushright}
\end{minipage}
\end{center}
\vspace{5mm}

This concept is well established among simulations practitioners in the form of (i) \textit{simulation results should not be extrapolated without a compelling reason} and (ii) \textit{uncertainties propagate through the calculations and determine the significance of the final outcome}. These propositions, often implicitly adopted as ``scientific common sense``, contribute to building a framework for interpreting and assess the predictive power of simulations, as well as their limitations. Results that expose these limitations inevitably raise skepticism. For instance, \cite{2023MNRAS.520.3164A} proved that EAGLE-like models of galaxy formation cannot reproduce the observed cool-core cluster population (failure in the validation); the results by \cite{2022arXiv221108442B} showed that modern galaxy formation simulations can lead up to a $\approx 20\%$ variability in the star-formation rate of a single object, when re-run with identical settings (limitation in reproducibility and determinism).

Skepticism, when combined with proactive initiative, can however expose some of the points of failure in numerical simulations, and help develop novel, robust, trustworthy models. This \textit{falsifiability} dynamics would then take place, just as predicted by the Popperian view of scientific progress \citep[see][and Section~\ref{sec:intro-philosophy}]{popper1934logic}.

\section{Thesis outlook}
\label{sec:thesis-layout}

The next chapters are divided into two categories: Chapters \ref{chapter:2} and \ref{chapter:3} provide a general \textit{introduction} to cosmology, the formation of galaxy clusters and modern numerical methods to simulate their evolution; Chapters \ref{chapter:4} -- \ref{chapter:6} present \textit{novel research} related to the study of groups and clusters of galaxies produced throughout the doctoral program. An overview of the content of each chapter is given below.

\subsection*{Second chapter}
Starting from a description of the standard model of cosmology, i.e. the $\Lambda$-cold dark matter model, we illustrate how matter-energy perturbations are thought to have been produced in the early Universe and grown to form the structures we observe today. Within this framework, particular importance is given to the astrophysical processes determining the properties of galaxy clusters and how these can be probed by observations.

Throughout this chapter, unnumbered subsections with dark-red titles include considerations pointing to numerical simulations, discussed in greater detail in later chapters.

\subsection*{Third chapter}
We describe the workflow used in the production of modern cosmological simulations of galaxy clusters, starting from initial conditions to the final scientific insights. Throughout this simulation pipeline, we illustrate the main physical processes implemented in modern cosmological simulations codes, mainly focusing on the \swift code, used in Chapters \ref{chapter:5} and \ref{chapter:6}, with comparisons to the older \textsc{Gadget-3} code, relevant to Chapter \ref{chapter:4}. The methodology of masking the refined region presented in this chapter was developed and implemented by the author in preparation to the simulations of Chapters \ref{chapter:5} and \ref{chapter:6}.

Finally, we present the computational performances of the \swift simulation code for solving known hydrodynamics problems. These benchmarks and the software used to perform them form part of the author's own contribution to the ExCALIBUR program.\footnote{ExCALIBUR is a UK nation-wide project focusing efforts to produce numerical simulations with future Exa-FLOP supercomputers. Further information is available at \href{https://excalibur.ac.uk/}{https://excalibur.ac.uk/}.} We also provide information on the usage of computational resources in the simulations in Chapters \ref{chapter:5} and \ref{chapter:6}.

\subsection*{Fourth chapter}
This chapter documents the research on the kinetic Sunyaev-Zeldovich (kSZ) effect due to the bulk rotation of galaxy clusters. We use the cluster sample from the MACSIS simulations \citep{macsis_barnes_2017} to investigate the alignment of the rotation axes of the hot gas in clusters, the galaxies and the dark matter halo. We found that the misalignment between these axes can lead to an underestimation of the rotational kSZ effect if the galaxies are used as proxies for the gas rotation. Moreover, our results suggest that the rotational kSZ signal from massive clusters could be much stronger than previously thought and that this contribution to the overall kSZ signal could be important to interpret data from upcoming CMB observations. This chapter is based on \cite{2023arXiv230207936A}.

\subsection*{Fifth chapter}
We investigate the possible causes of the \textit{entropy core problem}, representing the failure of many modern hydrodynamic simulations in reproducing the cool-core/non-cool-core cluster dichotomy found in X-ray observations. With different implementations of EAGLE\hyp{}like galaxy formation models in the \swift code, we show that the both groups and clusters of galaxies cannot maintain low\hyp{}entropy cores to low redshifts. These results indicate that the sub\hyp{}grid physics implemented in our models may not be the primary cause of high entropy levels in the core of our sample of objects. Instead, missing physical processes or poor numerical resolution could lead to large entropy cores. This research tests the limitations of the EAGLE model of galaxy formation and motivates further work to understand this shortcoming of hydrodynamic simulations. This chapter is based on \cite{2023MNRAS.520.3164A}.

\subsection*{Sixth chapter}
We extend the study of \cite{2023MNRAS.520.3164A} by probing the formation of high\hyp{}entropy cluster cores through cosmological redshift. By comparing entropy profiles associated with merger activity and quiescent periods, we aim to determine the connection between high levels of entropy in clusters and cosmological accretion. We then show results from the comparison with entropy profiles measured during high and low star\hyp{}formation rates, as well as high and low black hole activity. Finally, we provide a framework for studying the role of mergers, star formation and black hole activity in the persistence of cool cores. The research presented in this chapter will form the basis of an upcoming publication (Altamura et al., in preparation).

\subsection*{Seventh chapter}
We conclude this thesis with an outlook on the potential applications of the research conducted as part of the doctoral program. We assess the impact of these contributions in the context of cosmology and cluster astrophysics research, and we describe the scope for future work. 

%% file: Chapters/Chapter2.tex
\thispagestyle{plain}
\begin{spacing}{1.}
\topskip0pt
\vspace*{\fill}
\begin{flushright}
\begin{minipage}{0.5\textwidth}   
\begin{verbatim}
In the vast expanse of space,
Galaxies dance in their place,
Clustered together, they reveal,
Secrets the Universe can't conceal.

Gravity pulls them close,
As they spin in cosmic prose,
Their shapes and sizes diverse,
Help us understand the Universe.

In these clusters we can see,
Dark matter's mysterious decree,
And the ways in which it bends,
Light, time, and space until it ends.

Cosmology, a grand pursuit,
Expanding our knowledge to refute,
The mysteries of the universe,
And the role of galaxy clusters.
\end{verbatim}
\vspace{-2pt}
\par\noindent\rule{\textwidth}{0.5pt}
\vspace{-20pt}
\begin{flushright}
Chat-GPT by OpenAI\\
On the cosmology of galaxy clusters
\end{flushright}
\end{minipage}
\end{flushright}
\vspace*{\fill}
\end{spacing}

\chapter{Cosmology}
\label{chapter:2}

\section{Preface to the chapter}
\label{sec:ch1-preface}

This chapter is organised as follows. In Section \ref{sec:friedmann-equations}, we introduce the Friedmann equations for a homogeneous $\Lambda$CDM universe; in Section \ref{sec:early-perturbations} we provide an outlook on the framework for the evolution of linear perturbations in the early Universe; we then make connect the perturbations at early times with the non-linear perturbations at late times in Section \ref{sec:late-perturbations}, where we discuss the collapse of structures in an expanding Einstein-de Sitter universe and the formation of dark matter halos. In this context, we also present elements of tidal-torque theory for the generation of angular momentum in virialised structures, relevant for the work on galaxy cluster rotation in Chapter \ref{chapter:3}. Finally, in Section \ref{sec:cluster-observables}, we illustrate the physical mechanisms behind two of the main techniques in galaxy cluster observations: X-ray production and the Sunyaev-Zeldovich (SZ) effects.

\section{The homogeneous Universe}
\label{sec:friedmann-equations}

\subsection{The Big Bang}
\label{sec:flrw-big-bang}

The discovery of the expansion of the Universe made by \cite{1926ApJ....64..321H} using spectra of galaxies ruled out the concept of the static Universe, which had been considered the leading cosmology for four centuries. Together with his team at Mount Wilson Observatory (CA, USA), he captured the spectra of 46 galaxies, whose distance had already been determined with the period-luminosity relation of \textit{Cepheid variables} \citep[see e.g.][for details]{carroll_ostlie_2017}, and identified spectral lines that were shifted ($\delta \lambda$) from their reference wavelength $\lambda_0$. In the majority of cases, the lines were shifted towards lower wavelengths (i.e. red), suggesting that those galaxies were receding from the observer. For this reason, this particular shift is often referred to as cosmological \textit{redshift}, and is defined as
\begin{equation}
    z\equiv \frac{\delta \lambda}{\lambda_0} = \frac{v}{c},
\end{equation}
where $v$ is the recession speed along the line-of-sight and $c$ is the speed of light. Hubble identified a linear proportionality between $v$ (measured from spectral lines) and the distance $d$ (measured from Cepheids), known today as Hubble's law:
\begin{equation}
    v = H_0\, d = c\, z,
\end{equation}
with $H_0$ the Hubble constant. Hubble's research is an example of successful use of the \textit{distance ladder}, a method that is the foundation of today's endeavours to measure the distance of astrophysical objects at low- and medium-redshift \cite{carroll_ostlie_2017}.

This revolutionary finding, combined with the detection of the cosmic microwave background (CMB) radiation by \cite{1965ApJ...142..419P}, inspired the concept of a universe which originated from a extremely hot singularity and then cooled down as it expanded. This physical theory is known as the Big Bang, a term originally introduced by Fred Hoyle. To the present day, the Big Bang cosmology continues to find affinity with observations such as the the abundances of light elements as part of the Big Bang nucleosynthesis (BBN) and the stellar dating of globular clusters \citep[see e.g.][]{peebles1971physical}.

Furthermore, if the Universe had a beginning, then its age can be defined. If we were to reconstruct the expansion of the Universe back to its initial singular state, and define that instant as $t_i$, then the time elapsed since $t_i$ is commonly regarded as the age of the Universe, or $t_0$. Assuming a Big Bang cosmological model, this value can be computed by fitting observational data, as in \cite{2020A&A...641A...6P}, where it was found $t_0=(13.787 \pm 0.020)$ Gyr.

This value has important consequences for the astrophysics of galaxy clusters. Primarily, it dictates the number of massive clusters of galaxies, as matter subject to gravitational collapse for a longer time can form larger structures. Also, it determines the longest time that hot gas in galaxy clusters has available to cool down. Depending on its thermodynamic properties, hot gas can cool down in a characteristic timescale $t_{\rm cool}$; if $t_{\rm cool} > t_0$, then the gas can never cool within the timescale of the age of the Universe and it will remain hot, with significant consequences for the entropy (see Chapters \ref{chapter:4} and \ref{chapter:5}).

\subsection{The FLRW metric}
\label{sec:flrw-metric}

The FLRW metric describes the geometry of a homogeneous and isotropic space-time manifold. The condition of homogeneity implies that the properties of the space-time must the same at every point, and the condition of isotropy imposes that the properties of the universe must be the same in every direction. Historically, this metric was constructed from symmetry considerations \citep{1933RvMP....5...62R, 1972gcpa.book.....W, 1975ctf..book.....L}, assuming that the space-time geodesics are non-intersecting \citep[also known as Weyl's postulate, see][]{1922stm..book.....W}. Maximally symmetric manifolds are defined as isotropic and homogeneous and, following Schur's theorem, their Riemann tensor is

\begin{equation}
    R_{\mu\nu\alpha\beta} = k\, (g_{\mu\alpha}\, g_{\nu\beta} - g_{\mu\beta}\, g_{\nu\alpha}),
\end{equation}
where $g$ is the metric tensor and $k$ is the curvature of the manifold. Crucially, $k$ is constant throughout the manifold; then we can have $k>0$, corresponding to a spherical geometry, $k=0$ for flat, Euclidean space, or $k<0$ for hyperbolic space \citep{1973grav.book.....M}.

Now, the space-time metric $ds^2 = g_{\mu\nu}\, dx^\mu\, dx^\nu$ can be calculated in spherical polar coordinates $(r, \theta, \phi )$ to obtain the FLRW metric:
\begin{equation}
    ds^2 = c^2 dt^2 - a^2(t) \left[ \frac{d r^2}{1-kr^2} + r^2 d\Omega^2 \right],
\label{eq:flrw-metric}
\end{equation}
where the solid angle element $d\Omega^2 = d\theta^2 + \sin^2{\theta}d\phi^2$. Here, $c$ is the speed of light in vacuum and $a(t)$ is the (time-dependent) scale factor, which relates to the size of the universe at time $t$.\footnote{The scale factor at the present time $t_0$ is defined to be $a(t_0)=1$. Here, as well as in later sections, the subscript "0" indicates the present time for cosmological quantities varying over cosmic time, unless stated otherwise.} We note that the $(r, \theta, \phi )$ coordinates are co-moving with respect of the expansion of the universe, which is expressed by the scale-factor $a$. This quantity is defined in terms of cosmological redshift as
\begin{equation}
    a \equiv \frac{1}{z + 1}.
\end{equation}

The first term ($c^2 dt^2$) does not depend on the spatial coordinates $(r, \theta, \phi)$, indicating that time passes at the same rate for every point in the universe and that the FLRW metric is homogeneous with respect to time. If $k$ does not depend on spatial coordinates, then the manifold,  The second term $-a^2(t) \left[ dr^2 / (1-kr^2) + r^2 d\Omega^2 \right]$, does not \textit{explicitly} depend on time, meaning that it is the same in every space-time direction. This shows that the FLRW metric is isotropic with respect to space. Eq.~\eqref{eq:flrw-metric} expresses the evolution of a perturbation-free universe, where matter does not collapses locally, nor it forms clumps. For this reason, the FLRW metric and the evolution equations that originate from it (see Section \ref{sec:flrw-derived-friedmann-equations}) are often referred to as descriptors of the \textit{background} cosmology.

The particular choice of $k$ and $r$ determines the evolution of the metric. If we apply the coordinate transformation $r \longrightarrow \chi$
\begin{equation}
    d\chi \equiv \frac{dr}{\sqrt{1-kr^2}},
\end{equation}
then Eq.~\eqref{eq:flrw-metric} becomes 
\begin{equation}
    ds^2 = c^2 dt^2 - a^2(t) \left[ d\chi^2 + \Sigma^2 d\Omega^2 \right],
\label{eq:flrw-metric2}
\end{equation}
where we can identify three cases depending on the curvature:
\begin{equation}
    \Sigma=
    \begin{cases}
        \frac{1}{\sqrt{k}}\, \sin(\sqrt{k}\, \chi) & \text{if } k > 0 \quad : \quad \text{Closed universe}\\
        \chi & \text{if } k = 0 \quad : \quad \text{Flat universe (Euclidean)}\\
        \frac{1}{\sqrt{-k}}\, \sinh(\sqrt{-k}\, \chi) & \text{if } k < 0 \quad : \quad \text{Open universe}.
    \end{cases}
\end{equation}


\subsection{Equations of motion for a relativistic fluid}
\label{sec:flrw-field-equations}

One of the key results of GR is the relation between space-time topology and energy distribution, following Einstein's field equation:
\begin{equation}
    R_{\mu\nu} - \frac{1}{2}Rg_{\mu\nu} + \Lambda g_{\mu\nu} = \frac{8\pi G}{c^4}T_{\mu\nu},
\label{eq:einstein-field}
\end{equation}
where $R_{\mu\nu}$ is the Ricci tensor, which describes the curvature of space-time, $R$ is the Ricci scalar, which is a measure of the overall curvature of space-time, $g_{\mu\nu}$ is the metric tensor, which describes the properties of the space-time, $\Lambda$ is the cosmological constant, which is a term that was introduced by Einstein to tune the expansion of the universe, $G$ is the gravitational constant, which determines the strength of the gravitational force and finally $T_{\mu\nu}$ is the stress-energy tensor, which describes the matter and energy present in the space-time manifold.

So far, the FLRW metric can determine the functional form of the left-hand-side of Eq.~\eqref{eq:einstein-field}. Here, we outline the main considerations in the derivations of each term, and refer to e.g. Chapter 27 of \cite{1973grav.book.....M} for further details. The metric tensor corresponding to Eq.~\eqref{eq:flrw-metric}, in covariant and contravariant form, is

\begin{equation}
\begin{array}{lr}
    g_{\mu\nu}=\left[
    \begin{array}{cccc}
        c & 0 & 0 & 0 \\        
        0 & -\frac{a(t)^2}{1-kr^2} & 0 & 0 \\
        0 & 0 & -r^2 a(t)^2 & 0 \\
        0 & 0 & 0 & -r^2 a(t)^2\, \sin^2\theta
    \end{array}
    \right] &\quad \text{Covariant,}\\
    [5em]
    g^{\mu\nu}=\left[
    \begin{array}{cccc}
        \frac{1}{c} & 0 & 0 & 0 \\        
        0 & -\frac{1-kr^2}{a(t)^2} & 0 & 0 \\
        0 & 0 & -\frac{1}{r^2 a(t)^2} & 0 \\
        0 & 0 & 0 & -\frac{1}{r^2 a(t)^2\, \sin^2\theta},
    \end{array}
    \right]  &\quad \text{Contravariant.}
\end{array}
\end{equation}

These results can be used to compute the Ricci tensor and Ricci scalar as follows.
\begin{itemize}
    \item \textbf{Ricci tensor.} To compute $R_{\mu\nu}$, it is necessary to compute the Christoffel symbols $\Gamma^{\alpha}_{\mu\nu}$, then the Riemann tensor $R^\alpha_{\beta\mu\nu}$. Because of the symmetries of the FLRW metric and the properties of the connection,\footnote{In absence of torsion $T^{\alpha}_{\mu\nu} = 1/2\, (\Gamma^{\alpha}_{\mu\nu}-\Gamma^{\alpha}_{\nu\mu})=0$, the connection is symmetric: $\Gamma^{\alpha}_{\mu\nu}=\Gamma^{\alpha}_{\nu\mu}$. GR satisfies the torsion-less condition, which is a special case where the indices $(\mu, \nu)$ are commutative. Alternative theories of gravity, such as modified gravity, have $T^{\alpha}_{\mu\nu} \neq 0$, often linked to the production of beyond-standard-model (BSM) particles \citep[e.g.][]{2018PrPNP.102...89I, 2021JCAP...06..016S}.} most of the $\Gamma^{\alpha}_{\mu\nu}$ will also be either symmetric or null. Finally, the Ricci tensor can be found as a contraction of the Riemann tensor: $R_{\mu\nu} = R^m_{\mu m \nu}$. For the FLRW metric, one can find that $R_{\mu\nu}$ is diagonal, i.e. the off-diagonal terms $\mu \neq \nu$ are zero.

    \item \textbf{Ricci scalar.} $R$ is computed from the Ricci tensor as $R=g^{\mu\nu}R_{\mu\nu}$.
\end{itemize}

To solve the Eq.~\eqref{eq:einstein-field}, it is often convenient to use its trace-reversed form:
\begin{equation}
    \frac{8\pi G}{c^4}\left(T_{\mu\nu} - \frac{1}{2}Tg_{\mu\nu}\right) - \Lambda g_{\mu\nu} = R_{\mu\nu},
\label{eq:einstein-field-reversed}
\end{equation}
where $T$ is defined in analogy to the Ricci scalar: $T=g^{\mu\nu}T_{\mu\nu}$. For an ideal, homogeneous fluid without viscosity, the stress-energy tensor is
\begin{equation}
    T_{\mu\nu}=\text{diag}(\rho, P, P, P),
\end{equation}
where $\rho$ is the density and $P$ the pressure of the fluid, as specified in the equation of state. Here, we have set $c=1$, as often done when equating mass and energy \citep[see Section 3. of][]{2015imct.book.....L}. Finally, we can evaluate the term
\begin{equation}
    T_{\mu\nu} - \frac{1}{2}Tg_{\mu\nu} = \frac{1}{2}\, \text{diag}\left[\rho + 3P, -g_{ii}(\rho - P)\right],
\label{eq:einstein-field-reversed-solved}
\end{equation}
where $g_{ii}$ indicates the space components of the FLRW metric tensor, having $i=\{1, 2, 3\}$.

\subsection{The Friedmann equations}
\label{sec:flrw-derived-friedmann-equations}

The solution to the time component (00) of Eq.~\eqref{eq:einstein-field-reversed} leads to the first Friedmann equation. Therefore, we obtain
\begin{equation}
    \frac{\ddot{a}}{a} = -\frac{4\pi G}{3}\, (\rho + 3P) + \frac{1}{3}\, \Lambda,
\label{eq:friedmann-1}
\end{equation}
also known as the \textit{acceleration} equation because of the second time derivative in $\ddot{a}$.\footnote{Dotted notation indicates time derivatives, i.e. $\dot{x}\equiv dx/dt$.}

The space components ($ii$) of Eq.~\eqref{eq:einstein-field-reversed} lead to the second Friedmann equation:
\begin{equation}
    \left(\frac{\dot{a}}{a}\right)^2 \equiv H(t)^2 = \frac{8\pi G}{3}\, \rho - \frac{k}{a^2} + \frac{1}{3}\, \Lambda,
\label{eq:friedmann-2}
\end{equation}
where the \textit{Hubble parameter} is defined as $H(t)\equiv \dot{a}(t)/a(t)$. At the present time, the Hubble parameter is defined as Hubble constant:  $H(t_0) \equiv H_0$.

Finally, taking the covariant derivative ($\nabla_\mu$) of Einstein's equation, once can prove that the left-hand-side vanishes, and $\nabla_\mu\, T^\mu_{~~\nu} = 0$. This relation expresses the conservation of energy, which can be expanded as 
\begin{equation}
    \dot{\rho} + 3\, \frac{\dot{a}}{a}\, (\rho + P) = 0,
\label{eq:friedmann-3}
\end{equation}
subject to an equation of state $P = P(\rho)$. This relation is the third Friedmann equation, which is purely a conservation law.

We note that the three Friedmann equations are not independent: it is possible to combine any two of them to derive the third. For example, differentiating Eq.~\eqref{eq:friedmann-2} with respect to time and combining it with Eq.~\eqref{eq:friedmann-1}, it is possible to obtain Eq.~\eqref{eq:friedmann-3}. In summary, Friedmann's equations describe the evolution of the scale factor (and the Hubble parameter) of the universe in the context of GR. The scale factor is a measure of the size of the universe at any given time. These equations are derived from Einstein's field equation, which relates the curvature of space-time to its energy content.

\subsection{The evolution of the background}
\label{sec:flrw-background-evolution}

In absence of $\Lambda$, Eq.~\eqref{eq:friedmann-2} suggests that the density of the Universe is related to the space-time curvature. For a perfectly flat universe, $k=0$, the Friedmann equations require the density to be equal to its critical value
\begin{equation}
    \rho_{\rm crit}(t)=\frac{3H(t)^2}{8\pi G}
\label{eq:critical-density}
\end{equation}
at each time $t$. At the present time, we have $\rho_{\rm crit, 0}$. The energy content of the Universe can be composed of different fluids with different densities, which sum to the total density: $\rho=\sum_X \rho_X$. In cosmology, this value is often normalised to the present time $\rho_{\rm crit, 0}$ to obtain the density parameters, denoted with $\Omega$:
\begin{equation}
    \begin{array}{cl}
         \Omega_m \equiv \frac{\rho_{\rm m, 0}}{\rho_{\rm crit, 0}}, \qquad & \text{Matter} \\
         [1em]
         \Omega_r \equiv \frac{\rho_{\rm r, 0}}{\rho_{\rm crit, 0}}, \qquad & \text{Radiation} \\
         [1em]
         \Omega_\Lambda \equiv \frac{\Lambda_0}{3H_0^2}, \qquad & \text{Cosmological constant}  \\
         [1em]
         \Omega_k \equiv -\frac{k_0}{H_0^2}, \qquad & \text{Curvature}\\
    \end{array}
\label{eq:density-parameters}
\end{equation}
where we usually set $\Omega = \Omega_m + \Omega_r + \Omega_\Lambda$. In Eq.~\eqref{eq:density-parameters}, the subscript $m$ indicates the matter density, $r$ the radiation (photons), $\Lambda$ the effective contribution of the cosmological constant and $k$ the effective contribution of the curvature.

The density parameters $\Omega_X$ can be interpreted as the fraction of the total energy in the Universe determining its evolution. Often, the formulation of the background cosmology in terms of $\Omega_X$ is more instructive than using the densities $\rho_X$, and we can re-write the Friedmann's equation in Eq.~\eqref{eq:friedmann-2} after substituting the definitions of Eqs.~\eqref{eq:critical-density} and \eqref{eq:density-parameters}:
\begin{subequations}
\begin{align}
     H(t)^2 =& \, H_0^2\, \frac{\rho}{\rho_{\rm crit, 0}} - \frac{k}{a^2} + \frac{1}{3}\, \Lambda
\\[2mm]
     \left(\frac{H}{H_0}\right)^2 \equiv E(a)^2 = &\,  \underbrace{\frac{\Omega_r}{a^4}}_{=\, \Omega_r(a)} + \underbrace{\frac{\Omega_m}{a^3}}_{=\, \Omega_m(a)} + \underbrace{\frac{\Omega_k}{a^2}}_{=\, \Omega_k(a)} +  \underbrace{\Omega_\Lambda}_{\text{constant}},
\label{eq:friedmann-2-omega-parameters}
\end{align}
\end{subequations}
with $E(a)$ the \textit{dimensionless} Hubble parameter.

The equations of motion for the fluids derived from GR in Section \ref{sec:flrw-field-equations} do not assume equations of state. For the matter and radiation, we use the equation of state of ideal gases, $P(\rho) = w\, \rho$, where $w$ is a constant. By substituting this relation in Eq.~\eqref{eq:friedmann-3}, we can show that
\begin{subequations}
\begin{align}
    \dot{\rho} &=  -3\, \frac{\dot{a}}{a}\, (\rho + w\rho)
\\[2mm]
    \frac{\dot{\rho}}{\rho} &=  -3\, \frac{\dot{a}}{a}\, (1 + w)
\\[2mm]
    \int \frac{d\rho}{\rho} &=  -3(1 + w)\, \int \frac{da}{a}
\\[2mm]
    \rho &= C\, a^{-3(1 + w)},
\end{align}
\end{subequations}
where $C$ is an integration constant. This result highlights that the evolution of the density of an ideal gas with cosmological scale factor follows a power-law with constant index $n = -3(1 + w)$. 

\begingroup
\setlength{\tabcolsep}{6pt} 
\renewcommand{\arraystretch}{1.7} 
\begin{table}
    \centering
    \caption{Summary of the density evolution relations for three cosmological fluids in Friedmann's equations. The special case of a universe where only matter is present, i.e. $\Omega_m = 1$, is known as Einstein-de Sitter (EdS) universe.}
    \label{tab:fluid-evolution}
    \begin{tabular}{lccccl}
    \toprule
    \textbf{Fluid} & $w$ & $n$ & $\rho - a$ & \textbf{$a - $Time}  & \textbf{Limiting case} \\
    &  & & \textbf{dependence} & \textbf{dependence} &  \textbf{($X$-dominated)}\\
    \midrule
    Matter     & 0     & $-3$ & $\rho \propto a^{-3}$ & $a = \left(\frac{2}{3}H_0\, t \right)^{2/3}$ & $\Omega_m \approx 1$ \quad (EdS)\\
    Radiation  & $1/3$ & $-4$ & $\rho \propto a^{-4}$ & $a = \left(2 H_0\, t \right)^{1/2}$ & $\Omega_r \approx 1$\\
    Cosmological constant  & $-1$ & 0 & $\rho \approx \text{constant}$ & $a =  e^{H_0\, t}$ & $\Omega_\Lambda \approx 1$\\
    \bottomrule
    \end{tabular}
\end{table}
\endgroup

Up to this point, we have not chosen a functional form for the equation of state associated to the cosmological constant. This component can also be interpreted as cosmological fluid obeying the Einstein equation, and therefore we can write an effective energy-density $\rho_{\Lambda}$ associated with it:
\begin{equation}
         \rho_{\Lambda, 0} \equiv \Omega_\Lambda\, \rho_{\rm crit, 0} = \frac{\Lambda}{8\pi G}.
\end{equation}
We then use this relation to define the equation of state for $\Lambda$. If $\Lambda$ is constant throughout the Universe, it must satisfy the energy-conservation law in Eq.~\eqref{eq:friedmann-3}:
\begin{equation}
    \dot{\rho}_\Lambda = -3H\, (\rho_\Lambda + P_\Lambda) = 0,
\end{equation}
which is only satisfied for the equation of state $P_\Lambda = -\rho_\Lambda$.

We summarise the evolution of the fluid densities with scale factor in Tab.~\ref{tab:fluid-evolution}.

An additional corollary of Eq.~\eqref{eq:friedmann-2-omega-parameters} is the time evolution of the scale-factor dictated by each cosmological fluid. Assuming that the fluid $X$ makes up all the energy in the Universe, i.e. $\Omega_X \approx 1$ and all other density parameters are zero, then we can integrate Eq.~\eqref{eq:friedmann-2-omega-parameters} over cosmological time $t$, knowing that $H(t) = (da/dt)/a$ and $a = a(t)$. For a detailed derivation, we refer to \cite{1973grav.book.....M} and \cite{2015imct.book.....L}. We also report the equation for the evolution of $a$ with time in Tab.~\ref{tab:fluid-evolution}. These regimes characterised part of the history of the Universe, and referred to in the literature as 
\begin{itemize}
    \item \textit{radiation-dominated era}, with the radiation density parameter larger than all the others, $\Omega_r > \Omega_{X\neq r}$, and a scale-factor evolving as $a \propto t^{1/2}$;
    \item \textit{matter-dominated era}, with the matter density parameter larger than all the others, $\Omega_m > \Omega_{X\neq m}$, and a scale-factor evolving as $a \propto t^{2/3}$;
    \item \textit{$\Lambda$-dominated era}, with the $\Lambda$ density parameter larger than all the others, $\Omega_\Lambda > \Omega_{X\neq \Lambda}$, and a scale-factor increasing exponentially as $a \propto e^t$.
\end{itemize}
For a realistic, multi-component universe like ours, the evolution of the thermodynamic parameters and the scale-factor is weighted by the values of the density parameters measured experimentally, as discussed in Section \ref{sec:lambda-cdm} for the \cite{2020A&A...641A...6P} results.

\subsection{The Universe at the present day}
\label{sec:lambda-cdm}
One of the the most revolutionary achievements of modern cosmology is the measurement of the density parameters, which quantify the abundance of the constituents of the Universe. These parameters enabled the reconstruction of the thermodynamic history of the Universe, and even its fate. The density parameters, together with other metrics of interest to the scientific community, are often referred to as \textit{cosmological parameters}. If asked to compress the information describing the evolution of the Universe, the cosmological parameters would (nearly) be the minimum set of values that we would still capture the main features of cosmic history. This analogy to the JPEG compression algorithm may also lead to the question of what information would be lost. In fact, as the result of a ``lossy compression``, the cosmological parameters can only provide a crude estimate of the formation of structures of the Universe via the $\sigma_8$ parameter. The details of star formation, the growth of black holes or the feedback processes that shape the formation of galaxies and clusters would be lost if we had to re-build cosmology solely from the cosmological parameters. Conversely, these astrophysical mechanism strongly depend on the cosmological parameters. For instance, star formation and feedback depend on the amount of gas available (baryons) in the Universe, which in turn contributes to $\Omega_m$. For this reason, the cosmological parameters, which we will occasionally simply write as \textit{cosmology}, must be specified prior to running any hydrodynamic simulation similar to those presented in this work.

\begin{figure}
    \centering
    \includegraphics[width=0.88\textwidth]{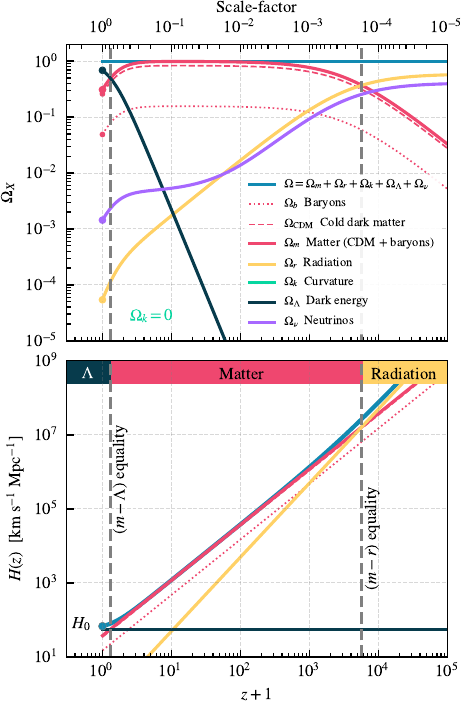}    
    \smallskip
    \caption[Planck 2018 cosmology]{\textit{Top.} The evolution with redshift (and scale-factor) of the density parameters $\Omega_X$ according to the \cite{2020A&A...641A...6P} cosmology. The values of the $\Omega_X$ at the present day are indicated with markers on the left-hand-side. \textit{Bottom.} The corresponding evolution of the Hubble parameter, with the contributions of the individual components $X$. The $\Lambda$, matter and radiation eras are indicated in the top panel and they are separated by dashed grey lines. The value of $H(z=0) \equiv H_0 = 67.66~{\rm km\,s}^{-1}\, {\rm Mpc}^{-1})$ is indicated with a marker, similarly to the plot above.}
    \label{fig:cosmological-parameters}
\end{figure}

Having stated the importance of the choice of cosmology when carrying out calculations and simulation work, we illustrate some of the key parameters often used for simulations and galaxy cluster studies. The currently accepted values of the cosmological parameters have been determined by \cite{2020A&A...641A...6P} and we report them in Tab.~\ref{tab:planck2018-cosmo-parameters}, with a short description of their meaning. We also quote some of the values from the Dark Energy Survey (DES) Year-3 release and the combined DES+Planck results \citep{2022PhRvD.105b3520A}. In Fig.~\ref{fig:cosmological-parameters}, we show how these parameters (top) and the the Hubble parameter (bottom) vary with scale factor and redshift, assuming a Planck 2018 cosmology and the evolution equations in Section \ref{sec:flrw-derived-friedmann-equations}. Crucially, we highlight the radiation, matter and $\Lambda$-dominated eras, separated by grey dashed lines at
\begin{itemize}
    \item $z = 3387 \pm 21$ at the matter-radiation ($m-r$) equality, when $\Omega_r \approx \Omega_m$;
    \item $z \approx 1.3$ at the matter-$\Lambda$ ($m-\Lambda$) equality, when $\Omega_\Lambda \approx \Omega_m$.
\end{itemize}

\begingroup
\setlength{\tabcolsep}{6pt} 
\renewcommand{\arraystretch}{1.7} 
\begin{sidewaystable}
\centering
\caption{Key cosmological parameters measured by \cite{2020A&A...641A...6P}. From their Tab. 2, we quote the results from the from Planck CMB power spectra (TT,TE,EE+lowE) CMB lensing reconstruction and baryon acoustic oscillations (BAO) modelling. The calculation assumes a $\Lambda$CDM model. The errors are the 68\% confidence intervals. For the $\Omega_k$ parameter measurement from Planck, we also refer to \cite{2020AnA...641A..10P}. We also show the measurements from the Dark Energy Survey Year-3 release and the combined DES+Planck results \citep{2022PhRvD.105b3520A}.}
\label{tab:planck2018-cosmo-parameters}
\begin{tabular}{lccll}
\toprule
\textbf{Parameter} & \textbf{Planck 2018} & \textbf{DES -Y3} & \textbf{Combined} & \textbf{Description} \\
\midrule
$h$                        & $0.6766 \pm 0.0042$   & -- & $0.680^{+0.004}_{-0.003}$ & Reduced Hubble constant, $\equiv\, H_0 / (100\, {\rm km\,s}^{-1}\, {\rm Mpc}^{-1})$ \\
$\Omega_b$                 & $0.04897 \pm 0.0003$  & $0.041^{+0.004}_{-0.010}$ & $0.0487^{+0.0005}_{-0.0003}$ & Baryonic matter density parameter today \\
$\Omega_{\rm CDM}$         & $0.2607 \pm 0.0020$   & -- & -- & Cold dark matter density parameter today\\
$\Omega_m$                 & $0.3111 \pm 0.0056$   & $0.318^{+0.020}_{-0.025}$  & $0.306^{+0.004}_{-0.005}$ & Matter (CDM + baryons) density parameter today\\
$\Omega_\Lambda$           & $0.6889 \pm 0.0056$   & -- & -- & Dark energy density parameter today\\
$\Omega_k$                 & $0.0006 \pm 0.0019$   & -- & -- & Curvature parameter today (consistent with a flat Universe) \\
Age [Gyr]                  & $13.787 \pm 0.020$    & -- & -- & Age of the Universe\\
$T_{\rm CMB, 0}$ [K]       & $2.7255 \pm 0.0006$   & -- & -- & Temperature of the CMB today \citep{2009ApJ...707..916F} \\
$\sigma_8$                 & $0.8102 \pm 0.0060$   & $0.756^{+0.037}_{-0.039}$ & $0.804^{+0.008}_{-0.005}$ & Amplitude of matter density fluctuations on 8 Mpc scales\\
$z_{\rm reion}$            & $7.82 \pm 0.71$       & -- & -- & Mid-point redshift of reionisation \\
$100\, \theta_\text{MC}$   & $1.04101 \pm 0.00029$ & -- & -- & Sound horizon angle for CMB anisotropies \\
$n_\text{s}$               & $0.9665 \pm 0.0038$   & -- & $0.969^{+0.004}_{-0.003}$ & Scalar spectral index \\
$\ln(10^{10}A_\text{s})$   & $ 3.047 \pm 0.014$    & -- & -- & Amplitude of primordial scalar perturbations \\
$\sum m_\nu$ [eV]          & $<0.12$               & -- & $<0.13$ & Sum of neutrino masses (must be $> 0.056$ eV)\\
\bottomrule
\end{tabular}
\end{sidewaystable}
\endgroup

\subsubsection*{Dark matter and baryons}
The matter content, $\Omega_m$ is composed of \textit{baryonic} and \textit{non-baryonic} matter. While both types of matter interact gravitationally, only baryons account for the observable Universe and are attributed to elementary particles in the standard model of particle physics, mostly constituting protons and neutrons. The amount of baryons in the Universe is quantified by the parameter $\Omega_b$. The non-baryonic matter, known as dark matter (DM), is thought to be composed by elementary particles outside the current standard model. Such particles have not yet been detected and extended searches for DM candidates are being undertaken by high-energy physics and cosmology collaborations \cite[e.g.][]{2011ARA&A..49..155P, PhysRevLett.121.081307, 2019PDU....24..249A}. Despite the fundamental nature of the DM is not yet known, research teams put constraints on some of its properties and determined its distribution in the Universe, because of its coupling to gravity. An important constraint involves the energy scale of the DM field. Models with \textit{warm} DM assume that DM particles move relativistically at early times, and have high energies ($\gtrsim$~GeV), while \textit{cold} DM predicts much lower energies ($\lesssim$~keV) and no relativistic species. Both warm and cold DM scenarios return non-relativistic species at late times, after decoupling from radiation.

The main effect of the high velocities of the WDM at early times is to wipe out density fluctuations on small scales. If proven, the deficit of small-scale structures, compared to the CDM scenario, could indeed  be the imprint of early WDM at the present time. 

Recent surveys, such as \cite{2020AnA...641A..10P} and \citep{2022PhRvD.105b3520A}, imposed tighter constraints on warm DM models and indicate that most of the DM in the Universe is in the cold phase. In this document, we will use $\Omega_{\rm CDM}$ to represent the cold dark matter (CDM) contribution to $\Omega_m = \Omega_{\rm CDM} + \Omega_{b}$.

In the computational cosmology paradigm, stars, gas and black holes are considered baryons. Recently, additional cosmological components have been introduced in the baryonic budget, such as neutrinos. These weakly interactive leptons were proved to play a crucial role in precision cosmology calculations, especially in the early Universe. \citep[see e.g.][]{2016JCAP...07..051D}. At early times ($z\sim 10^{3-4}$), the neutrino density follows the radiation density:
\begin{equation}
    \rho_\nu = \frac{7}{8}\left(\frac{4}{11}\right)^{4/3}\, N_{\rm eff}\, \rho_\gamma,
\end{equation}
where $N_{\rm eff}$ is the effective number of neutrino species ($N_{\rm eff} = 3.046$ according to the standard model of particle physics). This behaviour is shown by the evolution of the neutrino density parameter $\Omega_\nu(z)$ in Fig.~\ref{fig:cosmological-parameters}. At lower redshift, however, the neutrinos are subject to a phase transition, which causes them to decouple from the radiation field and behave like matter ($z\lesssim 10^2$). At this stage of the matter-dominated era, non-linear structure formation beginning to take place and, because of their stronger coupling to matter, neutrinos can affect the collapse of matter at small scales. Therefore, high-precision calculations on the matter power spectrum are sensitive to the effects of the neutrinos and motivated research on the neutrino sector from a cosmological standpoint. The measurement of the neutrino masses and their mass hierarchy (i.e. their order when sorted by rest mass) is an unsolved problem, tackled by particle physicists and cosmologists. The discovery of neutrino flavour oscillations proved that the neutrino masses must be non-zero and imposed a tight lower limit on the sum of the masses, i.e. $\sum m_\nu$ > 0.056 eV \citep{1998PhRvL..81.1562F, 2002PhRvL..89a1301A, GONZALEZGARCIA2016199}.\footnote{For the inverted mass hierarchy, $\sum m_\nu$ > 0.10 eV.} Cosmology experiments, on the other hand, constrained the upper limit on the sum of neutrino masses: $\sum m_\nu$ < 0.12 eV \cite{2020A&A...641A...6P, PhysRevD.106.043540}. The effect of massive neutrinos can be included in cosmological simulations in two ways.

\begin{enumerate}
    \item \textbf{Time-dependent power spectrum}, e.g. the BAHAMAS project \citep{2018MNRAS.476.2999M}. The initial matter power spectrum accounts $\Omega_\nu$ (for a particular choice of $\sum m_\nu$, often treated as a free parameter). Furthermore, the mutual gravitational reaction of CDM, baryons, and neutrinos is consistently evaluated through time in the simulations in Fourier space. This approach does not treat the neutrinos as particle-like mass tracers, as in SPH, but effectively as a time-dependent power spectrum which is sourced by the CDM and the baryons. This technique does not suffer from shot noise, which strongly affects standard neutrino particle simulations \citep[e.g.][]{2020ApJS..250....2V}. However, the Fourier method only introduces a linear response to the structure formation and does not capture the non-linear evolution of the neutrino field, which shapes the observationally interesting neutrino field on cluster-scales. 

    \item \textbf{Explicit neutrino particles}, e.g. the FLAMINGO project \citep{flamingo_schaye2023}. \cite{2021MNRAS.507.2614E} have devised a new particle implementation, called the $\delta f$ method, that suppresses shot noise and has the advantage of following the non-linear evolution of non-relativistic (i.e. slow-moving) neutrinos as well as the linear relativistic species on large scales.
\end{enumerate}

The upcoming flagship numerical simulations like FLAMINGO are set to include the effects of massive neutrinos in their physics models. Results from these simulations could greatly improve our understanding of neutrino fields in the Universe and, potentially, describe how the thermal SZ effect \citep{2020MNRAS.497.1332B} and the kinetic SZ effect \citep{2017MNRAS.467..985R} can be used as probes for neutrino cosmology.

\subsubsection*{Dark energy}
The role of the cosmological constant $\Lambda$ in the standard model of cosmology was redefined following the analysis of galaxy surveys in the 1990s with, e.g., the ADm galaxy\hyp{}redshift survey \citep{1992ApJ...390L...1D} and 10 type Ia supernovae (SNe) in the redshift range $0.16<z<0.62$ by \cite{1998AJ....116.1009R}. This result suggested a scenario where $\Omega_\Lambda > 0$, which is associated with a fluid of negative pressure (equation of state $P=-\rho$) referred to as \textit{dark energy}. The amount of dark energy in the Universe is also well constrained by CMB experiments, e.g. \cite{2020A&A...641A...6P} measured $\Omega_\Lambda \approx 0.7$, meaning that approximately 70\% of the matter-energy in the Universe must be in the form of dark energy. As for CDM, the fundamental nature of dark energy is still unknown and a vast branch of theoretical cosmology have proposed models which attempt to probe the mechanisms and phenomenology of the dark sector \citep[see][for a review]{2017JCAP...05..015H}. The Friedmann equations predict that $\Omega_\Lambda$ leads to an accelerating expansion rate of the Universe, i.e. $\ddot{a} > 0$, int the $\Lambda$-dominated era. 

\subsubsection*{Curvature}
The Planck mission has allowed the study of the CMB anisotropies with unprecedented detail, and imposed very tight constraints on the total density parameter $\Omega = \Omega_m + \Omega_r + \Omega_\Lambda \simeq 1$. This result implies that the curvature parameter $\Omega_k \simeq 0$, consistent with a flat Universe \citep{2021APh...13102607D}. Rearranging Eqs. \eqref{eq:friedmann-1} and \eqref{eq:friedmann-2-omega-parameters}, one can obtain the equation for the evolution of $\Omega$:
\begin{equation}
    \dot{\Omega} = H\Omega (\Omega - 1)(1 + 3w),
\end{equation}
which has a \textit{unstable} fixed point at $\Omega = 1$ (and $\Omega_k=0$), compatibly with the result from the Planck 2018 cosmology. Because of the instability of the $\Omega = 1$ solution, we can deduce that any deviation from a non-flat Universe must have been incredibly small at early times, in order for it to go undetected at the present day. Therefore, the $\Omega$ parameter must have been ``fine tuned`` at early times: a coincidence which is inexplicable in the context of classical FLRW cosmology. This puzzle is commonly referred to as the \textit{flatness problem}.

\section{The early Universe}
\label{sec:early-perturbations}

While the background cosmology continues to provide useful insights in today's experiments about the Universe \textit{in toto} (e.g. the measurements of the fundamental cosmological parameters), it fails to capture the formation of structures within it. The Friedmann equations show that a uniform metric will \textit{remain} uniform. In fact, the density $\rho$ and pressure $P$ are constant at any given time and do not depend on the spatial scale: we will always measure the same value regardless of the volume $V$ used to sample the matter distribution. For inhomogeneities to exist to the present time, they must have been seeded at very early times and then grown to form the structures that we observe today. To define the magnitude of the (matter) perturbations, we take the difference of the matter density at a specific point in space $\rho({\bf r})$ and the mean density of the Universe $\Bar{\rho}$ to obtain the density contrast
\begin{equation}
    \delta \equiv \frac{\rho({\bf r})}{\Bar{\rho}} = \frac{\delta \rho}{\Bar{\rho}},
\end{equation}
where $\delta \rho$ is the density perturbation. Assuming a flat Universe, we can write $\Bar{\rho} = \rho_{\rm crit}$. In galaxies and clusters, $\delta$ can be very large $\sim 10^{6-9}$, while in cosmic voids (under-dense regions) the density contrast can become negative. Such extreme values of $\delta$ measures in the nearby Universe are ad odds with the energy distribution in the CMB, which appears to be an extremely uniform black-body radiation at $T_{\rm CMB, 0} \approx 2.7255$ K, after correcting for the peculiar motion of the observer \citep{2009ApJ...707..916F}. Further research on the CMB in the 1990s lead to the discovery of small temperature \textit{anisotropies} from observations performed with the \textit{COBE} (Cosmic Background Explorer) satellite and its successors (e.g. \textit{WMAP} and \textit{Planck}). This result enabled scientists to establish a link between CMB anisotropies and the large-scale structures at low redshift. The development of this framework lead to two additional puzzles in combination with the flatness problem of Section \ref{sec:lambda-cdm}:
\begin{itemize}
    \item \textbf{Primordial perturbations problem}: what physical process seeded the perturbations that lead to the CMB anisotropies?
    \item \textbf{Horizon problem}: how could regions far apart, i.e. causally disconnected at the redshift of the CMB, be in thermal equilibrium and lead to such a uniform temperature distribution?
\end{itemize}

\subsection{Theory of cosmological inflation}

The fine-tuned state of the Universe when the photons decoupled from baryons to produce the CMB could be explained by a mechanism of rapid cosmological expansion at earlier times. This scenario is described by the theory of cosmological inflation, which postulates the existence of an \textit{inflaton} scalar field, $\phi$, which produces a negative pressure and leads to an exponential expansion of the scale-factor \citep{gorbunov2011introduction}. The modern theory of inflation was proposed by \cite{1982PhLB..108..389L} and aimed to solve the flatness, primordial perturbations and horizon problems\footnote{As well as the \textit{no-magnetic monopoles problem}, see also \cite{1978PhLB...79..239Z}.} via an inflaton field with a slowly varying potential $V(\phi)$. This \textit{slow roll} regime can guarantee a smooth transition (\textit{graceful exit}) from the inflationary expansion into the post-inflationary era \citep{1983PhRvD..28..679B}, allowing for the production of primordial perturbations from the vacuum state \citep{1970PhRvD...1.2726H, 1980MNRAS.192..663Z}. The quantum fluctuations of the inflaton field are stretched to cosmological scales by the end of inflation and are then frozen-in as they cross the particle horizon.

The rapid expansion of \textit{causal} regions during inflation causes distant \textit{physical} regions of the universe to come into causal contact well before the onset of recombination and the emission of the CMB. Hence, this scenario can explain the remarkable isotropy of the CMB radiation and resolve the horizon problem. The flatness problem is resolved in a similar fashion, as the expansion during inflation flattens the curvature of the Universe to an extraordinary degree leading to $\Omega_k \sim 10^{-60}$ at the end of inflation \citep{gorbunov2011introduction}. Finally, the theory of inflation allows for the production of primordial perturbations from vacuum states, which is a well-understood mechanism in quantum field theory.

Linear perturbations produced during inflation can be classified into scalar, vector, and tensor modes, depending on their transformation properties under spatial rotations. Inflation only generates scalar and tensor modes, since vector modes decay away during inflation due to the Hubble friction \citep{gorbunov2011introduction}. The scalar perturbations are associated with the seeds of the large-scale structure in the universe, while the tensor perturbations give rise to primordial gravitational waves. The ratio between the amplitudes of the scalar and tensor modes is known as the tensor-to-scalar ratio, which has been constrained by the combined \textit{Planck} and BICEP2/Keck Array BK15 data to obtain $r < 0.056$ \citep{2020A&A...641A..10P}. This result favours, though does not prove, slow-roll inflationary scenarios.

Despite being apparently free from critical shortcomings, and unconfirmed, inflation leaves a number of open questions. The fundamental nature of the inflaton is unconstrained. While numerous models are being checked for compatibility with the most up-to-date limits on $r$, the Higgs boson doublet \citep{2008PhLB..659..703B} appears to provide a suitable potential profile $V(\phi)$ to satisfy the slow-roll conditions in compliance with the current observational constraints. The \textit{graceful exit} out of the inflationary era must be associated with a mechanism to re-heat the Universe to the energy scales required for Big Bang nucleosynthesis and recombination at the onset of the radiation-dominated era. The details of the re-heating mechanism are subject of intense debate within the theoretical cosmology community \citep[e.g.][]{2020PhLB..81135888A, 2023JCAP...02..019A} and motivating the development of numerical simulation codes to model the evolution of inflaton-like vacuum fluctuations using lattice approaches \citep[e.g. the \texttt{CosmoLattice} code][]{2020arXiv200615122F}. As there are no constraints on re-heating in the post-inflationary era, there are preferential models which predict the production of dark matter candidates \citep[e.g.][]{2021PhRvD.103f3526S, 2023JCAP...02..034K}.

\subsection{Linear growth of perturbations}
After the exponential inflationary expansion of the Universe, the scalar perturbations which re-entered the horizon began a phase of linear growth through the radiation-dominated and matter-dominated eras.

\begin{figure}
    \centering
    \includegraphics[width=0.9\textwidth]{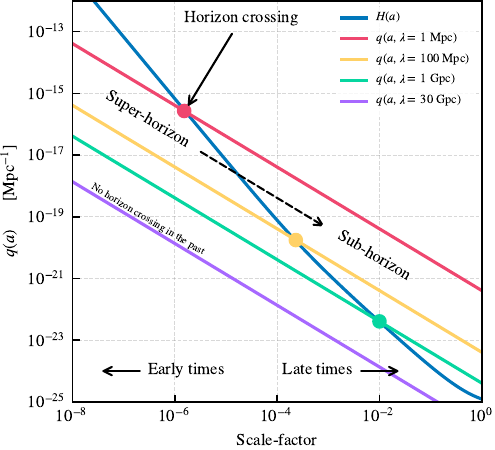}    
    \smallskip
    \caption[Horizon-crossing]{Evolution of the comoving momentum $q$ with scale-factor $a$ for three different scalar modes (wavelength $\lambda = $1 Mpc, 100 Mpc, 1 Gpc) crossing the horizon, indicated by the Hubble parameter $H(a)$ \citep{2020A&A...641A...6P}. We indicate the horizon-crossing scale factors with markers, assuming linear growth across the entire domain. Smaller perturbations enter the horizon at earlier times than larger-wavelength modes. Instead, the larger 30 Gpc mode (purple) never crosses the horizon for $a<1$.}
    \label{fig:horizon-crossing}
\end{figure}

To characterise the growth of these perturbations, we introduce fluctuations in the background metric $g_{\mu\nu}$ by $\delta g_{\mu\nu}$ and again for the temperature $\delta T$ and pressure $\delta P$. Here, we assume that these perturbations are small compared to the background, so that
\begin{equation}
    \begin{array}{cccc}
    \delta \rho \ll \Bar{\rho} \qquad & \qquad \delta T \ll T \qquad & \qquad  \delta P \ll P \qquad & \qquad \delta g_{\mu\nu} \ll g_{\mu\nu}.
    \end{array}
\end{equation}
In this regime, an equation for the evolution of linear perturbations can be derived \citep[see][for further details]{gorbunov2011introduction} in two regimes: scalar modes which did not yet re-enter the horizon, i.e. \textit{super-horizon} modes, and scalar modes which have re-entered the horizon, i.e. \textit{sub-horizon} modes. In Fourier space, these modes are often characterised by their wave-number $k = 2\pi / \lambda$, which depends on their wavelength $\lambda$. For adiabatic modes with comoving momentum $q(a) \sim k / a$, expressed in natural units, we define
\begin{equation}
    \begin{array}{cr}
    q(a) < H(a) & \textrm{Super-horizon} \\
    q(a_\times) \simeq H(a_\times) & \textrm{Horizon crossing} \\
    q(a) > H(a) & \textrm{Sub-horizon}, \\
    \end{array}
\end{equation}
where $a_\times$ indicates the scale factor at horizon crossing and $H(a)$ is the Hubble parameter in natural units. In Fig.~\ref{fig:horizon-crossing}, we illustrate an example for the horizon crossing of three perturbative modes of different wavelengths.
After a mode enters the horizon, $a>a_\times$, its growth phase begins. Small perturbations $\lambda \ll H^{-1} \sim 1$ Mpc entered the horizon at very early times. Hence, they were able to spend most of the cosmic time growing, linearly at first, and non-linearly towards late times. At the present day, these modes are associated with galaxies and clusters. Perturbations comparable to the Hubble parameter $\lambda \sim H^{-1} \sim 1$ Gpc entered the horizon recently and they are still growing linearly at the present day. Modes with $\lambda \gg H^{-1} \gtrsim 30$ Gpc (purple) have not yet entered the horizon, and their growth is inhibited because regions of space so far apart remain causally disconnected.




\subsection{Zeldovich approximation to first-order perturbations}
\label{sec:zeldovich-approximation}

In Section \ref{sec:early-perturbations}, we discussed that the scalar perturbations with $\lambda \lesssim 100$ Mpc that give rise to galaxies and clusters enter the particle horizon during the radiation-dominated. By the time the Universe reaches a radiation-matter equality state, these modes have already undergone linear growth and are beginning to collapse non-linearly. Here, we assume that the linear growth behaviour of these modes continues in the matter-dominated era, where $\Omega_m \approx 1$. This assumption is clearly an \textit{extrapolation} of the linear theory into the non-linear regime. While it provides an analytic formalism for the non-linear structure formation, this approximation is only robust until $z\sim 100$ (or very high redshifts from the perspective of numerical hydrodynamic simulations).

In an EdS universe, the evolution of sub-horizon modes is solely dictated by the gravitational force, which couples to the CDM collisionless fluid \cite{1969ApJ...155..393P, 1970Afz.....6..581D, 1984ApJ...286...38W}. The equations for the evolution of the matter field are derived from the continuity equation,
\begin{equation}
    \partial_t \rho + \mathbf{\nabla} \cdot (\rho \mathbf{v}) = 0,
    \label{eq:continuity-top-hat}
\end{equation}
Euler's equation
\begin{equation}
    \partial_t  \mathbf{v} + ( \mathbf{v} \cdot \mathbf{\nabla}) \cdot \mathbf{v} + \frac{1}{\rho} \mathbf{\nabla}P +  \mathbf{\nabla}\Psi = 0,
    \label{eq:euler-top-hat}
\end{equation}
the energy equation
\begin{equation}
    \partial_t u + ( \mathbf{v} \cdot \mathbf{\nabla}) u + \frac{p}{\rho} (\mathbf{\nabla} \cdot \mathbf{v}) = 0,
    \label{eq:energy-top-hat}
\end{equation}
and Poisson's equation
\begin{equation}
    \nabla^2 \Psi = 4\pi G \rho,
    \label{eq:poisson-top-hat}
\end{equation}
where $\partial_t$ denotes the partial time derivative, $\rho=\rho(\mathbf{x}, t)$ the density of the collisionless CDM fluid, $\mathbf{v}=\mathbf{v}(\mathbf{x}, t)$ its velocity field, $P=P(\mathbf{x}, t)$ the pressure field, $\Psi=\Psi(\mathbf{x}, t)$ is the scalar gravitational potential, and $u=u(\mathbf{x}, t)$ is the energy per unit mass. An important distinction must be made between the \textit{Eulerian} coordinate frame $\mathbf{r}$ and the \textit{Lagrangian} frame $\mathbf{x}$: $\mathbf{r}$ explicitly includes the background expansion of the Universe, while $\mathbf{x}$ is defined in the rest frame of the fluid (i.e. comoving). By differentiating the transformation relation between the comoving and physical frame: $\mathbf{r}=a(t) \mathbf{x}$ with respect ti time ($\partial_t$) we obtain

\begin{equation}
\label{eq:pecvelocity}
    \mathbf{v} = \dot{\mathbf{r}} = \underbrace{\dot{a}\mathbf{x}}_{\text{Hubble flow}} + \overbrace{a \dot{\mathbf{x}}}^{\text{Peculiar velocity}} \equiv \dot{a}\mathbf{x} + \mathbf{v}_\mathrm{p},
\end{equation}
with $\mathbf{v}_\mathrm{p}$ the peculiar velocity field (in the Lagrangian frame). Solving the evolution equations for an EdS background leads to $a \propto t^{2/3}$ and growing density modes parametrised by the \textit{growth factor} $D(t)=a(t)$ \citep[see e.g.][]{1980lssu.book.....P}.

As an additional corollary to the EdS solution, the $\mathbf{\dot{x}}$ field is irrotational, indicating that the velocity vector of fluid elements does not change direction over time due to cosmological expansion. This formalism is the foundation of the Zeldovich approximation \citep[ZA,][]{1970A&A.....5...84Z}, used to describe the linear growth of structures in a matter-dominated scenario.

The self-similar growth of CDM structures is described in terms of the density contrast $\delta\equiv\rho/\rho_b$, as defined in Section \ref{sec:early-perturbations}, yielding $\delta(\mathbf{x}, t) = D(t)\delta(\mathbf{x}_0)$. By substituting this relation into Poisson's equation, it is possible to find the $\Psi$ solution in a EdS background:
\begin{align}
    \nabla^2 \Psi(\mathbf{x}, t) &= 4\pi G\, \rho_b\, \delta(\mathbf{x}, t) a^2,\\
    \Psi(\mathbf{x}, t) &= \frac{D}{a} \Psi(\mathbf{x}_0). 
    \label{eq:poisson-EdS-ZA}
\end{align}

The latter expression suggests that in a EdS Universe with $D=a$ the gravitational potential due to the CDM field does not change over time. Eq.~\eqref{eq:poisson-EdS-ZA} allows to find solutions to Euler's equation and express the velocity
\begin{equation}
    \dot{\mathbf{x}}(t) = \frac{\dot{D}(t)}{4\pi G\, \rho_b} \nabla \Psi(\mathbf{x}_0),
\end{equation}
which can be further integrated over time to yield the final solution for $\mathbf{x}$:
\begin{equation}
\label{eq:edsmotion}
    \mathbf{x}(t) = -D(t)\frac{\nabla \Psi(\mathbf{x}_0)}{4\pi G\, \rho_b}  + \mathbf{x}_0 \equiv \mathbf{S}(\mathbf{x}_0, t) + \mathbf{x}_0.
\end{equation}
This expression represents the motion of a fluid element in a straight line, parametrised by the linear growth factor $D$ via the displacement vector $\mathbf{S}(\mathbf{x}_0, t)$.




\section{Non-linear perturbations}
\label{sec:late-perturbations}

We now provide an overview of the framework describing the formation of structures during the non-linear growth regime. We will then focus on the evolution of angular momentum in linear and first-order perturbation theory. The discussion of the angular momentum growth in an EdS universe introduces the concept of cluster halo rotation, which Chapter \ref{chapter:3} is built upon.

\subsection{The spherical top-hat collapse model}
\label{sec:top-hat-collapse}

Still in the context of an EdS universe, we now apply small perturbations to the continuity equation \eqref{eq:continuity-top-hat}, Euler's equation \eqref{eq:euler-top-hat} and Poisson's equation \eqref{eq:poisson-top-hat} as follows:
\begin{equation}
    \begin{array}{cr}
         \rho = \rho_b + \delta \rho & \qquad \text{Density perturbation},\\
         \phi = \phi_b + \delta \phi& \qquad \text{Potential perturbation},
    \end{array}
\end{equation}
where $\phi_b$ is the background gravitational potential. After defining the \textit{convetive derivative}
\begin{equation}
    \frac{d}{dt} \equiv \partial t + (\mathbf{v}_b\cdot\nabla)
\end{equation}
and the \textit{comoving gradient} operators
\begin{equation}
    \nabla_a \equiv a\nabla,
\end{equation}
one can derive the perturbed equations:
\begin{align}
         \frac{d\delta\rho}{dt} + \nabla_a \cdot \mathbf{\dot{x}} &= 0,\\   \vspace{3mm}
         \frac{d\mathbf{\dot{x}}}{dt} + 2H\mathbf{\dot{x}} + \frac{1}{a^2}\,\nabla_a\delta\phi &=0, \\  \vspace{3mm}
         \nabla_a^2 \delta\phi = 4\pi G\, a^2 \delta \rho &=0,
\end{align}

with $\mathbf{\dot{x}} \equiv d\mathbf{x}/dt$, i.e. the convective derivative of $\mathbf{x}$. In this formulation, we have neglected perturbative terms in second order or higher and retained only first-order terms. The perturbed equations can be then combined to yield the evolution equation for density perturbations:
\begin{equation}
    \frac{d^2}{dt^2}(\delta \rho) + 2H\, \frac{d}{dt}(\delta \rho) - 4\pi G\, \rho_b\, (\delta \rho) = 0, 
    \label{eq:sthc-perturbations-ode}
\end{equation}
which is a second-order ordinary differential equation (ODE) in $\delta\rho$. From Tab.~\ref{tab:fluid-evolution}, the time-dependence of the Hubble parameter is
\begin{equation}
    H(t) \equiv \frac{\dot{a}}{a} = \frac{2}{3t},
    \label{eq:sthc-hubble-parameter}
\end{equation}
and for the background density we have
\begin{equation}
    \rho_b \annoterel[\Big]{Assuming $\Omega_k = 0$}{=} \rho_{\rm crit} = \frac{1}{6\pi G t^2},
    \label{eq:sthc-background-density}
\end{equation}
assuming a flat universe. Substituting Eqs.~\eqref{eq:sthc-hubble-parameter} and \eqref{eq:sthc-background-density} into Eq.~\eqref{eq:sthc-perturbations-ode} gives
\begin{equation}
    \frac{d^2}{dt^2}(\delta \rho) + \frac{4}{3t}\, \frac{d}{dt}(\delta \rho) - \frac{2}{3t^2}\, (\delta \rho) = 0,
\end{equation}
which has real solutions of the form
\begin{equation}
    \delta \rho(t) = A\, t^{2/3} + B\, t^{-1} \approx A\, t^{2/3},
    \label{eq:sthc-background-density-evolution}
\end{equation}
where $A$ and $B$ are constants of integration. Neglecting the decaying mode $B/t$ for large $t$, we find that $\delta \rho(t) \approx A\, t^{2/3}$ in a matter-dominated universe. In terms of the growth factor $D(a) = a$, this relation can also be expressed as $\delta \rho(t) \propto D$, emphasising the linear proportionality between the density contrast and the growth factor.

\begin{figure}
    \centering
    \includegraphics[width=\textwidth]{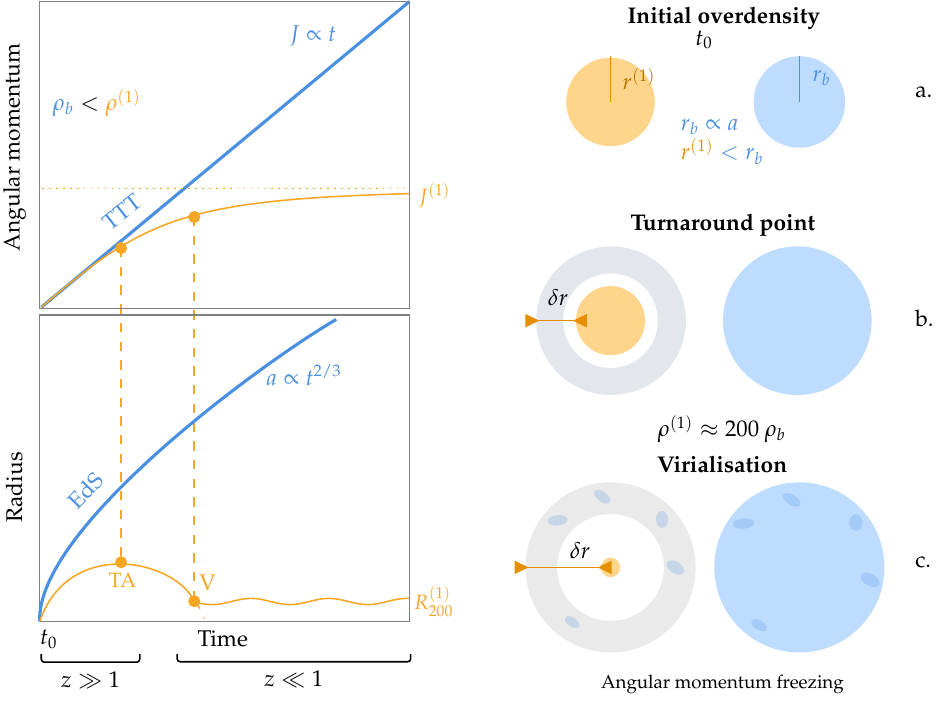}    
    \caption{Diagram illustrating the time-evolution of the radius and the angular momentum of a background region and an overdense region in the context of the STHC model (left panels). We show an idealised uniform CDM halo in three key moments of its collapse: initial state (\textit{a}), turnaround (\textit{b}) and  virialisation (\textit{c}). The lower left panel illustrates the time-evolution of the radius of two spherical regions (blue and orange). On the right (\textit{b}) and (\textit{c}), we superimpose the background to the overdensity as a visual guide to emphasize their increasing difference in radii $\delta r$.}
    \label{fig:tttdiagram}
\end{figure}

We now consider two spherically symmetric regions of space: one region is labelled as (\textit{b}) and is a region of space filled with matter at the background density $\rho_b$ and the other is labelled (\textit{1}) and has a slightly higher density than the background, $\rho^{(1)} = \rho_b + \delta \rho$. Their density distribution follows a top-hat profile and for this reason such a model is known as the spherical top-hat collapse (STHC) model. We define regions (\textit{b}) and (\textit{1}) to have the same mass, $M_b = M^{(1)}$. This scenario is illustrated in Fig.~\ref{fig:tttdiagram} (right panel), with the background region in blue and the overdense region (\textit{1}) in orange. We can write the evolution equation for the physical radius $r$ of the spherical overdensity using the Friedmann equation
\begin{equation}
    \left(\frac{dr}{dt}\right)^2 = \frac{8\pi G}{3}\, \frac{\rho}{r} = \frac{8\pi G}{3}\, \frac{\rho_b + \delta \rho}{r} = \frac{8\pi G \rho_b}{3}\, \frac{1 + \delta}{r} \equiv \alpha^2 \frac{1 + \delta}{r},
\end{equation}
where we have collected the constants in the $\alpha^2$ term. For the sphere with background density ($\delta = 0$), the radius evolves as
\begin{equation}
    r_b(t) = \left( \frac{3\alpha}{2}\, t\right)^{2/3} \propto t^{2/3},
\end{equation}
while for region (\textit{1}), with $\delta > 0$, we obtain an ODE which can be recast into $\dot{r}^2 = \alpha ^ 2 / r - \varepsilon^2$, where $\varepsilon^2$ can be interpreted as a positive-curvature term of the Friedmann equation. The solution for an overdensity is a cycloid, which can be parameterised as
\begin{equation}
\begin{array}{c}
     r(\theta) = A(1 - \cos \theta ), \\
     t(\theta) = B(\theta - \sin \theta).
\end{array}
\end{equation}
The cycloid solution for the radius of the overdense region is shown in the bottom left panel of Fig.~\ref{fig:tttdiagram}. It shows a maximum radius for $\theta = \pi$, called the turn-around point (TA), after which the radius stops increasing and the matter distribution collapses. While the analytic solution predicts that the radius vanishes at $\theta = 2\pi$, the virial theorem, and numerical N-body simulations, suggest that inhomogeneous overdense regions reach an equilibrium between internal kinetic energy and gravitational potential energy, which prevents further collapse. The overdensity is in a virialised state (V) and has a \textit{virial} radius $R_{\rm vir}$ which remains approximately constant. Assuming that the overdense halo is fully virialised after twice its turnaround point, then its density compared to the background is $\Delta = (r_b / r^{(1)})^3 = 18\pi^2\approx178$. Conservatively, this value is often approximated as $\Delta \approx 200$, which is the density contrast at which CDM halos are predicted to virialise according to the STHC model.
The mass enclosed in the spherical overdensity with $\Delta = 200$ is defined as $M_{200}$, or equivalently the virial mass $M_{\rm vir}$ and the radius of the spherical overdensity is defined as $r_{200}$, or virial radius $r_{\rm vir}$. Given that $\rho = 200\, \rho_b$ and, in a $\Lambda$CDM universe, $\rho_b=\rho_{\rm crit}$, we can show that
\begin{equation}
    M_{200} = \frac{4\pi}{3}\, r^3_{200} \cdot 200\, \rho_{\rm crit}. 
\end{equation}

\subsubsection*{\simulationinsight{Inhomogeneous spherical overdensities}}
Early N-body simulations have shown that matter perturbations collapsing under gravity are not homogeneous and do not have uniform density. This scenario arises from inhomogeneous initial conditions, which are often generated via perturbing a uniform particle lattice via Zeldovich displacement vectors (see the ZA in Section~\ref{sec:zeldovich-approximation}). Inner regions of the matter distributions were found to be denser than the outskirts. The density computed in concentric spherical shells of increasing radii, often known as \textit{density profile}, was found to follow the relation
\begin{equation}
    \rho(r) = \frac{\rho_c}{(r/r_s)(1 + r/r_s)^2},
\end{equation}
where $\rho_c$ is the density scale and $r_s$ is the scale-radius. This distribution is known as Navarro-Frenk-White \citep[NFW,][]{1996ApJ...462..563N} profile and was shown to provide a suitable description for the density profile of CDM halos in a $\Lambda$CDM Universe. The scale radius is related to the virial radius via the concentration parameter $c$ as $r_s= r_{200}/c$.

The definition of the virial radius can be generalised for different values of the density contrast, allowing the definition of $r_\Delta$ and $M_\Delta$. In microwave observations, galaxy cluster radii are often scaled as $r_{200}$, which indicates the radius where the mean density contained in the spherical aperture is $200\, \rho_{\rm crit}$. X-ray observations, instead, often use the value of $r_{2500}$ (common for \textit{Chandra} due to its narrow field of view) and $r_{500}$ (common for \textit{XMM-Newton}) as a radial scaling. Provided that the density profile decreases monotonically with radius, similarly to the NFW case, we must have $r_{2500} < r_{500} < r_{200}$, and equivalently $M_{2500} < M_{500} < M_{200}$.


\subsection{Angular momentum generation from tidal torques}
\label{sec:tidal-torque-theory}
After having developed the formalism of the STHC model and introduced the ZA, we now discuss the cosmological origin of angular momentum, a crucial quantity for the formation of clusters, galaxies, stars, down to the smallest scales of planetary systems.

Early works by \cite{1952PhRv...86..251G} and \cite{1969ApJ...155..393P} suggested that angular momentum on cosmological scales could originate from the collapse of dark matter structures, which themselves transfer it to smaller objects condensing within them, such as galaxies. The role of angular momentum in the formation of galaxies and clusters rapidly grew in interest within the research community.

Following the derivation outlined in \cite{1996MNRAS.282..436C}, the angular momentum $\mathbf{J}(t)$ of a CDM structure enclosed within a volume $a^3 V$ is given (in the Eulerian frame) by
\begin{equation}
   \mathbf{J}(t) = \int_{a^3V} d \mathbf{r}\rho (\mathbf{r} \times \mathbf{v}).
   \label{eq:angmomentumdef}
\end{equation}
In this formulation, it is important to note that $\mathbf{v}$ encapsulates both the Hubble flow and the peculiar velocity term. By expanding the vector product using the comoving transformation relation in equation~(\ref{eq:pecvelocity}), the Hubble flow vanishes, indicating that only $\mathbf{v}_\mathrm{p}$ contributes to the angular momentum of the CDM fluid. After writing $\rho$ in terms of the contrast $\delta$ and $\mathbf{r}=a\mathbf{x}$, the expression for $\mathbf{L}(t)$ becomes
\begin{equation}
\label{eq:angmomentumdef2}
    \mathbf{J}(t) =a^5\rho_b \int_{V}d\mathbf{x} (1+\delta)(\mathbf{x}\times\dot{\mathbf{x}}).
\end{equation}

In the Lagrangian formulation of the ZA, it is common to rewrite equation~(\ref{eq:edsmotion}) as $\mathbf{x}(t) = \mathbf{q}_0 + D(t)\mathbf{S}(\mathbf{q}_0)$, by remapping $\mathbf{x}\rightarrow\mathbf{q}$ \citep{1989RvMP...61..185S} via the Jacobian $(1+\delta)$, yielding

\begin{equation}
\label{eq:angmomentumdef3}
    \mathbf{J}(t) = a^5\rho_b \int_{\Gamma}d\mathbf{q} (\mathbf{q} + \mathbf{S})\times\dot{\mathbf{S}},
\end{equation}
where $\Gamma$ is now the initial Lagrangian volume. As stressed by \cite{1996MNRAS.282..436C}, equation~(\ref{eq:angmomentumdef3}) is exact in the EdS limit, since the displacement vector has not been expanded and made explicit yet. Performing a Taylor expansion on $\mathbf{S(\mathbf{q}, t)}$ and retaining only leading-order terms gives rise to semi-analytic perturbation theories, such as the canonical first-order tidal-torque theory \citep[TTT,][]{1984ApJ...286...38W, 1996MNRAS.282..436C} and higher-order Lagrangian formulations \citep[see the introduction by][and references therein]{1996MNRAS.282..455C}.

In this section, we only derive the first-order approximation and briefly discuss the benefits and limitations of more complex extensions pointing to the relevant literature works. The expansion of $\mathbf{S(\mathbf{q}, t)}$ in Taylor series is performed in the Lagrangian coordinate frame centred on the minimum of the potential of the CDM halo, referred to as centre of potential (CoP) and located at $\mathbf{q}=\mathbf{q}_0  = 0$. While maintaining an implicit time dependence through the linear growth factor, we expand $\mathbf{S}$ as   
\begin{equation}
\label{eq:taylorza}
    \mathbf{S}(\mathbf{q}, t) = \sum_{i=0}^{\infty}\mathbf{S}_i(\mathbf{q}, t) \annoterel[\Big]{Using equation~(\ref{eq:edsmotion})}{=}  -\frac{D(t)}{4\pi G_N \rho_b}\sum_{i=0}^{\infty} \frac{(\mathbf{q}-\mathbf{q}_0)^i}{i!}\frac{\partial^{i+1} \Psi(\mathbf{q})}{\partial q^{i+1}} \bigg\rvert_{\mathbf{q}=\mathbf{q}_0=0}.
\end{equation}

It is now straight-forward to write the first terms of the series and their non-zero partial time-derivatives:
\begin{align}
    \mathbf{S}_0(\mathbf{q}, t) &= -\frac{D(t)}{4\pi G_N \rho_b}\frac{\partial\Psi}{\partial q_\alpha} \bigg\rvert_{\mathbf{q}_0}\nonumber\\ 
    &= -\frac{D(t)}{4\pi G_N \rho_b}\nabla_\alpha\Psi(\mathbf{q}_0) \label{eq:displacement-nabla-psi-zero}\\
    &= 0 \quad\text{(In the centre-of-mass frame)}\nonumber\\
    \dot{\mathbf{S}}_0(\mathbf{q}, t) &= 0 \quad\text{(In the centre-of-mass frame)} \nonumber\\
    \nonumber \\
    \mathbf{S}_1(\mathbf{q}, t) &= -\frac{D(t)}{4\pi G_N \rho_b}q_\beta\frac{\partial^2\Psi}{\partial q_\alpha \partial q_\beta} \bigg\rvert_{\mathbf{q}_0}\nonumber\\ 
    &= -\frac{D(t)}{4\pi G_N \rho_b}q_\beta\mathcal{H}_{\alpha\beta}\Psi(\mathbf{q}_0)\nonumber\\
    &= D(t)\,q_\beta\mathcal{D}_{\alpha\beta} \label{eq:displacement-nabla-psi}\\
    \dot{\mathbf{S}}_1(\mathbf{q}, t) &= \dot{D}(t)\,q_\beta\mathcal{D}_{\alpha\beta} \label{eq:displacement-nabla-psi-dot}\\
    \nonumber \\
    \mathbf{S}_2(\mathbf{q}, t) &= D(t)\,q_\beta\, q_\gamma\, \partial_{\alpha\beta\gamma} \Psi = \mathcal{O}(\partial^3\Psi)\nonumber\\
    \dot{\mathbf{S}}_2(\mathbf{q}, t) &= \dot{D}(t)\,q_\beta\, q_\gamma\, \partial_{\alpha\beta\gamma} \Psi = \mathcal{O}(\partial^3\Psi).\nonumber
\end{align}
The $(-\nabla\Psi)$ term in Eq.~\eqref{eq:displacement-nabla-psi} expresses the force acting on the centre of mass (CoM) of the halo. By choosing to compute the \textit{proper} angular momentum, we evaluate this term in the CoM frame, which is assumed to coincide with the Lagrangian frame centred on the CoP. Note that the Taylor expansion was performed about the CoP, defined as the centre of the CDM halo, while $\mathbf{S}_0(\mathbf{q}, t)$ vanishes at the CoM. The two reference frames are conceptually distinct and the derivation in the linear regime gives no justification for them to coincide. However, we implicitly assume that the CoP and CoM do approximately coincide.\footnote{Formally, we assume that $|\mathbf{x}_\mathrm{CoP} - \mathbf{x}_\mathrm{CoM}|\ll R_{200}$, where $R_{200}$ is the CDM halo's virial radius in the Lagrangian frame. As will be described in later sections, the separation between the CoP and the CoM is used to quantify how perturbed a CDM halo is in response to the interaction with a close merger.} This approximation is valid for unperturbed CDM halos, which are subject to weak tidal gravitational fields, such as in the idealised TTT set-up. While $\dot{\mathbf{S}}_0(\mathbf{q}, t)$ trivially vanishes (in the CoM reference frame), the first-order terms do not. The explicit expansion of the $\Psi$-derivatives in the first order results in the Hessian $\mathcal{H}$ of the scalar potential, which quantifies the degree of anisotropy in the $\Psi$ field. Tidal forces originate from an anisotropic topology of $\Psi$ and cause the local matter distribution to deform. For this reason, $\mathcal{H}\Psi$ is commonly denoted as deformation tensor $\mathcal{D}$ (includes the $\mathbf{q}$-invariant pre-factor). In the Taylor expansion of the Zeldovich displacement vector, spatial derivatives of $\Psi$ do not have any explicit time dependence, while the time evolution is entirely expressed by the linear growth factor $D$. This feature, which can be regarded as part of the ZA approach, allows to easily compute the first-order $\dot{\mathbf{S}}_1(\mathbf{q}, t)$ from Eq.~\eqref{eq:displacement-nabla-psi-dot}. Recalling from above, only the expansion terms in $\mathbf{S}_1$ contribute to the TTT, while higher-order ones are shown above for demonstration purposes and are truncated \citep[see e.g.][for a full treatment of $\mathcal{O}(\partial^3\Psi)$ terms]{1996MNRAS.282..455C}.

Eq.~\eqref{eq:displacement-nabla-psi} and Eq.~\eqref{eq:displacement-nabla-psi-dot} can be combined with Eq.~\eqref{eq:angmomentumdef3} to deliver the perturbed form of the angular momentum to first-order ZA:
\begin{align}
    \mathbf{J}(t) &= a^5\rho_b \int_{\Gamma}d\mathbf{q} (\mathbf{q} + \mathbf{S})\times\dot{\mathbf{S}} \\
    &= a^5\rho_b \int_{\Gamma}d\mathbf{q} (\mathbf{q}\times\dot{\mathbf{S}}) + \underbrace{(\mathbf{S}\times\dot{\mathbf{S}})}_{=0} \\
    &\approx a^5\rho_b \dot{D}(t)\int_{\Gamma}d\mathbf{q} \, \mathbf{q}\times (\mathbf{q}\mathcal{D}) \qquad \text{(First-order ZA)} \\
    &\approx a^5 \rho_b \dot{D}(t) \int_{\Gamma}d q \,\epsilon_{\alpha\beta\gamma}\, q_\beta\,  q_\delta\,  \mathcal{D}_{\delta\gamma},
\end{align}
where $\epsilon_{\alpha\beta\gamma}$ is the Levi-Civita symbol, used in the expansion of the vector product with index notation. In the above equation, we note that the integral 
\begin{equation}
\label{eq:inertiatensor}
    (a^3 \rho_b) \int_{\Gamma}d q \, q_\beta\,  q_\delta \equiv \mathcal{I}_{\beta\delta},
\end{equation}
namely the inertia tensor of the CDM halo computed assuming a constant mass element $(a^3 \rho_b)$ throughout the volume $\Gamma$. By gathering the terms from equation~(\ref{eq:inertiatensor}) \citep{1984ApJ...286...38W, 1996MNRAS.282..436C}, we can write the time-dependent angular momentum as
\begin{equation}
\label{eq:tttmain}
    J_\alpha(t) = a^2 \dot{D}(t) \,\epsilon_{\alpha\beta\gamma} \, \mathcal{I}_{\beta\delta}\,  \mathcal{D}_{\delta\gamma}.
\end{equation}

Early studies by \citep{1984ApJ...286...38W, 1996MNRAS.282..436C, 1996MNRAS.282..455C} assumed $\mathcal{I}$ and $\mathcal{D}$ to be uncorrelated. The deformation tensor maps the asymmetric shape of the \textit{local potential well} and shows no explicit dependence on the inertia tensor, which instead represents the \textit{morphology of the matter distribution} within $\Gamma$. In fact, $\mathcal{I}$ and $\mathcal{D}$ are conceptually different tensors, with distinct eigenvalues $\lambda^{(i)}_\mathcal{D}$, $\lambda^{(i)}_\mathcal{I}$ and eigenvectors $\{\mathbf{a}_\mathcal{D}, \mathbf{b}_\mathcal{D}, \mathbf{c}_\mathcal{D}\}$, $\{\mathbf{a}_\mathcal{I}, \mathbf{b}_\mathcal{I}, \mathbf{c}_\mathcal{I}\}$ (the eigenvectors are sorted from the largest to the smallest corresponding eigenvalue). Since the CDM distribution itself introduces a gravitational field (which may dominate if the halo is relatively isolated), it is reasonable to disagree with the assumption of uncorrelated $\mathcal{I}$ and $\mathcal{D}$. \cite{2002MNRAS.332..325P} and \cite{2004ApJ...613L..41N} investigated this idea and found, as expected, a strong correlation between the eigenvectors of the two tensors, which however emerged from analysing N-body simulations in a statistical manner.

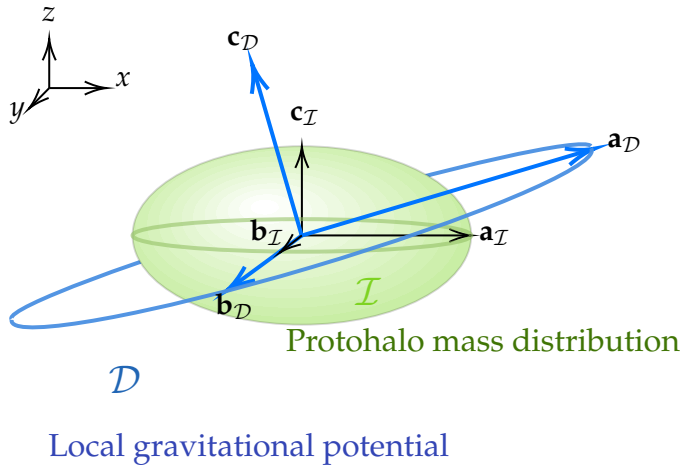
\begin{figure}
    \centering

    \tikzset {_84j66t27s/.code = {\pgfsetadditionalshadetransform{ \pgftransformshift{\pgfpoint{89.1 bp } { -128.7 bp }  }  \pgftransformscale{1.32 }  }}}
    \pgfdeclareradialshading{_0afup590s}{\pgfpoint{-72bp}{104bp}}{rgb(0bp)=(1,1,1);
    rgb(0bp)=(1,1,1);
    rgb(0bp)=(1,1,1);
    rgb(25bp)=(0.13,0.4,0.73);
    rgb(400bp)=(0.13,0.4,0.73)}
    \tikzset{_609aio8t5/.code = {\pgfsetadditionalshadetransform{\pgftransformshift{\pgfpoint{89.1 bp } { -128.7 bp }  }  \pgftransformscale{1.32 } }}}
    \pgfdeclareradialshading{_fojkviuxj} { \pgfpoint{-72bp} {104bp}} {color(0bp)=(transparent!0);
    color(0bp)=(transparent!0);
    color(0bp)=(transparent!47);
    color(25bp)=(transparent!17);
    color(400bp)=(transparent!17)} 
    \pgfdeclarefading{_b5wj3n7c1}{\tikz \fill[shading=_fojkviuxj,_609aio8t5] (0,0) rectangle (50bp,50bp); } 
    
      
    \tikzset {_0le17zmzh/.code = {\pgfsetadditionalshadetransform{ \pgftransformshift{\pgfpoint{89.1 bp } { -128.7 bp }  }  \pgftransformscale{1.32 }  }}}
    \pgfdeclareradialshading{_a56sonfst}{\pgfpoint{-72bp}{104bp}}{rgb(0bp)=(1,1,1);
    rgb(0bp)=(1,1,1);
    rgb(0bp)=(1,1,1);
    rgb(25bp)=(0.72,0.91,0.53);
    rgb(400bp)=(0.72,0.91,0.53)}
    \tikzset{every picture/.style={line width=0.75pt}} 
    
    \begin{tikzpicture}[x=0.75pt,y=0.75pt,yscale=-1,xscale=1]
    
    \draw  [draw opacity=0][line width=1.5]  (298.01,76.33) .. controls (293.96,71.56) and (228.86,86.23) .. (150.48,109.67) .. controls (70.54,133.58) and (7.07,157.55) .. (8.7,163.21) .. controls (8.7,163.22) and (8.7,163.23) .. (8.71,163.24) -- (153.44,119.93) -- cycle ; \draw  [color={rgb, 255:red, 74; green, 144; blue, 226 }  ,draw opacity=0.95 ][line width=1.5]  (298.01,76.33) .. controls (293.96,71.56) and (228.86,86.23) .. (150.48,109.67) .. controls (70.54,133.58) and (7.07,157.55) .. (8.7,163.21) .. controls (8.7,163.22) and (8.7,163.23) .. (8.71,163.24) ;
    \draw  [draw opacity=0][shading=_0afup590s,_84j66t27s,path fading= _b5wj3n7c1 ,fading transform={xshift=2}] (7.67,164.97) .. controls (-6.04,117.43) and (48.24,59.34) .. (128.91,35.21) .. controls (209.59,11.09) and (286.1,30.06) .. (299.81,77.6) .. controls (313.52,125.14) and (259.24,183.24) .. (178.57,207.36) .. controls (97.9,231.49) and (21.38,212.51) .. (7.67,164.97) -- cycle ;
    \path  [shading=_a56sonfst,_0le17zmzh] (69.22,121.29) .. controls (69.22,96.49) and (107.06,76.39) .. (153.74,76.39) .. controls (200.42,76.39) and (238.26,96.49) .. (238.26,121.29) .. controls (238.26,146.08) and (200.42,166.19) .. (153.74,166.19) .. controls (107.06,166.19) and (69.22,146.08) .. (69.22,121.29) -- cycle ; 
     \draw  [color={rgb, 255:red, 0; green, 0; blue, 0 }  ,draw opacity=0.15 ] (69.22,121.29) .. controls (69.22,96.49) and (107.06,76.39) .. (153.74,76.39) .. controls (200.42,76.39) and (238.26,96.49) .. (238.26,121.29) .. controls (238.26,146.08) and (200.42,166.19) .. (153.74,166.19) .. controls (107.06,166.19) and (69.22,146.08) .. (69.22,121.29) -- cycle ; 
    
    \draw    (153.74,121.29) -- (236.26,121.29) ;
    \draw [shift={(238.26,121.29)}, rotate = 180] [color={rgb, 255:red, 0; green, 0; blue, 0 }  ][line width=0.75]    (10.93,-3.29) .. controls (6.95,-1.4) and (3.31,-0.3) .. (0,0) .. controls (3.31,0.3) and (6.95,1.4) .. (10.93,3.29)   ;
    \draw    (153.74,121.29) -- (153.74,78.39) ;
    \draw [shift={(153.74,76.39)}, rotate = 450] [color={rgb, 255:red, 0; green, 0; blue, 0 }  ][line width=0.75]    (10.93,-3.29) .. controls (6.95,-1.4) and (3.31,-0.3) .. (0,0) .. controls (3.31,0.3) and (6.95,1.4) .. (10.93,3.29)   ;
    \draw  [color={rgb, 255:red, 145; green, 191; blue, 98 }  ,draw opacity=0.57 ][line width=1.5]  (69.22,121.29) .. controls (69.22,116.77) and (107.06,113.1) .. (153.74,113.1) .. controls (200.42,113.1) and (238.26,116.77) .. (238.26,121.29) .. controls (238.26,125.81) and (200.42,129.47) .. (153.74,129.47) .. controls (107.06,129.47) and (69.22,125.81) .. (69.22,121.29) -- cycle ;
    \draw [color={rgb, 255:red, 0; green, 117; blue, 255 }  ,draw opacity=1 ][line width=1.5]    (153.65,121.37) -- (296.93,78.47) ;
    \draw [shift={(299.81,77.61)}, rotate = 523.3299999999999] [color={rgb, 255:red, 0; green, 117; blue, 255 }  ,draw opacity=1 ][line width=1.5]    (14.21,-4.28) .. controls (9.04,-1.82) and (4.3,-0.39) .. (0,0) .. controls (4.3,0.39) and (9.04,1.82) .. (14.21,4.28)   ;
    \draw [color={rgb, 255:red, 0; green, 117; blue, 255 }  ,draw opacity=1 ][line width=1.5]    (153.65,121.37) -- (129.73,38.1) ;
    \draw [shift={(128.91,35.21)}, rotate = 433.98] [color={rgb, 255:red, 0; green, 117; blue, 255 }  ,draw opacity=1 ][line width=1.5]    (14.21,-4.28) .. controls (9.04,-1.82) and (4.3,-0.39) .. (0,0) .. controls (4.3,0.39) and (9.04,1.82) .. (14.21,4.28)   ;
    \draw [color={rgb, 255:red, 0; green, 117; blue, 255 }  ,draw opacity=1 ][line width=1.5]    (153.65,121.37) -- (126.09,141.76) -- (118.78,147.17) ;
    \draw [shift={(116.37,148.95)}, rotate = 323.5] [color={rgb, 255:red, 0; green, 117; blue, 255 }  ,draw opacity=1 ][line width=1.5]    (14.21,-4.28) .. controls (9.04,-1.82) and (4.3,-0.39) .. (0,0) .. controls (4.3,0.39) and (9.04,1.82) .. (14.21,4.28)   ;
    \draw  [draw opacity=0][line width=1.5]  (298.06,77.44) .. controls (298.21,77.64) and (298.32,77.86) .. (298.38,78.09) .. controls (300.93,86.91) and (238.19,113.45) .. (158.26,137.36) .. controls (78.32,161.26) and (11.45,173.48) .. (8.91,164.66) .. controls (8.83,164.37) and (8.81,164.06) .. (8.87,163.73) -- (153.65,121.37) -- cycle ; \draw  [color={rgb, 255:red, 74; green, 144; blue, 226 }  ,draw opacity=0.95 ][line width=1.5]  (298.06,77.44) .. controls (298.21,77.64) and (298.32,77.86) .. (298.38,78.09) .. controls (300.93,86.91) and (238.19,113.45) .. (158.26,137.36) .. controls (78.32,161.26) and (11.45,173.48) .. (8.91,164.66) .. controls (8.83,164.37) and (8.81,164.06) .. (8.87,163.73) ;
    \draw    (153.74,121.29) -- (144.19,128.76) ;
    \draw [shift={(142.61,129.99)}, rotate = 321.98] [color={rgb, 255:red, 0; green, 0; blue, 0 }  ][line width=0.75]    (10.93,-3.29) .. controls (6.95,-1.4) and (3.31,-0.3) .. (0,0) .. controls (3.31,0.3) and (6.95,1.4) .. (10.93,3.29)   ;
    \draw    (27.62,47.17) -- (54.27,47.17) ;
    \draw [shift={(56.27,47.17)}, rotate = 180] [color={rgb, 255:red, 0; green, 0; blue, 0 }  ][line width=0.75]    (10.93,-3.29) .. controls (6.95,-1.4) and (3.31,-0.3) .. (0,0) .. controls (3.31,0.3) and (6.95,1.4) .. (10.93,3.29)   ;
    \draw    (27.62,47.17) -- (27.62,24.27) ;
    \draw [shift={(27.62,22.27)}, rotate = 450] [color={rgb, 255:red, 0; green, 0; blue, 0 }  ][line width=0.75]    (10.93,-3.29) .. controls (6.95,-1.4) and (3.31,-0.3) .. (0,0) .. controls (3.31,0.3) and (6.95,1.4) .. (10.93,3.29)   ;
    \draw    (27.62,47.17) -- (18.74,56.22) ;
    \draw [shift={(17.33,57.65)}, rotate = 314.49] [color={rgb, 255:red, 0; green, 0; blue, 0 }  ][line width=0.75]    (10.93,-3.29) .. controls (6.95,-1.4) and (3.31,-0.3) .. (0,0) .. controls (3.31,0.3) and (6.95,1.4) .. (10.93,3.29)   ;
    
    \draw (178.96,140.97) node [anchor=north west][inner sep=0.75pt]  [font=\Large,color={rgb, 255:red, 126; green, 211; blue, 33 }  ,opacity=1 ]  {$\mathcal{I}$};
    \draw (55.61,185.37) node [anchor=north west][inner sep=0.75pt]  [font=\Large,color={rgb, 255:red, 0; green, 0; blue, 0 }  ,opacity=1 ]  {$\mathcal{\textcolor[rgb]{0.13,0.4,0.71}{D}}$};
    \draw (58.99,37.77) node [anchor=north west][inner sep=0.75pt]    {$x$};
    \draw (6.51,52.42) node [anchor=north west][inner sep=0.75pt]    {$y$};
    \draw (22.59,5.02) node [anchor=north west][inner sep=0.75pt]    {$z$};
    \draw (142.44,166.1) node [anchor=north west][inner sep=0.75pt]  [font=\large,color={rgb, 255:red, 0; green, 117; blue, 255 }  ,opacity=1 ] [align=left] {
    \begin{minipage}[lt]{150.35684pt}\setlength\topsep{0pt}
    \begin{center}
        \textcolor[rgb]{0.25,0.46,0.02}{Protohalo mass distribution}
        \end{center}    
        \end{minipage}
    };
    \draw (304.54,68.62) node [anchor=north west][inner sep=0.75pt]    {$\mathbf{a}_{\mathcal{D}}$};
    \draw (108.93,148.8) node [anchor=north west][inner sep=0.75pt]    {$\mathbf{b}_{\mathcal{D}}$};
    \draw (114.71,16.94) node [anchor=north west][inner sep=0.75pt]    {$\mathbf{c}_{\mathcal{D}}$};
    \draw (146.26,54.62) node [anchor=north west][inner sep=0.75pt]    {$\mathbf{c}_{\mathcal{I}}$};
    \draw (126.17,111.74) node [anchor=north west][inner sep=0.75pt]    {$\mathbf{b}_{\mathcal{I}}$};
    \draw (240.52,115.33) node [anchor=north west][inner sep=0.75pt]    {$\mathbf{a}_{\mathcal{I}}$};
    \draw (25,219) node [anchor=north west][inner sep=0.75pt]   [align=left] {
    \begin{minipage}[lt]{151.062pt}\setlength\topsep{0pt}
        \begin{center}
        \textcolor[rgb]{0.2,0.27,0.71}{{\large Local gravitational potential}}
        \end{center}    
        \end{minipage}
    };
    
    \end{tikzpicture}
    \vspace{5pt}
    \caption{Schematic of the triaxial CDM distribution of a model proto-halo (green ellipsoid) in TTT, embedded within an anisotropic triaxial gravitational potential (blue ellipsoid). The equipotential surface is labelled with the deformation tensor $\mathcal{D}$, while the mass distribution is marked with the inertia tensor $\mathcal{I}$. The eigenvectors of $\mathcal{D}$ and $\mathcal{I}$ follow the notation earlier introduced and their 3-dimensional sets are de-rotated by an angle in the $x-y$ plane. As a consequence, $\mathbf{b}_\mathcal{I}\parallel\mathbf{b}_\mathcal{D}\parallel \hat{y}$, intentionally constructed to illustrate the vanishing component of $\tau_\alpha\bot\hat{y}$. The centres of the equipotential ellipsoid and the mass distribution coincide, as do the CoP and the CoM in the first-order TTT set-up.}
    \label{fig:ttttensors}
\end{figure}

The results earlier obtained for the evolution of a CDM background in a EdS universe are useful for the interpretation of equation~(\ref{eq:tttmain}). Given $D(t)=a(t)$ and $a\propto t^{2/3}$, the $a^2 \dot{D}(t)$ factor evaluates as being $\propto t$. In full,

\begin{equation}
\label{eq:ttteds}
    J_\alpha(t) \propto \overbrace{t}^\text{Time dependence} \cdot \underbrace{(\epsilon_{\alpha\beta\gamma} \, \mathcal{I}_{\beta\delta}\,  \mathcal{D}_{\delta\gamma})}_\text{Spatial dependence}.
\end{equation}
Equation~(\ref{eq:ttteds}) is the main expression in first-order TTT and its results are outlined in the following corollary.
\begin{itemize}
    \item \textbf{Time sector}. Independently of the dynamics of the spatial sector in equation~(\ref{eq:ttteds}), the first-order TTT predicts $\mathbf{J}$ to increase linearly with cosmological time, as represented in the top left panel of Fig.~(\ref{fig:tttdiagram}). In this schematic, the time-evolution of $J_\alpha(t)$ is juxtaposed to the evolution of the EdS background in the STHC model, visible in the lower panel. Both these solutions are based on the results from first-order theory and do ignore the highly non-linear nature of the collapse of structures at late times, i.e. $z \ll 1$. The TTT formulation thus provides an accurate picture for the angular momentum growth of linearly-behaving structures, which can either be protohalos in their early stages of collapse ($z \gg 1$) or newly emerged sub-horizon CDM modes at low redshifts.\footnote{Not only is linear theory accurate for describing the formation of CDM protohalos at high redshift, but also it is useful for studying the collapse of large scale modes that have recently entered the horizon. These structures, with a typical size $k^{-1}\sim100\, h^{-1}$ Mpc, have not yet reached the turnaround point in the picture of the STHC model and are just starting to play a role in the hierarchical growth of structure formation, alongside with smaller modes (e.g. cluster-sized halos with $k^{-1}\sim1\, h^{-1}$ Mpc), most of which are already fully collapsed.}
    
    \item \textbf{Spatial sector}. The cross product between $\mathcal{I}$ and $\mathcal{D}$ in equation~(\ref{eq:ttteds}) only acts on the off-diagonal terms of the tensors, i.e. their traceless parts. The gravitation tidal torque is, in fact, greater the more misaligned the eigenvectors of $\mathcal{I}$ and $\mathcal{D}$ are. This can be easily observed by writing the torque $\mathbf{\tau} = d\mathbf{J}/dt$ examining the index-components of $\epsilon_{\alpha\beta\gamma}$. By definition, the Levi-Civita symbol is zero if $\alpha=\beta$ or $\beta=\gamma$ or $\alpha=\gamma$, which implies that the diagonal terms in $\mathcal{I}$ and $\mathcal{D}$ do not contribute to $J_\alpha$. Fig.~(\ref{fig:ttttensors}) visualises the geometry of both tensors in 3-dimensions, with the elongation of the ellipsoids much more amplified that in a realistic scenario. In this particular case, the set of $\mathcal{I}$ eigenvectors is chosen to be aligned with the reference orthogonal axes on the top left corner, while the set of $\mathcal{D}$ eigenvectors is rotated about the $y$ axis by an angle. The misalignment of $\mathbf{a}_{\mathcal{I}, \mathcal{D}}$ and $\mathbf{c}_{\mathcal{I}, \mathcal{D}}$ generates a torque in the direction perpendicular to the $x-z$ plane, while the perfectly aligned $\mathbf{a}_{\mathcal{I}, \mathcal{D}}$ do not produce a component $\tau_\alpha\bot\hat{y}$.
    
    \item \textbf{Null torque}. Two possible scenario exist where no angular momentum is transferred to a CDM overdensity by the local gravitational field. As explained above, the protohalo is not subject to any tidal torque if the eigenvectors of $\mathcal{I}$ are aligned with those of $\mathcal{D}$. Alternatively, if either the local scalar potential or the matter distribution (or both) is spherically symmetric, it follows that $\tau_\alpha=0$. A spherically symmetric field can be described by purely diagonal tensors, whose off-diagonal elements are zero. Given the form of the the Levi-Civita tensor, only traceless parts of $\mathcal{I}$ and $\mathcal{D}$ propagate to $\mathbf{J}$, which implies that if at least one of them has vanishing traceless elements, then  $\tau_\alpha=0$. In order to simplify the manipulation of the tensors in equation~(\ref{eq:ttteds}), several works in the literature \citep[e.g.][]{2002MNRAS.332..325P} define the velocity shear tensor (also known as tidal field tensor) $\mathcal{T}_{\alpha\beta}$ as the traceless deformation tensor and the quadrupolar inertia tensor $\mathcal{I^\prime}_{\alpha\beta}$ as the traceless inertia tensor, i.e.
\begin{align}
     \mathcal{T}_{\alpha\beta} &= \mathcal{D}_{\alpha\beta} - \frac{1}{3}\mathcal{D}_{\alpha\alpha}\, \delta_{\alpha\beta}  \\
     \mathcal{I^\prime}_{\alpha\beta} &= \mathcal{I}_{\alpha\beta} - \frac{1}{3}\mathcal{I}_{\alpha\alpha}\, \delta_{\alpha\beta},
\end{align}
where $\delta_{\alpha\beta}$ is the Kronecker delta symbol.    
\end{itemize}

The assumptions introduced in the derivations of the canonical TTT and the Zeldovich approximation provide a reliable description for the generation of angular momentum in linearly growing structures embedded within an EdS background. To probe the dynamics of tidal torques in the non-linear regimes of structure formation, several extensions of first\hyp{}order TTT have been developed. This section introduces some of them, illustrating the main results which are relevant to the present project work.
\begin{itemize}
    \item \textbf{Higher\hyp{}order perturbative theories}. Equation~(\ref{eq:taylorza}) shows the definition of the Taylor expansion of the displacement vector in the ZA regime. As encountered in the derivation for the first-order TTT, each term in the series introduces an additional perturbative term in the final form of $\mathbf{J}$ \citep[see e.g.][and references therein for an overview]{1996MNRAS.282..436C, 1996MNRAS.282..455C, 1997MNRAS.287..753M, 1997MNRAS.290..439M}. The advantage of including higher-order terms in the $\mathbf{S}$-expansion lies in the possibility to gain accuracy in the predictions at lower redshift in comparison to the canonical TTT formulation. In addition, as shown by \cite{1996MNRAS.282..436C, 1996MNRAS.282..455C}, second-order terms may affect the results of TTT ensemble statistics and correlations between $\mathbf{J}$, $\mathcal{I}$ and $\mathcal{D}$. The mathematical formalism of such correlations could be relevant for the present work and will be explored in later stages in the project.
    
    \item \textbf{Relativistic Zeldovich approximation}. The canonical TTT, as well as most N-body simulations, is developed using a Newtonian approach to the formation of structures. This choice is usually motivated by the relatively shallow gravitational potential wells that diffuse CDM proto\hyp{}halos produce. From a theoretical point of view, however, the full general-relativistic treatment of structure formation can be of great interest, particularly when comparing its predictions with those from the Newtonian limit. In the context of tidal torques, the ZA constitutes a pivotal stage in the formulation of TTT. Similarly, most theories, alternative to the Newtonian limit, aim in the first instance at developing a formulation of the ZA within their paradigm. An example of this procedure, presented as the foundation of a general-relativistic Lagrangian theory of structure formation, can be found in \cite{2012PhRvD..86b3520B}. An in-depth discussion of the relativistic ZA and its consequences does not fall within the scope of this document and will not be included. The relativistic generalisation of TTT based on this formulation has not been developed yet; however, deviations from the Newtonian limit are expected to be negligible and unimportant for most studies of CDM structure formation.
\end{itemize}

\section{Galaxy cluster observables}
\label{sec:cluster-observables}

Cutting edge studies of the astrophysics of galaxy clusters today seldom rely on observations in a single region of the electromagnetic spectrum. Multi-wavelength observations, i.e. combining data from different instruments at different wavelengths, have become a powerful tool for investigating the properties of these structures.

Multi-wavelength observations of galaxy clusters provide a wealth of information, such as the distribution of their baryonic and dark matter, the temperature and density of the ICM, the presence of shocks and turbulence, and the role of cosmic rays, high-energy particles that are accelerated via magneto-hydrodynamic processes. In addition, these observations allow us to study the complex interplay between galaxies, gas, and dark matter in galaxy clusters and to test theoretical models of structure formation and evolution. Some of the most important recent advances in the field of multi-wavelength observations of galaxy clusters include the use of \textit{X-ray}, \textit{optical}, and \textit{radio} observations \citep[see][for a review]{2011ARA&A..49..409A}. In this section, we will present the most recent advances of observations in these three windows of the electromagnetic spectrum.

\subsection{Visible light}
\label{sec:cluster-observables:visible}

Visible light has been the only electromagnetic messenger used for astronomy throughout history until the 20$^{\rm th}$ century. Visible-light observations through photographic plates were used by \cite{1933AcHPh...6..110Z, 1937ApJ....86..217Z} to estimate the mass of a handful of galaxy clusters in two ways:
\begin{itemize}
    \item the spectra of individual galaxies allowed the determination of their radial velocity, which can be used to compute the velocity dispersion $\sigma_v = \sqrt{\langle v^2 \rangle - \langle v \rangle^2}$, where the angle-brackets indicate the mean value of the ensemble. Then, $\sigma_v$ can be used to compute the \textit{virial} mass of the cluster via the virial theorem, $U+2K = 0$, which links the gravitational potential energy $U$ and the kinetic energy $K$ of a system in equilibrium.
    \item photometry on the galaxies can be used to infer the \textit{luminous} mass attributed to stars and diffuse sources within the galaxies.
\end{itemize}
The systematic discrepancy between the virial and luminous mass of galaxy clusters uncovered the first piece of evidence for the existence of dark matter.

Recent galaxy cluster surveys, such as the Sloan Digital Sky Survey (SDSS), adopt more advanced imaging technique and more rigorous analysis methods, but their main objective remains the spectro\hyp{}photometry of the member galaxies \citep[e.g.][]{2002ApJ...571..172Z, 2011ApJ...736...59Z}. In Chapter \ref{chapter:4}, we will refer extensively to the studies by \cite{2017MNRAS.465.2616M} and \cite{2019JCAP...06..001B}, which used radial-velocity information to probe the rotational dynamics of galaxy clusters from SDSS galaxy spectra.

Most galaxy clusters host a bright central galaxy (BCG) which, as the denomination suggests, is the brightest objects located in centre of clusters. Most observed BCGs are red elliptical galaxies, with very little to no star formation ongoing \cite[e.g.][]{10.1093/mnras/stv1271}. The star-formation rate (SFR) and the specific star-formation rate (sSFR) of the BCG were shown to be correlated to the state of the active galactic nucleus (AGN) and the activity of the central SMBH. In fact, both observations and hydrodynamic simulations confirmed the role of AGN feedback in removing the cold gas from the centre of the cluster, ultimately stopping star formation \citep[e.g.][]{2018ApJ...863...62P, 2022MNRAS.tmp.1955N, 2023MNRAS.520.3164A}. The BCG then becomes quenched and its sSFR drops typically below $10^{-2}$ Gyr$^{-1}$. This example illustrates how the properties of the BCG can be useful metrics to probe the physics of the cosmic \textit{baryon cycle} and we will discuss this topic further in Chapters \ref{chapter:5} and \ref{chapter:6}.

In optical bands, galaxies are the brightest components of clusters, however, the intra\hyp{}cluster light (ICL) has rapidly gained interest among researchers. The ICL is a very diffuse, low surface-brightness stellar component that surrounds the BCG and, occasionally, other massive members of galaxy clusters. Although still debated, the formation of the ICL is often associated with merger events, tidal stripping and pre-processing of stellar components from other objects \citep[see][for a review]{2021Galax...9...60C}.

Finally, an additional observational probe, often conducted in visible and infrared bands, but also in radio \citep[e.g.][]{2020MNRAS.495.1706B}, is gravitational lensing, which uses the deflection of background light due to massive objects in the foreground to study the distribution of the total matter (baryonic and dark matter) in galaxy clusters. This method already produced high-quality results and has become of the main targets of the \textit{Euclid} space mission, operated by the European Space Agency and launched on a Falcon 9 on 1 July 2023 \citep{2011arXiv1110.3193L, 2022A&A...662A.112E}.\footnote{Further information about the \textit{Euclid} mission is available at \href{https://www.euclid-ec.org/}{https://www.euclid-ec.org/}.} We will not discuss gravitational weak lensing in this thesis, and we refer to \cite{2020A&ARv..28....7U} for a review of the topic.

\subsection{X-rays}
\label{sec:cluster-observables:xrays}

Clusters emit X\hyp{}rays due to their hot, diffuse intra\hyp{}cluster medium (ICM), which has temperatures in the range of $10^7 - 10^8$ K. The X-ray emission from clusters can be divided into two components: continuum emission and spectral lines.
\begin{itemize}
    \item The continuum emission is due to thermal Bremsstrahlung. This process occurs when charged particles, such as those in the electron plasma of cluster atmospheres, are deflected by the electric fields of other particles. As electrons experience an acceleration, e.g. in the form of a change in the velocity direction, some of the kinetic energy is lost via the emission of high\hyp{}energy photons.

    \item The spectral lines are due to energy transitions in atoms in the ICM, which have been ionized by the high temperature.
\end{itemize}
  
The continuum emission from clusters of galaxies can be described by the equation
\begin{equation}
S_\nu = j_\nu n_e n_i V,
\end{equation}
where $S_\nu$ is the specific intensity of the radiation, $j_\nu$ is the thermal Bremsstrahlung emissivity, $n_e$ and $n_i$ are the electron and ion number densities, respectively, and $V$ is the volume of the emitting region. The thermal Bremsstrahlung emissivity, i.e., the energy emitted per time and volume, is given by
\begin{equation}
j_\nu \propto n_e n_i T^{-1/2} \exp\left(-h\nu/k_{\rm B}T\right),
\end{equation}
where $T$ is the temperature of the gas, $h$ is Planck's constant, and $k_{\rm B}$ is Boltzmann's constant. The continuum emission from clusters is sensitive to the density and temperature of the ICM, as well as the metallicity \citep[see e.g.][for further details]{1986RvMP...58....1S}.

The spectral lines in the X-ray emission from clusters of galaxies are due to transitions between ionization states of heavy elements in the ICM, such as iron and oxygen. The strength and shape of the spectral lines depend on the ionization state of the emitting atoms, as well as the temperature and density of the gas.

Assuming  collisional ionisation equilibrium (CIE, $n_e \approx n_i$) and optically thin cluster atmospheres, the energy released by thermal emission and atomic transitions escapes the system without being re-absorbed. As a consequence of this energy loss, some gas cools down with a characteristic cooling rate, expressed by
\begin{equation}
\dot{E}_{\rm cool} = n_e^2 \Lambda(T,Z),
\end{equation}
where $\Lambda(T,Z)$ is the cooling function, which describes the rate at which the gas loses thermal energy per unit volume. The cooling function depends on the temperature $T$ and metallicity $Z$ of the gas. At low temperatures ($T\lesssim 10^6$ K), the contribution from spectral lines emission dominates the profile of the cooling function, while at high temperatures ($T\gtrsim 10^8$ K) the Bremsstrahlung (continuum) emission dominates over the spectral lines.

In absence of external heating mechanisms, we can define a cooling time in terms of the cooling rate $\dot{E}_{\rm cool}$ and the specific energy $E_{\rm cool} = 3/2\, k_{\rm B} T\, n_e$, as
\begin{equation}
t_{\rm cool} \equiv \frac{E_{\rm cool}}{|\dot{E}_{\rm cool}|} = \frac{3}{2}\, \frac{k_{\rm B}T}{n_e \Lambda(T,Z)}.
\end{equation}
The cooling time can provide a description of the thermodynamic evolution of the gas, in comparison to the Hubble time $t_{H}$ and the $r_{200}$ crossing time $t_{\rm cross}$:\footnote{We define the crossing time $t_{\rm cross} \sim r/\sigma_v \sim \sqrt{r^3/GM}$, evaluated at $r=r_{200}$ as the timescale for a collisionless DM mass-element to complete one orbit around the centre of mass of the cluster at an $r_{200}$ distance.}
\begin{itemize}
    \item for $t_{\rm cool} < t_{\rm cross}$, the gas can cool before one crossing time; as it cools, it becomes denser and falls towards the centre. This process is crucial for accreting material onto SMBHs and their role in the baryon cycle can influence the formation of cool-core clusters.

     \item For $t_{\rm cool} > t_{H}$, the gas cannot cool down within a Hubble time, so that it always remains hot throughout cosmic time. This scenario occurs in the gravitationally\hyp{}heated ICM and, most importantly in the gas which is heated during feedback events, as discussed in \cite{2023MNRAS.520.3164A}.

     \item Intermediate cases where $t_{\rm cross} < t_{\rm cool} < t_{H}$ represent gas which slowly cools down during the evolution of the cluster, forming a \textit{cooling flow}. Similarly to the the case where $t_{\rm cool} < t_{\rm cross}$, cooling flows contribute to the SMBH feeding and the resulting self-regulation of the baryon cycle in the cluster environment.
\end{itemize}

The radius at which $t_{\rm cool} = t_{\rm cross}$ is known as the cooling radius $r_{\rm cool}$. For a cluster atmosphere satisfying the assumptions above, this radius is $r_{\rm cool} \approx 1-3$\% of $r_{200}$ at $z=0$, suggesting that most of the ICM gas is in the hot, X-ray emitting phase for clusters of mass $M_{200}\gtrsim 10^{14}$ M$_\odot$.

In hydrodynamic simulations of realistic clusters, as well as in real ones, the assumptions of CIE, optically thin gas cannot often be neglected. The emissivity of dense, metal-rich gas in the centre of galaxy clusters can heavily depend on self-shielding, the presence of dust and molecular abundances \citep{2020MNRAS.497.4857P}, which are included in the state-of-the-art hydrodynamic simulations presented in later chapters.

Finally, X-ray observations often reconstruct the gas mass distribution under the assumption of (i) spherical symmetry and (ii) hydrostatic equilibrium, described by
\begin{equation}
    \frac{dP}{dr} = - \frac{G M(<r)}{r^2}\, \rho,
    \label{eq:hse-equilibrium-equation}
\end{equation}
where $M(<r)$ indicates the total mass within radius $r$. Most clusters, especially mergers, are characterised by large internal motions, which are not captured by the hydrostatic equilibrium equation \citep[e.g.][]{2021MNRAS.506.2533B}. When integrating the density profile in Eq.~\ref{eq:hse-equilibrium-equation} to derive the cluster mass (e.g. $M(<r_{500})$), the departure from hydrostatic equilibrium introduces a \textit{hydrostatic mass bias}, defined as
\begin{equation}
    b \equiv 1 - \frac{M(<r_{500})}{M_{500}},
\end{equation}
where we label $M_{500}$ as the \textit{true} mass and $M(<r_{500})$ the observed mass. While all observational methods introduce a mass bias, the parameter $b$ can be computed independently from simulations by comparing the \textit{true} halo mass with mock observations of synthetic galaxy clusters.

\subsection{SZ effects}
\label{sec:cluster-observables:sz}

\begin{figure}

	\begin{center}	
	\tikzset{every picture/.style={line width=0.75pt}} 

    \begin{tikzpicture}[x=0.75pt,y=0.75pt,yscale=-.8,xscale=.8]
    
    \draw  [draw opacity=0][line width=1.5]  (31.84,68.8) .. controls (68.78,41.03) and (128.34,23) .. (195.53,23) .. controls (261.72,23) and (320.52,40.5) .. (357.57,67.58) -- (195.53,135.1) -- cycle ; \draw  [line width=1.5]  (31.84,68.8) .. controls (68.78,41.03) and (128.34,23) .. (195.53,23) .. controls (261.72,23) and (320.52,40.5) .. (357.57,67.58) ;
    \draw   (168.5,166.2) .. controls (195.5,164.2) and (225,133.2) .. (243.5,149.2) .. controls (262,165.2) and (257,208.2) .. (240.5,229.2) .. controls (224,250.2) and (136.5,247.2) .. (116.5,217.2) .. controls (96.5,187.2) and (141.5,168.2) .. (168.5,166.2) -- cycle ;
    \draw [color={rgb, 255:red, 255; green, 0; blue, 0 }  ,draw opacity=1 ]   (78,94.2) .. controls (80.36,94.19) and (81.54,95.37) .. (81.54,97.73) .. controls (81.54,100.09) and (82.72,101.27) .. (85.08,101.26) .. controls (87.44,101.25) and (88.62,102.43) .. (88.62,104.79) .. controls (88.63,107.15) and (89.81,108.33) .. (92.17,108.32) .. controls (94.53,108.31) and (95.71,109.49) .. (95.71,111.85) .. controls (95.71,114.21) and (96.89,115.39) .. (99.25,115.38) .. controls (101.61,115.37) and (102.79,116.55) .. (102.79,118.91) .. controls (102.79,121.27) and (103.97,122.45) .. (106.33,122.44) .. controls (108.69,122.43) and (109.87,123.61) .. (109.87,125.97) .. controls (109.87,128.33) and (111.05,129.51) .. (113.41,129.5) .. controls (115.77,129.49) and (116.95,130.67) .. (116.95,133.03) .. controls (116.96,135.39) and (118.14,136.57) .. (120.5,136.56) .. controls (122.86,136.55) and (124.04,137.73) .. (124.04,140.09) -- (127.34,143.38) -- (133.01,149.03) ;
    \draw [shift={(134.43,150.44)}, rotate = 224.91] [color={rgb, 255:red, 255; green, 0; blue, 0 }  ,draw opacity=1 ][line width=0.75]    (10.93,-3.29) .. controls (6.95,-1.4) and (3.31,-0.3) .. (0,0) .. controls (3.31,0.3) and (6.95,1.4) .. (10.93,3.29)   ;
    
    \draw [color={rgb, 255:red, 0; green, 0; blue, 255 }  ,draw opacity=1 ]   (184.5,62.2) .. controls (186.17,63.87) and (186.17,65.53) .. (184.5,67.2) .. controls (182.83,68.87) and (182.83,70.53) .. (184.5,72.2) .. controls (186.17,73.87) and (186.17,75.53) .. (184.5,77.2) .. controls (182.83,78.87) and (182.83,80.53) .. (184.5,82.2) .. controls (186.17,83.87) and (186.17,85.53) .. (184.5,87.2) .. controls (182.83,88.87) and (182.83,90.53) .. (184.5,92.2) .. controls (186.17,93.87) and (186.17,95.53) .. (184.5,97.2) .. controls (182.83,98.87) and (182.83,100.53) .. (184.5,102.2) .. controls (186.17,103.87) and (186.17,105.53) .. (184.5,107.2) .. controls (182.83,108.87) and (182.83,110.53) .. (184.5,112.2) .. controls (186.17,113.87) and (186.17,115.53) .. (184.5,117.2) .. controls (182.83,118.87) and (182.83,120.53) .. (184.5,122.2) -- (184.5,125.2) -- (184.5,133.2) ;
    \draw [shift={(184.5,135.2)}, rotate = 270] [color={rgb, 255:red, 0; green, 0; blue, 255 }  ,draw opacity=1 ][line width=0.75]    (10.93,-3.29) .. controls (6.95,-1.4) and (3.31,-0.3) .. (0,0) .. controls (3.31,0.3) and (6.95,1.4) .. (10.93,3.29)   ;
    
    \draw  [fill={rgb, 255:red, 0; green, 0; blue, 0 }  ,fill opacity=1 ][line width=0.75]  (179.5,199.35) .. controls (179.5,197) and (181.4,195.1) .. (183.75,195.1) .. controls (186.1,195.1) and (188,197) .. (188,199.35) .. controls (188,201.7) and (186.1,203.6) .. (183.75,203.6) .. controls (181.4,203.6) and (179.5,201.7) .. (179.5,199.35) -- cycle ;
    \draw    (183.75,199.1) -- (129.49,204.98) ;
    \draw [shift={(127.5,205.2)}, rotate = 353.81] [color={rgb, 255:red, 0; green, 0; blue, 0 }  ][line width=0.75]    (10.93,-3.29) .. controls (6.95,-1.4) and (3.31,-0.3) .. (0,0) .. controls (3.31,0.3) and (6.95,1.4) .. (10.93,3.29)   ;
    
    \draw [color={rgb, 255:red, 0; green, 0; blue, 255 }  ,draw opacity=1 ]   (154.5,224.2) .. controls (154.17,226.53) and (152.83,227.53) .. (150.5,227.2) .. controls (148.17,226.87) and (146.83,227.88) .. (146.5,230.21) .. controls (146.17,232.54) and (144.84,233.54) .. (142.51,233.21) .. controls (140.18,232.88) and (138.84,233.88) .. (138.51,236.21) .. controls (138.18,238.54) and (136.84,239.55) .. (134.51,239.22) .. controls (132.18,238.89) and (130.84,239.89) .. (130.51,242.22) .. controls (130.18,244.55) and (128.85,245.55) .. (126.52,245.22) .. controls (124.19,244.89) and (122.85,245.9) .. (122.52,248.23) .. controls (122.19,250.56) and (120.85,251.56) .. (118.52,251.23) .. controls (116.19,250.9) and (114.85,251.9) .. (114.52,254.23) .. controls (114.19,256.56) and (112.86,257.57) .. (110.53,257.24) .. controls (108.2,256.91) and (106.86,257.91) .. (106.53,260.24) .. controls (106.2,262.57) and (104.86,263.57) .. (102.53,263.24) .. controls (100.2,262.91) and (98.86,263.91) .. (98.53,266.24) -- (95.62,268.43) -- (89.22,273.24) ;
    \draw [shift={(87.63,274.44)}, rotate = 323.08000000000004] [color={rgb, 255:red, 0; green, 0; blue, 255 }  ,draw opacity=1 ][line width=0.75]    (10.93,-3.29) .. controls (6.95,-1.4) and (3.31,-0.3) .. (0,0) .. controls (3.31,0.3) and (6.95,1.4) .. (10.93,3.29)   ;
    
    \draw [color={rgb, 255:red, 255; green, 0; blue, 0 }  ,draw opacity=1 ]   (184,230.2) .. controls (185.67,231.87) and (185.67,233.53) .. (184,235.2) .. controls (182.33,236.87) and (182.33,238.53) .. (184,240.2) .. controls (185.67,241.87) and (185.67,243.53) .. (184,245.2) .. controls (182.33,246.87) and (182.33,248.53) .. (184,250.2) .. controls (185.67,251.87) and (185.67,253.53) .. (184,255.2) .. controls (182.33,256.87) and (182.33,258.53) .. (184,260.2) .. controls (185.67,261.87) and (185.67,263.53) .. (184,265.2) .. controls (182.33,266.87) and (182.33,268.53) .. (184,270.2) .. controls (185.67,271.87) and (185.67,273.53) .. (184,275.2) .. controls (182.33,276.87) and (182.33,278.53) .. (184,280.2) .. controls (185.67,281.87) and (185.67,283.53) .. (184,285.2) .. controls (182.33,286.87) and (182.33,288.53) .. (184,290.2) .. controls (185.67,291.87) and (185.67,293.53) .. (184,295.2) -- (184,303.2) ;
    \draw [shift={(184,305.2)}, rotate = 270] [color={rgb, 255:red, 255; green, 0; blue, 0 }  ,draw opacity=1 ][line width=0.75]    (10.93,-3.29) .. controls (6.95,-1.4) and (3.31,-0.3) .. (0,0) .. controls (3.31,0.3) and (6.95,1.4) .. (10.93,3.29)   ;
    
    \draw  [dash pattern={on 0.84pt off 2.51pt}]  (132.5,149.35) -- (179.5,196.35) ;

    \draw  [dash pattern={on 0.84pt off 2.51pt}]  (183.75,195.1) -- (184.5,135.2) ;

    \draw  [dash pattern={on 0.84pt off 2.51pt}]  (183.75,330.2) -- (183.75,203.6) ;

    \draw  [draw opacity=0] (249.43,358.07) .. controls (235.64,376.78) and (211.38,389.2) .. (183.75,389.2) .. controls (156.75,389.2) and (132.96,377.34) .. (119.02,359.33) -- (183.75,322.2) -- cycle ; \draw   (249.43,358.07) .. controls (235.64,376.78) and (211.38,389.2) .. (183.75,389.2) .. controls (156.75,389.2) and (132.96,377.34) .. (119.02,359.33) ;
    \draw   (117.17,356.84) .. controls (117.17,350.39) and (146.78,345.16) .. (183.3,345.16) .. controls (219.82,345.16) and (249.43,350.39) .. (249.43,356.84) .. controls (249.43,363.29) and (219.82,368.52) .. (183.3,368.52) .. controls (146.78,368.52) and (117.17,363.29) .. (117.17,356.84) -- cycle ;
    \draw [line width=1.5]    (183.3,331.2) -- (183.3,368.52) ;
    
    \draw [shift={(183.3,331.2)}, rotate = 90] [color={rgb, 255:red, 0; green, 0; blue, 0 }  ][line width=1.5]      (6.71,-6.71) .. controls (3.01,-6.71) and (0,-3.7) .. (0,0) .. controls (0,3.7) and (3.01,6.71) .. (6.71,6.71) ;
    \draw  [dash pattern={on 0.84pt off 2.51pt}]  (183.5,199.35) -- (154.5,224.2) ;

    \draw    (269.5,131.2) .. controls (258.83,130.23) and (254.75,129.26) .. (237.17,139.24) ;
    \draw [shift={(235.5,140.2)}, rotate = 329.93] [color={rgb, 255:red, 0; green, 0; blue, 0 }  ][line width=0.75]    (10.93,-3.29) .. controls (6.95,-1.4) and (3.31,-0.3) .. (0,0) .. controls (3.31,0.3) and (6.95,1.4) .. (10.93,3.29)   ;
    
    \draw    (238.5,314.2) .. controls (227.83,313.23) and (218.1,317.91) .. (200.19,328.23) ;
    \draw [shift={(198.5,329.2)}, rotate = 329.93] [color={rgb, 255:red, 0; green, 0; blue, 0 }  ][line width=0.75]    (10.93,-3.29) .. controls (6.95,-1.4) and (3.31,-0.3) .. (0,0) .. controls (3.31,0.3) and (6.95,1.4) .. (10.93,3.29)   ;
    
    \draw    (103.5,70.2) .. controls (107.5,88.2) and (111.5,86.2) .. (92.5,108.2) ;
    \draw [shift={(92.5,108.2)}, rotate = 130.82] [color={rgb, 255:red, 0; green, 0; blue, 0 }  ][fill={rgb, 255:red, 0; green, 0; blue, 0 }  ][line width=0.75]      (0, 0) circle [x radius= 3.35, y radius= 3.35]   ;
    
    \draw    (142.5,58.2) .. controls (146.5,76.2) and (163.5,72.2) .. (184.5,72.2) ;
    \draw [shift={(184.5,72.2)}, rotate = 0] [color={rgb, 255:red, 0; green, 0; blue, 0 }  ][fill={rgb, 255:red, 0; green, 0; blue, 0 }  ][line width=0.75]      (0, 0) circle [x radius= 3.35, y radius= 3.35]   ;
    
    \draw    (106.5,312.2) .. controls (125.5,300.2) and (132.5,269.2) .. (114.5,254.2) ;
    \draw [shift={(114.5,254.2)}, rotate = 219.81] [color={rgb, 255:red, 0; green, 0; blue, 0 }  ][fill={rgb, 255:red, 0; green, 0; blue, 0 }  ][line width=0.75]      (0, 0) circle [x radius= 3.35, y radius= 3.35]   ;
    
    \draw    (106.5,312.2) .. controls (125.5,300.2) and (158.5,265.2) .. (184,267.7) ;
    \draw [shift={(184,267.7)}, rotate = 5.6] [color={rgb, 255:red, 0; green, 0; blue, 0 }  ][fill={rgb, 255:red, 0; green, 0; blue, 0 }  ][line width=0.75]      (0, 0) circle [x radius= 3.35, y radius= 3.35]   ;

    \draw (115,54.2) node [scale=1,rotate=-342.43] [align=left] {CMB photons};
    \draw (332,135) node  [align=left] {Hot electron gas};
    \draw (103,137) node [color={rgb, 255:red, 255; green, 0; blue, 0 }  ,opacity=1 ]  {$1$};
    \draw (171,121) node [color={rgb, 255:red, 0; green, 0; blue, 255 }  ,opacity=1 ]  {$2$};
    \draw (199,294) node [color={rgb, 255:red, 255; green, 0; blue, 0 }  ,opacity=1 ]  {$1'$};
    \draw (81,260) node [color={rgb, 255:red, 0; green, 0; blue, 255 }  ,opacity=1 ]  {$2'$};
    \draw (152,191) node   {$\mathbf{v}$};
    \draw (201,187) node   {$\mathrm{e^{-}}$};
    \draw (312,313) node  [align=left] {Microwave detector};
    \draw (65,333) node  [align=left] {Comptonised\\photons};
    \draw (331,21) node   {$z\ \approx 1100$};
    \draw (321,161) node   {$z\ \sim 0.5$};
    \draw (321,364) node   {$z\ =0$};

    \end{tikzpicture}
    \end{center}
 
	\caption{Schematic of the inverse Compton scattering of two CMB photons, labelled 1 and 2, off electrons in a hot ionised gas. The velocity $\mathbf{v}$ of the electrons is assumed to follow a non-relativistic Maxwell-Boltzmann distribution. Although being previously aligned with the observer's line of sight, as a consequence of the scattering photon 2 is deflected into 2', while being replaced by photon 1', which had entered the IGM from a different angle. The process illustrated in this figure does not involve the simultaneous interaction of two photons with an electron. Double Comptonisation is not discussed in this work, but further details can be found in \cite{1976tper.book.....J, 1979rpa..book.....R}, and \cite{2018Ap&SS.363..224F}.}
	\label{fig:szphotons}
\end{figure}
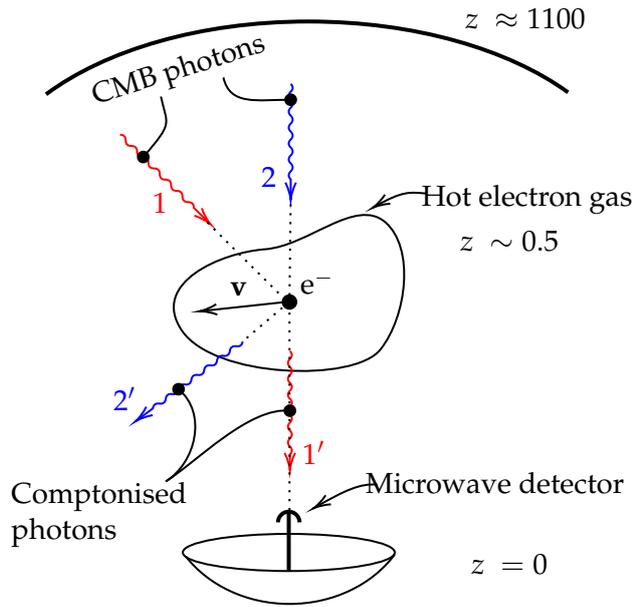

The third observational probe for galaxy clusters discussed in this chapter is through the CMB emission, whose spectrum is distorted by clusters in the foreground via the Sunyaev-Zeldovich (SZ) effect.

The free electrons in the ICM, having energies of order $\sim 10$ keV for the most massive clusters, produce processes of inverse Compton scattering with CMB photons, as well as the diffuse free-free X-ray emission discussed in Section~\ref{sec:cluster-observables:xrays} \citep{2009MNRAS.400..705P, 2011ARA&A..49..409A}. The average of the inverse Compton scattering processes in the ICM, giving rise to the SZ effect, can be observed in microwave bands as a deficit of low-energy CMB photons, which are scattered towards higher energies as a result of the momentum transfer from the electrons (see Fig.~\ref{fig:szphotons}).

By applying the equation for inverse Compton scattering to a cloud of hot electrons described by a Maxwell-Boltzmann distribution (in the non-relativistic limit) and using a generalised form of the \citeauthor{1956ZhETF..31..876K} equations (\citeyear{1956ZhETF..31..876K}), \cite{1970Ap&SS...7...20S} derived for the first time the distortion $\Delta I_{v}$ in the spectrum of the 
\begin{figure}
	\centering	
	\includegraphics[width=0.7\textwidth]{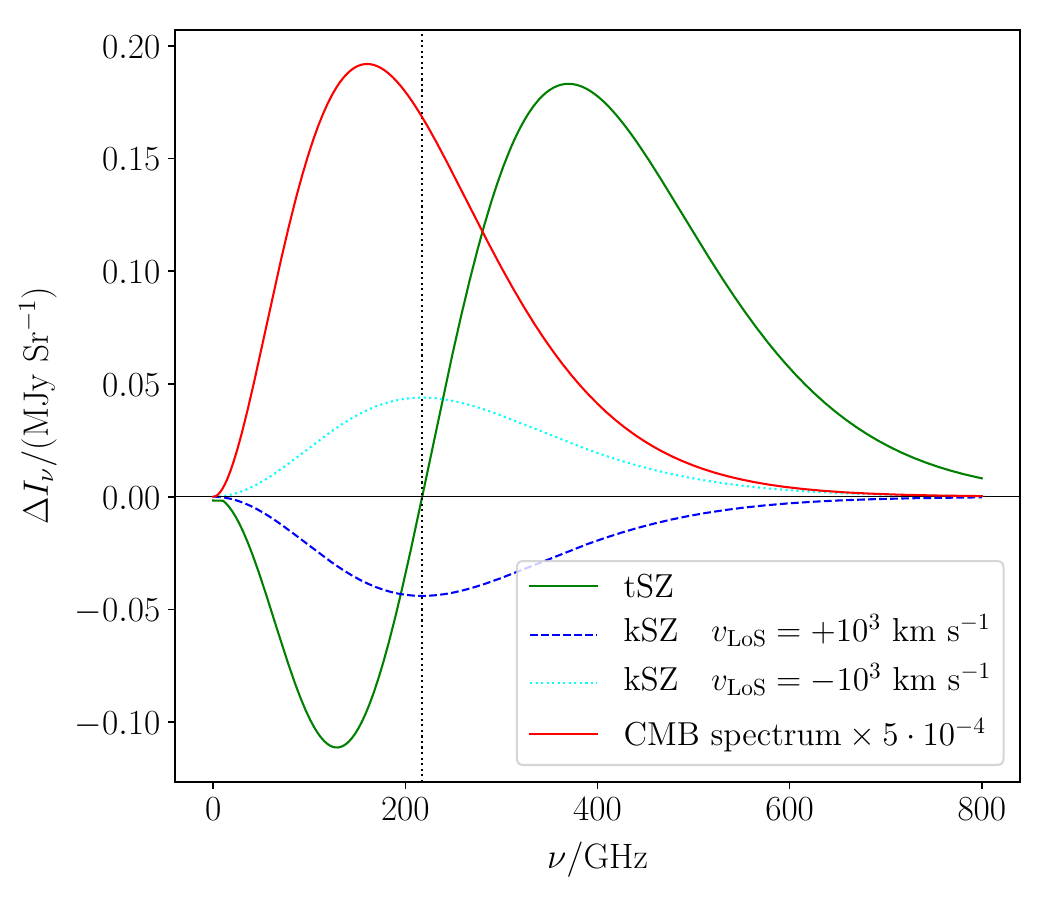}
	\caption{Plot displaying the comparison of the spectral distortions for the tSZ (green solid curve) and kSZ (blue dashed and dotted lines) against the spectrum of the CMB down-scaled by a factor $5 \times 10^{-4}$ (red solid curve). The tSZ spectrum was computed using $y_{\mathrm{tSZ}} = 10^{-4}$. The kSZ instead was simulated with $y_{\mathrm{kSZ}} = 10^{-5}$ for an object moving with velocity along the line of sight (LoS) $v_{\mathrm{LoS}} = \pm 10^3 \mathrm{km\ s^{-1}}$ (positive is defined as towards the observer). The vertical dotted line marks the turn-over of the tSZ at $\nu \approx 217$ GHz, where the tSZ deviation from the CMB spectrum is expected to vanish.}
	\label{fig:szspec}
\end{figure}
CMB caused by the energy transfer of its photons during Comptonisation. This spectral signature, known as thermal SZ (tSZ) effect, is predicted to cause a decrement in photons below $\nu \approx 217$ GHz, complemented by an increment in photon count above this threshold. The shape of tSZ distortion is displayed by the green solid line in figure \ref{fig:szspec} and corresponds to the equation
\begin{equation}
	\Delta I_{v} \approx I_{0} y\, \frac{x^{4} e^x}{\left(e^x-1\right)^{2}}\left(x\,  \frac{e^x+1}{e^x-1}-4\right) \equiv I_{0}\, y\, g(x),
\label{eq:tsz1}
\end{equation}
where
\begin{equation}
	I_0 = 2\frac{(k_{\mathrm{B}} T_{\mathrm{CMB}})^3}{(hc)^2}=270.33\left[\frac{T_{\mathrm{CMB}}}{2.7255\ \mathrm{K}}\right]^{3}\ \mathrm{MJy} \ \mathrm{sr^{-1}}
\label{eq:tsz2}
\end{equation}
\noindent and $x$ is the dimensionless frequency parameter
\begin{equation}
		x \equiv \frac{h\nu}{k_{\mathrm{B}} T_{\mathrm{CMB}}},
\label{eq:tsz3}
\end{equation}
\noindent with $T_{\mathrm{CMB}}$ the CMB temperature, $k_{\mathrm{B}}$ the Boltzmann constant and $y_{\mathrm{tSZ}}$ the tSZ Compton parameter
\begin{align}
y_{\mathrm{tSZ}} &\equiv \int_{\mathrm{LoS}} \frac{k_{\mathrm{B}} T_{\mathrm{e}}}{m_{\mathrm{e}} c^{2}}\, \mathrm{d} \tau_{\mathrm{e}} \label{eq:tsz4}\\ \vspace{2mm}
&=\int_{\mathrm{LoS}} \frac{k_{\mathrm{B}} T_{\mathrm{e}}}{m_{\mathrm{e}} c^{2}} n_{\mathrm{e}} \sigma_{\mathrm{T}}\, \mathrm{d} l \label{eq:tsz5}\\ \vspace{2mm}
&=\frac{\sigma_{\mathrm{T}}}{m_{\mathrm{e}} c^{2}} \int_{\mathrm{LoS}} P_{\mathrm{e}}\, \mathrm{d} l\ . \label{eq:tsz6}
\end{align}

The equations above show that $y_{\mathrm{tSZ}}$ can be recast in terms of the IGM optical depth $\tau_{\mathrm{e}}$ or by integrating the (classical) electron gas pressure $P_{\mathrm{e}}$ of the line elements $\mathrm{d} l$ along the direction of observation, or alternatively by considering the electron temperature $T_{\mathrm{e}}$, the electron number density $n_{\mathrm{e}}$ and the Stefan–Boltzmann constant $\sigma_{\mathrm{T}}$ \citep{1969Ap&SS...4..301Z, 1969Natur.223..721S, 1970CoASP...2...66S, 1970Ap&SS...7....3S, 1970Ap&SS...7...20S, 1972CoASP...4..173S, 1999PhR...310...97B, 2019SSRv..215...17M}.

The largest contribution to the SZ effect in clusters arises from the thermal motion of the free electrons. However, the galaxy cluster may also exhibit a \textit{bulk traslational motion} in the rest frame of the CMB, which introduces an additional constant velocity term $\boldsymbol{\beta}_{\mathrm{p}}$ in the inverse Compton scattering calculations \citep[see e.g.][]{1980ARA&A..18..537S}.
This scenario now includes an additional monopole component to the overall SZ effect, known as \textit{kinetic} SZ (kSZ) distortion. Similarly to $y_{\mathrm{tSZ}}$, the derivation for the kSZ effect leads to a frequency-dependent distortion given by
 \begin{equation}
\Delta I_{v} \approx I_{0}\, \frac{x^{4}\, \mathrm{e}^{x}}{\left(\mathrm{x}^{2}-1\right)^{2}}\, y_{\mathrm{ksZ}}
\end{equation}
\noindent and a $y_{\mathrm{kSZ}}$ parameter
\begin{equation}
y_{\mathrm{ksz}} \equiv \int_{\mathrm{LoS}} \sigma_{\mathrm{T}} n_{\mathrm{e}} \boldsymbol{n} \cdot \boldsymbol{\beta}_{\mathrm{p}} \mathrm{d} l=\int_{\mathrm{LoS}} \boldsymbol{n} \cdot \boldsymbol{\beta}_{\mathrm{p}} \mathrm{d} \tau_{\mathrm{e}},
\label{eq:kszdefinition}
\end{equation}
\noindent which is sensitive to the component of $\boldsymbol{\beta}_{\mathrm{p}}$ along the line of sight (LoS), i.e. $\boldsymbol{n} \cdot \boldsymbol{\beta}_{\mathrm{p}}$ \citep{1980ARA&A..18..537S, 1972CoASP...4..173S, 1999PhR...310...97B, 2013MNRAS.432.3508R, 2017A&A...598A.115A, 2019SSRv..215...17M}.

Unlike the tSZ effect, the kSZ signature is direction-dependent and can lead either to a deficit or a boost in CMB photons, according to the sign of peculiar velocity of the gas along the LoS (see figure \ref{fig:szspec}). From an observational standpoint, both the tSZ and kSZ distortions are measured as fluctuations $\Delta T_{\mathrm{CMB}}$ in the temperature of the CMB caused by the foreground galaxy cluster, which can be expressed as 
\begin{equation}
\frac{\Delta T_{\mathrm{CMB}}}{T_{\mathrm{CMB}}} \approx y_{\mathrm{tSZ}}\Bigg(x\,\underbrace{\frac{\mathrm{e}^{x}+1}{\mathrm{e}^{x}-1}}_{\equiv \, \coth(x/2)}-4\Bigg)=y_{\mathrm{tSZ}}\ f(x)
\end{equation}
for the tSZ distortion and
\begin{equation}
\frac{\Delta T_{\mathrm{CMB}}}{T_{\mathrm{CMB}}} \approx - y_{\mathrm{kSZ}}
\end{equation}
for the kSZ signal.

In addition to the traslational kSZ signal, the \textit{bulk rotation} of galaxy clusters may also contribute as a dipole-like signature in their $y_{\mathrm{ksz}}$ maps, as displayed in Chapter \ref{chapter:3}. This effect, known as \textit{rotational} kSZ (rkSZ), was first predicted by \cite{2002A&A...396..419C, 2002ApJ...573...43C}, who estimated a maximum $y_{\mathrm{ksz}}$ contribution of $\sim 10^{-6}$. Further analyses have been recently conducted by \cite{2017MNRAS.465.2616M, 2018ApJ...869..124S, 2019MNRAS.485.3909L, 2019arXiv190904690Z} and in particular by \cite{ 2017MNRAS.465.2584B, 2018mgm..confE...8B, 2018MNRAS.479.4028B, 2019JPhCS1226a2003B, 2020arXiv200101739M} using hydrodynamic simulations.

Over the last two decades, several techniques have been developed to disentangle the microwave components of the tSZ and kSZ effects from the CMB and other contaminating foregrounds \citep{2004A&A...424..389C, 2007astro.ph..3541S, 2012MNRAS.426..510C, 2013JCAP...03..012S, 2017A&A...598A.115A}. However, the extremely weak kSZ signal from cluster rotation poses the difficulty of detecting a $\Delta T_{\mathrm{rkSZ}}$ variation in CMB temperature which is $\approx 3\ \mu\mathrm{K}$ in the most favourable scenario \citep[rotation axis perpendicular to the LoS,][]{CC02}. \cite{2019JCAP...06..001B} have used clusters identified in the Sloan Digital Sky Survey \citep[SDSS][]{2014ApJS..211...17A} to infer the orientation of their rotation axes from spectroscopic observations of galaxies \citep[the detailed procedure is reported by][]{2017MNRAS.465.2616M}. Subsequently, the positions of the SDSS clusters were mapped onto the Planck data-set and the resulting cut-outs have been stacked after aligning the projected rotation axes. The use of optical observations as priors for stacking and maximising the rkSZ signal allowed \cite{2019JCAP...06..001B} to claim a $2\, \sigma$ detection of galaxy clusters rotation. Using the galaxy clusters in the MACSIS simulations \citep{macsis_barnes_2017}, \cite{2023arXiv230207936A} showed that the rkSZ contribution can reach values $\approx 30-50$ times larger than initially estimated by \cite{2002ApJ...573...43C}. We will investigate the sensitivity of the kSZ signal from cluster rotation to the kinematics of the galaxies and other cluster properties in Chapter \ref{chapter:4}.


%% file: Chapters/Chapter3.tex
\thispagestyle{plain}
\begin{spacing}{1.}
\topskip0pt
\vspace*{\fill}
\begin{flushright}
\begin{minipage}{0.61\textwidth}   
\begin{verbatim}
Galaxy clusters, simulations in view,
A tool to explore, the Universe anew,
From dark matter to gas and light,
Intricate interactions, we must get right.

Hydrodynamics, the physics of flow,
Recreating cosmos, with fidelity, we sow,
Virtual galaxies, we bring to life,
Their dynamics, evolution, in virtual strife.

The challenge is great, no easy feat,
Simulating reality, the laws we must meet,
But with each iteration, we push ahead,
Advancing cosmology, our curiosity fed.

Galaxy clusters, a window to the past,
A wealth of knowledge, we seek to amass,
In simulations we see, their hidden ways,
Revealing secrets, that forever stays.
\end{verbatim}
\vspace{-2pt}
\par\noindent\rule{\textwidth}{0.5pt}
\vspace{-20pt}
\begin{flushright}
Chat-GPT by OpenAI\\
On cosmological hydrodynamic simulations
\end{flushright}
\end{minipage}
\end{flushright}
\vspace*{\fill}
\end{spacing}

\chapter{Numerical simulations}
\label{chapter:3}


\section{Simulation pipeline}
\label{sec:simulation-pipeline}

The goal of most cosmological simulations is to produce synthetic digital copies of our Universe which are (i) \textit{representative} of the vast population of astrophysical objects we can observe and (ii) \textit{physically motivated}, meaning that the calculations must be based on known physical processes, such as star formation and black hole accretion, or fundamental laws of physics, such as gravity or thermodynamics. The EAGLE-type simulations were designed using these conceptual guidelines. However, the journey to achieving a suitable agreement with the observational findings is challenging, and establishing what defines a match \textit{suitable} can also hide difficulties and caveats.

\begin{figure}
    \centering
    \includegraphics[width=\textwidth]{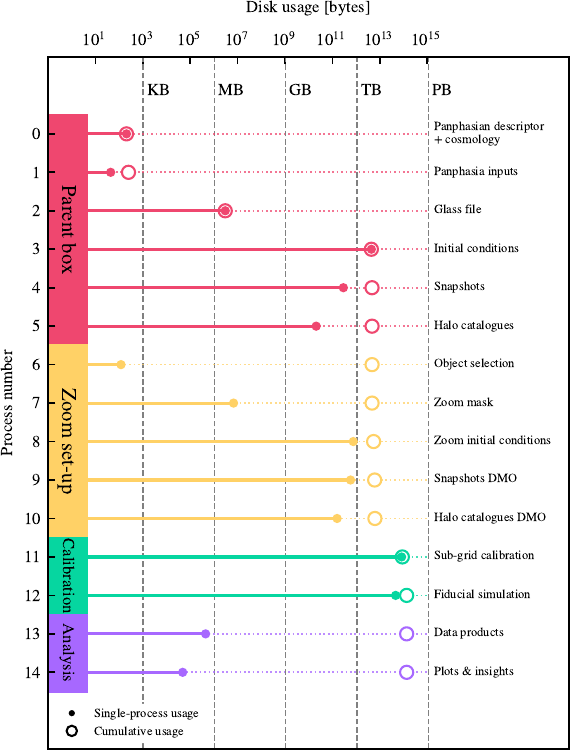}
    \caption{Disk usage at different stages of the zoom-in simulation project in chapters \ref{chapter:5} and \ref{chapter:6}. The usage, in bytes, is shown in the $x$-axis and the processes are numbered sequentially, as indicated in the $y$-axis. The pipeline starts at the top (0) and ends at bottom (14). We divide the simulation steps into four categories: parent simulation (red), the object selection and zoom-in simulation set-up (yellow), the simulation model calibration (green) and the final analysis (purple). The filled markers indicate the disk usage for each specific process, and the empty markers are the cumulative usage. The vertical lines indicate common file sizes used by the human-readable format.}
    \label{fig:simulation-pipeline}
\end{figure}

This chapter provides an overview of the computational pipeline used to generate modern zoom-in simulations of groups and clusters of galaxies. Before describing each step individually, we discuss the logic behind the workflow used produce the simulations in chapters \ref{chapter:5} and \ref{chapter:6} \citep[see also][]{2023MNRAS.520.3164A}. To simulate groups and clusters with high resolution, we use the zoom-in technique \citep{1993ApJ...412..455K, 1997MNRAS.286..865T}, which allows to identify individual objects in a low-resolution volume, increase the the resolution selectively in the region of interest and re-simulate the high-resolution region. Fig.~\ref{fig:simulation-pipeline} shows a summary of the pipeline, starting from the generations of the initial conditions used to generate the low-resolution (pathfinder) volume, down to the production of scientific figures and insights published in journals.

In the workflow, we identify 14 steps, grouped in four categories:
\begin{itemize}
    \item generating the \textit{parent box} from a set of integers as parameters, a random seed, and cosmological parameters (red);
    
    \item selecting the objects of interest, masking and \textit{re-simulating} the high-resolution \textit{zoom} regions (yellow);
    
    \item iterating over \textit{calibration} cycles to tune the simulation model with full physics to match observations or other metrics until a \textit{fiducial model} is found (green);
    
    \item performing the final analysis on the data products and producing publication-ready figures and scientific insights (purple).    
\end{itemize}
We will describe each step individually in the following sections of this chapter.

The only input required to produce (or reproduce) a unique cosmological simulation is a limited set of integers and floats, specifying seed and phases of the Gaussian random field seeding the large-scale structures and the cosmological parameters (see Section \ref{sec:methods:initial-conditions}). These data can be stored in a space of $\approx 100$ bytes and they are used to generate the initial state of the simulation at $z=127$, which requires $\approx 1$ TB of disk storage. The selection of zooms, based on a few tens of integer indexes from a halo catalogue ($\sim 1$ KB), can once again increase to $>1$ TB once the simulation code produced dark-matter only (DMO) snapshots for each object, with their respective new halo catalogues. The inclusion of full physics and the necessity to calibrate the free parameters, e.g. the AGN thermal feedback temperature, leads to the production of tens more realisation of each simulation, totalling to $\approx 300$ TB in our case. Finally, only one of these models is usually selected as \textit{fiducial}. Then, the subset of data from this simulation contracts into $\sim 1$ MB database and the final plot with the insights for scientific publications is usually $\sim 100$ KB.

Throughout the pipeline the data-set sizes vary dramatically from less than 1 KB to hundreds of TB, and then back to under 1 MB. The increase in size reflects the attempt to reproduce the complexity of the formation of cosmic structures and their internal processes numerically, while the down-sizing occurs in the attempt to distill a limit, manageable set of values from a complex problem. This process requires large amounts of computational and human resources to produce an small-sized output from a small-sized input. The result, however, encodes the new insights driving the scientific progress.

\section{Initial conditions}
\label{sec:methods:initial-conditions}

Similarly to solving any ordinary differential equation, hydrodynamic simulations require an initial state which represents the Universe at early times. While we could begin the simulation with direct numerical integration (DNI) methods from arbitrarily early times, most projects use the Zeldovich approximation and higher-order Lagrangian perturbation theory (LPT) to compute the preliminary displacement of the particles. This procedure, referred to as the derivation of the \textit{pre-initial conditions}, starts from a \textit{glass file}, which contains the coordinates a uniform distribution of particles, and has the aim of perturbing that uniform lattice from early times ($z\sim 10^3$) to $z\sim 100$. Then, the DNI simulation would start from $z_{\rm start} \lesssim 100$. The advantage of this strategy is to reduce: 
\begin{itemize}
    \item \textbf{Computational cost}. Because DNI simulations are resource-intensive, we should attempt to evolve our synthetic universe for as long as possible using the computationally cheaper LPT instead of DNI. The initial conditions are then produced to represent the state of the Universe as much into the non-linear regime as possible. In terms of redshift, the start redshift for the simulations, $z_{\rm start}$, should be made as small as LPT allows without significant errors \citep[see][and references therein]{2021MNRAS.500..663M}.

    \item \textbf{Discretisation noise}. Pre-initial conditions approximate a smooth distribution of matter with a set of disrcete particles. The particle-particle interactions (or, equivalently, the lattice self-interactions) can cause random noise which manifests as spurious decaying modes. As shown in the ODE solution of Section \ref{eq:sthc-perturbations-ode}, decaying modes are large at early times and can propagate through the DNI simulation if $z_{\rm start}$ is large \citep[see][and references therein]{2021MNRAS.500..663M}.
\end{itemize}

We discussed in Section \ref{sec:zeldovich-approximation} how perturbation theory can accurately describe the collapse of structures after the perturbations enter the non-linear regime. The Zeldovich approximation is a formulation of first-order Lagrangian perturbation theory (1LPT), which is accurate down to $z\approx 127$ and is often used to evolve an ensemble pf particles analytically before starting the DNI, saving a large number of resource-intensive integration time-steps.

\begin{figure}
\centering
\begin{minipage}{0.8\textwidth}   
\tikzset{every picture/.style={line width=0.75pt}} 

\begin{tikzpicture}[x=0.85pt,y=0.85pt,yscale=-1,xscale=0.8]
\centering

\draw    (206,154) -- (291.67,154) ;
\draw [shift={(293.67,154)}, rotate = 180] [color={rgb, 255:red, 0; green, 0; blue, 0 }  ][line width=0.75]    (10.93,-3.29) .. controls (6.95,-1.4) and (3.31,-0.3) .. (0,0) .. controls (3.31,0.3) and (6.95,1.4) .. (10.93,3.29)   ;
\draw    (206,185.67) -- (335.67,185.67) ;
\draw [shift={(337.67,185.67)}, rotate = 180] [color={rgb, 255:red, 0; green, 0; blue, 0 }  ][line width=0.75]    (10.93,-3.29) .. controls (6.95,-1.4) and (3.31,-0.3) .. (0,0) .. controls (3.31,0.3) and (6.95,1.4) .. (10.93,3.29)   ;
\draw    (206,217.33) -- (389.67,217.33) ;
\draw [shift={(391.67,217.33)}, rotate = 180] [color={rgb, 255:red, 0; green, 0; blue, 0 }  ][line width=0.75]    (10.93,-3.29) .. controls (6.95,-1.4) and (3.31,-0.3) .. (0,0) .. controls (3.31,0.3) and (6.95,1.4) .. (10.93,3.29)   ;
\draw    (206,249) -- (366.33,249) ;
\draw [shift={(368.33,249)}, rotate = 180] [color={rgb, 255:red, 0; green, 0; blue, 0 }  ][line width=0.75]    (10.93,-3.29) .. controls (6.95,-1.4) and (3.31,-0.3) .. (0,0) .. controls (3.31,0.3) and (6.95,1.4) .. (10.93,3.29)   ;
\draw   (40,143) -- (153,143) -- (153,271.67) -- (40,271.67) -- cycle ;
\draw   (471.33,134.33) -- (584.33,134.33) -- (584.33,263) -- (471.33,263) -- cycle ;

\draw (46,187.33) node [anchor=north west][inner sep=0.75pt]   [align=center] {Linear theory\\$z\sim 10^3$};
\draw (160,147) node [anchor=north west][inner sep=0.75pt]   [align=left] {1LPT};
\draw (160,178.11) node [anchor=north west][inner sep=0.75pt]   [align=left] {2LPT};
\draw (160,209.22) node [anchor=north west][inner sep=0.75pt]   [align=left] {3LPT};
\draw (160,240.33) node [anchor=north west][inner sep=0.75pt]   [align=left] {3LPT*};
\draw (195,270) node [anchor=north west][inner sep=0.75pt]   [align=center] {$\vdots$\\$n$LPT with $n>3$ is\\less efficient than DNI};
\draw (309.83,146.57) node [anchor=north west][inner sep=0.75pt]    {$z\approx 127~^\dagger$};
\draw (376.5,239.9) node [anchor=north west][inner sep=0.75pt]    {$z\approx 24~^\ddagger$};
\draw (401.17,209.23) node [anchor=north west][inner sep=0.75pt]    {$z\approx 12$};
\draw (349.17,178.57) node [anchor=north west][inner sep=0.75pt]    {$z\approx 50$};
\draw (485,169.67) node [anchor=north west][inner sep=0.75pt]   [align=center] {Direct\\Numerical\\Integration};

\end{tikzpicture}
\end{minipage}
    \caption{Timeline of the generation of high\hyp{}order $n$LPT for generating initial conditions with sub-percent errors in the matter power spectrum using different $\mathcal{O}(n)$ in $n$LPT. The linear theory, on the left, can describe the growth of structures only at high redshift, while (1-3)LPT can capture the non\hyp{}linear formation of structures down to $z\approx 12$ for a single\hyp{}fluid universe, e.g. matter-dominated EdS universe. Beyond this redshift, the DNI simulations are started and the initial conditions are evolved to $z=0$. \textsuperscript{*}3LPT formulation for a 2-fluid universe including baryons and CDM \citep[e.g. the \textsc{MUSIC2\hyp{}monofonIC} code,][]{2021MNRAS.503..426H}. $^\dagger$The starting redshift we use for the simulations throughout this thesis. We use the \textsc{IC\_Gen} code by \cite{2010MNRAS.403.1859J}. $^\ddagger$The recommended starting redshift for 2-fluid 3LPT initial conditions in \textsc{MUSIC2\hyp{}monofonIC}.}
    \label{fig:ics-process}
\end{figure}
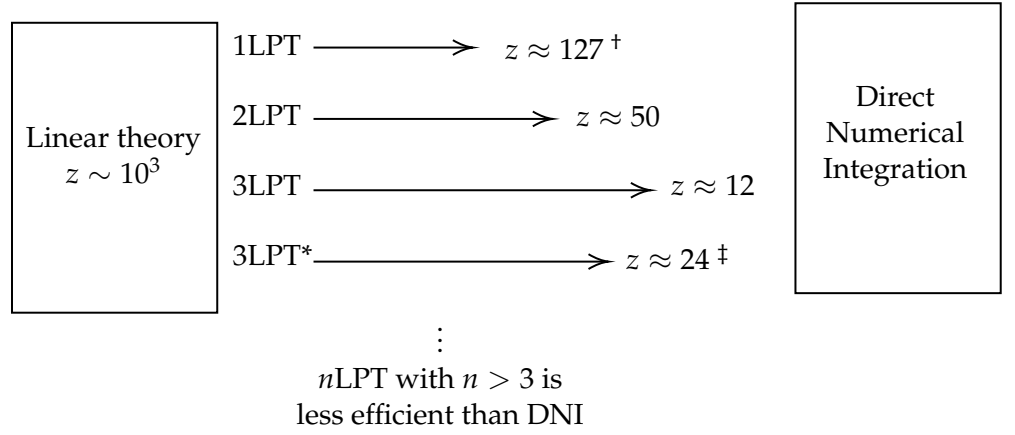

In principle, extending the 1LPT formalism to higher orders ($n$LPT) can capture the non-linear collapse with enough precision to even later times, which would reduce the computational footprint of DNI even further. Parallel codes with these capabilities have been developed by \cite{2006MNRAS.373..369C}, who compared the evolution of the matter power spectrum from initial conditions generated with linear theory, the ZA (i.e. 1LPT) and finally including second-order terms (2LPT) with significantly improved results. The later implementation of \cite{2010MNRAS.403.1859J}, used in these projects, found that the 1LPT could be used to generate initial conditions as late as $z=127$ with sub-percent errors in the matter power spectrum, while 2LPT could reach $z\approx50$ before the errors would grow over 1\%. A diagram of the timeline and the $z_{\rm start}$ for different PT formulations is shown in Fig.~\ref{fig:ics-process}. We will provide further details on the method by \cite{2010MNRAS.403.1859J} later in this section.

A formulation of 3LPT which could, in principle, provide sub-percent errors down to $z=12$ was introduced by \cite{2021MNRAS.500..663M} and is currently used by the Virgo Consortium working teams in the upcoming COLIBRE flagship project (COLIBRE Collaboration, in preparation). However, including higher-order terms in the perturbation theory leads to a steeply rising computational cost associated with the generation of the initial conditions. Beyond 3LPT, the late start of the DNI simulations can no longer be justified based on the efficient allocation of computational resources, given the current hardware and software technologies. Nevertheless, future precision cosmology could demand initial conditions accurate to 0.1 \% or lower, for which the frameworks to produce arbitrarily precise initial conditions will be an invaluable asset \citep[see Section 6 of][]{2021MNRAS.500..663M}. Finally, recent developments on the \textsc{MUSIC2-monofonIC} code allow the use of 3LPT calculation of perturbations with separate transfer functions for CDM and baryons, which can drastically reduce the small-scale discretisation noise \citep{2021MNRAS.503..426H}.

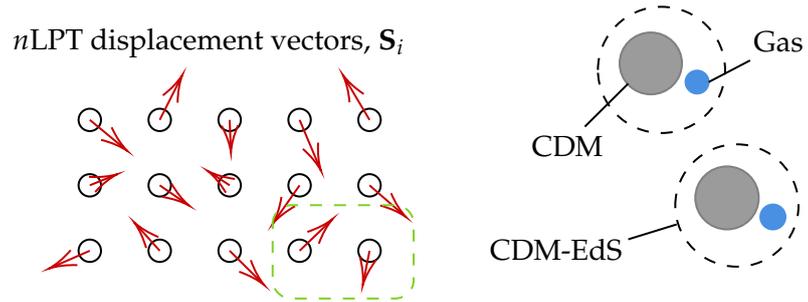
\begin{figure}
\centering
\begin{minipage}{0.9\textwidth}   

\tikzset{every picture/.style={line width=0.75pt}} 
\centering
\begin{tikzpicture}[x=0.75pt,y=0.75pt,yscale=-1.25,xscale=1.25]

\draw   (64.83,78.5) .. controls (64.83,76.01) and (66.84,74) .. (69.31,74) .. controls (71.79,74) and (73.79,76.01) .. (73.79,78.5) .. controls (73.79,80.99) and (71.79,83) .. (69.31,83) .. controls (66.84,83) and (64.83,80.99) .. (64.83,78.5) -- cycle ;
\draw   (92.76,78.5) .. controls (92.76,76.01) and (94.76,74) .. (97.24,74) .. controls (99.71,74) and (101.72,76.01) .. (101.72,78.5) .. controls (101.72,80.99) and (99.71,83) .. (97.24,83) .. controls (94.76,83) and (92.76,80.99) .. (92.76,78.5) -- cycle ;
\draw   (148.61,78.5) .. controls (148.61,76.01) and (150.61,74) .. (153.09,74) .. controls (155.56,74) and (157.57,76.01) .. (157.57,78.5) .. controls (157.57,80.99) and (155.56,83) .. (153.09,83) .. controls (150.61,83) and (148.61,80.99) .. (148.61,78.5) -- cycle ;
\draw   (176.54,78.5) .. controls (176.54,76.01) and (178.54,74) .. (181.02,74) .. controls (183.49,74) and (185.5,76.01) .. (185.5,78.5) .. controls (185.5,80.99) and (183.49,83) .. (181.02,83) .. controls (178.54,83) and (176.54,80.99) .. (176.54,78.5) -- cycle ;
\draw   (120.68,78.5) .. controls (120.68,76.01) and (122.69,74) .. (125.16,74) .. controls (127.64,74) and (129.64,76.01) .. (129.64,78.5) .. controls (129.64,80.99) and (127.64,83) .. (125.16,83) .. controls (122.69,83) and (120.68,80.99) .. (120.68,78.5) -- cycle ;

\draw   (64.83,104.83) .. controls (64.83,102.34) and (66.84,100.33) .. (69.31,100.33) .. controls (71.79,100.33) and (73.79,102.34) .. (73.79,104.83) .. controls (73.79,107.32) and (71.79,109.33) .. (69.31,109.33) .. controls (66.84,109.33) and (64.83,107.32) .. (64.83,104.83) -- cycle ;
\draw   (92.76,104.83) .. controls (92.76,102.34) and (94.76,100.33) .. (97.24,100.33) .. controls (99.71,100.33) and (101.72,102.34) .. (101.72,104.83) .. controls (101.72,107.32) and (99.71,109.33) .. (97.24,109.33) .. controls (94.76,109.33) and (92.76,107.32) .. (92.76,104.83) -- cycle ;
\draw   (148.61,104.83) .. controls (148.61,102.34) and (150.61,100.33) .. (153.09,100.33) .. controls (155.56,100.33) and (157.57,102.34) .. (157.57,104.83) .. controls (157.57,107.32) and (155.56,109.33) .. (153.09,109.33) .. controls (150.61,109.33) and (148.61,107.32) .. (148.61,104.83) -- cycle ;
\draw   (176.54,104.83) .. controls (176.54,102.34) and (178.54,100.33) .. (181.02,100.33) .. controls (183.49,100.33) and (185.5,102.34) .. (185.5,104.83) .. controls (185.5,107.32) and (183.49,109.33) .. (181.02,109.33) .. controls (178.54,109.33) and (176.54,107.32) .. (176.54,104.83) -- cycle ;
\draw   (120.68,104.83) .. controls (120.68,102.34) and (122.69,100.33) .. (125.16,100.33) .. controls (127.64,100.33) and (129.64,102.34) .. (129.64,104.83) .. controls (129.64,107.32) and (127.64,109.33) .. (125.16,109.33) .. controls (122.69,109.33) and (120.68,107.32) .. (120.68,104.83) -- cycle ;

\draw   (64.83,131.16) .. controls (64.83,128.67) and (66.84,126.66) .. (69.31,126.66) .. controls (71.79,126.66) and (73.79,128.67) .. (73.79,131.16) .. controls (73.79,133.65) and (71.79,135.66) .. (69.31,135.66) .. controls (66.84,135.66) and (64.83,133.65) .. (64.83,131.16) -- cycle ;
\draw   (92.76,131.16) .. controls (92.76,128.67) and (94.76,126.66) .. (97.24,126.66) .. controls (99.71,126.66) and (101.72,128.67) .. (101.72,131.16) .. controls (101.72,133.65) and (99.71,135.66) .. (97.24,135.66) .. controls (94.76,135.66) and (92.76,133.65) .. (92.76,131.16) -- cycle ;
\draw   (148.61,131.16) .. controls (148.61,128.67) and (150.61,126.66) .. (153.09,126.66) .. controls (155.56,126.66) and (157.57,128.67) .. (157.57,131.16) .. controls (157.57,133.65) and (155.56,135.66) .. (153.09,135.66) .. controls (150.61,135.66) and (148.61,133.65) .. (148.61,131.16) -- cycle ;
\draw   (176.54,131.16) .. controls (176.54,128.67) and (178.54,126.66) .. (181.02,126.66) .. controls (183.49,126.66) and (185.5,128.67) .. (185.5,131.16) .. controls (185.5,133.65) and (183.49,135.66) .. (181.02,135.66) .. controls (178.54,135.66) and (176.54,133.65) .. (176.54,131.16) -- cycle ;
\draw   (120.68,131.16) .. controls (120.68,128.67) and (122.69,126.66) .. (125.16,126.66) .. controls (127.64,126.66) and (129.64,128.67) .. (129.64,131.16) .. controls (129.64,133.65) and (127.64,135.66) .. (125.16,135.66) .. controls (122.69,135.66) and (120.68,133.65) .. (120.68,131.16) -- cycle ;

\draw [color={rgb, 255:red, 201; green, 0; blue, 0 }  ,draw opacity=1 ]   (69.31,78.5) -- (82.8,89.72) ;
\draw [shift={(84.33,91)}, rotate = 219.77] [color={rgb, 255:red, 201; green, 0; blue, 0 }  ,draw opacity=1 ][line width=0.75]    (10.93,-3.29) .. controls (6.95,-1.4) and (3.31,-0.3) .. (0,0) .. controls (3.31,0.3) and (6.95,1.4) .. (10.93,3.29)   ;
\draw [color={rgb, 255:red, 201; green, 0; blue, 0 }  ,draw opacity=1 ]   (69.31,104.83) -- (78.48,101.09) ;
\draw [shift={(80.33,100.33)}, rotate = 157.8] [color={rgb, 255:red, 201; green, 0; blue, 0 }  ,draw opacity=1 ][line width=0.75]    (10.93,-3.29) .. controls (6.95,-1.4) and (3.31,-0.3) .. (0,0) .. controls (3.31,0.3) and (6.95,1.4) .. (10.93,3.29)   ;
\draw [color={rgb, 255:red, 201; green, 0; blue, 0 }  ,draw opacity=1 ]   (97.24,104.83) -- (107.36,111.86) ;
\draw [shift={(109,113)}, rotate = 214.78] [color={rgb, 255:red, 201; green, 0; blue, 0 }  ,draw opacity=1 ][line width=0.75]    (10.93,-3.29) .. controls (6.95,-1.4) and (3.31,-0.3) .. (0,0) .. controls (3.31,0.3) and (6.95,1.4) .. (10.93,3.29)   ;
\draw [color={rgb, 255:red, 201; green, 0; blue, 0 }  ,draw opacity=1 ]   (97.24,78.5) -- (105.44,62.12) ;
\draw [shift={(106.33,60.33)}, rotate = 116.59] [color={rgb, 255:red, 201; green, 0; blue, 0 }  ,draw opacity=1 ][line width=0.75]    (10.93,-3.29) .. controls (6.95,-1.4) and (3.31,-0.3) .. (0,0) .. controls (3.31,0.3) and (6.95,1.4) .. (10.93,3.29)   ;
\draw [color={rgb, 255:red, 201; green, 0; blue, 0 }  ,draw opacity=1 ]   (125.16,78.5) -- (125.59,89.67) ;
\draw [shift={(125.67,91.67)}, rotate = 267.81] [color={rgb, 255:red, 201; green, 0; blue, 0 }  ,draw opacity=1 ][line width=0.75]    (10.93,-3.29) .. controls (6.95,-1.4) and (3.31,-0.3) .. (0,0) .. controls (3.31,0.3) and (6.95,1.4) .. (10.93,3.29)   ;
\draw [color={rgb, 255:red, 201; green, 0; blue, 0 }  ,draw opacity=1 ]   (125.16,104.83) -- (118,100.1) ;
\draw [shift={(116.33,99)}, rotate = 33.44] [color={rgb, 255:red, 201; green, 0; blue, 0 }  ,draw opacity=1 ][line width=0.75]    (10.93,-3.29) .. controls (6.95,-1.4) and (3.31,-0.3) .. (0,0) .. controls (3.31,0.3) and (6.95,1.4) .. (10.93,3.29)   ;
\draw [color={rgb, 255:red, 201; green, 0; blue, 0 }  ,draw opacity=1 ]   (69.31,131.16) -- (54.83,137.53) ;
\draw [shift={(53,138.33)}, rotate = 336.26] [color={rgb, 255:red, 201; green, 0; blue, 0 }  ,draw opacity=1 ][line width=0.75]    (10.93,-3.29) .. controls (6.95,-1.4) and (3.31,-0.3) .. (0,0) .. controls (3.31,0.3) and (6.95,1.4) .. (10.93,3.29)   ;
\draw [color={rgb, 255:red, 201; green, 0; blue, 0 }  ,draw opacity=1 ]   (97.24,131.16) -- (88.33,121.16) ;
\draw [shift={(87,119.67)}, rotate = 48.31] [color={rgb, 255:red, 201; green, 0; blue, 0 }  ,draw opacity=1 ][line width=0.75]    (10.93,-3.29) .. controls (6.95,-1.4) and (3.31,-0.3) .. (0,0) .. controls (3.31,0.3) and (6.95,1.4) .. (10.93,3.29)   ;
\draw [color={rgb, 255:red, 201; green, 0; blue, 0 }  ,draw opacity=1 ]   (125.16,131.16) -- (137.59,143.59) ;
\draw [shift={(139,145)}, rotate = 225.01] [color={rgb, 255:red, 201; green, 0; blue, 0 }  ,draw opacity=1 ][line width=0.75]    (10.93,-3.29) .. controls (6.95,-1.4) and (3.31,-0.3) .. (0,0) .. controls (3.31,0.3) and (6.95,1.4) .. (10.93,3.29)   ;
\draw [color={rgb, 255:red, 201; green, 0; blue, 0 }  ,draw opacity=1 ]   (153.09,131.16) -- (164.9,119.72) ;
\draw [shift={(166.33,118.33)}, rotate = 135.92] [color={rgb, 255:red, 201; green, 0; blue, 0 }  ,draw opacity=1 ][line width=0.75]    (10.93,-3.29) .. controls (6.95,-1.4) and (3.31,-0.3) .. (0,0) .. controls (3.31,0.3) and (6.95,1.4) .. (10.93,3.29)   ;
\draw [color={rgb, 255:red, 201; green, 0; blue, 0 }  ,draw opacity=1 ]   (153.09,104.83) -- (143.47,118.69) ;
\draw [shift={(142.33,120.33)}, rotate = 304.75] [color={rgb, 255:red, 201; green, 0; blue, 0 }  ,draw opacity=1 ][line width=0.75]    (10.93,-3.29) .. controls (6.95,-1.4) and (3.31,-0.3) .. (0,0) .. controls (3.31,0.3) and (6.95,1.4) .. (10.93,3.29)   ;
\draw [color={rgb, 255:red, 201; green, 0; blue, 0 }  ,draw opacity=1 ]   (153.09,78.5) -- (160.89,97.16) ;
\draw [shift={(161.67,99)}, rotate = 247.29] [color={rgb, 255:red, 201; green, 0; blue, 0 }  ,draw opacity=1 ][line width=0.75]    (10.93,-3.29) .. controls (6.95,-1.4) and (3.31,-0.3) .. (0,0) .. controls (3.31,0.3) and (6.95,1.4) .. (10.93,3.29)   ;
\draw [color={rgb, 255:red, 201; green, 0; blue, 0 }  ,draw opacity=1 ]   (181.02,78.5) -- (171.35,62.06) ;
\draw [shift={(170.33,60.33)}, rotate = 59.54] [color={rgb, 255:red, 201; green, 0; blue, 0 }  ,draw opacity=1 ][line width=0.75]    (10.93,-3.29) .. controls (6.95,-1.4) and (3.31,-0.3) .. (0,0) .. controls (3.31,0.3) and (6.95,1.4) .. (10.93,3.29)   ;
\draw [color={rgb, 255:red, 201; green, 0; blue, 0 }  ,draw opacity=1 ]   (181.02,104.83) -- (194.16,116.35) ;
\draw [shift={(195.67,117.67)}, rotate = 221.23] [color={rgb, 255:red, 201; green, 0; blue, 0 }  ,draw opacity=1 ][line width=0.75]    (10.93,-3.29) .. controls (6.95,-1.4) and (3.31,-0.3) .. (0,0) .. controls (3.31,0.3) and (6.95,1.4) .. (10.93,3.29)   ;
\draw [color={rgb, 255:red, 201; green, 0; blue, 0 }  ,draw opacity=1 ]   (181.02,131.16) -- (178.71,143.04) ;
\draw [shift={(178.33,145)}, rotate = 280.98] [color={rgb, 255:red, 201; green, 0; blue, 0 }  ,draw opacity=1 ][line width=0.75]    (10.93,-3.29) .. controls (6.95,-1.4) and (3.31,-0.3) .. (0,0) .. controls (3.31,0.3) and (6.95,1.4) .. (10.93,3.29)   ;
\draw  [color={rgb, 255:red, 126; green, 211; blue, 33 }  ,draw opacity=1 ][dash pattern={on 4.5pt off 4.5pt}] (142.33,120.33) .. controls (142.33,116.18) and (145.7,112.82) .. (149.85,112.82) -- (191.13,112.82) .. controls (195.28,112.82) and (198.65,116.18) .. (198.65,120.33) -- (198.65,142.87) .. controls (198.65,147.02) and (195.28,150.38) .. (191.13,150.38) -- (149.85,150.38) .. controls (145.7,150.38) and (142.33,147.02) .. (142.33,142.87) -- cycle ;
\draw  [color={rgb, 255:red, 128; green, 128; blue, 128 }  ,draw opacity=1 ][fill={rgb, 255:red, 155; green, 155; blue, 155 }  ,fill opacity=1 ] (280.89,55.78) .. controls (280.89,48.9) and (286.46,43.33) .. (293.33,43.33) .. controls (300.21,43.33) and (305.78,48.9) .. (305.78,55.78) .. controls (305.78,62.65) and (300.21,68.22) .. (293.33,68.22) .. controls (286.46,68.22) and (280.89,62.65) .. (280.89,55.78) -- cycle ;
\draw  [color={rgb, 255:red, 128; green, 128; blue, 128 }  ,draw opacity=1 ][fill={rgb, 255:red, 155; green, 155; blue, 155 }  ,fill opacity=1 ] (311.11,109.89) .. controls (311.11,102.89) and (316.78,97.22) .. (323.78,97.22) .. controls (330.77,97.22) and (336.44,102.89) .. (336.44,109.89) .. controls (336.44,116.88) and (330.77,122.56) .. (323.78,122.56) .. controls (316.78,122.56) and (311.11,116.88) .. (311.11,109.89) -- cycle ;
\draw  [dash pattern={on 4.5pt off 4.5pt}] (272.22,58.39) .. controls (272.22,44.37) and (283.59,33) .. (297.61,33) .. controls (311.63,33) and (323,44.37) .. (323,58.39) .. controls (323,72.41) and (311.63,83.78) .. (297.61,83.78) .. controls (283.59,83.78) and (272.22,72.41) .. (272.22,58.39) -- cycle ;
\draw  [dash pattern={on 4.5pt off 4.5pt}] (303.11,112.94) .. controls (303.11,99.35) and (314.13,88.33) .. (327.72,88.33) .. controls (341.31,88.33) and (352.33,99.35) .. (352.33,112.94) .. controls (352.33,126.54) and (341.31,137.56) .. (327.72,137.56) .. controls (314.13,137.56) and (303.11,126.54) .. (303.11,112.94) -- cycle ;
\draw  [color={rgb, 255:red, 74; green, 144; blue, 226 }  ,draw opacity=1 ][fill={rgb, 255:red, 74; green, 144; blue, 226 }  ,fill opacity=1 ] (307.22,63.28) .. controls (307.22,60.73) and (309.29,58.67) .. (311.83,58.67) .. controls (314.38,58.67) and (316.44,60.73) .. (316.44,63.28) .. controls (316.44,65.82) and (314.38,67.89) .. (311.83,67.89) .. controls (309.29,67.89) and (307.22,65.82) .. (307.22,63.28) -- cycle ;
\draw  [color={rgb, 255:red, 74; green, 144; blue, 226 }  ,draw opacity=1 ][fill={rgb, 255:red, 74; green, 144; blue, 226 }  ,fill opacity=1 ] (337.11,117.56) .. controls (337.11,114.79) and (339.35,112.56) .. (342.11,112.56) .. controls (344.87,112.56) and (347.11,114.79) .. (347.11,117.56) .. controls (347.11,120.32) and (344.87,122.56) .. (342.11,122.56) .. controls (339.35,122.56) and (337.11,120.32) .. (337.11,117.56) -- cycle ;
\draw    (316.44,63.28) -- (333,54.56) ;
\draw    (284.33,63.89) -- (267,81.22) ;
\draw    (303.67,120.56) -- (284.33,126.56) ;

\draw (333.33,41.67) node [anchor=north west][inner sep=0.75pt]  [xscale=1,yscale=1] [align=left] {Gas};
\draw (244.67,83.67) node [anchor=north west][inner sep=0.75pt]  [xscale=1,yscale=1] [align=left] {CDM};
\draw (228,125) node [anchor=north west][inner sep=0.75pt]  [xscale=1,yscale=1] [align=right] {CDM-EdS};
\draw (37.33,40.33) node [anchor=north west][inner sep=0.75pt]  [xscale=1,yscale=1] [align=left] {$n$LPT displacement vectors, $\mathbf{S}_i$};
\end{tikzpicture}
\end{minipage}
    \caption{\textit{Left:} diagram of particles (black circles) arranged in a uniform square lattice. Each particle is then displaced with vectors $\mathbf{S}_i$ (red) computed using $n$LPT. The particles in the initial state are tracers of the CDM density field, defined in a smoothed-particle hydrodynamics fashion. We assume that the $n$LPT is formulated for a single (CDM) fluid in an EdS universe. In production simulation projects, particle loads are arranged in a glass-like structure and not in a square lattice; in this figure, we use the latter purely for illustration purposes. \textit{Right:} zoomed view of the green dashed region, containing two perturbed CDM particles (dashed circles), labelled CDM-EdS. The CDM-EdS particles are then split into CDM particles and gas particles with masses weighted by $\Omega_{\rm CDM}$ and $\Omega_b$ respectively before the start of a hydrodynamic simulation.}
    \label{fig:ics-gas-split}
\end{figure}

LPT is used to calculate the vectors $\mathbf{S}$, in analogy to the Zeldovich approximation, used to displace particles from a regular lattice arrangement. An example of uniform square lattice pattern in 2D is shown in Fig.~\ref{fig:ics-gas-split} and the displacement vectors are shown in red. Including high-order terms in the LPT increases the precision on the length and direction of the $\mathbf{S}_i$ vectors. Non-linear systems, such as late-times structure formation, are sensitive to the initial conditions and small errors in the initial state can propagate through the DNI simulation and grow significantly over millions of time-steps.

\subsubsection*{Glass-like particle distribution}
The uniform, unperturbed distribution of particles used in our simulations is glass-like, generated using the method of \cite{white1996schaeffer}. In a glass-like distribution, the particles are not arranged in a square Cartesian lattice. Arranging particles in a glass-like structure is shown to minimise the Fourier modes within the volume which would interfere with the power spectrum injected using the LPT \citep[e.g.][]{2007MNRAS.380...93W}. This method of generating the pre-initial conditions is well converged with resolution and provides similar results to the more recent capacity-constrained Voronoi tessellation, proposed by \cite{2018MNRAS.481.3750L}. We refer to Fig. 3 and 4 of \cite{2018MNRAS.481.3750L} for a visual comparison between the square lattice and glass-like distributions, and the Fourier spectrum respectively.

\begin{figure}
    \centering
    \includegraphics[width=0.9\textwidth]{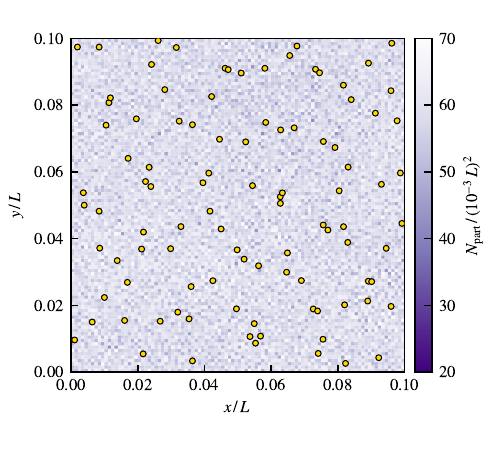}
    \caption{Glass particles (yellow) and 2LPT-displaced dark matter particle density (purple) in the initial conditions of the parent simulations of \cite{2023MNRAS.520.3164A}. The original cubic volume is $L=300$ Mpc in size and $564^3$ particles. Its EAGLE-like identifier is \texttt{EAGLE-XL\_L0300N0564}. We selected only the glass and dark matter particles in a cubic subset volume of side $L/10$, projected along the $z$-axis. The glass particles are used as interpolation points to compute the positions and LPT displacement vectors of the higher-resolution initial conditions.}
    \label{fig:particle-load}
\end{figure}

3D initial conditions are generated from a 3D glass-like particle distribution. In Fig. \ref{fig:particle-load}, we show a portion of such arrangement used to produce the initial conditions of the parent box in Fig. \ref{fig:dmo_box_with_halos}. Following the method of \cite{2010MNRAS.403.1859J}, the glass-like distribution is not necessarily mapped to the particles to displace. The ``glass-particles`` are used as interpolation points to generate a denser (or sparser) particle distribution, depending on the required resolution. The new particles identified in this manner are also known as \textit{particle load}. Crucially, the refined glass-like distribution has the same Fourier properties of the original one \citep[see Appendix A of][for a discussion on the interpolation methods]{2010MNRAS.403.1859J}. Finally, the LPT displacement vectors are also interpolated using the glass-particles and evaluated at the particle load positions.

\subsubsection*{\textsc{Panphasia} and \textsc{IC\_Gen}}
The next step involves computing the displacement vectors to perturb the particle load, consistently with the chosen cosmology. This goal is achieved via (i) a discrete realization of a Gaussian white noise field that has a hierarchical structure, based on octrees, and (ii) a matter power spectrum which modulates the amplitude of perturbations at different spatial scales. In our work, we use the \textsc{Panphasia} code \citep{2013arXiv1306.5771J}, which implements this algorithm in \texttt{Fortran} language.
\textsc{Panphasia} is integrated in \textsc{IC\_Gen} the framework for generating initial conditions used by the Virgo Consortium, described in \cite{2013MNRAS.434.2094J}. \textsc{IC\_Gen} can be configured to produce two types initial conditions:
\begin{itemize}
    \item \textbf{Single-resolution volumes}, such as the \texttt{EAGLE-XL\_L0300N0564} parent simulation in \cite{2023MNRAS.520.3164A}. These volumes are suitable for studies of large-scale structures and galaxy populations \citep[e.g. the EAGLE simulations,][]{eagle.schaye.2015}.

    \item \textbf{Multi-resolution volumes}, such as zoom-in set-ups, e.g., for VR18 and VR2915 in \cite{2023MNRAS.520.3164A}. These cases are used to boost the resolution in one specific region of interest, e.g. a galaxy cluster, for detailed studies.
\end{itemize}

The parent box in the pipeline of Fig. \ref{fig:simulation-pipeline} is a single-volume simulation, while the zoom set-up uses \textsc{IC\_Gen} with a multi-resolution configuration (see Section \ref{sec:zoom-setup}). The parameter file supplied to \textsc{IC\_Gen} is reported in Appendix \ref{appendix:1} (Listing \ref{app1:parent-sim-ics}).

\subsubsection*{Periodic boundary conditions}
Following once again the analogy of an ODE problem, we must specify the \textit{boundary conditions} at the edges of the cubic simulation volume. To reproduce realistic large-scale structures, we need to emulate the gravitational tidal fields that would be generated outside our limited-volume box. Capturing this effect is required because gravity is a long-range force and can influence the collapse of structures even on $\sim 100 - 1000$ Mpc scales (provided they map to sub-horizon modes, see Section \ref{sec:early-perturbations}). To emulate an infinite space using a finite volume, periodic boundary conditions are usually adopted in cosmological and astrophysical simulations. This choice of boundaries makes the space a topological 3-torus (in 3D). The power spectrum calculation used in \textsc{Panphasia} assumes periodic boundary conditions, and so does every other DNI or analysis code used used to produce this work. 

\subsubsection*{Initial particle mass}
Initial conditions generated with \textsc{IC\_Gen} for our projects are dark matter-only (DMO) and do not automatically generate gas particles. This set-up can be used without modification to run DMO simulations (i.e. an EdS universe), but hydrodynamic simulations require gas particles, which prompt a cascade of baryonic effects, such as star formation. In our simulations, the dark matter particles in DMO initial conditions are split -- after being drifted with 2LPT -- into a (smaller) dark matter particle and a gas particle. We illustrate this technique in the right panel of Fig.~\ref{fig:ics-gas-split}. The CDM-EdS particle, generated with \textsc{IC\_Gen} is split into two smaller CDM+gas particles, with masses weighted by the cosmological density parameters $\Omega_{\rm CDM}$ and $\Omega_b$ provided. Since the masses must be conserved, we must have $m_{\rm CDM-EdS} = m_{\rm CDM} + m_{\rm gas}$. Given that $\Omega_m = \Omega_{\rm CDM} + \Omega_b$, we obtain
\begin{align}
     m_{\rm CDM} &= \frac{\Omega_{\rm CDM}}{\Omega_m}~ m_{\rm CDM-EdS},  \\
     m_{\rm gas} &= \frac{\Omega_b}{\Omega_m}~ m_{\rm CDM-EdS}.
\end{align}
Assuming a $\Lambda$CDM (flat) cosmology, the mean density of the Universe is $\rho_{\rm crit,0}$. Since the comoving volume of the simulation box is $V_{\rm box} = L^3$, containing $N_{\rm part}$, the mean density can be derived as
\begin{equation}
    m_{\rm CDM-EdS} ~ \frac{N_{\rm part}}{V_{\rm box}} = \rho_{\rm m,0} = \Omega_m\, \rho_{\rm crit,0},
\end{equation}
which can be rearranged to give the DMO particle mass as a function of cosmological parameters and simulation parameters: 
\begin{equation}
    m_{\rm CDM-EdS} = \rho_{\rm m,0}~\frac{L^3}{N_{\rm part}}.
\end{equation}

The split CDM and gas particles are assigned the same initial velocity as the CDM-EdS particle. At early times, baryons track the dark matter distribution, justifying this method of seeding gas particles in hydrodynamic initial conditions. In our simulations, the DMO initial conditions generated with \textsc{IC\_Gen} are parsed into the \swift (DNI) simulation code, which performs the gas seeding. This functionality can be enabled with the parameter 
\begin{verbatim}
    InitialConditions: generate_gas_in_ics: 1
\end{verbatim}
in the \swift parameter file.

For initial conditions generated with CDM and gas \textit{ab initio} by computing the transfer function separately \citep[e.g. \textsc{MUSIC2-monofonIC},][]{2021MNRAS.503..426H}, this functionality is not needed. This set-up was selected by the Virgo Consortium for the upcoming simulations of single-resolution boxes and multi-resolution zooms with \textsc{MUSIC2-polyfonIC} (Hahn et al., in preparation).

\section{Numerical gravity}
\label{sec:numerical-gravity}

The next step in the simulation pipeline consists in evolving the DMO initial conditions using direct integration. For our simulations, we used the \swift code configured in DMO mode. The MACSIS simulations \citep{macsis_barnes_2017}, on the other hand, were run with \textsc{Gadget-3}. In this section and in Section \ref{sec:numerical-sph} we will mainly focus on the \swift code, used as the main simulation tool for running simulations anew, highlighting the main differences with \textsc{Gadget-3}.

\subsubsection*{Introducing \swift}
We begin by providing a general overview of the \swift code, before focusing the discussion its gravity solver. \swift \citep{schaller_2018_swift} is an open-source code used for numerous astrophysics simulations, ranging from planetary collisions to large-scale structures. Written in \texttt{C} language and designed with a modular structure, it was developed by members of the Virgo Consortium, external computer-science researchers and software engineers. The source code can be downloaded from \href{www.swiftsim.com}{www.swiftsim.com} and the most up-to-date documentation is hosted at \href{https://swift.dur.ac.uk/docs/index.html}{this website}.

\swift and \textsc{Gadget-3} differ in the paradigm they adopt to scale to large number of computing nodes for massively parallel HPC simulations. \textsc{Gadget-3} uses \textit{data-based} parallelism, where the computing workload is split among CPU cores in a sequence of collective operations; instead, \swift uses \textit{task-based} parallelism, where calculations are bundled into tasks and allocated to threads \textit{asynchronously} as soon as they become available. By avoiding CPU cores to remain idle while waiting for all the other cores to complete tasks and execute collective operations, task-based parallelism can reduce idle-time and is often described as a more resource-effective method for running large-scale simulations on the most powerful HPC systems in use. The asynchronicity of communications, however, can make complex algorithms particularly difficult to implement and execute reliably. For instance, when one task requires the output of a different one, it must wait until the first is complete before starting. This requirement can impact the run-time efficiency, but is often needed to prevent the algorithm from breaking causality. We will return to this topic in Section \ref{sec:computational-performance} on the scaling efficiency of the \swift default hydrodynamic solver.

\subsubsection*{Gravitational force softening}
The gravitational force dictates the formation of structures in the Universe and can be modelled by discretising a continuous matter distribution with particles. These particles trace the underlying matter distribution and are evolved over time using Lagrangian equations of motion (see Section \ref{sec:early-perturbations}). Using Newtonian mechanics, the gravitational potential $\phi(\mathbf{r}_i)$ at the comoving position $\mathbf{r}_i$ can be computed by summing the contributions over all possible particle pairs $(i,j)$ without repetition:
\begin{equation}
    \phi(\mathbf{r}_i) = - G~\sum_{j\neq i} \frac{m_j}{|\mathbf{r}_i - \mathbf{r}_j |}.
    \label{eq:gravitational-potential-newton}
\end{equation}
This formulation assumes that the gravity particles are described by a Dirac-delta in 3D, $m(\mathbf{r}_i) = m_i\, \delta(\mathbf{r}_i)$, which is prone to generating unphysically strong interactions where two particles in the systems approach each other, making the denominator singular: $\mathbf{r}_i \approx \mathbf{r}_j$. Such unrealistically large two-body scattering effects can be mitigated by convolving the Dirac-delta with a spline, causing the gravity particles to no longer be point-like, but \textit{softened}, allowing them to even superimpose with each other without returning extremely high values of $|\phi|$.

The simplest prescription of gravitational softening invokes a \cite{1911MNRAS..71..460P} softening length $\epsilon_{\rm Plummer}$, which modified Eq.~\eqref{eq:gravitational-potential-newton} as
\begin{equation}
    \phi_{\rm Plummer}(\mathbf{r}_i) = - G~\sum_{j\neq i} \frac{m_j}{\left(|\mathbf{r}_i - \mathbf{r}_j |^2 + \epsilon_{\rm Plummer}^2 \right)^{1/2}}
    \label{eq:gravitational-potential-plummer}
\end{equation}
to give the Plummer potential. While reducing artificial two-body scattering, the $\epsilon_{\rm Plummer}^2$ term causes deviations from the Newtonian potential even at moderately large distances, affecting the long-range forces.

For this reason, cosmological simulations codes in the past two decades adopted a different spline to soften the gravity particles. In \swift, we use a C2 kernel
\citep{wendland1995piecewise}, $W(|\mathbf{r}|, h) \equiv W(r, h)$, defined as
\begin{align}
    W(r,h) &= \frac{21}{2\pi h^3} \cdot \left\lbrace\begin{array}{ccl}
    4u^5 - 15u^4 + 20u^3 - 10u^2 + 1 & \text{if} & u < 1,\\
    0 & \text{if} & u \geq 1,
    \end{array}
    \right.
    \end{align}
    where $u = r/h$ is the distance scaled by a smoothing length $h = 3\, \epsilon_{\rm Plummer}$. The potential $\phi_i(r,h)$ corresponding to this density distribution of particle $i$ is therefore
    \begin{align}
    \phi_i^{\rm (C2)}(r,h) = G\, m_i\, \cdot\, 
    \left\lbrace\begin{array}{ccl}
    \left(-3u^7 + 15u^6 - 28u^5 + 21u^4 - 7u^2 + 3\right) / h & \text{if} & r < h,\\
    1 / r & \text{if} & r \geq h.
    \end{array}
    \right.
    \label{eq:fmm:potential}
\end{align}
In the limit of $r \longrightarrow 0$,
\begin{equation}
    \lim_{r \longrightarrow 0}  - G~\sum_{j\neq i} \phi_j^{\rm (C2)}(r,h) =  - G~\sum_{j\neq i} \frac{3\, m_j}{3\, \epsilon_{\rm Plummer}} = \phi_{\rm Plummer}(\mathbf{r}_i = \mathbf{r}_j),
\end{equation}
proving that the C2 kernel tends to a Plummer sphere in the centre of the gravity particle and to a Newtonian potential for large $r$ by construction.

\begin{figure}
    \centering
    \includegraphics[width=0.8\textwidth]{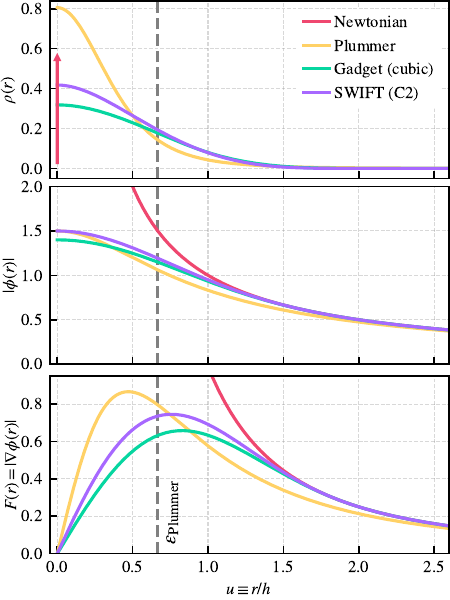}
    \caption{Gravitational softening kernels corresponding to the \cite{1911MNRAS..71..460P} potential (yellow), the cubic spline used in \textsc{Gadget} (green) and the C2 spline used in \swift (purple). These kernels are compared to the Newtonian potential (red), which is asymptotically large at $u=0$. From tow to bottom, we show (in arbitrary units) the density profile ($\rho(r)$) for an isolated particle, the potential profile in absolute value ($|\varphi(r)|$) and the force $F(r)$. The dashed line indicates the Plummer-equivalent smoothing length $\epsilon_{\rm Plummer}$. The density profile of an isolated particle with a Newtonian potential is a Dirac delta, represented in the top panel with an arrow at $u=0$.}
    \label{fig:gravitational-softening}
\end{figure}

In contrast, \textsc{Gadget} uses the \cite{Monaghan1985} cubic spline kernel, which is more computationally expensive on modern CPU architectures while retaining a similar profile and properties to the C2 kernel. We compare the Newtonian, Plummer, cubic and C2 softening rules in Fig. \ref{fig:gravitational-softening}. 


This calculation summarises the \textit{direct summation} algorithm and is implemented with two nested \texttt{for} loops over $n(n-1)$ particle pairs. Therefore, this calculations scales quadratically with the number $n$ of particles in the system, i.e. it has a run-time complexity of $\mathcal{O}(n^2)$. The direct summation method returns the most accurate results for the calculation of the gravitational forces between particles, $F_i = - \nabla \phi_i$, but is computationally expensive. For long-range forces (and large spatial scales) this method is replaced by slightly less accurate, but faster methods, such as the fast multipole method (FMM) or the tree-particle mesh method (TreePM).

\subsubsection*{Fast multipole method}
Fast multipoles, originally introduced by \cite{GREENGARD1987325}, are used by default in \swift to compute the gravitational forces between particles. 
This method relies on a hierarchical decomposition of the simulation domain into smaller sub-regions, and uses a series of low-rank approximations to evaluate the interactions between distant particles.

In FMM, the simulation domain is recursively subdivided into a tree-like structure (oc-tree), which is composed of cubic cells containing particles. The interactions between particles are computed at multiple levels of the tree, where each level corresponds to a different granularity of the simulation domain. At each level of the tree, the particles are grouped into clusters, and the gravitational interactions between clusters are approximated using a multipole expansion. By approximating the gravitational field of distant particles using a low-order multipole expansion, FMM can greatly reduce the number of pairwise interactions that need be computed. The run-time complexity of the FMM algorithm is $\mathcal{O}(n \log n)$, which outperforms the $\mathcal{O}(n^2)$ complexity of the direct summation method.

In all our simulations, we use the \textit{adaptive} multipole acceptance criterion (MAC) from \cite{2014ComAC...1....1D}, which uses the particle information to estimate the level at which the multipole expansion can be truncated.

The most recent release of \textsc{Gadget-4} also adopts the FMM method \citep[see Section 2.4 of][]{2021MNRAS.506.2871S}, albeit with a different MAC specification than \swift. Instead, \textsc{Gadget-2} and \textsc{Gadget-3} rely on the hybrid TreePM method for compute short- and long-range gravitational forces, explicitly separated in Fourier space by a characteristic threshold scale. This approach combines the oct-tree method \citep{1986Natur.324..446B}, more accurate for short-range forces because of its hierarchical structure, and the particle-mesh method \citep{1981csup.book.....H}, preferred for long-range interactions. Overall, the TreePM method has an algorithmic complexity of $\mathcal{O}(n\log n)$.

Both TreePM and FMM methods include a hierarchical tree, and for this reason they are adaptive to domains where the density of particles varies over a vast dynamic range, e.g. in zoom-in simulations or late-time large-scale structures. Their implementation is based on the \textit{Fastest Fourier Transform in the West} library \citep[FFTW,][]{frigo1998fftw}.\footnote{The FFTW library is available at \href{https://www.fftw.org/}{https://www.fftw.org/}. For our simulations, we use versions of FFTW greater than 3.0.} The FFTW library can be compiled with support for threads and the message-passing interface \citep[MPI,][]{gropp1996high}, which are required for running massive parallel simulations with in \swift or \textsc{Gadget} on HPC systems.

\section{Object selection methods}
\label{sec:object-selection}

Following the generation of the initial conditions of the \textit{parent volume} (step 3 in Fig.~\ref{fig:simulation-pipeline}), we run \swift configured in DMO mode, i.e. only using the gravity solver. The $z=0$ snapshot of the simulation (step 4) was then post-processed with \velociraptor \citep{2019PASA...36...21E}, a halo-finding code written in \texttt{C++} and adopted by the Virgo Consortium for collaboration projects such as EAGLE-XL and the \swift-based simulations in this thesis. To identify the centre of potential (CoP) and the spatial extent of halos, \velociraptor uses both the positions and velocity information of the particles in the simulation. This method, referred to as 6D\footnote{The total Lagrangian phase-space has 6 dimensions: 3 for the Cartesian coordinates of the particles, and 3 for the Cartesian velocities.} friends-of-friends (FOF), is an extension of the original FOF methodology introduced by \cite{1985ApJ...292..371D}, which only included the coordinate information.

\subsubsection*{Halo-finding codes}
Most halo finders utilised in production runs today use algorithms analogous to FOF to locate the centre of halos. In addition, they include algorithms tailored to finding specific structures or applications. For instance,
\begin{itemize}
    \item the \textbf{\textsc{HBT+}} code \citep{2018MNRAS.474..604H} leverages the hierarchical bound-tracing (HBT) algorithm \citep{2012MNRAS.427.2437H} to produce particularly clean merger trees between main halos and their substructure components. The results enabled precise reconstructions of the assembly history of a Local Group-like DMO simulation through constrained realisations of the SIBELIUS project \citep{2022MNRAS.512.5823M}.
    \item \textbf{\velociraptor}, instead, is a specialised tool which uses 6DFOF to find halos and substructures reliably, even in environments with low density contrast \citep{2019PASA...36...21E}. This capability makes it suitable for locating stream-like structures and separating \textit{dynamically distinct} particles in halos about to merge.\footnote{Distinguishing halos near a close encounter (e.g. mergers) is a particularly challenging task for FOF-based algorithms, which grow a geometrical network between the particles that would likely connect the two halos \citep[e.g.][]{2015MNRAS.454.3020B}. In this example, the kinematic information used in the 6DFOF prescription \citep[see Section 2 of][]{2019PASA...36...21E}} Thanks to its modular \texttt{C++11} code structure and HPC support via the OpenMP and message-passing interface (MPI) libraries, \velociraptor can be configured to compute a wide range of halo properties and has is being used by most Virgo Consortium scientists for simulation projects \citep[e.g.][]{2022MNRAS.514..249C, 2023MNRAS.520.3164A}. \swift can also be coupled with \velociraptor to run the halo finder on-the-fly. However, this functionality requires large amounts of free memory, since \velociraptor creates a deep copy of the particle data, and cannot be used in large simulations (e.g. SIBELIUS-DARK, \citealt{2022MNRAS.512.5823M} and FLAMINGO, \citealt{flamingo_schaye2023}) that require most of the HPC system's memory capacity to run.
\end{itemize}
Several other halo-finding codes, not mentioned in this document, have been successful in producing production-quality halo catalogues; for a more comprehensive overview of halo-finding algorithms, we recommend recent code-comparison project series by \cite{2011MNRAS.415.2293K, 2012MNRAS.423.1200O, 2013MNRAS.436..150S, 2013MNRAS.433.1537E, 2013MNRAS.428.2039K, 2014MNRAS.441.3488A, 2015MNRAS.454.3020B}.

\subsubsection*{Selection criteria in \cite{2023MNRAS.520.3164A}}
Using the \velociraptor halo catalogue from the $z=0$ snapshot of the parent DMO simulation, we selected \textit{isolated} field halos (i.e. not bound to larger objects) in the mass range $M_{500} / {\rm M}_\odot \in [10^{12.9}, 10^{14.5}]$. We define a field halo of mass $M$ isolated if no other field halos with a mass larger than $0.1\, M$ are located within a $10\, r_{500}$ distance from its centre. This selection returned $\sim 100$ objects satisfying those conditions.

Finally, we divided the mass range into 3 log-spaced bins and selected 9 halos from each bin randomly. This sample of 27 halos (referred to as the extended sample in Chapter \ref{chapter:4}) will then be re-simulated individually using the zoom-in technique (step 7 and 9 in Fig.~\ref{fig:simulation-pipeline}). Spanning approximately 2 order of magnitude in mass, this simulation suite was designed to have enough resolution and number of objects to provide reliable estimates of the thermodynamic profiles.

\subsubsection*{Selection criteria in MACSIS}
Each set of selection rules are tailored to the scope of the project. For MACSIS \citep{macsis_barnes_2017}, the criteria for selecting halos from $(2.3~{\rm Gpc})^3$ DMO parent volume are different and aimed at extending the halo-mass function (HMF) of BAHAMAS \citep{2018MNRAS.476.2999M} at the high-mass end. The 390 galaxy clusters in MACSIS simulation suite constitute a large statistical sample of massive clusters run with the same resolution as BAHAMAS, and the were selected as follows:
\begin{itemize}
    \item the FOF halos with $M_{\rm FOF} / {\rm M}_\odot \in [10^{15}-10^{16}]$ were divided into 5 equally-spaced log-bins (0.2 dex wide).
    \item Starting from the highest-mass bin, the first subset contains 7 objects and the second bin 83. All these halos are promoted to be part of the re-simulated sample. Since all the clusters at the high-mass end were selected without additional filtering, this subset is \textit{mass-limited} in the range $M_{\rm FOF} / {\rm M}_\odot \in [10^{15.6}-10^{16}]$.
    \item For each of the remaining mass bins, numbered 3-5, 100 halos were selected from a larger sample, making the interval $M_{\rm FOF} / {\rm M}_\odot \in [10^{15}-10^{15.6}]$ \textit{not} mass-limited. The halos selected in this mass range add up to 300, which combine with the 90 objects selected in the mass-limited interval.
    \item To avoid over-representing the low-mass halos in each bin 3-5, each of these 0.2 dex-wide bins was further split into 10 sub-bins (0.02 dex wide) and 10 halos were randomly sampled from each sub-bin.
\end{itemize}

Selecting individual objects to be re-simulated with the zoom-in technique has another advantage in addition to the resolution boost at fixed computational cost. While single-resolution volumes need to be run at once, zoom re-simulations of selected objects can be run independently of each other, making the computational endeavour much more ``liquid`` and flexible. For instance, the C-EAGLE cluster sample \citep{ceagle.barnes.2017} was partly run on the DiRAC Cosma-5 machine in Durham (UK) and partly \citep{2017MNRAS.470.4186B} on the HPC system in Munich (Germany); this arrangement would not have been possible within the limits of current software stacks for a cosmological volume by running communicating cells across HPC systems.

A disadvantage of selecting a large number of objects to re-simulate comes from tasks which cannot (yet) be automated, such as assessing the topology of the initial refinement mask (see Section \ref{sec:zoom-setup:lagrangian-region}). In this scenario, operating on a large sample of objects can become particularly laborious and resource-intensive. The final outcome may also be affected by selection biases, making the comparison with different simulation data-sets and observations more complex. Other than the 390 zoom-in MACSIS simulations, The Three Hundred project also strives to provide the same rich statistical sample of MACSIS by zoom re-simulating 324 galaxy clusters with full-physics \cite{2018MNRAS.480.2898C}. At the time of writing, MACSIS and The Three Hundred are the only zoom-in simulation projects striving to provide a large statistical data-set with many simulated clusters and they have made an impact in cosmology research with $\approx 200$ citations combined. Limitations in automation, and human resources, are encouraging computational researchers to re-shape the future MACSIS-like projects, propending towards Gpc-size single-resolution volumes, run with increasing resolution and physics complexity as new HPC hardware and software technologies are deployed. An example of this strategy is found in the FLAMINGO simulations project \citep{flamingo_schaye2023}.

\section{Zoom-in simulations}
\label{sec:zoom-setup}

The next step of the zoom-in simulation pipeline (steps 7 and 8 in Fig.~\ref{fig:simulation-pipeline}) involves defining the portion of the parent box to refine and re-simulate. In our runs, we choose the high-resolution region to be a sphere centres in the centre of potential of a particular halo in the parent box at $z=0$. The geometry of the selected region is chosen based on the specific scientific requirements; for instance, studies of baryon acoustic oscillations (BAO) use a \textit{slab}, i.e. a cuboid with one dimension much larger than the other two, to enhance the resolution along a $\sim$ Gpc-long line of sight \citep[e.g.][]{2020arXiv200601039L}. A simple spherical region is sufficient to produce a high-quality re-simulation of relatively isolated groups and clusters.

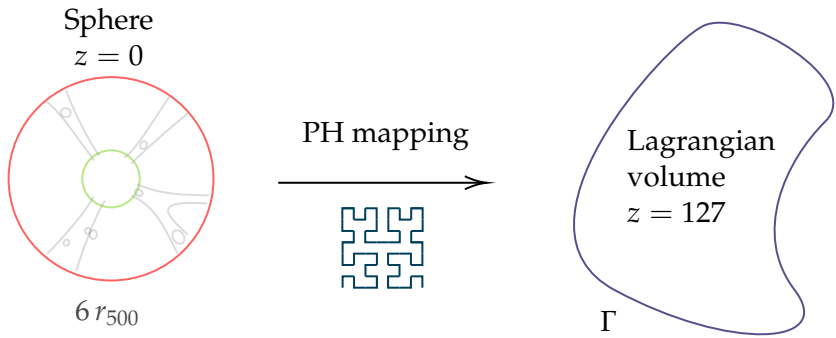
\begin{figure}
\centering
\begin{minipage}{\textwidth}   

  
\tikzset {_2q7ocq1ux/.code = {\pgfsetadditionalshadetransform{ \pgftransformshift{\pgfpoint{0 bp } { 0 bp }  }  \pgftransformscale{1 }  }}}
\pgfdeclareradialshading{_33890c2tz}{\pgfpoint{0bp}{0bp}}{rgb(0bp)=(1,1,1);
rgb(12.410714285714286bp)=(1,1,1);
rgb(25bp)=(0.76,1,0);
rgb(400bp)=(0.76,1,0)}
\tikzset{_3czgovdl3/.code = {\pgfsetadditionalshadetransform{\pgftransformshift{\pgfpoint{0 bp } { 0 bp }  }  \pgftransformscale{1 } }}}
\pgfdeclareradialshading{_fr4fh0dvj} { \pgfpoint{0bp} {0bp}} {color(0bp)=(transparent!100);
color(12.410714285714286bp)=(transparent!100);
color(25bp)=(transparent!38);
color(400bp)=(transparent!38)} 
\pgfdeclarefading{_baq963wyn}{\tikz \fill[shading=_fr4fh0dvj,_3czgovdl3] (0,0) rectangle (50bp,50bp); } 
\tikzset{every picture/.style={line width=0.75pt}} 
    
\centering     
\begin{tikzpicture}[x=0.75pt,y=0.75pt,yscale=-1,xscale=1]

\draw  [color={rgb, 255:red, 255; green, 99; blue, 97 }  ,draw opacity=1 ] (68.69,130.16) .. controls (68.69,101.91) and (91.58,79.02) .. (119.83,79.02) .. controls (148.07,79.02) and (170.97,101.91) .. (170.97,130.16) .. controls (170.97,158.4) and (148.07,181.3) .. (119.83,181.3) .. controls (91.58,181.3) and (68.69,158.4) .. (68.69,130.16) -- cycle ;
\path  [shading=_33890c2tz,_2q7ocq1ux,path fading= _baq963wyn ,fading transform={xshift=2}] (105.5,130.16) .. controls (105.5,122.23) and (111.91,115.8) .. (119.83,115.8) .. controls (127.74,115.8) and (134.15,122.23) .. (134.15,130.16) .. controls (134.15,138.09) and (127.74,144.52) .. (119.83,144.52) .. controls (111.91,144.52) and (105.5,138.09) .. (105.5,130.16) -- cycle ; 
 \draw  [color={rgb, 255:red, 184; green, 233; blue, 134 }  ,draw opacity=1 ] (105.5,130.16) .. controls (105.5,122.23) and (111.91,115.8) .. (119.83,115.8) .. controls (127.74,115.8) and (134.15,122.23) .. (134.15,130.16) .. controls (134.15,138.09) and (127.74,144.52) .. (119.83,144.52) .. controls (111.91,144.52) and (105.5,138.09) .. (105.5,130.16) -- cycle ; 

\draw [color={rgb, 255:red, 155; green, 155; blue, 155 }  ,draw opacity=0.35 ]   (87.63,91.45) .. controls (94.29,95.85) and (105.94,113.85) .. (111.46,121.01) ;
\draw [color={rgb, 255:red, 155; green, 155; blue, 155 }  ,draw opacity=0.35 ]   (97.46,85.51) .. controls (99.73,92.99) and (108.44,111.35) .. (113.96,118.51) ;
\draw [color={rgb, 255:red, 155; green, 155; blue, 155 }  ,draw opacity=0.35 ]   (87.96,168.01) .. controls (92.66,162.96) and (107.87,146.65) .. (110.96,139.01) ;
\draw [color={rgb, 255:red, 155; green, 155; blue, 155 }  ,draw opacity=0.35 ]   (102.46,176.51) .. controls (104.24,170.16) and (112.09,147.35) .. (116.96,141.01) ;
\draw [color={rgb, 255:red, 155; green, 155; blue, 155 }  ,draw opacity=0.35 ]   (152.96,167.51) .. controls (146.47,155.31) and (143.07,144.91) .. (129.96,138.51) ;
\draw [color={rgb, 255:red, 155; green, 155; blue, 155 }  ,draw opacity=0.35 ]   (168.64,138.44) .. controls (154.36,138.28) and (144.64,137.89) .. (132.96,133.01) ;
\draw [color={rgb, 255:red, 155; green, 155; blue, 155 }  ,draw opacity=0.35 ]   (126.46,119.51) .. controls (132.46,115.6) and (143.06,97.48) .. (148.9,88.86) ;
\draw [color={rgb, 255:red, 155; green, 155; blue, 155 }  ,draw opacity=0.35 ]   (129.96,122.51) .. controls (134.82,117.46) and (149.62,104.07) .. (158.39,97.08) ;
\draw  [color={rgb, 255:red, 155; green, 155; blue, 155 }  ,draw opacity=0.35 ] (131.83,136.1) .. controls (132.22,135.35) and (133.35,135.16) .. (134.35,135.68) .. controls (135.35,136.21) and (135.84,137.25) .. (135.44,138) .. controls (135.05,138.76) and (133.92,138.95) .. (132.92,138.42) .. controls (131.92,137.9) and (131.43,136.86) .. (131.83,136.1) -- cycle ;
\draw  [color={rgb, 255:red, 155; green, 155; blue, 155 }  ,draw opacity=0.35 ] (135.3,115.04) .. controls (134.51,114.72) and (134.22,113.61) .. (134.64,112.56) .. controls (135.06,111.51) and (136.05,110.93) .. (136.84,111.25) .. controls (137.63,111.57) and (137.92,112.68) .. (137.5,113.73) .. controls (137.07,114.77) and (136.09,115.36) .. (135.3,115.04) -- cycle ;
\draw  [color={rgb, 255:red, 155; green, 155; blue, 155 }  ,draw opacity=0.35 ] (94.83,96.13) .. controls (95.2,94.96) and (96.56,94.34) .. (97.87,94.75) .. controls (99.18,95.16) and (99.95,96.45) .. (99.58,97.62) .. controls (99.21,98.8) and (97.85,99.42) .. (96.54,99) .. controls (95.23,98.59) and (94.46,97.31) .. (94.83,96.13) -- cycle ;
\draw  [color={rgb, 255:red, 155; green, 155; blue, 155 }  ,draw opacity=0.35 ] (109.25,158.1) .. controls (109.32,156.87) and (110.17,155.93) .. (111.14,155.99) .. controls (112.11,156.05) and (112.83,157.1) .. (112.75,158.33) .. controls (112.67,159.55) and (111.82,160.5) .. (110.86,160.44) .. controls (109.89,160.38) and (109.17,159.33) .. (109.25,158.1) -- cycle ;
\draw  [color={rgb, 255:red, 155; green, 155; blue, 155 }  ,draw opacity=0.35 ] (106.96,155.83) .. controls (107.26,155.02) and (108.08,154.58) .. (108.79,154.84) .. controls (109.5,155.11) and (109.84,155.97) .. (109.54,156.78) .. controls (109.25,157.59) and (108.43,158.03) .. (107.71,157.77) .. controls (107,157.5) and (106.66,156.64) .. (106.96,155.83) -- cycle ;
\draw  [color={rgb, 255:red, 155; green, 155; blue, 155 }  ,draw opacity=0.35 ] (96.25,161.85) .. controls (96.55,161.04) and (97.37,160.6) .. (98.08,160.86) .. controls (98.79,161.12) and (99.13,161.99) .. (98.83,162.8) .. controls (98.54,163.61) and (97.72,164.05) .. (97,163.79) .. controls (96.29,163.52) and (95.95,162.65) .. (96.25,161.85) -- cycle ;
\draw  [color={rgb, 255:red, 155; green, 155; blue, 155 }  ,draw opacity=0.35 ] (151.42,156.41) .. controls (152.66,155.58) and (154.58,156.26) .. (155.69,157.94) .. controls (156.81,159.62) and (156.7,161.66) .. (155.45,162.49) .. controls (154.21,163.32) and (152.29,162.63) .. (151.18,160.95) .. controls (150.06,159.27) and (150.17,157.24) .. (151.42,156.41) -- cycle ;
\draw [color={rgb, 255:red, 155; green, 155; blue, 155 }  ,draw opacity=0.35 ]   (167.87,142.55) .. controls (142.24,141.01) and (145.01,146.47) .. (158.06,158.14) ;
\draw  [color={rgb, 255:red, 88; green, 80; blue, 141 }  ,draw opacity=1 ] (415.37,53.28) .. controls (435.37,43.28) and (496.37,81.28) .. (476.37,101.28) .. controls (456.37,121.28) and (439.63,157.72) .. (461,186) .. controls (482.37,214.28) and (428.37,217.28) .. (371,186) .. controls (313.63,154.72) and (395.37,63.28) .. (415.37,53.28) -- cycle ;
\draw    (203.28,132.14) -- (305.37,132.14) ;
\draw [shift={(307.37,132.14)}, rotate = 180] [color={rgb, 255:red, 0; green, 0; blue, 0 }  ][line width=0.75]    (10.93,-3.29) .. controls (6.95,-1.4) and (3.31,-0.3) .. (0,0) .. controls (3.31,0.3) and (6.95,1.4) .. (10.93,3.29)   ;
\draw  [color={rgb, 255:red, 0; green, 63; blue, 92 }  ,draw opacity=1 ][line width=0.75]  (241.12,173.5) -- (235.41,173.5) -- (235.41,167.79) ;
\draw  [color={rgb, 255:red, 0; green, 63; blue, 92 }  ,draw opacity=1 ][line width=0.75]  (241.12,173.5) -- (241.12,167.79) -- (246.83,167.79) ;
\draw  [color={rgb, 255:red, 0; green, 63; blue, 92 }  ,draw opacity=1 ][line width=0.75]  (252.54,173.5) -- (252.54,167.79) -- (246.83,167.79) ;
\draw  [color={rgb, 255:red, 0; green, 63; blue, 92 }  ,draw opacity=1 ][line width=0.75]  (252.54,173.5) -- (246.83,173.5) -- (246.83,179.21) ;
\draw  [color={rgb, 255:red, 0; green, 63; blue, 92 }  ,draw opacity=1 ][line width=0.75]  (252.54,184.92) -- (252.54,179.21) -- (246.83,179.21) ;
\draw  [color={rgb, 255:red, 0; green, 63; blue, 92 }  ,draw opacity=1 ][line width=0.75]  (241.12,179.21) -- (235.41,179.21) -- (235.41,184.92) ;
\draw  [color={rgb, 255:red, 0; green, 63; blue, 92 }  ,draw opacity=1 ][line width=0.75]  (241.12,179.21) -- (241.12,184.92) -- (246.83,184.92) ;
\draw  [color={rgb, 255:red, 0; green, 63; blue, 92 }  ,draw opacity=1 ][line width=0.75]  (235.41,167.79) -- (235.41,162.08) -- (241.12,162.08) ;
\draw  [color={rgb, 255:red, 0; green, 63; blue, 92 }  ,draw opacity=1 ][line width=0.75]  (252.54,179.21) -- (252.54,184.92) -- (246.83,184.92) ;
\draw  [color={rgb, 255:red, 0; green, 63; blue, 92 }  ,draw opacity=1 ][line width=0.75]  (246.7,156.39) -- (246.7,162.1) -- (252.41,162.1) ;
\draw  [color={rgb, 255:red, 0; green, 63; blue, 92 }  ,draw opacity=1 ][line width=0.75]  (246.7,156.39) -- (252.41,156.39) -- (252.41,150.68) ;
\draw  [color={rgb, 255:red, 0; green, 63; blue, 92 }  ,draw opacity=1 ][line width=0.75]  (246.7,144.97) -- (252.41,144.97) -- (252.41,150.68) ;
\draw  [color={rgb, 255:red, 0; green, 63; blue, 92 }  ,draw opacity=1 ][line width=0.75]  (246.7,144.97) -- (246.7,150.68) -- (240.99,150.68) ;
\draw  [color={rgb, 255:red, 0; green, 63; blue, 92 }  ,draw opacity=1 ][line width=0.75]  (235.28,144.97) -- (240.99,144.97) -- (240.99,150.68) ;
\draw  [color={rgb, 255:red, 0; green, 63; blue, 92 }  ,draw opacity=1 ][line width=0.75]  (240.99,156.39) -- (240.99,162.1) -- (235.28,162.1) ;
\draw  [color={rgb, 255:red, 0; green, 63; blue, 92 }  ,draw opacity=1 ][line width=0.75]  (240.99,156.39) -- (235.28,156.39) -- (235.28,150.68) ;
\draw  [color={rgb, 255:red, 0; green, 63; blue, 92 }  ,draw opacity=1 ][line width=0.75]  (240.99,144.97) -- (235.28,144.97) -- (235.28,150.68) ;
\draw  [color={rgb, 255:red, 0; green, 63; blue, 92 }  ,draw opacity=1 ][line width=0.75]  (269.52,173.5) -- (275.23,173.5) -- (275.23,167.79) ;
\draw  [color={rgb, 255:red, 0; green, 63; blue, 92 }  ,draw opacity=1 ][line width=0.75]  (269.52,173.5) -- (269.52,167.79) -- (263.81,167.79) ;
\draw  [color={rgb, 255:red, 0; green, 63; blue, 92 }  ,draw opacity=1 ][line width=0.75]  (258.1,173.5) -- (258.1,167.79) -- (263.81,167.79) ;
\draw  [color={rgb, 255:red, 0; green, 63; blue, 92 }  ,draw opacity=1 ][line width=0.75]  (258.1,173.5) -- (263.81,173.5) -- (263.81,179.21) ;
\draw  [color={rgb, 255:red, 0; green, 63; blue, 92 }  ,draw opacity=1 ][line width=0.75]  (258.1,184.92) -- (258.1,179.21) -- (263.81,179.21) ;
\draw  [color={rgb, 255:red, 0; green, 63; blue, 92 }  ,draw opacity=1 ][line width=0.75]  (269.52,179.21) -- (275.23,179.21) -- (275.23,184.92) ;
\draw  [color={rgb, 255:red, 0; green, 63; blue, 92 }  ,draw opacity=1 ][line width=0.75]  (269.52,179.21) -- (269.52,184.92) -- (263.81,184.92) ;
\draw  [color={rgb, 255:red, 0; green, 63; blue, 92 }  ,draw opacity=1 ][line width=0.75]  (275.23,167.79) -- (275.23,162.08) -- (269.52,162.08) ;
\draw  [color={rgb, 255:red, 0; green, 63; blue, 92 }  ,draw opacity=1 ][line width=0.75]  (258.1,179.21) -- (258.1,184.92) -- (263.81,184.92) ;
\draw  [color={rgb, 255:red, 0; green, 63; blue, 92 }  ,draw opacity=1 ][line width=0.75]  (263.94,156.39) -- (263.94,162.1) -- (258.23,162.1) ;
\draw  [color={rgb, 255:red, 0; green, 63; blue, 92 }  ,draw opacity=1 ][line width=0.75]  (263.94,156.39) -- (258.23,156.39) -- (258.23,150.68) ;
\draw  [color={rgb, 255:red, 0; green, 63; blue, 92 }  ,draw opacity=1 ][line width=0.75]  (263.94,144.97) -- (258.23,144.97) -- (258.23,150.68) ;
\draw  [color={rgb, 255:red, 0; green, 63; blue, 92 }  ,draw opacity=1 ][line width=0.75]  (263.94,144.97) -- (263.94,150.68) -- (269.65,150.68) ;
\draw  [color={rgb, 255:red, 0; green, 63; blue, 92 }  ,draw opacity=1 ][line width=0.75]  (275.36,144.97) -- (269.65,144.97) -- (269.65,150.68) ;
\draw  [color={rgb, 255:red, 0; green, 63; blue, 92 }  ,draw opacity=1 ][line width=0.75]  (269.65,156.39) -- (269.65,162.1) -- (275.36,162.1) ;
\draw  [color={rgb, 255:red, 0; green, 63; blue, 92 }  ,draw opacity=1 ][line width=0.75]  (269.65,156.39) -- (275.36,156.39) -- (275.36,150.68) ;
\draw  [color={rgb, 255:red, 0; green, 63; blue, 92 }  ,draw opacity=1 ][line width=0.75]  (269.65,144.97) -- (275.36,144.97) -- (275.36,150.68) ;
\draw [color={rgb, 255:red, 0; green, 63; blue, 92 }  ,draw opacity=1 ][line width=0.75]    (252.41,162.1) -- (258.23,162.1) ;

\draw (118.82,58.68) node   [align=center] {Sphere\\$z=0$};
\draw (213.82,100) node [anchor=north west][inner sep=0.75pt]   [align=left] {PH mapping};
\draw (376,104) node [anchor=north west][inner sep=0.75pt]   [align=left] {Lagrangian \\volume\\$z=127$};
\draw  [draw opacity=0][fill={rgb, 255:red, 255; green, 255; blue, 255 }  ,fill opacity=1 ]  (95.11,185.11) -- (141.17,185.11) -- (141.17,210.22) -- (95.11,210.22) -- cycle  ;
\draw (118.14,197.66) node  [color={rgb, 255:red, 74; green, 74; blue, 74 }  ,opacity=1 ,rotate=-0.16]  {$6\, r_{500}$};
\draw (362.5,194.9) node [anchor=north west][inner sep=0.75pt]    {$\Gamma$};

\end{tikzpicture}
\end{minipage}
    \caption{\textit{Left:} diagram of a dark matter (DM) halo at $z=0$ selected from the parent simulation and centred on a $6\, r_{500}$ sphere (red). The $r_{500}$ sphere is shown in green and schematic of filaments and substructures in grey. Each DM particle in the $6\, r_{500}$ sphere is then mapped onto its position in the initial conditions via a Peano-Hilbert (PH) space-filling curve (a 2D example of PH curve is shown in the middle). \textit{Right:} the positions of the particles in the initial conditions at $z=127$ define a Lagrangian volume $\Gamma$, which is then masked and prepared for the resolution refinement.}
    \label{fig:ics-map-particles}
\end{figure}

As shown in the diagram of Fig.~\ref{fig:ics-map-particles}, we consider all particles within a sphere of radius $6\, r_{500}$ around a halo at $z=0$ in the parent simulation. Our goal is now to track the particles back in time to the initial conditions (ICs), and mark their original positions before the onset non-linear collapse. This task can be performed in two ways:
\begin{itemize}
    \item marking the unique identifier (ID) of all particles in the soon-to-be high-resolution sphere, $p^{\{z=0\}}_{\rm ID}{(r<6\, r_{500})}$, and then search for the particles in the ICs that match the same IDs. The search operation can be performed using a \textit{lookup table} algorithm, such as the one implemented in    \begin{center} \href{https://numpy.org/doc/stable/reference/generated/numpy.isin.html}{\texttt{numpy.isin($*$, assume\_unique$=$True, kind$=$"table")}} \end{center} (using the \texttt{Python} language), which can be sped up by assuming unique IDs. Nevertheless, this strategy can become problematic in zooms with large number of particles, $N_{\rm DM} \sim 10^9$, where there is insufficient random-access memory (RAM) for the search algorithm to operate.

    \item The second method, used in our simulations and most of the zoom-based projects run within the Virgo Consortium, bypasses the search algorithm by \textit{encoding} the initial ($z=127$) 3D coordinates in the particle IDs, and then \textit{decoding} this information from any snapshot in a deterministic fashion. The encoding is performed by \textsc{IC\_Gen} during the generation of the ICs of the parent simulation (step 3 of Fig.~\ref{fig:simulation-pipeline}) using a space-filling Peano-Hilbert (PH) curve.\footnote{The PH curve is a one-dimensional function which can be constructed iteratively by progressively bisecting the domain. In the middle of Fig.~\ref{fig:ics-map-particles}, we show the third iteration of the PH construction. As the number $k$ of bisections increases, the number of distinct segments also increases, until it fills the domain as $k\rightarrow \infty$. The PH curve can be generalised to map a domain of $n>1$ dimensions while retaining its properties.} Given that a sufficiently dense PH curve can map each point in the domain to a unique 1D position along the curve, we can use this method to associate each particle to a unique cell centred in one of the vertices of the PH curve. Under these conditions, the integer identifier of each vertex also identifies the unique particle in that cell, establishing a bijective relation between 3D coordinates $\mathbf{r}_i$ and integer particle ID\footnote{The relation if one-to-one, i.e. bijective, only as long as there is at most one particle per PH cell. EAGLE-type simulations traditionally stopped after 14 iterations generated 14-bit PH indices (see also the \texttt{nbit} parameter in the Listing \ref{app1:parent-sim-ics}), while we use 21-bit PH indices, which can hold more particles and can identify them uniquely even when two particles are particularly close together.}
    \begin{equation}
        p^{\{z=0\}}_{\rm ID}{(r<6\, r_{500})} \Longleftrightarrow\mathbf{r}_i^{\{z=127\}}\in \Gamma,
    \end{equation}
    where $\Gamma$ is the Lagrangian volume in the ICs at $z=127$ that collapses into a sphere by $z=0$. Unlike the previous method, this operation does not require the copy of particle arrays in memory and is therefore more computationally efficient.
\end{itemize}

\subsection{Defining the Lagrangian volume with topological closure}
\label{sec:zoom-setup:lagrangian-region}

\begin{figure}
    \centering
    \includegraphics[width=\textwidth]{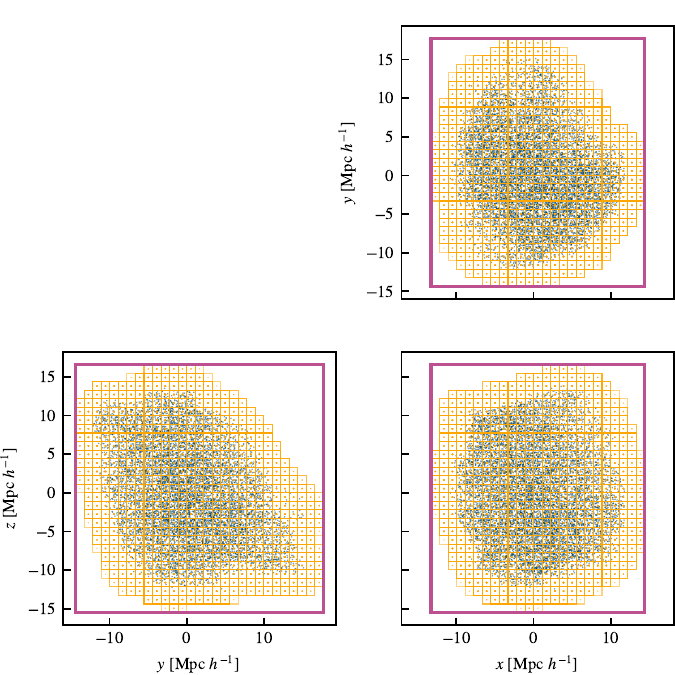}
    \caption{2D projections of the DM particles (blue) mapped onto the initial conditions starting from a $6\, r_{500}$ sphere at $z=0$ around \texttt{L0300N0564\_VR18} (abbreviated as \texttt{VR18}) at a resolution 8 times lower than EAGLE. This object is the \textit{cluster} referred to in \cite{2023MNRAS.520.3164A} and Chapters \ref{chapter:5} and \ref{chapter:6}. The Lagrangian volume around the DM particles is masked using cubic bins (yellow) and expanded using the \textit{topological closure} technique. The cuboid enclosing the mask is shown in purple. The dimensions are quoted in units of $h^{-1}$, with $h=0.6766$ \citep{planck.2018.cosmology}, and are re-scaled to the centre of mass of the particle ensemble.}
    \label{fig:zoom-mask-VR18}
\end{figure}

\begin{figure}
    \centering
    \includegraphics[width=\textwidth]{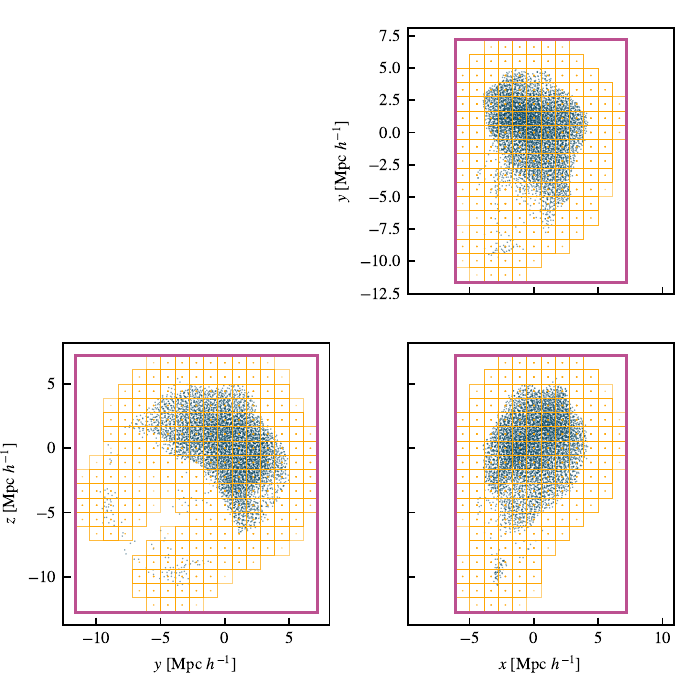}
    \caption{As in Fig.~\ref{fig:zoom-mask-VR18}, but for \texttt{L0300N0564\_VR2915} (abbreviated as \texttt{VR2915}) at a resolution 8 times lower than EAGLE. This object is the \textit{group} referred to in \cite{2023MNRAS.520.3164A} and Chapters \ref{chapter:5} and \ref{chapter:6}.}
    \label{fig:zoom-mask-VR2915}
\end{figure}

Once the ensemble of particles $i = \{\mathbf{r}_i\in \Gamma\}$ has been obtained by decoding the PH indices, it is necessary to determine the extent of the volume $\Gamma$. Strictly, point particles do not define a bounding surface, but can be used to construct one that contains them. Our strategy involves binning the 3D domain and counting the number of particles $N'$ in each cubic bin. An example of this binning procedure is shown in Fig.~\ref{fig:zoom-mask-VR18}, where we decoded the initial positions of DM particles producing the cluster \texttt{VR18} (blue) and binned the 3D domain with adjacent cubes with side $1.1$ Mpc $h^{-1}$ (we show the same for the group \texttt{VR2915} in Fig.~\ref{fig:zoom-mask-VR2915}). The outline of the bins containing particles defines the volume $\Gamma$, which will be enriched with higher-resolution particles for the zoom re-simulation.

By inspecting these examples of mask, we note that (i) the masked volume covered by the cubic bins extends further than the ensemble $\mathbf{r}_i$ and (ii) some of the particles (particularly in Fig.~\ref{fig:zoom-mask-VR2915}) are not covered by the mask. Adding a \textit{padding} region around the particle ensemble, i.e. increasing the volume of $\Gamma$, can be advantageous for creating a slightly larger high-resolution region during the collapse of the object from $z=127 \rightarrow 0$. Such a scenario is desirable for studies of galaxies or subhalos in the suburbs of large objects \citep[e.g. the Hydrangea simulations,][]{2017MNRAS.470.4186B}. However, this choice would also result in a larger particle load, and ultimately the computational cost would become higher than originally estimated.

In an attempt to balance the benefits of the higher resolution with the run-time costs, masks for Langrangian volumes of zoom simulations should follow the bounds of the particle ensemble as closely as possible, so that the buffer region would be reduced to a minimum. The mask of \texttt{VR18} in Fig.~\ref{fig:zoom-mask-VR18} is an example of a mask where the padding could have been reduced further while avoiding particles in the periphery to be outside the mask. The Lagrangian volume of \texttt{VR18} is, in fact, largely convex and this geometrical property facilitates the definition of a \textit{fully-connected} mask which captures most of, if not all, the particles.

The case of \texttt{VR2915} in Fig.~\ref{fig:zoom-mask-VR2915} is different. Firstly, the particles are distributed in a concave space; secondly, particles are not clustered in a single region of space: one disconnected region is visible at the bottom of the $x-z$ projection, with an extended arc-like structure in the $y-z$ projection. Minimising the padding of this mask would either (i) omit the particles in secondary structures and only retain the central ensemble or (ii) include the particles in the periphery by forming a satellite mask around them and leaving the satellite mask disconnected from the central mask. Previous tests (unpublished) conducted by members of the EAGLE-XL Collaboration on similar zoom-in set-ups showed low-resolution DM particles being loaded either (i) in place of the omitted high-resolution particles or (ii) in the gaps between the disconnected mask domains. The low-resolution particles, whose sole role is to provide the surrounding tidal gravitational field, instead actively participate in the collapse of the halo and are often found inside the high-resolution sphere ($6\, r_{500}$ in our case) defined at the start. The high-resolution region is therefore said to be \textit{contaminated} by low-resolution DM particles.

\begin{figure}
    \centering
    \includegraphics[width=\textwidth]{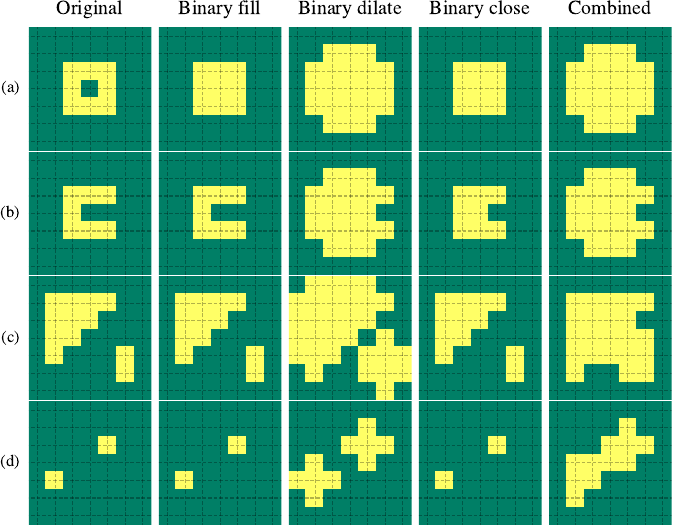}
    \caption{Illustration of the three masking algorithms (\textit{binary fill}, \textit{binary dilate} and \textit{binary close}) used to define the Lagrangian volume in the initial conditions. The diagram are 2D binary matrices, where green corresponds to 0 and yellow to 1. The rows (a-d) show four different initial mask topologies 2D (\textit{original}) and on the right the corresponding results when each algorithm is applied individually. The last column shows the result of applying the three algorithms combined in the order fill-dilate-close. The latter choice was found to perform best in resolving topologically disconnected regions (case c and d), despite increasing the spatial coverage. In this example and in the ICs masking pipeline we use a structuring element with connectivity $l=1$.}
    \label{fig:masking-algorithm}
\end{figure}

The particular definition of the Lagrangian volume and the mask affects the level of contamination of the high-resolution region. Here, we present a novel approach, referred to as \textit{topological closure}, to mitigate contamination by adding more cubic cells to the mask while minimising the computational overhead associated with the increased particle load. This method acts on the initial 3D mask, treated as a discrete binary object; zeros are associated with cubic elements with no or too few particles, and ones to cubic elements with enough particles. Our method consists of three steps, implemented with \texttt{scipy} \citep{virtanen2020scipy}:
\begin{enumerate}
    \item we run a \textbf{binary filling} operation, which runs a cubic (or square if in 2D) structuring element\footnote{A structuring element is a mathematical object representing a particular geometrical shape in matrix form. As an example, a square structuring element in 2D with a connectivity-length $l=2$ can be represented by $\left[\begin{array}{cc}
       1 & 1 \\
       1 & 1 \\
    \end{array}\right]$. The rules according to which the structuring element interacts with the mask (or input image) defines a \textit{morphological operation}. These techniques are applications of concepts in topology and are commonplace in image processing \citep{dougherty1992introduction}.} along the boundary of the mask to detect empty regions (zeros) surrounded by masked regions (ones). If such empty region exists, the central zero is switched to one, filling the topological hole. This strategy is effective in filling holes with size comparable to the structuring element, such as the example shown in row \textit{a} of Fig. \ref{fig:masking-algorithm}. However, binary filling is less effective on concave boundaries (row \textit{b}), larger holes or wider gaps (rows \textit{c} and \textit{d}). 

    \item Where the gaps are larger than the cell size, we run a \textbf{binary dilation} algorithm to produce a padding around the mask boundary. This technique produces an 1-cell-wide extrusion of the outer boundary, which is the primary cause of the increased particle load. However, the same extrusion in inner boundaries, if present, can fill small holes which may have been missed by the binary filling or reduces the size of larger holes. The examples in Fig.~\ref{fig:masking-algorithm} confirm that this method is highly effective in smoothing concave boundaries and reducing gaps, but leads to an increase in volume, which may impact the run-time of a zoom-in simulation. This step is crucial to reduce the gaps between topologically disconnected regions (rows \textit{c} and \textit{d}).

    \item Finally, the \textbf{binary close} operation targets concave boundaries specifically (row \textit{b} in Fig. \ref{fig:masking-algorithm}). Mathematically, it is defined as the erosion of a dilation and it is an effective method for softening sharp boundaries and closing small gaps.
\end{enumerate}

In our pipeline, we apply the fill, dilate and close operations in this order (determined heuristically) and produced excellent results, as demonstrated in the \textit{Combined} column of Fig. \ref{fig:masking-algorithm}. Not only is the scheme capable of closing small gaps in simple topological configurations (rows a and b), but it can also connect disconnected domains (rows c and d) which cannot be resolved with either of the operations individually. We expect these scenario to be representative of initial Lagrangian volumes of small DM halos, such as \texttt{VR2915} (see Fig.~\ref{fig:zoom-mask-VR2915}).

By default, the topological closure method is configured with a single iteration of fill, dilate and close. The first step is only applied once, but the dilation and closing operations can be repeated by specifying the number of iterations in \texttt{topology\_fill\_holes}, \texttt{topology\_dilation\_niter} and \texttt{topology\_closing\_niter} under the \textsc{ topological closure} section (see also the parameter file for \texttt{VR18} in Listing \ref{app1:vr18-mask}).

The topological closure technique is integrated in the zoom\hyp{}simulation pipeline used by members of the Virgo Consortium. By leveraging MPI functionalities, our new algorithm can scale to large particle loads ($N_{\rm DM} \gtrsim 10^{10}$), as well as Lagrangian volumes with complex morphology and disconnected domains. The latest stable version of the \textit{masking} pipeline can be accessed and downloaded from the following GitHub repository:
\begin{center} \href{https://github.com/stuartmcalpine/zoom_mask_creator}{https://github.com/stuartmcalpine/zoom\_mask\_creator}. \end{center} 
A different private shared repository was used for the development of the topological closure method. Topological closure has been successfully used in production runs as part of the Sibelius project \citep{2022MNRAS.512.5823M} and most zoom\hyp{}in simulations assisting the model-calibration phase in the EAGLE-XL project \citep{2023MNRAS.520.3164A}.

\subsection{Particle-resolution refinement}
\label{sec:zoom-setup:refinement}

Masking the Lagrangian volume $\Gamma$ leads to the next step, number 8 in Fig.~\ref{fig:simulation-pipeline}, which involves populating $\Gamma$ with high-resolution particles and setting up the low-resolution tidal field around the high-resolution volume. At the end of this procedure, the initial conditions for the zoom simulation will be produced, ready to be run with a numerical integration code, such as \swift.

\begin{figure}
    \centering
    \includegraphics[width=\textwidth]{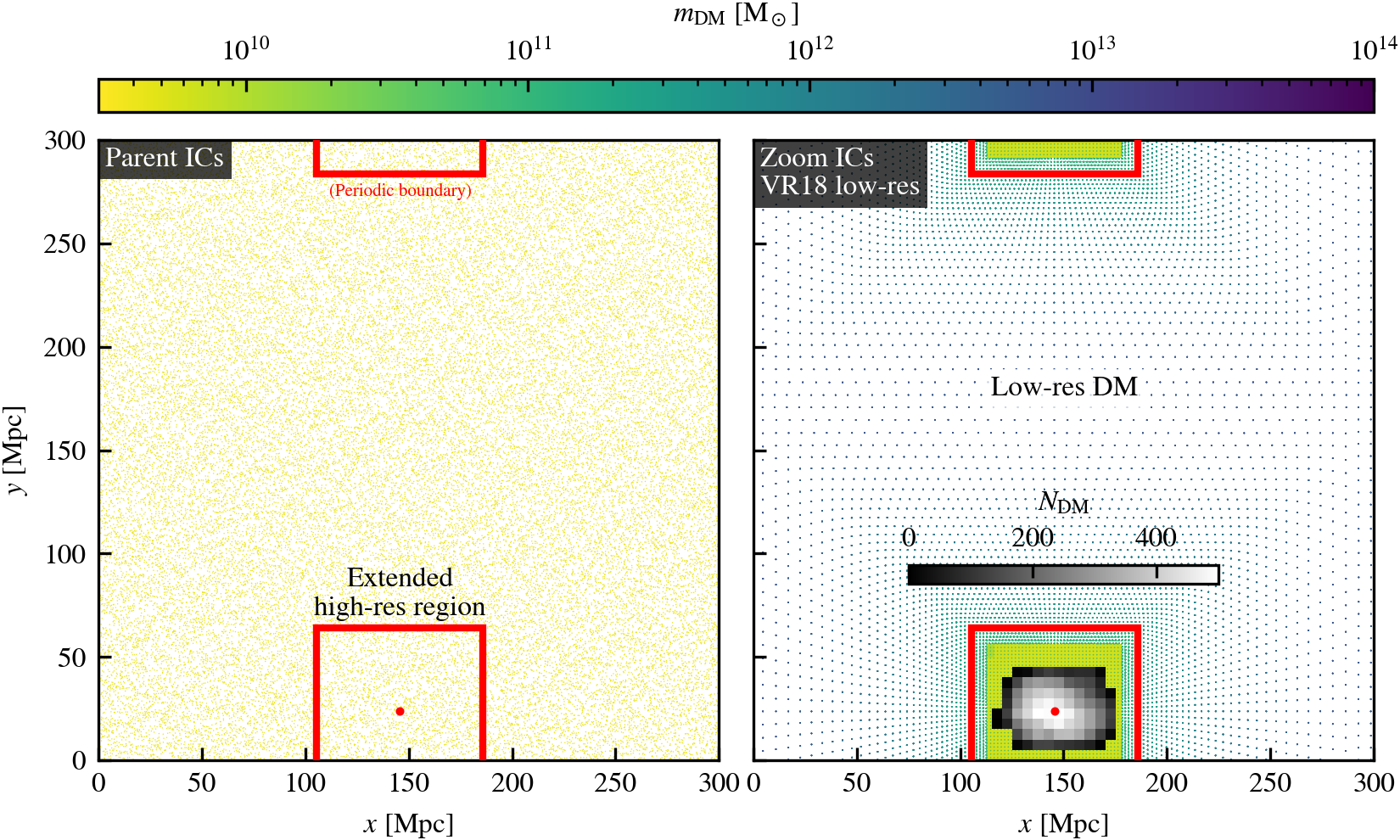}
    \caption{\textit{Left:} projected $x-y$ positions of the DM particles in the initial conditions (ICs) of the parent simulation. In this space, we indicate the centre and the perimeter (red) of the \textit{extended} high-resolution region from the \textsc{Panphasia} parameter file (Listing~\ref{app1:vr18-ics}). The high-resolution region extends beyond the periodic boundary at $y=0$ and appears on the top of the space. In both panels, the DM particles are coloured according to their mass $m_{\rm DM}= 5.95 \times 10^9$ M$_\odot$, and in the parent simulation, such mass resolution is constant throughout the volume. \textit{Right:} projected $x-y$ positions of the DM particles in the initial conditions (ICs) of the zoom-in simulation. The number $N_{\rm DM}$ of high-resolution DM particles is shown by the black-white colour map within the extended high-res region. The other particles represent the low-resolution \textit{background} dark matter tidal field. As we move further away from the refined region, the low-res DM particles become more massive and more sparse, indicating a progressively decreasing resolution.}
    \label{fig:zoom-setup-vr18}
\end{figure}

\begin{figure}
    \centering
    \includegraphics[width=\textwidth]{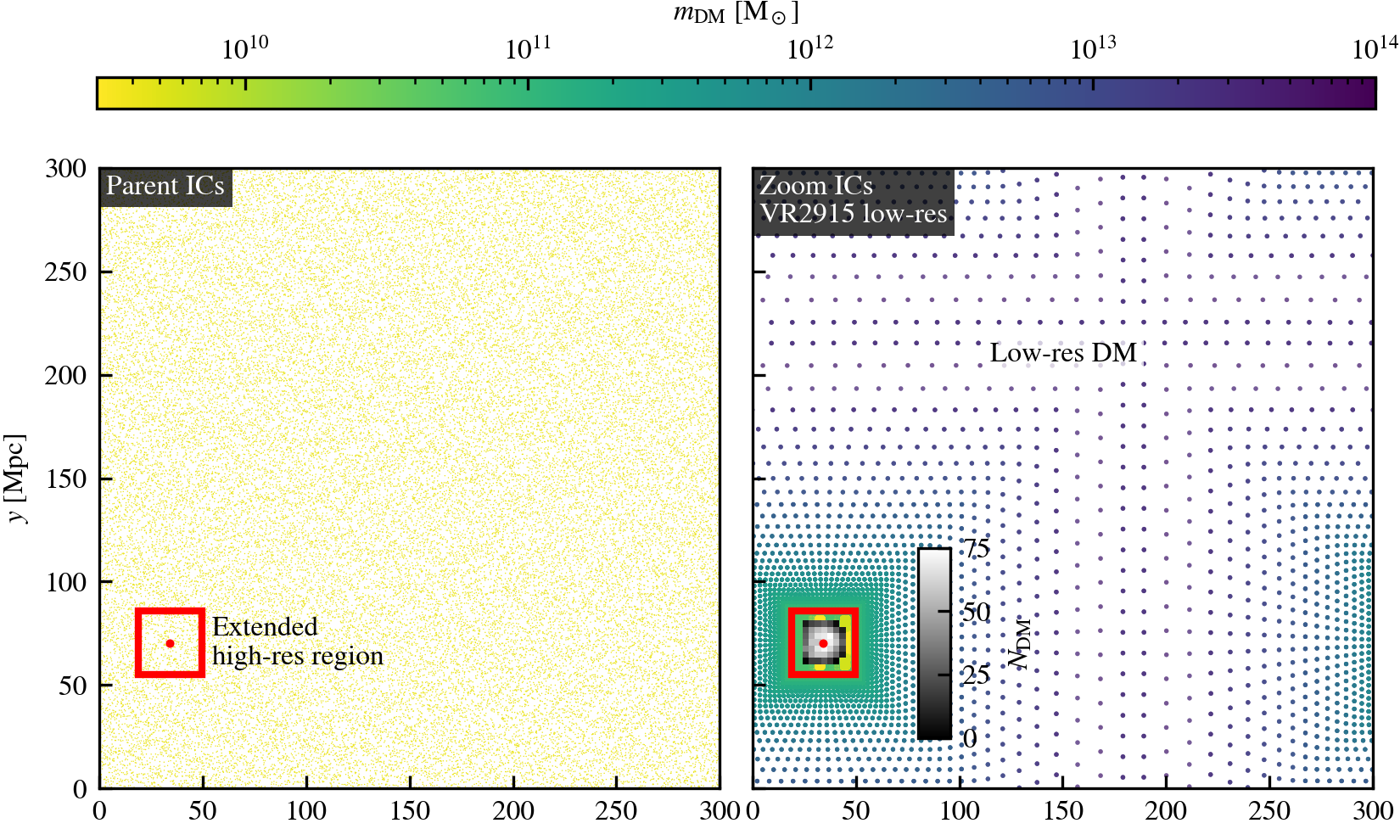}
    \caption{As in Fig.~\ref{fig:zoom-setup-vr18}, but for the group \texttt{VR2915}.}
    \label{fig:zoom-setup-vr2915}
\end{figure}

To generate the particle load used in our simulations, we used a pipeline based on the code in the following repository \begin{center}
    \href{https://github.com/stuartmcalpine/zoom_particle_load_creator}{https://github.com/stuartmcalpine/zoom\_particle\_load\_creator}.
\end{center}
The particle load is applies in two phases, starting from (1) the high-resolution region and following with (2) the low-resolution boundary particles.
\begin{enumerate}
    \item The particle resolution in $\Gamma$ is controlled by the parameter \texttt{n\_particles} specified in a \texttt{YAML} file. This value expresses the total number of particles if the entire volume were to be simulated at high resolution. In Tab.~\ref{tab:resolutions-particle-load}, we report the number of particles used to generate our groups and cluster sample at three resolution. We also compare them to the EAGLE Ref resolution. Not all the high-resolution particles are allocated to the simulation volume: only those within $\Gamma$ are kept, while those outside $\Gamma$ are masked out.

    \item The low-resolution DM particles, by definition, have a larger mass than the high-resolution DM particles. Outside of the masked region, the low-resolution (or boundary) particles are allocated in cubic layers, commonly referred to as \textit{skins}, starting from the outer boundary of $\Gamma$ and moving further away to fill the entire simulation volume. The first skin extends just outside the cuboid enclosing the mask, shown in purple in Figs.~\ref{fig:zoom-setup-vr18} and \ref{fig:zoom-setup-vr2915}. The particles in this this region are a factor of a few more massive than the high-resolution ones. As more skins are added at greater distances from the high-resolution region, their thickness increases, and so do the volume and the mass of the DM particles within that skin.    
\end{enumerate}

\begin{table}
    \centering
    \caption{Summary of the particle resolutions in the refined region used to generate the particle load. We also specify the equivalent resolution with respect to EAGLE Ref, an alias used in our database and the number of CPU cores requested to generate the particle displacements with \textsc{IC\_Gen}. $^\star$For these calculations, we used the Cosma-7 HPC system, equipped with compute nodes running 28 CPU cores each.}
    \label{tab:resolutions-particle-load}
    \begin{tabular}{ccccc}
    \toprule
        Resolution & \texttt{n\_particles} & Equivalent to & Alias & \textsc{IC\_Gen} CPU cores$^\star$\\
        \midrule
        Parent & $546^3$ & -- & -- & 24 \\
        Low-res & $2254^3$ & $1/8 \times$ EAGLE & \texttt{-8res} &56\\ 
        Mid-res & $4508^3$ & $1 \times$ EAGLE & \texttt{+1res} &112\\ 
        High-res & $9016^3$ & $8 \times$ EAGLE & \texttt{+8res} &392\\ 
    \bottomrule
    \end{tabular}
\end{table}

In Figs.~\ref{fig:zoom-setup-vr18} and \ref{fig:zoom-setup-vr2915} we illustrate the particle load set-up for the small and larger halos which we used as case studies in this section. Firstly, we consider cluster \texttt{VR18} in Fig.~\ref{fig:zoom-setup-vr18}. On the left, we show a fraction ($1/1000^{\rm th}$) of the particles in the DM parent box at the initial redshift ($z=127$), projected along the $z$-axis and we highlight the cubic boundary enclosing (red) enclosing the high-resolution region. We also indicate the position of the centre of mass of the high-resolution region. In this case, the bounding region overlaps with the periodic boundary of the simulation volume and, therefore, the high-resolution boundary wraps around the box, as specified. On the right, we show the same high-resolution cubic boundary containing the \textit{true} high-resolution domain ($\Gamma$) at $z=127$. The number of high-resolution DM particles $N_{\rm DM}$ is shown as a grey-scale colour map. Just outside $\Gamma$, the first cubic boundary skin of low-resolution DM particle is visible in a green-yellow colour, corresponding to $m_{\rm DM}\approx8\times 10^9$ M$_\odot$. The red boundary, drawn based on the value specified in the \textsc{IC\_Gen} parameter file, is slightly larger than $\Gamma$ because the code requires a buffer region to apply a smooth transition when computing the LPT displacements between the high- and low-resolution regimes. Outside the high-resolution region, we only project DM particles within a slice along the $z$-axis as thick as the side of the high-resolution boundary. The mass of these particles, represented by the color, increases with their distance from the high-resolution region, and so does the spacing between them. The group \texttt{VR2915} in Fig.~\ref{fig:zoom-setup-vr2915}, instead, shows a smaller high-resolution region, which does not intersect any periodic boundary of the original simulation box.

\section{Numerical hydrodynamics}
\label{sec:numerical-sph}

Most of the observables which can be predicted from simulations are associated with baryonic physics. These processes are largely driven by the gas and, together with gravity, they are one of the foundations nearly all cosmological simulations today. Modelling the hydrodynamics of the gas is a computational challenge that attracted the interest of cosmologists and astrophysicists since the 1980s. A vast number of methods to run hydrodynamic simulations have been devised, all with their strengths and shortcomings. In a broad context, numerical methods which tend to be more stable, converged and rigorous, such as the particle-in-cell \citep{eastwood1981computer} or finite volume \citep{versteeg2007introduction} methods, require large computational resources, which hinder their capability to scale to large number of particles and highly inhomogeneous fluids.

Among slightly less stable, but computationally cheaper methods to model the dynamics of a continuous fluid with discrete elements, smoothed particle hydrodynamics (SPH) is the technique of choice for numerous simulation projects across disciplines. The SPH method, initially formulated by \cite{1977AJ.....82.1013L} and \cite{1977MNRAS.181..375G}, became one of the schemes of choice for computational cosmologists because of one advantage it provides: adaptability. Unlike grid-based methods, SPH does not require a mesh, whose definition is often as laborious and resource-intensive as the time integration itself. It being a mesh-less method, allows SPH to retain its computational stability in low-density regions, as well as in high densities ones, even when the gas properties vary over several orders of magnitude. Such vast dynamic ranges are often found in astrophysical models of galaxies or clusters of galaxies; this natural adaptability of the SPH method makes it a key resource for many large-scale simulation projects \citep[see e.g.][for a review]{2012JCoPh.231..759P}.

In SPH, the fluid is modeled as a collection of discrete particles, each of which represents an \textit{interpolation point} of the fluid properties in 3D space. The point-like particles are convolved with a smoothing kernel $W(\mathbf{r} - \mathbf{r}_b, h)$, which depends on the position of the particle $\mathbf{r}$ with respect to its neighbour (often abbreviated \textit{ngb}) particles $\mathbf{r}_b$ and a smoothing length $h$, which sets the spatial scale of the kernel are  over a certain distance.

The first step of the SPH calculation involves defining the density $\rho(\mathbf{r})$ of the fluid at the position of a particle. We can compute this value by summing the mass $m$ of the particles within the smoothing kernel weighted by $W$ as follows
\begin{equation}
    \rho(\mathbf{r}) = \sum_{b}^{N_{\rm ngbs}} m_b\, W(\mathbf{r} - \mathbf{r}_b, h),
    \label{eq:sph-density}
\end{equation}
where $N_{\rm ngbs}$ is the number of neighbours. Notably, the profile of $W$ must follow the normalisation condition \citep{1992ARA&A..30..543M}:
\begin{equation}
    \int_V\, W(\mathbf{r} - \mathbf{r}_b, h)\, dV = 1 \qquad \forall ~ (|\mathbf{r} - \mathbf{r}_b|, h),
    \label{eq:sph-normalisation}
\end{equation}
so that the integral of the mass over the volume $V$ is equal to the total mass enclosed:
\begin{equation}
    \int_V\, \rho\, dV = \sum_b m_b.
\end{equation}

Extending the concept of the SPH density in Eq.~\eqref{eq:sph-density}, we can use the smoothing kernel to compute any quantity $Q(\mathbf{r})$ in an SPH-like fashion:
\begin{equation}
    Q(\mathbf{r}) = \sum_{b}^{N_{\rm ngbs}} Q_b\, W(\mathbf{r} - \mathbf{r}_b, h),
    \label{eq:sph-quantity-general}
\end{equation}
where $Q_b$ could represent, for instance, the temperature $T_b$ of a gas particle, its entropy $K_b$, or its metallicity $Z_b$. Because value of $Q$ at a position $\mathbf{r}$ only depends on a set number of neighbours, the properties of SPH particles are \textit{local}. The hydrodynamic forces are also local, in contrast with gravity, which affects the matter field also over long ranges. 

After defining the fundamental scheme for computing the gas properties, we discuss how these evolve over time by introducing the equations of motion for a fluid represented by the continuity equation (introduced in Section~\ref{sec:zeldovich-approximation}, Eq.~\ref{eq:continuity-top-hat})
\begin{equation}
    \partial_t \rho + \mathbf{\nabla} \cdot (\rho \mathbf{v}) = 0,
    \label{eq:continuity-sph}
\end{equation}
Euler's equation,
\begin{equation}
    \partial_t  \mathbf{v} + \frac{1}{\rho} \mathbf{\nabla}P +  \mathbf{\nabla}\Psi = 0,
    \label{eq:euler-sph}
\end{equation}
and the conservation of energy, encapsulated by the first law of thermodynamics
\begin{equation}
    \partial_t u = - \frac{P}{\rho} (\mathbf{\nabla} \cdot \mathbf{v}) - \Lambda\, \rho,
    \label{eq:energy-sph}
\end{equation}
where we recast the term containing the specific internal energy $u$ as a function of the density $\rho$ and a \textit{cooling function} $\Lambda = \Lambda(u, \rho)$ \citep{1992ARA&A..30..543M}. The particular functional form of $\Lambda$ dictates how the gas cools and dissipates energy over time and in most simulations it is calculated independently based on the chemical elements included in the sub-grid scheme. For non-radiative (also known as \textit{adiabatic}) simulations \citep[e.g.][]{vkb_2005}, $\Lambda = 0~\forall~(T, \rho)$ by definition. In Lagrangian formulations of SPH, such as in our case, the particles move with the flow, so that the advection term is set to zero, $ ( \mathbf{v} \cdot \mathbf{\nabla}) \cdot \mathbf{v} = 0$.

Finally, imposing an (artificial) equation of state allows to relate the three key thermodynamic quantities, namely the pressure $P$, the internal energy $u$ (related to the temperature $T$) and the density $\rho$. In our simulations, we assume an ideal gas law, with equation of state
\begin{equation}
    P = (\gamma - 1)\, u\, \rho,
    \label{eq:eos-sph}
\end{equation}
with the adiabatic index $\gamma=5/3$ for a mono-atomic gas. SPH schemes which use the smoothed density (Eq.~\ref{eq:sph-density}) and the equation of state (Eq.~\ref{eq:eos-sph}) to compute the pressure $P$ are labelled \textit{density-SPH}. In contrast, schemes which reconstruct the smoothed pressure field directly, $P=\sum_b (\gamma - 1) m_b u_b W(\mathbf{r} - \mathbf{r}_b, h)$, are labelled \textit{pressure-SPH}.

In the simulations introduced in \cite{2023MNRAS.520.3164A} and Chapters \ref{chapter:5} and \ref{chapter:6}, we model the hydrodynamics of the gas with \textsc{Sphenix} \citep{2022MNRAS.511.2367B}, the new default SPH scheme for the \swift code. Instead, the MACSIS simulations, produced with \textsc{Gadget-3} code, used the \textsc{Gadget-SPH} implementation. Although \textsc{Gadget-SPH} and \textsc{Sphenix} are both density-SPH schemes, they use different approaches to determine the equations of motions. We will now summarise the key differences.

\subsubsection*{The \textsc{Sphenix} hydrodynamics scheme}
The first formulation of SPH was derived from the equation of state in Eq.~\eqref{eq:eos-sph}, which links the density and the internal energy. SPH schemes which adopt this formulation are often referred to as \textit{density-energy} or \textit{traditional}, and \textsc{Sphenix} is among these. As in other traditional SPH approaches, 

Following the recommendation of \cite{2022MNRAS.511.2367B}, we select the quartic spline (M5) SPH kernel, which was found to be a suitable compromise between the accuracy in reconstructing forces and computational cost.\footnote{SPH kernels with greater powers tend to provide a more accurate force reconstruction, at the expense of larger computational cost. In simulations of galaxy formation, spline orders greater than M5 were shown not to provide significant enough improvements in the hydrodynamics modelling to justify the increase in run-time \citep{2022MNRAS.511.2367B}.} The quartic spline is defined as
\begin{align}
    w(q) = \left\lbrace\begin{array}{lcr}
    \left(\frac{5}{2} - q\right)^4 - 5\, \left(\frac{3}{2} - q\right)^4 - 10\, \left(\frac{1}{2} - q\right)^4 & \text{if} & q < \frac{1}{2}\\ [2mm]
    \left(\frac{5}{2} - q\right)^4 - 5\, \left(\frac{3}{2} - q\right)^4 & \text{if} & \frac{1}{2} \leq q < \frac{3}{2}\\ [2mm]
    \left(\frac{5}{2} - q\right)^4 & \text{if} & \frac{3}{2} \leq q < \frac{5}{2}\\ [2mm]
    0 & \text{if} & q \geq \frac{5}{2}.\\
    \end{array}
\right.
\end{align}
Here, $q=r/h$ and is used to define the smoothing kernel
\begin{equation}
    W(r, h) = \kappa_{\zeta}\, w(r/h)/h^\zeta,
\end{equation}
with $\zeta$ the number of spatial dimensions. We configure \swift with $\zeta=3$ in all cosmological simulations, and $\zeta=2$ for the 2D hydrodynamics tests of Section~\ref{sec:computational-performance}. For $\zeta=3$, we have  $\kappa_{3} = 7/478\pi$. $\kappa_{\zeta}$ is a normalisation constant, required to satisfy the condition in Eq.~\eqref{eq:sph-normalisation} and depends on the number of dimensions \citep[see][for further details on the normalisation and other properties of spline kernels]{10.1111/j.1365-2966.2012.21439.x}.

The smoothing length $h$ is computed by solving the following equation
\begin{equation}
    \sum_b^{N_{\rm ngbs}}  W(\mathbf{r} - \mathbf{r}_b, h) = \left( \frac{\eta}{h} \right)^\zeta,
\end{equation}
where the parameter setting the resolution scale $\eta=1.2348$ is chosen to minimise the density reconstruction errors (see Section 3 of \citealt{2022MNRAS.511.2367B} and \citealt{10.1111/j.1365-2966.2012.21439.x} for a discussion). Under these conditions, the average number of neighbours is $N_{\rm ngbs}\approx58$. In full-physics simulations, however, $N_{\rm ngbs}$ can vary: for instance, we find that our cluster \texttt{VR18\_-8res} run with the Ref model of \cite{2023MNRAS.520.3164A} has $N_{\rm ngbs}\approx65$.

In the density-energy formalism, the evolution equation for $u$ can be derived from the Lagrangian \citep[see][for further details]{1992ARA&A..30..543M, 2012JCoPh.231..759P} as
\begin{equation}
    \frac{du_i}{dt}= -\sum_j m_j\, f_{ij}\, \frac{P_i}{\rho_i^2} \mathbf{v}_{ij}\cdot \nabla_i W_{ij},
\end{equation}
where we contracted the notation for the smoothing kernel as $W_{ij} \equiv W(\mathbf{r}_i - \mathbf{r}_j, h_i)$, and $f_{ij}$ is a factor representing the spatial asymmetry of the kernel, and is usually set to $f_{ij}\approx 1$. Here, we have used $\nabla_i\equiv \partial/\partial x_i$. The velocity $\mathbf{v}_{ij} = \mathbf{v}_{j} - \mathbf{v}_{i}$ is the velocity of particle $j$ (the neighbour) in the rest frame of particle $i$ (the central one). In addition, the equation of motion for particle $i$ is given by
\begin{equation}
    \frac{d\mathbf{v}_i}{dt} = -\sum_j m_j\, \left[\frac{f_{ij}P_i}{\rho_i^2}\, \nabla_i W_{ij}\, + \frac{f_{ji}P_j}{\rho_j^2}\, \nabla_j W_{ji} \right],
    \label{eq:sph-acceleration}
\end{equation}
which, for a uniform smoothing kernel ($f_{ij}=f_{ji}=1$ and $W_{ij}=W_{ji}$), simplifies to the original equation of motion (Eq.~3.3) in \cite{1992ARA&A..30..543M}. Eq.~\eqref{eq:sph-acceleration} expresses the acceleration of particle $i$, which is equivalent to the force per unit mass. For this reason, it is known as the \textit{momentum equation} and can be interpreted in physical terms by the following consideration. Given a SPH-like smoothed quantity $Q$
\begin{equation}
    \langle Q(\mathbf{r}) \rangle = \int \frac{Q(\mathbf{r'}}{\rho(\mathbf{r'})}\, W(\mathbf{r}-\mathbf{r'}, h)\, \rho(\mathbf{r'})\,  d\mathbf{r'} \approx \sum_j^{N_{\rm ngbs}} m_j\, \frac{Q_j}{\rho_j}\, W(\mathbf{r}-\mathbf{r}_j, h),
\end{equation}
then the gradient of $Q$ would be written as
\begin{align}
     \nabla Q &= \frac{\partial}{\partial \mathbf{r}} \left\{ \int \frac{Q(\mathbf{r'})}{\rho(\mathbf{r'})}\, \delta(\mathbf{r}-\mathbf{r'}, h)\, \rho(\mathbf{r'})\,  d\mathbf{r'} \right\} \\
     &= \frac{\partial}{\partial \mathbf{r}} \left\{ \int \frac{Q(\mathbf{r'})}{\rho(\mathbf{r'})}\, W(\mathbf{r}-\mathbf{r'}, h)\, \rho(\mathbf{r'})\,  d\mathbf{r'} + \mathcal{O}(h^2) \right\} \\
      &\approx \int \frac{Q(\mathbf{r'})}{\rho(\mathbf{r'})}\, \frac{\partial W(\mathbf{r}-\mathbf{r'}, h)}{\partial \mathbf{r}}\, \rho(\mathbf{r'})\,  d\mathbf{r'}  \\
      &= \int \frac{Q(\mathbf{r'})}{\rho(\mathbf{r'})}\, \big[\nabla W(\mathbf{r}-\mathbf{r'}, h) \big]\, \rho(\mathbf{r'})\,  d\mathbf{r'}  \\
      &\approx \sum_j^{N_{\rm ngbs}} m_j\, \frac{Q_j}{\rho_j}\, \nabla W(\mathbf{r}-\mathbf{r}_j, h).
      \label{eq:sph-density-identity}
\end{align}
Therefore, assuming a uniform smoothing kernel \citep[e.g.][]{1992ARA&A..30..543M} and setting $Q=P / \rho$, we obtain
\begin{equation}
    \frac{d\mathbf{v}_i}{dt} \approx -\sum_j m_j\, \left[\frac{P_i}{\rho_i^2}\, + \frac{P_j}{\rho_j^2} \right] \nabla W_{j} = \underbrace{-\sum_j m_j\, \frac{P_i}{\rho_i^2}\, \nabla W_{j}}_{\text{Term (I)}} \underbrace{-\sum_j m_j\, \frac{P_j}{\rho_j^2}\, \nabla W_{j}}_{\text{Term (II)}}.
\end{equation}
For the first term (I) we can use the identity in Eq.~\eqref{eq:sph-density-identity} with $Q=\rho$, while for the second term (II) we can set $Q=P/\rho$. By recasting the discretised form of the identity into its continuum equivalent, we have
\begin{align}
    \frac{d\mathbf{v}_i}{dt} &\approx - \frac{P_i}{\rho_i^2}\, \underbrace{\sum_j m_j\, \nabla W_{j}}_{=\nabla\rho} - \underbrace{\sum_j m_j\, \frac{P_j}{\rho_j^2}\, \nabla W_{j}}_{=\nabla \left( \frac{P}{\rho}\right)}\\
    &\approx - \frac{P}{\rho^2} \nabla \rho - \nabla \left( \frac{P}{\rho}\right) \\
    &= - \frac{1}{\rho} \left[ \left(\frac{P}{\rho}\right) \nabla \rho + \rho \nabla \left( \frac{P}{\rho}\right) \right]\\
    &= - \frac{1}{\rho} \nabla\left(\frac{P}{\rho}\, \rho \right)\\
    &= - \frac{\nabla P}{\rho},
\end{align}
which demonstrates a direct relationship between the acceleration and the pressure gradient \citep[see e.g.][]{2012JCoPh.231..759P}. By inverting the order of this derivation, we would recover the logic used in, e.g., \citep{1992ARA&A..30..543M} to prove the SPH equations of motion from Euler's equation (i.e. Eq.~\ref{eq:euler-sph} without gravity). In fact, the term $\nabla P/\rho$ plays indeed the role of a force in Euler's equation, as expected.

Defined in differential form, Euler's equation is suitable for continuous media, but returns non-defined solutions for sharp discontinuities, such as shocks and contact discontinuities. Numerically, this formulation of SPH would fail to capture these features and lead to errors. This shortcoming occurs because the entropy function $A(s)$\footnote{In the SPH literature, the entropy function of the fluid particles, referred to as $A$, is assumed to only a function of the specific (per unit mass) thermodynamic entropy $s$: $A=A(s)$. Notably, this quantity does not directly represent the thermodynamic entropy $S$, but it is the \textit{adiabat} corresponding to $S$ and satisfies the bijective (one-to-one) relation $A(s) \Leftrightarrow s \Leftrightarrow S$.} is not explicitly evolved, so that $dA(s)/dt = 0$ while implicitly assuming the absence of dissipation and shocks. In these conditions, however, the Rankine-Hugoniot jump conditions\footnote{The Rankine-Hugoniot jump conditions are obtained from the \textit{integral} form of Euler's equation, which is stable around non-differentiable points \citep{LANDAU1959310}.} predict an increase in entropy across the shock front. In SPH, shocks are captured by introducing an \textbf{artificial viscosity} term. The \textsc{Sphenix} scheme use the artificial viscosity implementation of \cite{2010MNRAS.408..669C}, which adds the following term to the equation of motion (Eq.~\ref{eq:sph-acceleration})
\begin{equation}
    \frac{d\mathbf{v}_i}{dt} \longrightarrow \frac{d\mathbf{v}_i}{dt} + \frac{d\mathbf{v}_i}{dt} \Big|_{\rm visc}.
\end{equation}
Here, the second term is the artificial viscosity
\begin{equation}
    \frac{d\mathbf{v}_i}{dt} \Big|_{\rm visc} = -\sum_j m_j\, \pi_{ij}\, \left[f_{ij} \nabla_i W_{ij}\, + f_{ji} \nabla_j W_{ji} \right],
    \label{eq:sph-viscosity}
\end{equation}
where the $\pi_{ij}$ is defined as
\begin{equation}
    \pi_{ij} = -\alpha_V\, \frac{\overbrace{c_i + c_j - 3\mu_{ij}}^{\text{Signal velocity}}}{\rho_i + \rho_j}\, \mu_{ij},
    \label{eq:viscosity-tensor}
\end{equation}
with $\mu_{ij}$ being a limiter which switches off the viscosity term for diverging flows:
\begin{equation}
    \mu_{ij} = \left\lbrace\begin{array}{lcr}
    \frac{\mathbf{v}_{ij} \cdot \mathbf{r}_{ij}}{|\mathbf{r}_{ij}|} & \text{if} & \nabla \cdot \mathbf{v}_{ij} < 0\\ [2mm]
    0 & \text{if} & \nabla \cdot \mathbf{v}_{ij} \geq 0,
    \end{array}
\right.
\end{equation}
 and $c_i=\sqrt{P_i/\rho_i}=\sqrt{(\gamma - 1)\gamma u_i}$ is the speed of sound associated with particle $i$.

In \textsc{Sphenix}, the viscosity coefficient $\alpha_V$ is not set to a constant, but depends on the iterations between particle $i$ and its neighbours, i.e. $\alpha_V =\alpha_{V, ij}$. This definition implies that each SPH gas particle needs to carry a value of $\alpha_{V, i}$ in order to compute the viscosity coefficient, increasing the memory footprint of the SPH scheme, but enhancing the accurate of the shock-capturing algorithm by (i) making $\alpha_{V, i}$ dependent on the \cite{1989BAAS...21Q1093B} switch and (ii) weighting by a shock indicator \cite[see Section 3.1 of][for further details]{2022MNRAS.511.2367B}.

Velocity discontinuities can be captured in SPH using artificial viscosity. To capture jumps in internal energy (or entropy), instead, it is necessary to add an \textbf{artificial conduction} term to Eq.~\eqref{eq:energy-sph}:
\begin{equation}
     \frac{du_i}{dt} \longrightarrow \frac{du_i}{dt} + \frac{du_i}{dt}\Big|_{\rm cond},
\end{equation}
where
\begin{equation}
    \frac{du_i}{dt}\Big|_{\rm cond} = -\sum_j m_j\, v_{D, ij}\, (u_i - u_j)\, \underbrace{\frac{\mathbf{r}_{ij}}{|\mathbf{r}_{ij}|}}_{\text{Unit vector}\, \mathbf{\hat{r}}_{ij}} \cdot \left[ \frac{f_{ij}}{\rho_i} \nabla_i W_{ij}\, + \frac{f_{ji}}{\rho_j} \nabla_j W_{ji}\right],
\end{equation}
and $v_{D, ij}(\alpha_{D, ij})$ is the pairwise conductivity speed, which is a function of the conduction coefficient $\alpha_{D, ij}$ and the average of two speeds: (i) a velocity-dependent term used in the \textsc{Phantom} code \citep{2018PASA...35...31P} which improves the modelling of fluid mixing and (ii) a pressure-dependent term introduced by \cite{2018PASA...35...31P} which targets targets contact discontinuities. Moreover, the pairwise artificial conduction coefficient is computed as a pressure-weighted average of the $\alpha_{D, i}$ of the individual particles as follows
\begin{equation}
    \alpha_{D, ij} = \frac{P_i\, \alpha_{D, i} + P_j\, \alpha_{D, j}}{P_i + P_j}.
\end{equation}
Throughout the simulation, the SPH scheme evolves the values of $\alpha_{D, i}$ over time. However, these values are capped to a maximum value $\alpha_{D,{\rm max}}$ via a \textit{conduction limiter} to prevent thermal feedback in the sub-grid model from creating unphysical discontinuities in internal energy \cite[see Section 4 of][for further details]{2022MNRAS.511.2367B}. In Section \ref{sec:results_model_variations:conduction}, we will discuss the effects of switching the ceiling $\alpha_{D,{\rm max}} = \{0, 1\}$ on the thermodynamic properties of cosmological simulations \citep[see also][]{2023MNRAS.520.3164A}.

\subsubsection*{The \textsc{Gadget-SPH} hydrodynamics scheme}
The \textsc{Gadget-SPH} implementation was used model the hydrodynamics in BAHAMAS and MACSIS and is based on the \textit{density-entropy} formulation described in \cite{2002MNRAS.333..649S}. Unlike the \textit{traditional} density-energy scheme of \textsc{Sphenix}, here the equation of state is recast as
\begin{equation}
    P=A(s)\, \rho^\gamma,
\end{equation}
where $A(s)$ denotes the entropy function. Here, $s$ is the specific entropy, which can be written as
\begin{equation}
    s\, (1-\gamma) = \log (P/\rho^\gamma),
\end{equation}
consistently with the equation of state \citep{1987ApJ...323..614B}.

Differently from \textsc{Sphenix}, \textsc{Gadget-SPH} uses a cubic spline (M4) kernel, expressed as
\begin{align}
    w(q) = \left\lbrace\begin{array}{lcr}
    1 - 6q^2 + 6q^3 & \text{if} & q < \frac{1}{2}\\ [2mm]
    2\, \left(1 - q\right)^3 & \text{if} & \frac{1}{2} \leq q < 1\\ [2mm]
    0 & \text{if} & q \geq 1,\\
    \end{array}
\right.
\end{align}
with $\kappa_{3} = 8/\pi$, so that $W(r, h) = 8/\pi\, w(r/h)/h^3$. Similarly to \textsc{Sphenix}, the smoothing length $h$ is determined as the value that satisfied the equation (for $\zeta=3$)
\begin{equation}
    N_{\rm ngbs} = \frac{4}{3}\pi h^3  \sum_b^{N_{\rm ngbs}}  W(\mathbf{r} - \mathbf{r}_b, h),
\end{equation}
with $N_{\rm ngbs} \approx 48$. Although this choice of kernel is expected to lead to a slightly less accurate reconstruction of the forces when $d\mathbf{v}/dt \approx 0$ (e.g. in Gresho-Chan vortex-like structures), we do not expect any large qualitative impacts on the hydrodynamic modelling in cosmological simulations.

The equation of motion for the acceleration is the same as in \textsc{Sphenix} (Eq.~\ref{eq:sph-acceleration}, see also Eq.~18 in \citealt{2002MNRAS.333..649S}), however the dimensionless factor $f_{ij}$ is not computed pairwise, but it only depends on the central particle, i.e. $f = f_i$. This factor disappears in the definition of the artificial viscosity term for the momentum equation, which becomes
\begin{equation}
    \frac{d\mathbf{v}_i}{dt} \Big|_{\rm visc} = - \frac{1}{4}\,\sum_j m_j\, \pi_{ij}\, \underbrace{ \left[ \nabla_i W_{ij}\, + \nabla_j W_{ji} \right] }_{\equiv \nabla \bar{W}_{ij}}.
\end{equation}

Together with the momentum equation, the evolution equation for the entropy function is the second pillar of the \textsc{Gadget-SPH} scheme. Combining explicitly the terms for dissipation and artificial viscosity, we can express $dA_i/dt$ as
\begin{align}
    \frac{dA_i}{dt} &= 0 +\frac{dA_i}{dt} \Big|_{\rm diss} + \frac{dA_i}{dt} \Big|_{\rm visc} \\ \vspace{3mm}
                    &=  \underbrace{- \frac{\gamma - 1}{\rho_i^\gamma}\, \Lambda(\rho_i, u_i)}_{\text{Cooling-heating}}~+~ \underbrace{\frac{1}{8}\, \frac{\gamma - 1}{\rho_i^{\gamma - 1}}\, \sum_j m_j\, \pi_{ij}\, \mathbf{v}_{ij} \cdot \nabla \bar{W}_{ij}.}_{\text{Artificial viscosity}}
\end{align}
In this equation, $\pi_{ij}$ has the same form as in Eq.~\eqref{eq:viscosity-tensor}, but with a fixed $c_V=2$. $\mu_{ij}$ which still encapsulates the \cite{1989BAAS...21Q1093B} switch. Finally, artificial conduction is not implemented in \textsc{Gadget-SPH} for versions of \textsc{Gadget-2} \citep{2005MNRAS.364.1105S}, but it is included in the \textsc{Anarchy-SPH} in versions of \textsc{Gadget-3} (see \citealt{2015MNRAS.451.1247S} and Appendix A of \citealt{eagle.schaye.2015}).

\subsubsection*{Time integration scheme}
To describe the time-evolution of an SPH fluid, the equations of motions must be integrated over time. This operation evolves the particles by a time interval $\Delta t$ (as $t_0 \rightarrow t_0 + \Delta t$) and is often referred to as \textit{time-stepping}. Numerous time-stepping methods have been developed within the field of numerical particle-based simulations. Most industrial SPH applications, such as for the aerospace or hydraulics sectors, use a constant time-step duration for all the particles throughout the simulations. In cosmological simulations, however, only a fraction of the particles will experience a significant displacement over a given $\Delta t$, while all the others would remain largely unchanged. Evolving the \textit{inactive} particles would therefore use memory space and CPU cycles which could, instead, be allocated to different tasks. Cosmological codes such as \swift and \textsc{Gadget} use an \textit{adaptive} time-stepping, which evolves only the \textit{active} particles and skips the others, until they are no longer passive as they require updating. Crucially, the requirement of adaptive time-stepping, where $\Delta t$ can vary over orders of magnitude, is rather unique to cosmological simulations; optimising multi-time-step integration methods is a computational challenge that attracted the interest of a large portion of the SPH community, fostering inter-disciplinary collaborations among computational astrophysicists, computer scientists and engineers.

While managing the timestep hierarchy in a task-based paradigm, such as the \textsc{SWIFT} back-end or in \textsc{Gadget}'s data-based parallelism, can be a complex, the integration over a single timestep for one single particle follows a simpler prescription. \swift, and therefore \textsc{Sphenix}, encapsulates a velocity\hyp{}verlet scheme based on the kick-drift-kick approach \citep{1997astro.ph.10043Q}. This method initially evolves the velocity $\mathbf{v}$ (kick) over half timestep ($\Delta t / 2$), then \textit{drifts} the particle over the time-step, and finally updates $\mathbf{v}$ again (kick) over the second half of the time-step, as outlined below:
\begin{align}
    \mathbf{v}\left(t + \frac{\Delta t}{2} \right) = \mathbf{v}(t) + \frac{\Delta t}{2}\, \mathbf{a} & \qquad \text{Kick}\\
    \mathbf{r}(t + \Delta t) = \mathbf{r}(t) + \mathbf{v}\left(t + \frac{\Delta t}{2} \right)\, \Delta t & \qquad \text{Drift}\\
    \mathbf{v}(t + \Delta t) = \mathbf{v}\left(t + \frac{\Delta t}{2} \right)\, \mathbf{a}(t + \Delta t) & \qquad \text{Kick.}
\end{align}
The time-step duration is computed for each particle, according to the Courant\hyp{}Friedrichs\hyp{}Lewy \citep[CFL][]{1928MatAn.100...32C} condition:
\begin{equation}
    \Delta t_i = C_{\rm CFL}\, \frac{2\gamma_K h_i}{v_{{\rm sig},i}},
\end{equation}
where $v_{{\rm sig},i}$ is the signal velocity (see Eq.~\ref{eq:viscosity-tensor}), $\gamma_K$ is a dimensionless factor which relates the smoothing length and the cut-off distance of the smoothing kernel and $C_{\rm CFL}$ is a constant set to 0.2 \citep[see also Section 6 of][]{2021MNRAS.505.2316B}.

Crucially, this method, also known as \textit{leapfrog}, is a second-order method correct to $\mathcal{O}(\Delta t^2)$, even containing only first-order terms. This feature ensures a great numerical stability of this method, making it particularly suited for simulations of non-linear structure formation. It is used in a range of cosmology simulation codes, including \textsc{Gadget-2/3} \citep[see Section 4 of][for a Hamiltonian formulation of the leapfrog algorithm]{2005MNRAS.364.1105S} and \textsc{Gadget-4} \citep{2021MNRAS.506.2871S}.

\section{Sub-grid models and calibration}
\label{sec:sub-grid-calibration}
Although the numerical methods to model collisionless gravity (Section \ref{sec:numerical-gravity}) and collisional hydrodynamics (Section \ref{sec:numerical-sph}) are well-constrained and tested against idealised problems with known analytic solutions \citep[e.g.][]{2022MNRAS.511.2367B}, full-physics cosmological simulations include sub-grid processes which couple to both hydrodynamics and gravity. This coupling can be direct, such as the direct heating of gas particles due to black hole feedback (see Chapter \ref{chapter:5} and \citealt{2023MNRAS.520.3164A}), or indirect, such as the shocks generated by the feedback and their impact on the hot gas distribution in groups and clusters of galaxies.

If the inclusion of sub-grid processes, i.e. anything other than gravity or hydrodynamics, augments the predictive power of simulations, their couplings, on the other hand, can affect the simulation metrics in a complex, highly non-trivial way. For instance, varying the strength of the supernova feedback can have the immediate effect to halt local outflows, but can also lead to a drastic increase in star formation and can even affect the global environment of small groups of galaxies (see the example of the \textit{group}, \texttt{VR\_2915} in  Section \ref{fig:profiles_nofeedback} and \citealt{2023MNRAS.520.3164A}). Varying one single parameter in the sub-grid model can therefore lead to a series of events. If such parameter is not constrained observationally or theoretically, then it becomes a degree of freedom of the sub-grid model. For ``general-purpose`` simulations of galaxy formation, such as EAGLE (and EAGLE-XL in its original form), the \textit{free parameters} in the sub-grid model must be tuned in order to find at least one combination which reproduces given observational data. This procedure is known as \textbf{sub-grid calibration} and the observational data to test the model against are referred to as \textbf{calibration metrics}.

Since the effects of a change in free parameters can lead to non-linear changes in calibration metrics, the most direct method to search for an optimal combination of parameters involves running a large number of low-volume, low-resolution simulations until the metrics are matched and a portion of the parameter space is ruled out \citep{2010MNRAS.407.2017B}. To further constrain the parameter space, a new set of simulations with larger volume and typically higher resolution is designed, this time with parameters varying within the newly constrained region. According to the requirements of the final production run, more iterations of this procedure may become necessary before reaching the target volume and resolution \citep{2015MNRAS.450.1937C}. For most such galaxy formation projects, this step (11 in Fig.~\ref{fig:simulation-pipeline}) is associated with a large computational cost and the production of large intermediate data sets before the selection of a \textit{fiducial} sub-grid model (step 12 in Fig.~\ref{fig:simulation-pipeline}).

To minimise the number of sub-grid model realisations required in each intermediate step of the calibration, the EAGLE-XL Collaboration developed \textsc{SWIFT-emulator} \citep{2022JOSS....7.4240K}, a \texttt{Python} tool designed to automate the process and available at \begin{center}
    \href{https://github.com/SWIFTSIM/emulator}{https://github.com/SWIFTSIM/emulator}.
\end{center} Crucially, it uses the Gaussian Process Regression library \textsc{George} \citep{2015ITPAM..38..252A} to train a Gaussian emulator model, which estimates the likelihood of the test simulations to match the target calibration metrics and recommends new sets of free parameters around the maximum likelihood region \citep[see also][for a constrained parameter space using the principle component analysis, PCA, method]{2017MNRAS.466.2418R}. \textsc{SWIFT-emulator} was successfully used to determine the fiducial model for the FLAMINGO project \citep{flamingo_calibration_kugel}.

The calibration strategy to find for the EAGLE-XL fiducial model was split in two parts:
\begin{itemize}
    \item calibration of \textbf{cosmological volumes}. Powered by the \textsc{SWIFT-emulator}, five \textit{waves} of simulations were designed and run. The calibration metrics included (i) the \textit{stellar mass function} with observational data from \cite{2009MNRAS.398.2177L, 2013A&A...556A..55I, 2015MNRAS.454.4027D, 2017MNRAS.470..283W}, (ii) the \textit{galaxy sizes} (stellar mass in 100 kpc-to-the projected stellar half-mass radius) with data from \cite{2015MNRAS.447.2603L, 2021A&A...649A..39L} and (iii) the stellar mass-to-black hole mass relation, with data from \cite{2013ApJ...764..184M, 2019ApJ...876..155S}.

     \item Calibration of \textbf{cosmological zoom-in simulations} (including this work). We designed heuristically motivated sets of simulations varying the AGN heating temperature and feedback energy-distribution schemes to match the observed thermodynamic profiles of groups and clusters \citep[see][or later chapters for a detailed discussion]{2023MNRAS.520.3164A}. 
\end{itemize}

By including a suite of zoom simulations and managing a sub-grid calibration in parallel to the cosmological boxes, the strategy adopted within the EAGLE-XL model was the first attempt in the literature to tune the free parameters to match galaxy and cluster properties simultaneously and explicitly. While small (25-50 Mpc) calibration-type boxes can represent the properties of galaxies, they cannot reliably predict the effects of, e.g., AGN feedback, which largely acts on super-galactic scales. For this reason, calibrating the AGN feedback schemes tuned on zoom simulations of galaxy clusters provides complementary information which could avoid common shortcomings of recent cluster simulations projects, such as high-entropy cores in the C-EAGLE objects \citep{ceagle.barnes.2017}. These considerations motivate future studies of these novel \textit{zoom-assisted calibration} techniques.  

\section{Computational performance}
\label{sec:computational-performance}

\subsection{Hydrodynamics weak-scaling}

Numerical approaches to problem solving have revolutionised most areas of quantitative sciences since the development and adoption of distributed high\hyp{}performance computing (HPC) systems. The benefits of using simulations to build models are driving the efforts from both academic and industry research teams towards the Exa\hyp{}FLOPs HPC era (see e.g. \citealt{2021ApJ...908...11L} for examples in the field of computational cosmology and medicine and \citealt{mcinnes2021community} for an overview on computer science and multidisciplinary potential of future HPC development).
Beside the expected hardware challenges related to clustering and interconnecting thousands of compute nodes in one working facility, many research programs, such as the UK\hyp{}centric \href{https://excalibur.ac.uk/excalibur}{ExCALIBUR}, the European \href{https://www.etp4hpc.eu/euexascale.html}{ETP4HPC} or the US-lead \href{https://www.exascaleproject.org/}{Exascale Computing Project}, invested over 50\% of their budget in software development. ExCALIBUR is a multi\hyp{}disciplinary project for the development of numerical methods designed to scale to large numbers of computing units with minimal overhead. Within the program, different working groups lead the development of codes such as ExaHyPE, mainly used in finite element methods, and \href{http://swift.dur.ac.uk/}{SWIFT}, configured to solve smoothed\hyp{}particle hydrodynamics (SPH) problems and assess its performance.

This section outlines details on benchmark tests of SPH simulations using the SWIFT code. Due to its simplicity and ease of implementation, we selected the well\hyp{}known Kelvin\hyp{}Helmholtz (KH) instability problem as the main test case for the first part of the project. 
The KH instability occurs at the interface between two fluids moving with different velocities relative to each other. A formal treatment of the physics governing the time-evolution of the KH instability can be found, e.g., in \cite{ferrari1978relativistic} and will not be reported here in its full form. Crucially, the evolution of this problem in 2 dimensions is linear at early times and can be solved analytically without the need of expensive numerical methods. In this regime, the KH problem becomes an excellent unit-test to assess the reliability of various numerical hydrodynamic simulation codes by comparing their output with analytic solutions (see e.g. the SPHENIX hydrodynamics scheme for SWIFT in \citealt{2022MNRAS.511.2367B} or the \textsc{Gizmo} code first presented by \citealt{2015MNRAS.450...53H}).

\subsection{Initial conditions and tiling scheme}
Throughout the development stages of the SWIFT code, \cite{2016arXiv160602738S} produced a \texttt{Python} script to generate initial conditions for a KH instability simulation in 2 dimensions. In this work, SWIFT's original KH script was extended with useful functionalities for benchmark tests. The parameters used to generate initial conditions can be found in table \ref{tab:khparams} and are kept unchanged throughout the testing campaign, with the exception of $L_2$, which can be varied to obtain initial conditions at different resolutions. The smoothing lengths associated with high\hyp{} and low\hyp{}density particles respectively are $h_{1, 2}=1.2348 / L_{1, 2}$, where the value of 1.2348 is computed by imposing the Courant\hyp{}Friedrichs\hyp{}Lewy convergence condition on a 3D cubic spline kernel with approximately 48 neighbours \citep{10.1111/j.1365-2966.2012.21439.x}.

\begin{sidewaystable}
\centering
\caption{Simulation parameters used for generating initial conditions for all the KH simulations in the benchmarking suite. The resolution of the simulation is set by the number of particles $L_2$, which results in $\approx L_2^2$ particles in 2D problems or $\approx L_2^3$ particles in the 3D case.}
\label{tab:khparams}
\begin{tabular}{cl}
\toprule
Quantity & \multicolumn{1}{c}{Description}    \\
{[}dimensionless or arbitrary units{]} &   \\ \midrule
$L_2$ (user-defined)           & Particles along one edge in the low-density region  \\
$L_1 = L_2^2 / (\rho_1 \rho_2)$& Particles along one edge in the high-density region  \\
$\gamma = 5/3$                 & Gas adiabatic index                                 \\
$P_1 = 2.5$                    & Central region pressure                             \\
$P_2 = 2.5$                    & Outskirts pressure                                  \\
$v_1 = 0.5$                    & Central region velocity (along $\hat{\mathbf{x}}$)  \\
$v_2 = -0.5$                   & Outskirts velocity (along $\hat{\mathbf{x}}$)       \\
$\rho_1 = 2$                   & Central density                                     \\
$\rho_2 = 1$                   & Outskirts density                                   \\ \bottomrule
\end{tabular}
\end{sidewaystable}

\begin{figure}
    \centering
    \includegraphics[width=\textwidth]{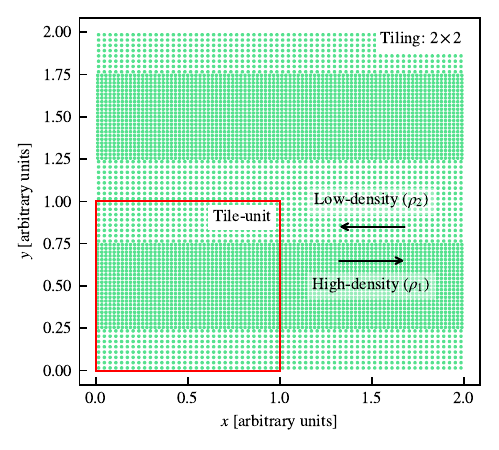}
    \caption{Example of a 2D KH initial conditions showing a $2\times2$ tiling scheme. A tile-unit, as indicated by the red square, is replicated twice in both dimensions, leaving to a 4-tile particle system. Each tile used in this image was defined by $L_2=32$, with particles arranged in a lattice-like structure between $x = (0, 1)$ [arbitrary units] and $y = (0, 1)$ [arbitrary units]. The arrows indicate the direction of the (unperturbed) fluid's velocity in the high- and low-density regions.}
    \label{fig:hk_setup}
\end{figure}

The original script in SWIFT's KH test package produces 2D initial conditions, where particles are positioned in a lattice-like configuration as shown by the markers enclosed by the red square in Fig.~\ref{fig:hk_setup}. In this configuration, where $L_2=32$, the high-density region (within the highlighted tile-unit) was defined as a band spanning the $x = (0, 1)$ [arbitrary units] and $y = (0.25, 0.75)$ [arbitrary units] ranges, while the rest of the space was filled with low-density gas. The fluid velocity along $\hat{\mathbf{x}}$ was conventionally chosen to be positive for high-density gas particles and negative for low-density ones, as reported in table \ref{tab:khparams}. Throughout the KH idealised simulation tests, we assume SWIFT's internal arbitrary unit system and units referring to specific quantities (such as $x$) will be dropped hereafter.

The velocity perturbation is applied along $\hat{\mathbf{y}}$ and its modulus can be expressed as
\begin{equation}
    \delta v = \omega_0 \sin\left(4\pi x \right) \times \left\{\exp \left[-\frac{(y-y_0)^2}{2\sigma^2}\right] + \exp \left[-\frac{(y-y_1)^2}{2\sigma^2}\right]\right\},
\end{equation}
where $x, y$ are dimensionless coordinates scaled by the tile box\hyp{}size, $\omega_0=0.1$, $\sigma = 0.05/\sqrt{2}$ and the constants $y_0=0.25$, $y_1=0.75$ are the $y$\hyp{}coordinates of the density discontinuities.

In order create initial conditions for progressively larger KH problems to simulate with SWIFT, we adopted a square tiling scheme, consisting in generating copies of the original tile-unit, shifting the coordinates of copied particles by a multiple of the box-size, and leaving other parameters (e.g. velocity components, mass, smoothing lengths) unchanged. For a 2D test-case, this operation is repeated for the same number of times in both dimensions, resulting in $n^2$ tiles for a $n\times n$ tiling choice. Fig.~\ref{fig:hk_setup} shows an example of this scheme with $n=2$, containing over $(nL_2)^2$ particles.\footnote{Note that $L_2$, set to define the tile's resolution, is the number of particles along the low-density side. However, the average separation of the particles in the high-density region is lower than in the low-density, bringing the total number of particles in one tile above $L_2^2$. The exact total number of particles in each simulation is recorded in the \texttt{\textbackslash Header\textbackslash NumPart\_Total} attribute in the initial conditions files (\texttt{HDF5} format).}

While weak\hyp{}scaling tests performed on KH set\hyp{}ups in two dimensions may offer insights into I/O and peek performance, SWIFT's hydrodynamics modules and back\hyp{}end engine are not optimised for running 2D hydrodynamic simulations. In addition, most production runs in astrophysics and engineering are performed in three dimensions, motivating the choice of building 3D KH initial conditions for the weak\hyp{}scaling performance tests in object.

\begin{figure}
    \centering
    \includegraphics[trim={1cm 0 0 0},clip,width=\textwidth]{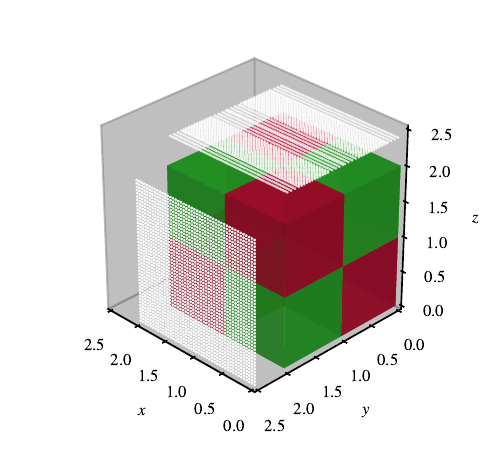}
    \caption{Example of $n=2$ (or $2 \times 2 \times 2$) tiling of 3D KH tile-units. Tiles are indicated with cubes of alternating colour for clarity. The white markers illustrate the projection of the particle assembly onto two planes (these initial conditions were generated with $L_2=32$, although only half of the particles are displayed). The top particles represent the $x-y$ projection, with two high-density regions visible. The markers on the left-hand side are the $x-z$ projection, showing a uniform distribution of $L_2 \times L_2$ particles originated from stacking 2D initial conditions along $z$.}
    \label{fig:tiles}
\end{figure}

Although the KH instability is formulated in two dimensions (i.e. $x, y$ in Fig.~\ref{fig:hk_setup}), it is possible to stack copies of this particle assembly along the perpendicular (i.e. $z$) dimension. The number of planes stacked in a tile-unit is $L_2$, defined such that the separation between planes is equal to the separation between particles along the low-density edge of a 2D KH set-up. Constructed with this method, the tile unit has $\approx L_2^3$ particles arranged in a cubic structure. Similarly to the 2D case, the tiling scheme in three dimensions involves periodic boundary conditions and $n^3$ copies of the tile-unit, themselves arranged to form a larger cubic ensemble of particles. Fig.~\ref{fig:tiles} shows an example of a 3D tiling scheme for $n=2$ (or $2 \times 2 \times 2$, to be compared to the 2D case in Fig.~\ref{fig:hk_setup}).

\subsection{Weak-scaling performance}

\begin{figure}
    \centering
    \includegraphics[width=\textwidth]{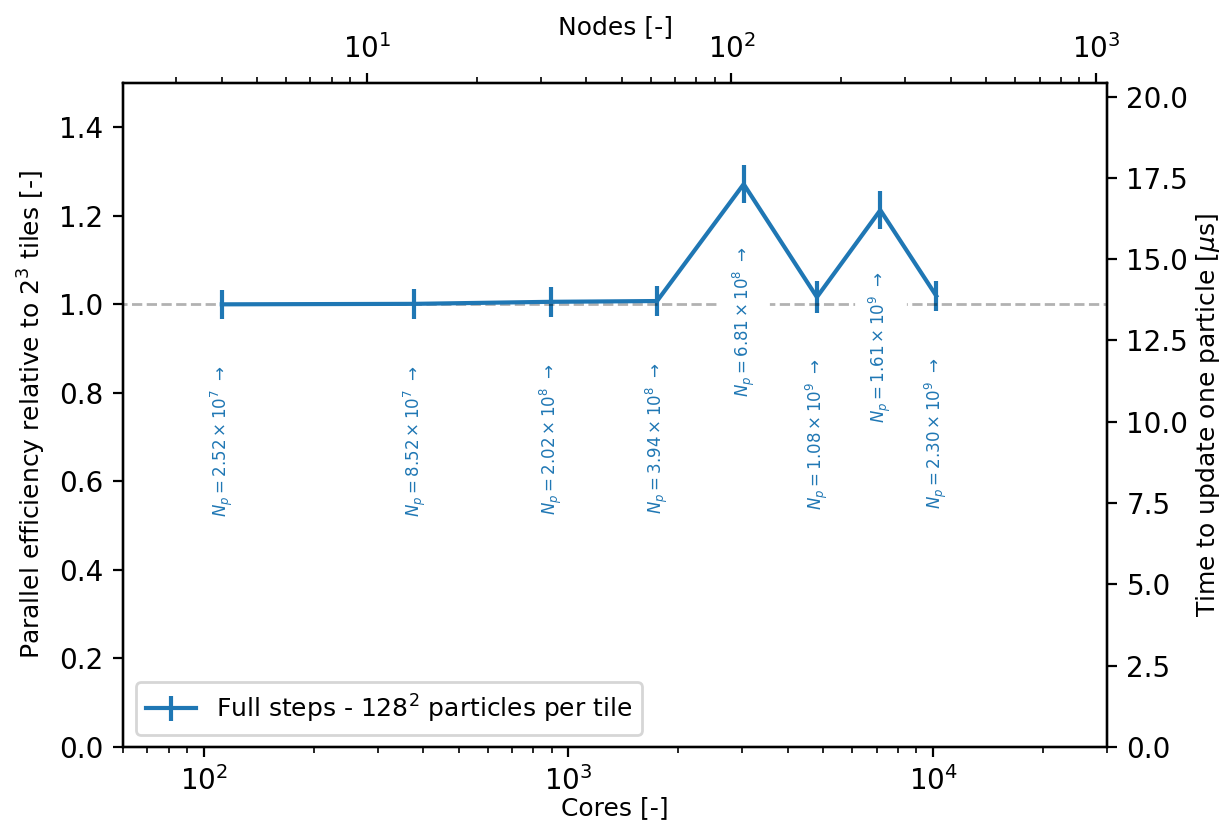}
    \caption{Weak-scaling efficiency for the 3D KH problem using tiles with $L_2=128$ and the Cosma-7 HPC system. The parallel efficiency is computed by averaging the wall-clock time of complete simulation steps and normalising by the average time-step duration of the $n=2$ test.}
    \label{fig:weak_scaling_128}
\end{figure}

\begin{figure}
    \centering
    \includegraphics[width=\textwidth]{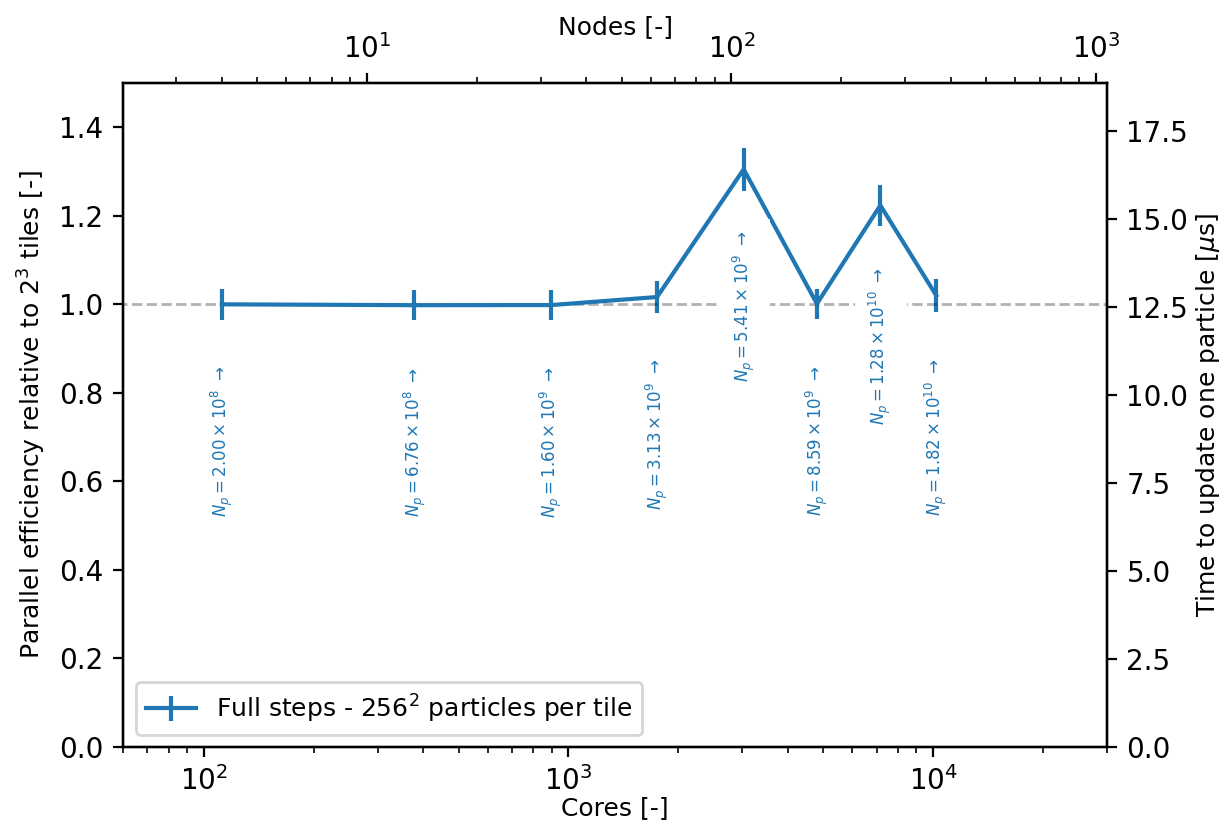}
    \caption{As in Fig.~\ref{fig:weak_scaling_128}, but using $L_2=256$.}
    \label{fig:weak_scaling_256}
\end{figure}

\begin{figure}
    \centering
    \includegraphics[width=\textwidth]{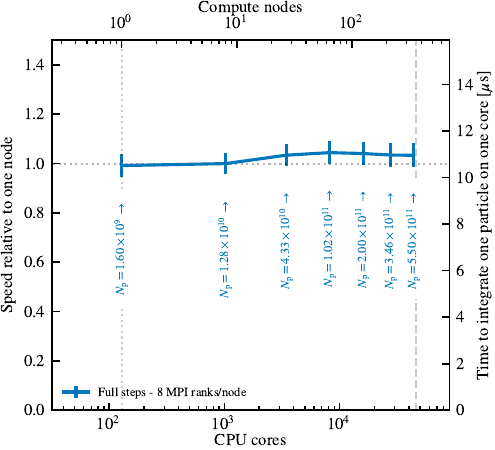}
    \caption{As in Fig.~\ref{fig:weak_scaling_128}, but using $L_2=512$ and the Cosma-8 HPC system. The vertical dotted line indicates the cores in one node (128 for Cosma-8), the vertical dashed line marks the full capacity of Cosma-8 (44000 cores during the commissioning phase) and the horizontal line sets the normalisation baseline for the run-time on one node.}
    \label{fig:weak_scaling_512}
\end{figure}

The ability to extend the problem size by increasing the tiling order $n$, while proportionally allocating more and more HPC resources, allows to test the weak-scaling performance of \swift when configured in hydrodynamics-only mode (using SPHENIX and a quintic SPH spline). Weak-scaling tests performed on the Cosma-7 HPC cluster included 3D KH simulations ranging from $n=2$ to $n=9$, progressively using between 2 and 365 compute nodes. Communications across CPUs are managed by Intel's Message Passing Interface (MPI) library, while processes within the same CPU are allocated in a threaded-based manner by the \textsc{Quicksched} library \citep{2016arXiv160105384G}. Cosma-7 compute nodes are equipped with a dual-socket motherboard, with one 14-core Intel Xeon Gold 5120 CPU per socket. As communications between NUMA regions in the same CPU are faster than inter-socket or inter-node communications, we allocated one MPI rank per socket, while performing asynchronously threaded operations within the rank. One tile-unit is associated with each MPI rank, leading to the configuration of two 3D tiles per node (i.e. one per socket). Scaling the number of tiles as $n^3$ implies that $n^3$ MPI ranks need to be allocated across the HPC system, or equivalently ${\rm ceil}(n^3/2)$ compute nodes. Note that in case $n^3$ is an odd integer, the last node will only use one of the two sockets, although the node is exclusively reserved by the SLURM batch management system (hence the use of the ${\rm ceil}$ function).

Benchmarks are performed by running the simulations for one hour (after which they are forced to stop at the SLURM level) and then counting the number of (completed) integration time-steps operated by \swift within this time-window. The weak-scaling parallel efficiency of a KH test with $n^3$ tiles is defined as the ratio between its number of time-steps completed in one hour and the number of time-steps completed by a $2^3$-tile simulation in the same time. By normalising the benchmarks to the $n=2$ test, rather than a single-tile simulation, we ensured to include MPI-related latency due to communications across $2^3=8$ MPI ranks (allocated across 4 nodes and a total of 8 sockets). Figures \ref{fig:weak_scaling_128}, \ref{fig:weak_scaling_256} and \ref{fig:weak_scaling_512} show the weak-scaling parallel efficiency of KH problems using low-resolution ($L_2=128$) and high-resolution ($L_2=256$) 3D tiles respectively. Starting from the left, the first point represents the $n=2$ run (4 nodes and 112 CPU threads), having a parallel efficiency of 1 by construction. Moving to the right, subsequent points represent integer increments of the tiling order $n$ up to $n=9$, corresponding to a simulations with 729 cubic tiles, allocated across the same number of MPI ranks. The Cosma-7 system having a total of 448 compute nodes limited the tiling order to $n=9$, as the $n=10$ test would have required 500 nodes to be allocated with the chosen tiling configuration.

Pure SPH simulations only involve loops over neighbours of each particle, unlike gravity-based schemes which need to account for long-range interactions. Thank to this property, particle-particle interactions in KH simulations are local, meaning that most communications are expected to take place between adjacent sockets or nodes. In the ideal case where no MPI overhead or memory-bandwidth limitations can affect the performance of simulations and 100\% of the run-time is spent in SPH calculations, we predict a constant parallel efficiency of 1 as the size of the problem is increased proportionally with the HPC resources allocated. The grey horizontal dashed lines in figures \ref{fig:weak_scaling_128} and \ref{fig:weak_scaling_256} indicate the theoretical limit for the weak-scaling efficiency. Points lying close to this level represent simulations with good weak-scaling efficiency, while an efficiency greater than 1 can be caused by communication overheads or a sub-optimal simulation set-up.

In figures \ref{fig:weak_scaling_128} and \ref{fig:weak_scaling_256}, KH simulations run at both resolutions show an excellent weak-scaling behaviour, with mean efficiency deviating from the theoretical limit by just a few percents across nearly two order of magnitude in number of CPU cores. However, the simulations having tiling order $n=6$ and $n=8$ display a $\approx30\%$ performance loss at both resolutions. To investigate the origin of this behaviour, we run each simulation three times with the aim of quantifying the variance in the network load. When MPI is invoked between two nodes which are located at the far ends of the system and are not connected to the same network switch, additional latency may arise from the communication needing to propagate through various switches across the interconnect fabric. If no particular network configuration is specified, SLURM optimally allocates resources based on available nodes. By considering three independent simulation run-times, the mean and standard deviation were computed and displayed in Figs.~\ref{fig:weak_scaling_128} - \ref{fig:weak_scaling_512} as error bars. The efficiency of simulations with $n=6$ and $n=8$ in Figs.~\ref{fig:weak_scaling_128} and \ref{fig:weak_scaling_256} are clearly incompatible with the overall trend within one standard deviation and therefore the discrepancy is unlikely to originate from network noise or resolution effect. However, the combination of the initial particle geometry and the tiling scheme could potentially lead to the observed performance losses. In fact, the original formulation of the KH problem is developed in 2D, while the initial conditions were generated by tiling planar (2D) particle assemblies along the perpendicular direction. The use of an unconventional particle configuration could lead to a few particles requiring longer integration times, hence causing SWIFT to compute a significantly smaller number of time-steps within one hour compared to the $n=7$ and $n=9$ cases. In order to perform a conclusive analysis on the origin of this effect, these weak-scaling tests could be compared to similar simulations running SWIFT (SPH-only mode) on a volume with particles uniformly distributed. An isotropic particle distribution would not be affected by geometry-related issues, despite not offering insights into the performance of the code in solving SPH calculations in realistic scenarios.

\begin{figure}
    \centering
    \includegraphics[width=\textwidth]{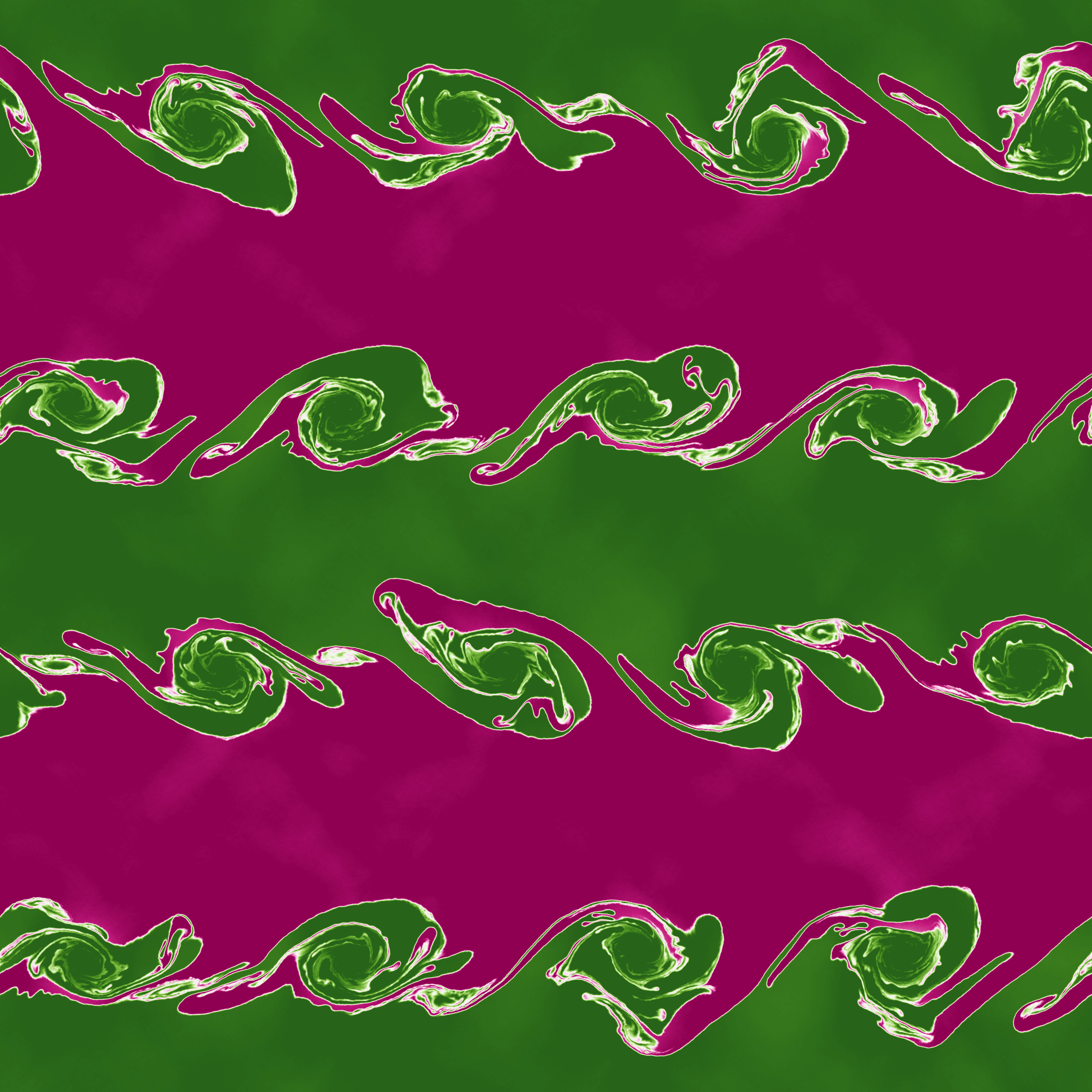}
    \caption{Snapshot of a KH simulation with $2\times2$ tiling scheme and $4096^2$ overall gas particles. The image is colour-coded according to the SPH density of the fluid, ranging from high density (purple regions) to low-density (green regions). The hydrodynamic scheme used in this example was SPHENIX, configured with a quintic SPH spline.}
    \label{fig:khrun}
\end{figure}

\subsection{Outlook and further development}
The benchmark results presented in this document were conducted as part of the ExCALIBUR \textit{phase-1} portfolio. They demonstrate the ability of \swift to perform pure SPH calculations with up to $5.5 \times 10^{11}$ particles, while maintaining a parallel weak-scaling efficiency deviating $\approx5\%$ from the ideal case. Tests run with two different particle resolutions show consistent weak-scaling behaviour across two orders of magnitude in number of cores using the Cosma-7/8 HPC systems.

Further performance tests have been also conducted on the Cosma-8 system, comprising 350 compute nodes with dual-socket AMD EPYC 7542 32-core processors at the time of commissioning. Despite the challenges posed by the topology of AMD NUMA domains (different from Intel domains in terms of core-bus layout and L-cache access protocols), Cosma-8 allowed to scale KH simulations up to $14^3$ tiles using $\approx44$ thousand logical threads, and up to $5.5 \times 10^{11}$ particles (using $L_2 = 512$). We delegate further profiling and weak-scaling tests with other hydrodynamics initial conditions to future work.

%% file: Chapters/Chapter4.tex
\chapter{Cluster rotation in the MACSIS simulations}
\label{chapter:4}

\section{Preface to the chapter}
\label{sec:ch4-preface}

\subsubsection*{Declaration of contributions}
The MACSIS simulations were produced by \cite{macsis_barnes_2017}, while the analysis framework and the figures were developed by the first author. Most of the text, except for Section \ref{sec:power-spectrum}, was produced by Edoardo Altamura, with comments from Scott Kay and Jens Chluba in parts of the document. The discussion on the temperature power-spectrum (Section \ref{sec:power-spectrum}) was lead by Jens Chluba, with comments from Edoardo Altamura. The values of the non-thermal pressure fraction in Fig.~\ref{fig:corner-plot-dynamical-state} were provided by Imogen Towler. The proof of the correlation between the kinetic-to-thermal ratio and the the non-thermal pressure fraction (see Section \ref{app:correlation:alpha-beta}) was prompted by Scott Kay and re-structured by Edoardo Altamura.

\newpage

\begin{spacing}{1.}

\topskip0pt
\vspace*{\fill}
\begin{center}

    {\Large{Galaxy cluster rotation revealed in the MACSIS simulations with the kinetic Sunyaev-Zeldovich effect}}

    \vspace{1cm}
    
    \textbf{Edoardo Altamura}, Scott T Kay, Jens Chluba, Imogen Towler, Monthly Notices of the Royal Astronomical Society, Volume 524, Issue 2, September 2023, Pages 2262–2289, \href{https://doi.org/10.1093/mnras/stad1841}{https://doi.org/10.1093/mnras/stad1841}
    
    \vspace{1.3cm}
    
    \begin{minipage}{0.85\textwidth}
        \textsc{Abstract}
        \vspace{0.5cm}
        
        The kinetic Sunyaev-Zeldovich (kSZ) effect has now become a clear target for ongoing and future studies of the cosmic microwave background (CMB) and cosmology. Aside from the bulk cluster motion, internal motions also lead to a kSZ signal. In this work, we study the rotational kSZ effect caused by coherent large-scale motions of the cluster medium using cluster hydrodynamic cosmological simulations. To utilise the rotational kSZ as a cosmological probe, simulations offer some of the most comprehensive data sets that can inform the modelling of this signal. In this work, we use the MACSIS data set to investigate the rotational kSZ effect in massive clusters specifically. Based on these models, we test stacking approaches and estimate the amplitude of the combined signal with varying mass, dynamical state, redshift and map-alignment geometry. We find that the dark matter, galaxy and gas spins are generally misaligned, an effect that can cause a sub-optimal estimation of the rotational kSZ effect when based on galaxy motions. Furthermore, we provide halo-spin-mass scaling relations that can be used to build a statistical model of the rotational kSZ. The rotational kSZ contribution, which is largest in massive unrelaxed clusters ($\gtrsim$100 $\mu$K), could be relevant to studies of higher-order CMB temperature signals, such as the moving lens effect. The limited mass range of the MACSIS sample strongly motivates an extended investigation of the rotational kSZ effect in large-volume simulations to refine the modelling, particularly towards lower mass and higher redshift, and provide forecasts for upcoming cosmological CMB experiments (e.g. Simons Observatory, SKA-2) and X-ray observations (e.g. \textit{Athena}/X-IFU).

        \vspace{-5pt}
        \par\noindent\rule{\textwidth}{1pt}
        Published in \mnras~and referenced in this thesis as \citet{2023arXiv230207936A}. 
        
    \end{minipage}
    
\end{center}
\vspace*{\fill}

\end{spacing}

\newpage

\section{Introduction}
The kinetic Sunyaev-Zeldovich (kSZ) effect is {related to} a late-{time} Doppler boost of the cosmic microwave background (CMB) which manifests itself when ionised gas moves with a non-zero velocity in the CMB rest-frame \citep{ksz_sunyaev_1980}. The kSZ has been investigated theoretically and observationally \citep[e.g.][]{2020PhRvL.125k1301C, 2022MNRAS.510.5916C} {on both cosmological and astrophysical levels.} In the cosmological context, {the kSZ contribution from astrophysical CMB foregrounds can be used to track the prokected velocity field in large-scale structures \citep[e.g.][]{2023JCAP...03..039B} or} subtracted to retrieve the signal from CMB anisotropies \citep[e.g.][]{2020A&A...641A...4P}. In the astrophysical context, the kSZ signal is a direct and unique probe for the dynamical state of the intra-cluster medium (ICM) in galaxy clusters \citep{mroczkowski2019_review}. {In addition to} direct measurements of the kSZ effect due to the {pair-wise momentum of thousands of} clusters \citep{2012PhRvL.109d1101H},\footnote{In \cite{2012PhRvL.109d1101H}, the objects were selected based on 27291 luminous galaxies from the Baryon Oscillation Spectroscopic Survey Data Release 9 \citep[BOSS-DR9,][]{2011ApJ...728..126W}. Atacama Cosmology Telescope \citep[ACT,][]{2011ApJS..194...41S} observations provided the microwave data to measure the temperature distortions.} more sensitive instruments and sophisticated post-processing pipelines have enabled the detection of peculiar motion of individual substructures within {an individual} cluster \citep{2017A&A...598A.115A}.

While galaxy clusters in quasi\hyp{}hydrostatic equilibrium are predominantly pressure\hyp{}supported, they gain angular momentum from the surrounding matter during their gravitational collapse and maintain a residual rotational support, typically accounting for {$\simeq 5\%$} of their total kinetic energy \citep{1995MNRAS.272..570S, 1996MNRAS.281..716C}. The presence of ordered motions in the ICM affects the assumption of hydrostatic equilibrium, often adopted to estimate cluster masses \citep[see e.g.][]{2005ApJ...628..655V} and the ability to quantify this discrepancy may improve the current estimates on hydrostatic mass bias and the angular momentum distribution of large assemblies of galaxies. Using the galaxy cluster Abell 2107 as a case study, \cite{2005MNRAS.359.1491K} illustrated that accounting for bulk rotation can lead to a few 10s percentile differences in the mass estimates and can help reconstruct the recent dynamical history of the system \citep{2019MNRAS.485.3909L} and the connection to the surrounding large\hyp{}scale structures \citep{2018ApJ...869..124S}. 

Using very similar assumptions, \citet{CC02} [henceforth, \citetalias{CC02}] and \citet{Chluba2001Diploma, CM02} [henceforth, \citetalias{CM02}] estimated the additional kSZ signal deriving from ordered cluster rotation, which combines with that from the cluster's bulk peculiar velocity. This effect, known as rotational kSZ [or rkSZ, not to be confused with the \textit{relativistic} SZ, see e.g. \citet{Sazonov1998, Challinor1998, Itoh98, Chluba2012SZpack, 2020MNRAS.493.3274L}], was initially estimated to produce a temperature variation $\Delta T_{\rm rkSZ}$ over the CMB ranging from $\simeq 3.5\, \mu$K for a relaxed cluster to $\simeq 146\, \mu$K for a recent merger \citepalias{CM02}, assuming a halo $\beta$-model from \cite{1976A&A....49..137C} and solid body rotation. The rkSZ signal has not yet been observed in individual clusters due to its remarkably small amplitude and its dependence on the orientation of the rotation axis relative to the line of sight (LoS).

The ICM gas moving towards the observed produces a temperature increment over the CMB, while gas moving in the opposite direction leads to a  temperature {decrement}. In the presence of cluster rotation, these patterns are adjacent and produce a dipole-like signature, which in the general case is superimposed to the monopole-like signal due to the cluster's bulk motion along the LoS. {Such} dipolar patterns in the CMB temperature map can also arise from the gravitational moving-lens effect {\citep{1986Natur.324..349G, 2007MNRAS.380.1023S, 2021PhRvD.104h3529H, 2021PhRvD.103d3536H}}: the CMB photons, deflected by the deep gravitational potential well of clusters, cause the anisotropies to be re-mapped and imprint an additional dipole-like feature in the temperature distribution. The dipole-like pattern from weak lensing has a temperature and angular scale comparable to that from the rkSZ effect {(see \citealt{2000ApJ...538...57S}, Section 4.3.1 of \citealt{2015ApJ...806..247B}, and \citealt{2019PhRvL.123r1301R})}, making it challenging to distinguish the two effects.

Recently, attempts to isolate the the rkSZ signal from the hot circumgalactic medium of 2000 galaxies were performed by \citet{2020PhRvD.101h3016Z}. {Slightly earlier,} \citet{2019JCAP...06..001B} presented a similar analysis of {\textit{Planck} data using 13 galaxy clusters from the SDSS-DR10 \citep{2014ApJS..211...17A} showing indications of bulk rotation \citep{2017MNRAS.465.2616M}}. In both works, the authors state the importance of \textit{aligning} and \textit{stacking} the kSZ maps from multiple objects to retrieve the rotational signal with sufficient signal-to-noise ratio. Because the rkSZ signal produces a dipolar pattern in the observed $\Delta T$ field, the maps must be {oriented} such that the {projected} rotation axis of the objects in the sample is aligned to maximise the rotational signal. The scale of the maps is then normalised to the objects' self-similar scale radii and the results are finally stacked. 

Synthetic galaxy clusters produced in hydrodynamic simulations offer unique test-cases for predicting the kSZ signal from bulk motion and rotation of the ICM. Crucially, simulations model the formation of clusters from cosmological accretion and therefore can capture the angular momentum transfer during gravitational collapse, mergers and substructures, all of which are not included in the analytic models used by \citetalias{CM02}. Using six clusters selected from the MUSIC simulations \citep{2013MNRAS.429..323S}, \citet{2018MNRAS.479.4028B} found that the rkSZ signal can account for up to 23\% of the kSZ component purely from bulk motion. They extend the study by showing the rkSZ signal variation at different orientations of the rotation axis relative to the LoS. Using a 6-parameter \citet{2006ApJ...640..691V} model fit to the electron number density profile of each halo and a parametric tangential velocity profile from \cite{2017MNRAS.465.2584B}, they could recover the tangential scale-velocity and bulk velocity by fitting the analytic model to the synthetic kSZ maps. Since the work by \citet{2018MNRAS.479.4028B}, \citet{2021MNRAS.504.4568M} have produced rkSZ maps to improve the accuracy of halo spin bias estimates using $5\times10^4$ halos with virial mass between $1.48 \times 10^{11}$ M$_\odot$ and $4.68 \times 10^{14}$ M$_\odot$ drawn from the IllustrisTNG simulation \citep{2018MNRAS.473.4077P}.

In this paper, we focus our discussion on the rotation of clusters, with particular reference to the work by \citet{2019JCAP...06..001B}. To obtain the rotation axis of the galaxy clusters in their sample, \citet{2019JCAP...06..001B} matched the {\textit{Planck} SZ maps with SDSS-DR10 galaxies in clusters showing evidence of coherent rotation based on the {LoS} velocity of the sources \citep{2017MNRAS.465.2616M}}. {Under the assumption that the galaxies and the ICM rotate about} the same axis, they appropriately oriented, scaled and then stacked the \textit{Planck} SZ maps to maximise the amplitude of the dipole-like signature of the kSZ effect from cluster rotation. Using the rkSZ analytic model from \citetalias{CM02}, they estimated the model parameters using the maximum likelihood estimation method, {yielding $\simeq 2\, \sigma$ evidence for the presence of the effect}.

In a $\Lambda$-cold dark matter (CDM) cosmology, the collapse of structures and the shape of the gravitational potential at low redshift is dominated by dark matter, with the baryonic matter following the same evolution. This concept implies that the different components of galaxy clusters (ICM gas, dark matter and stars in galaxies) are expected to co-rotate and to have their total angular momenta aligned. Early simulations found that the spin of dark matter halos is usually well-aligned with that of the central galaxy and the ICM, except for a non-negligible fraction of the population of objects showing misalignment \citep{2002ApJ...576...21V, 2010MNRAS.404.1137B}. The same works found that the angular momentum orientation of matter in the inner and outer halo is often vastly different, suggesting the importance in choosing an appropriate aperture when defining the spin of a galaxy cluster. More recently, these results have been corroborated using the Illustris simulation, which produced a surprisingly large ($30-50^{\circ}$ considering particles within the virial radius) median misalignment between gas and galaxies in cluster-sized objects, explained by the old-type stars being subject to the gravitational potential of the dark matter field, while being relatively unaffected by the gas hydrodynamics in the ICM in the late-time halo assembly \citep[see section 5.5 of][]{2017MNRAS.466.1625Z}. Although this estimate for the gas-stars misalignment includes \textit{all stars} within the virial radius instead of just the satellite galaxies as in the set-up used by \citet{2019JCAP...06..001B}, the result from the Illustris simulation suggests that the co-rotation of stars and gas, critical for recovering the elusive rkSZ signal, requires further inspection.

In this work, we aim to investigate the assumption of co-rotating galaxies and ICM gas used by \citet{2019JCAP...06..001B}, modelling the rkSZ signal of massive clusters in different alignment and stacking scenarios. The MACSIS simulations \citep{macsis_barnes_2017} provide an excellent suite of synthetic galaxy clusters simulated with the the \textsc{Gadget-3} code and the BAHAMAS sub-grid physics model. The MACSIS clusters were selected to have FoF mass above $10^{15}$ M$_\odot$, which extends the sample in \citet{2017MNRAS.466.1625Z} and \citet{2018MNRAS.479.4028B} to higher masses by one order of magnitude. Crucially, the amplitude of the SZ effects is larger in massive halos, meaning that observational surveys are most likely to detect rotational features in the SZ sky by selecting massive and merger-prone MACSIS-like clusters \citep{CM02, 2003AstL...29..783S}.

This work is organised as follows. Section \ref{sec:datasets} introduces the key features of the MACSIS cluster sample and  the theoretical framework for the rkSZ effect; in Section \ref{sec:cluster-properties}, we examine the alignment of the angular momenta of cluster components; in Section \ref{sec:image-processing}, we reproduce the rkSZ map-stacking method used in \citet{2019JCAP...06..001B} and in Section \ref{sec:results} explore the rotational signal with different selection criteria. We then fit a analytic model to the profiles as discussed in Section~\ref{sec:results:fitting} and, {starting from the prescription of \citetalias{CC02},} in Section~\ref{sec:power-spectrum} we illustrate how {our} results can be used to predict the contribution of the cluster rotation to the kSZ power spectrum. Finally, in Section \ref{sec:summary}, we discuss prospects for future models and observations of the kSZ effect from cluster rotation.

Throughout this work, we adopt the cosmology used in MACSIS \citep{macsis_barnes_2017}, with parameters: $\Omega_{\rm b} = 0.04825$, $\Omega_{\rm m} = 0.307$, $\Omega_\Lambda = 0.693$, $h \equiv H_0/(100 ~{\rm km~s^{-1}Mpc^{-1}}) = 0.6777$, $\sigma_8 = 0.8288$, $n_s = 0.9611$ and $Y = 0.248$ \citep{2014A&A...571A...1P}.

\section{Overview of the MACSIS simulations}
\label{sec:datasets}

\begin{figure}
	\includegraphics[width=\textwidth]{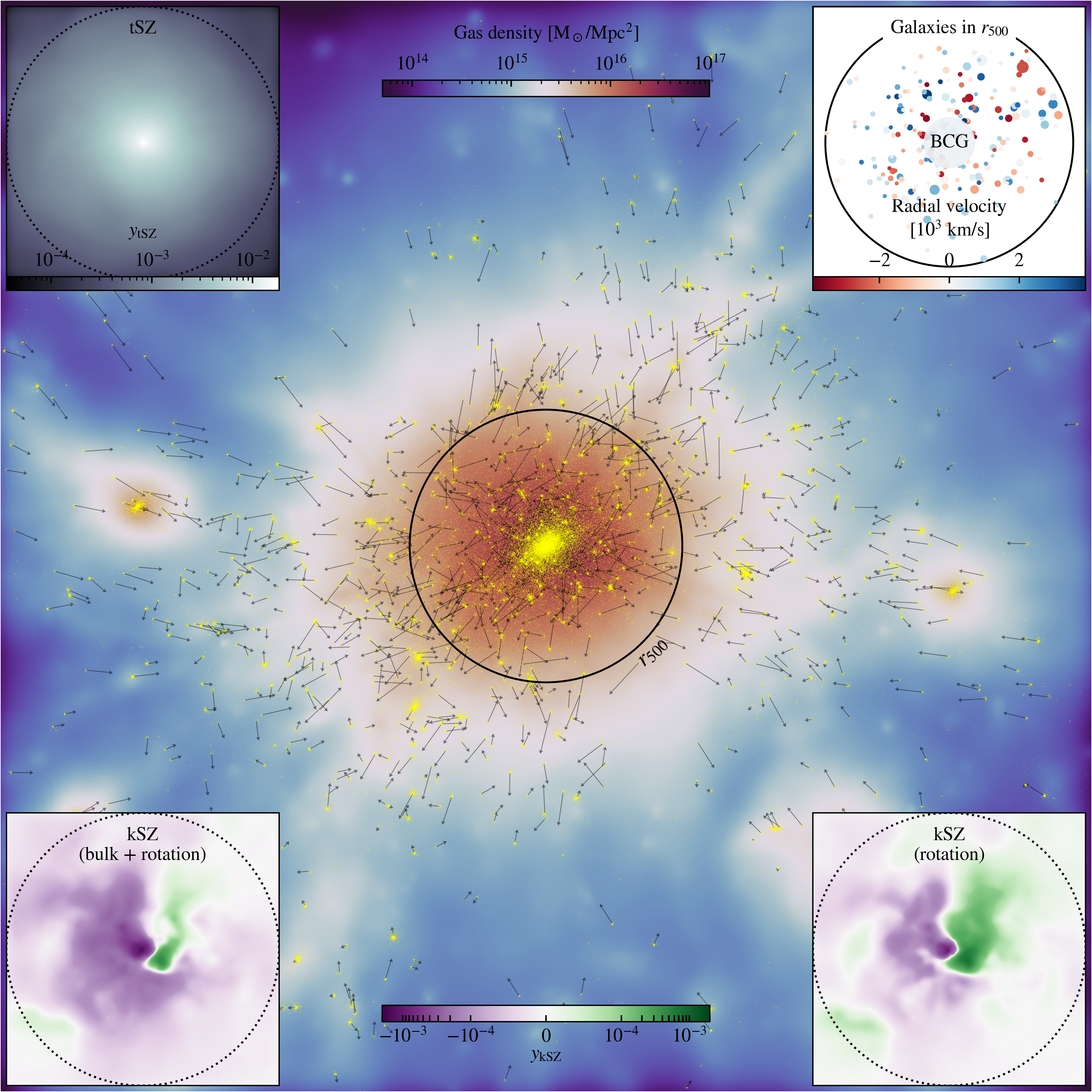}
    \caption{Illustration of the hot gas and stars in the MACSIS 0 cluster environment at $z=0$. The central image shows the projected gas density in the background (colour-coded), with superimposed the star particles marked in yellow. For the substructures with stellar mass above $10^{10}$ M$_\odot$, we also show the projected velocity vectors as black arrows. The spatial extent of the central map is 8 $r_{500}$, and the $r_{500}=2.39$ Mpc circle is drawn to guide the eye. The cluster is rotated such that the angular momentum of the hot gas in the ICM points vertically upwards in the plane of the page. The maps in the insets all have an extent of 2 $r_{500}$; the dotted circles in the three SZ maps indicate $r_{500}$. In the top-left, we show a map of the tSZ Compton-$y$ parameter; in the bottom-left, we show the kSZ Compton-$y$ parameter for the hot gas in the rest frame of the CMB; in the bottom-right is the same kSZ map, but without the cluster's bulk motion, as indicated by the label. The kSZ (bulk + rotation) and rotation-only amplitudes are comparable, however, we note that a large component of the bulk velocity of the cluster is oriented tangentially and, therefore, it does not contribute to the kSZ signal {at order $\simeq v/c$}. Finally, the top-right plot shows the position of the galaxies inside $r_{500}$ (3D, not projected), with markers colour-coded based on the {LoS} velocity and with size proportional to the logarithm of their stellar mass. The BCG is indicated in the centre of the plot, as well as the $r_{500}$ radius.}
    \label{fig:cluster_display}
\end{figure}

\begin{figure}
	\includegraphics[width=\textwidth]{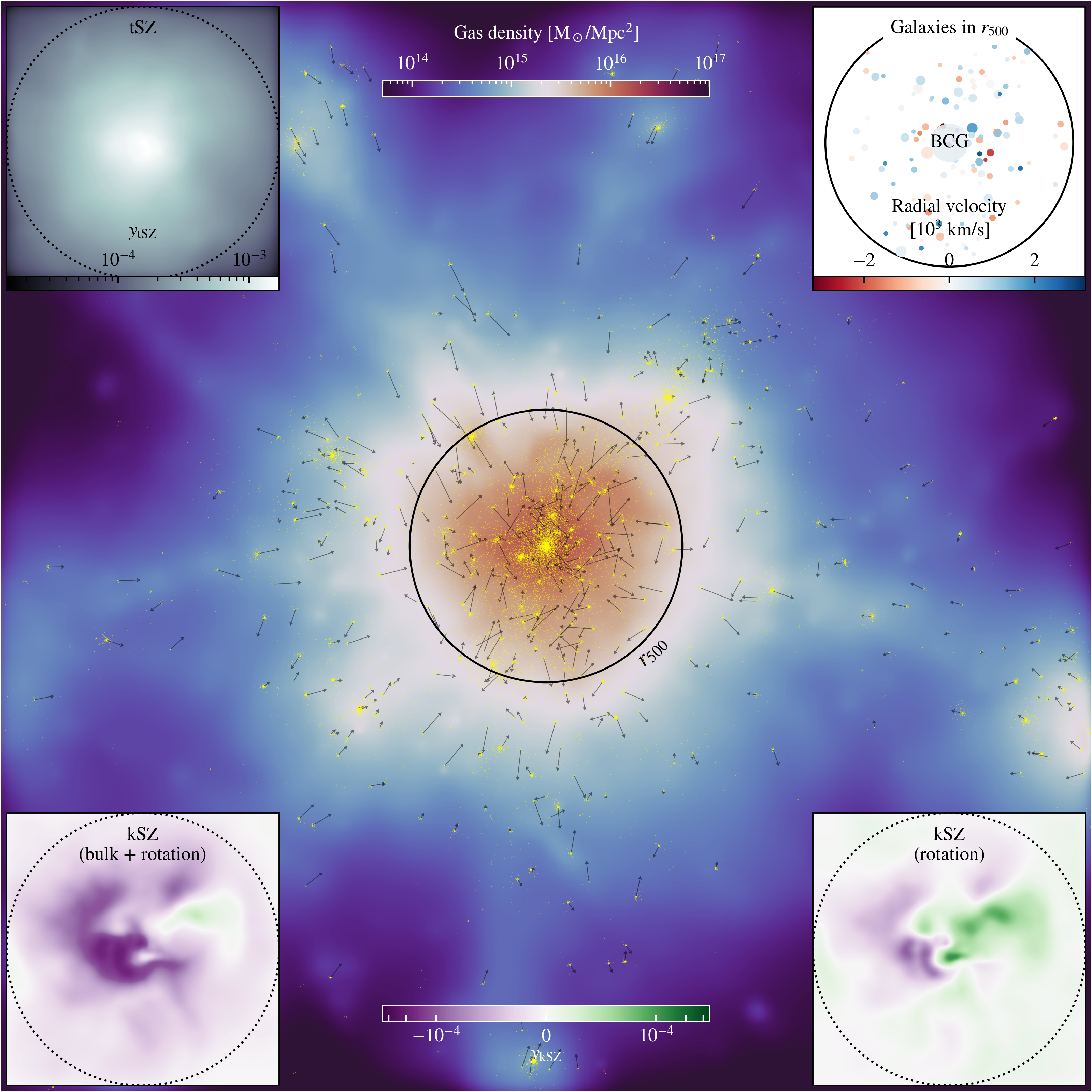}
    \caption{As in Fig. \ref{fig:cluster_display}, but showing the MACSIS 100 cluster at $z=0$. This smaller system, with $M_{500}=1.77 \times 10^{15}$ M$_\odot$ and $r_{500}=1.88$ Mpc, shows a lower number of galaxies with stellar mass above $10^{10}$ M$_\odot$. Kinematically, we report a significant bulk motion (bottom left) along the {LoS} and a residual dipolar kSZ pattern due to bulk rotation (bottom right). {The dipolar rotational signature is less pronounced due to the presence of moving substructures.}}
    \label{fig:cluster_display_100}
\end{figure}

\begin{figure}
	\includegraphics[width=\textwidth]{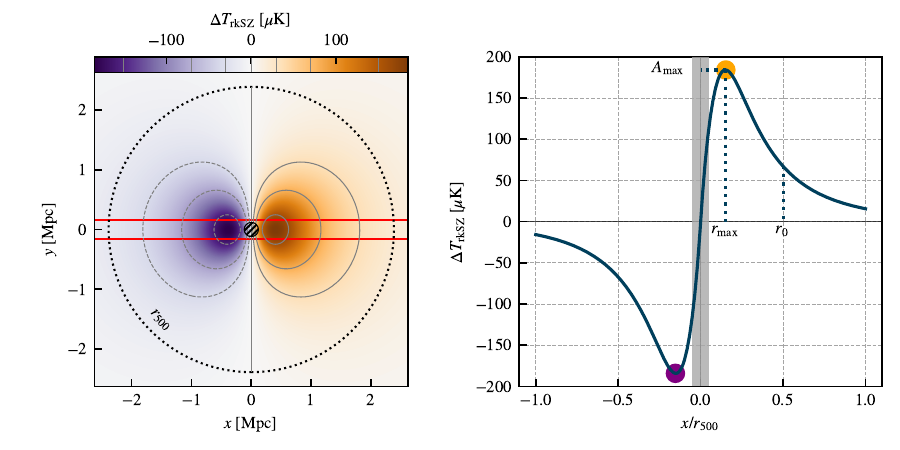}
    \caption{Model of the rkSZ signal map (left) and equatorial slice (right) based on the MACSIS 0 cluster at $z=0$. The resolution of the map is 20.5 kpc/pixel (256 pixels) and the horizontal red slice has a width of 657 kpc (32 pixels). In the left panel, we show line contours for $\Delta T_{\rm rkSZ}=0~\mu {\rm K},\, \pm 30 ~\mu {\rm K},\, \pm 70~\mu {\rm K},\, \pm 150~\mu {\rm K}$. We also indicate $r_{500}$ with a dotted circle. The inner hatched circle represents the $0.05\,r_{500}$ region excluded in the analysis. In the right panel, this region is represented as a vertical grey band between $\pm 0.05\,r_{500}$. The maximum (and minimum) rkSZ amplitude is $A_{\rm max}$, occurring at a radius $r_{\rm max}$. We also show the position of $r_0$ as a guideline, and we mark the extreme values in the profile with the same colours as in the map.}
    \label{fig:rksz_model}
\end{figure}

The MACSIS hydrodynamic simulations \citep{macsis_barnes_2017} are a suite of 390 galaxy clusters simulated with dark matter, gas and sub-grid physics using the \textsc{Gadget-3} smoothed\hyp{}particle hydrodynamics (SPH) code \citep[updated from \textsc{Gadget-2}, see][for details]{2005MNRAS.364.1105S}. These objects were initially selected from a (3.2 Gpc)$^3$ dark-matter-only parent volume based on their friends-of-friends (FoF) mass ($10^{15}<M_{\rm FoF}/{\rm M}_\odot<10^{16}$) at redshift $z=0$. The FoF groups within this mass range were then placed in logarithmic mass bins with a constant width of 0.2 dex. The two largest mass bins contained 7 and 83 halos, while 100 halos were selected from each of the three lowest mass bins.\footnote{The mass bins with $\log_{10}(M_{\rm FoF}/{\rm M}_\odot)$ in the interval 15.0-15.2, 15.2-15.4 and 15.4-15.6 were further divided into 10 sub-bins each (constant logarithmic spacing of 0.02 dex), and 10 halos were selected from each sub-bin, amounting to 100 halos per bin. This method was performed to minimise bias selection towards low masses, caused by the steep slope of the halo mass function.} The 390 halos were then re-simulated individually using the zoom-in technique \citep{1993ApJ...412..455K, 1997MNRAS.286..865T}, firstly in dark-matter-only mode and then with full physics. The MACSIS project used the same sub-grid model, particle-mass resolution and softening as in the BAHAMAS project \citep{2017MNRAS.465.2936M}: for the hydro-simulations, the dark matter particles had a mass of $6.49 \times 10^9$~M$_\odot$ and the gas particles had an initial mass of $1.18 \times 10^9$~M$_\odot$. \cite{macsis_barnes_2017} showed that the MACSIS (and the combined BAHAMAS+MACSIS) sample are in good agreement with the mass dependence of the observed hot gas fraction, the X-ray luminosity and the SZ Compton-$y$ parameter at $z=0$. To define the halos and substructures, we use the products of the SUBFIND code \citep{2001MNRAS.328..726S, 2009MNRAS.399..497D}. We also define galaxies as self-bound substructures with an associated stellar mass contained in a 70 kpc-radius spherical aperture above $10^{10}$ M$_\odot$, as in \cite{2019MNRAS.484.1526A}. The selection rules used in the definition of the MACSIS sample lead to an underrepresented low-mass halo population, biased towards low concentrations and high dark-matter spin parameter \citep{2017MNRAS.465.3361H}. Since our work focuses on the rotational dynamics of the cluster population, we discard MACSIS clusters below $M_{200} = 10^{14.5} h^{-1}$ M$_\odot$ at $z=0$ to mitigate this bias.\footnote{$M_{200}$ is defined as the total mass within a spherical overdensity of radius $r_{200}$, centred in the gravitational potential minimum. $r_{200}$ is the radius at which the internal mean density exceeds the critical density by a factor of 200. Similarly, this approach is used to define $r_{500}$ and $M_{500}$.} We further discard a small number (13) of clusters with abnormally low $f_{\rm gas}<0.05$ at $z=0$, likely caused by an AGN feedback event at high redshift. After the selection, the MACSIS sample is reduced from 390 to 377 clusters.

In Fig.~\ref{fig:cluster_display}, we show the most massive cluster in the sample, MACSIS 0 ($M_{500}=3.6\times10^{15}$ M$_\odot$), at $z=0$ with hot gas and stellar components. The map in the background encodes the gas density along the LoS and the star particles are shown as yellow markers in the foreground.
{In bottom-right corner of Fig.~\ref{fig:cluster_display}, we show} a map of the kSZ amplitude without the bulk motion. {This highlights the contribution from cluster rotation to the kSZ morphology, {which should become strongly visible when added to the approximately null tSZ signal at $\nu\simeq 217\,{\rm GHz}$} (CM02 and upper left corner of Fig.~\ref{fig:cluster_display}).} We note that for this particular projection, the bulk velocity of the cluster is perpendicular to the LoS, and only a smaller radial component boosts the negative part of the rkSZ dipole. When the bulk velocity is oriented radially, its contribution to the kSZ signal is larger, as in the visualisation of MACSIS 100 in Fig. \ref{fig:cluster_display_100}. Using this comparison, we stress that the kSZ amplitude strongly depends on the geometry of the observer, the cluster bulk velocity and bulk angular momentum.

\subsection{Modelling the rkSZ signal}
\label{sec:rksz-modelling}
{We will now describe the technique used to obtain this projected map and introduce the formalism that provides a analytic model for the rkSZ profiles.}
The kSZ Compton-$y$ parameter, given by the temperature difference $\Delta T_{\rm kSZ}$ over the CMB blackbody temperature, $T_{\rm CMB}$, is proportional to the mass-weighted velocity integrated along the LoS \citep{ksz_sunyaev_1980, mroczkowski2019_review}:
\begin{equation}
\label{eq:kSZ_definition}
     y_{\rm kSZ} \equiv \frac{\Delta T_{\rm kSZ}}{T_{\rm CMB}} = -\frac{\sigma_T}{c}\int_{\rm LoS} n_e \mathbf{v}\cdot {\rm d}\mathbf{l}.
\end{equation}
Here, $n_e$ is the free-electron number density and the product $\mathbf{v}\cdot d\mathbf{l}$ is the radial component of the velocity of the electron cloud element.
To construct a map of the projected Compton-$y$ parameter from SPH simulations, we discretise the integral above and sum the contributions of each particle $j$ to pixel $i$ as
\begin{equation}
    y_{{\rm kSZ,}i} = -\frac{\sigma_T}{c\, m_{\rm P}\, \mu_{\rm e}} \sum_{j} m_j v_{{\rm LoS,}j}\, W_{ij}(h_j),
\end{equation}
where $m_{\rm P}$ is the mass of the proton, $\mu_{\rm e}=1.14$ is the mean molecular weight per free electron, $v_{{\rm LoS,}j}$ is the velocity along the LoS, $h_j$ is the SPH smoothing length of the gas particle $j$ and $W_{ij}$ is the Wendland-C2 kernel \citep{wendland1995piecewise}, as implemented in \textsc{SWIFTsimIO} \citep{Borrow2020}.\footnote{The calculation of the smoothed projection maps is performed using the \textsc{SWIFTsimIO} \textit{subsampled} backend, which guarantees converged results by evaluating each kernel 32 times or more. The overlaps between pixels are taken into account for every particle \citep[see][for further details]{Borrow2021}}

We now derive the scaling relation of the kSZ amplitude with mass and redshift, predicted from self-similar cluster properties. Assuming fully ionised primordial gas, we obtain that $n_e$ is proportional to the critical density of the Universe, 
\begin{equation}
    \rho_{\rm crit}(z) = E^2(z)\, \frac{3 H_0^2}{8 \pi G},
\end{equation}
where $E^2(z)\equiv H^2(z) / H_0^2 =\Omega_{\rm m}(1+z)^3 + \Omega_\Lambda$, yielding $n_e \propto E(z)^2$. Similarly, assuming that the motion scales with the circular velocity of the cluster, we have 
{$|\mathbf{v}| \propto E(z)^{1/3}\, M_{500}^{1/3}$} 
and the electron column along the line-of-sight $\int |d\mathbf{l}| \propto r_{500} \propto  M_{500}^{1/3}\, E(z)^{-2/3}$. By combining these scaling relations in {Eq.~\eqref{eq:kSZ_definition}}, we obtain the predicted scaling for the kSZ amplitude:
\begin{equation}
    y_{\rm kSZ} \propto \Delta T_{\rm kSZ} \propto M_{500}^{2/3}\, E(z)^{5/3},
    \label{eq:kSZ-scaling}
\end{equation}
which is expected to moderately increase with cluster mass, and significantly increase with redshift. We note that this relation refers to the Compton-$y$ measured along the LoS, and not the Compton-$y$ integrated over the solid angle of the cluster. Compared to {the tSZ scaling} ($y_{\rm tSZ} \propto \int n_e T\, {\rm d}l \propto M_{500}\, E(z)^{2}$), the kSZ scaling has a weaker dependence on both mass and redshift.

{Assuming spherical symmetry and} following the formulation of \cite{2017MNRAS.465.2584B} for the rotating ICM, $\Delta T_{\rm kSZ}$ can be expressed in polar coordinates $(R, \phi)$ by the integral
\begin{equation}
\label{eq:kSZ-projection}
     y_{\rm rkSZ}(R,\phi) =-\frac{\sigma_T}{c}R\cos \phi \sin i \int_{R}^{r_{500}} n_e(r)~\omega(r)~\frac{2r~{\rm d}r}{\sqrt{r^2-R^2}},
\end{equation}
{with $R\in [0, r_{500}]$}. Here, the radial electron number density profile $n_e(r)$ and the angular velocity profile $\omega(r)$ are clearly separated in the integrand.\footnote{The $\frac{2r}{\sqrt{r^2-R^2}}$ part of the integrand is a projection factor computed by applying a direct Abell transform. The integration limits also reflect this mapping.} The inclination angle $i$ has the effect of reducing the overall amplitude of the rkSZ effect for $|\sin i|<1$.
This model shows that the rkSZ pattern is fully described by the choice of number density and angular velocity profiles. In our study, we model the rkSZ signal by assuming a \cite{2006ApJ...640..691V} $n_e(r)$ model {with the functional form}:
\begin{equation}
n_e= n_0\, \frac{\left(r / r_{\rm c}\right)^{-\alpha / 2}}{\left(1+r^{2} / r_{\rm c}^{2}\right)^{3 \beta / 2-\alpha / {4}}} \frac{1}{\Big(1+{[r/r_{\rm s}]^\gamma}\Big)^{\varepsilon / {2}\gamma}},
\label{eq:V06:emissionmeasure}
\end{equation}
where $\gamma = 3$ and $\varepsilon < 5$ are constrained and the other parameters are free to vary in the positive real interval.  {Following \cite{2017MNRAS.465.2584B} we found that the angular rotational velocity profile is well matched to the MACSIS data by}
\begin{equation}
\label{eq:omega-profile}
    \omega(r) = \frac{v_{\rm t0}}{r_0 \left[1 + (r/r_0)^\eta\right]},
\end{equation}
with the velocity scale radius $r_0$, the tangential velocity scale $v_{\rm t0}$, and the dimensionless slope parameter $\eta$ are to be determined by fitting the profile. Eq.~\eqref{eq:omega-profile} is a generalisation of the profile used to describe the rkSZ profiles of six relaxed MUSIC clusters \citep[fixing $\eta = 2$,][]{2017MNRAS.465.2584B}. Since most MACSIS clusters are dynamically unrelaxed and span over a wide range of halo masses, we allow $\eta$ to vary in the range $[1, 3]$ to match the slope of the decaying profile outside the peak radius. We illustrate the effect of changing the $r_0$, $v_{\rm t0}$ and $\eta$ parameters on the angular velocity and rkSZ profiles in Section~\ref{app:profile-fits}.

In Fig. \ref{fig:rksz_model}, we show an example an rkSZ map (top) modelled on the same cluster as above. For the density profile, we fit the $n_e(r)$ profile in Eq.~\eqref{eq:V06:emissionmeasure} with best-fit parameters
$$\left\{\frac{n_0}{\rm cm^{-3}}, \frac{r_{\rm c}}{r_{500}}, \frac{r_{\rm s}}{r_{500}}, \alpha,  \beta, \varepsilon\right\} = \{3.7 \times 10^{-3}, 0.17, 0.75, 1.5, 0.59, 2\}.$$
{We provide details on the profile fitting strategy in Section \ref{sec:results:fitting}.}

For the $\omega(r)$ profile, we assume a tangential velocity scale equal to the circular velocity of the cluster at $r_{500}$, given by $v_{\rm t0} \simeq v_{\rm circ} \simeq \sqrt{G M_{500}/r_{500}}$ and $r_0=r_{500} / 5$. We use the $n_e(r)$ fit from MACSIS~0 just to construct an example of rotation map and we do not fit for $v_{\rm t0}$ and $r_0$ at this stage. The top panel of Fig. \ref{fig:rksz_model} also shows the $r_{500}$ radius, and in solid grey the $0.05~r_{500}$ radius excluded from the \cite{2006ApJ...640..691V} fit. This threshold radius was determined by the particle-softening scale (4 $h^{-1}$ physical-kpc at $z < 3$) normalised to the value of $r_{500}$ for the smallest cluster in the MACSIS sample at $z=0$. To measure the amplitude of the rkSZ effect, we consider a horizontal slice through the centre of the halo and we average the pixel values in each column to give the rotation profile in the bottom panel of Fig. \ref{fig:rksz_model}. There, we also indicate the radius excluded from the density profile fit and the $r_{500}$ radius.

\section{Cluster properties and alignment between halo components}
\label{sec:cluster-properties}
Massive clusters form deep gravitational potential wells, causing the gas density to increase towards the centre. We therefore expect cluster atmospheres in large systems to produce more intense {SZ} signals. Moreover, MACSIS-like clusters acquire mass through a long history of mergers and accretion of matter from the surrounding filaments. For this reason, they present a complex dynamics even at $z=0$, making them ideal for future \textit{blind} rkSZ detection, {i.e. without prior knowledge of the cluster rotation axis}. In this paper, not only do we study the dependence of the rkSZ signal amplitude to the halo mass, $M_{500}$, but we also investigate its variations {with} other cluster properties, which we group in two categories: basic cluster properties and metrics for dynamical state. We will now investigate the MACSIS data set using these two classes of metrics and their correlations. 

\begin{figure}
    \centering
	\includegraphics[width=\textwidth]{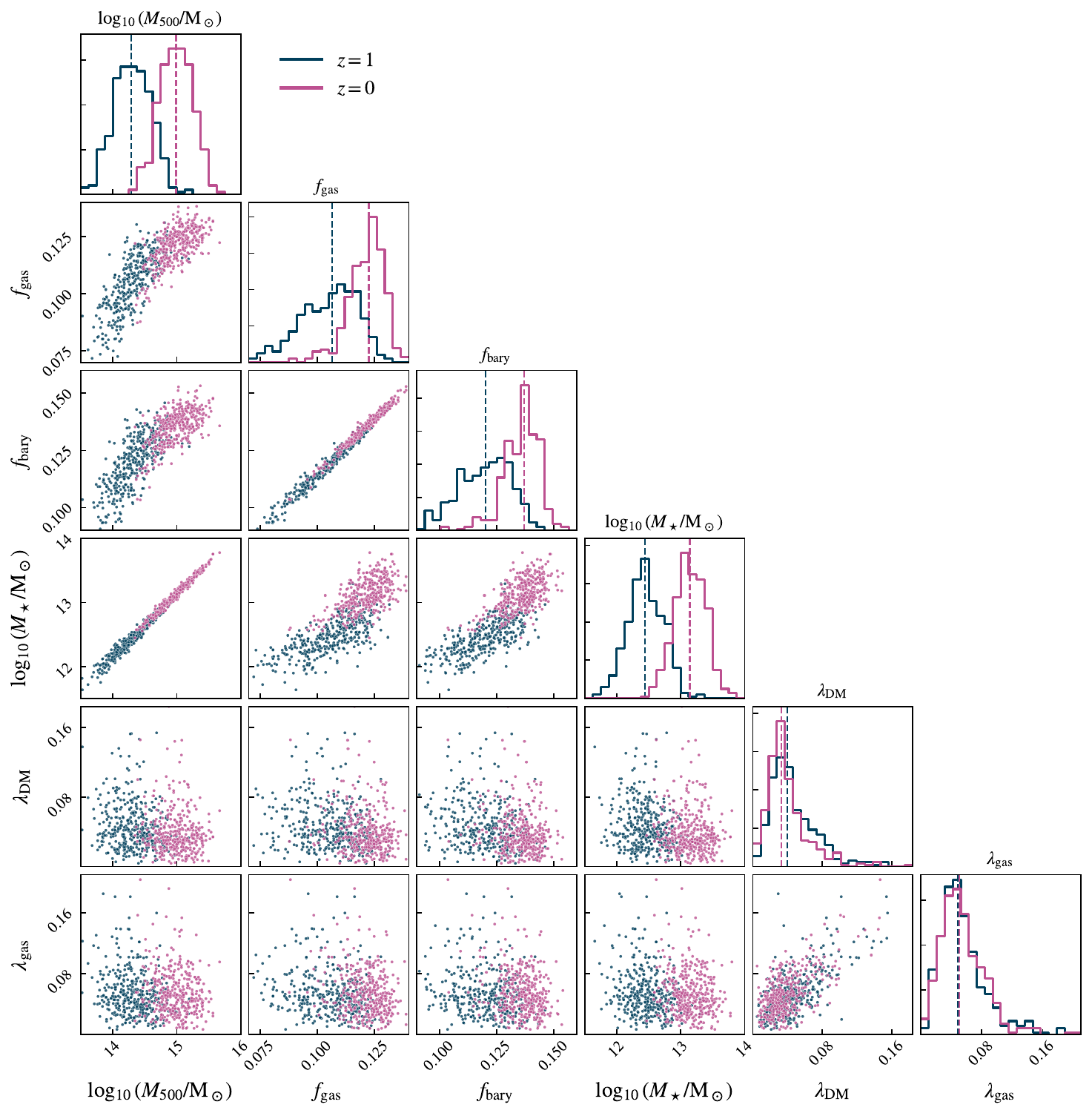}
    \caption{Corner plot with distributions of the basic cluster properties at $z=0$ (pink) and $z=1$ (blue). From left to right and top to bottom, we show the halo mass $M_{500}$, the hot gas fraction $f_{\rm gas}$, the baryon fraction $f_{\rm bary}$, the stellar mass $M_\star(r<r_{500})$, the dark matter spin parameter $\lambda_{\rm DM}$ and the hot gas spin parameter $\lambda_{\rm gas}$. Along the principal diagonal, we show the p.d.f. histograms of the cluster population, after removing the anomalous objects as specified in Section \ref{sec:datasets}. In each histogram, we indicate the sample median with vertical dashed lines. For clarity, each histogram in the diagonal plots is shown with a title indicating the quantity represented by the x-axis.}
    \label{fig:corner-plot-basic-properties}
\end{figure}

\subsection{Basic cluster properties}
\label{sec:alignment:properties}
We begin by defining the global cluster properties used in the rest of this analysis. The hot gas fraction is $f_{\rm gas}\equiv M_{\rm gas}(r<r_{500})/M_{500}$, where $M_{\rm gas}(r<r_{500})$ is the total mass of the gas above the hydrogen ionisation temperature (we apply a temperature cut of $T\geq 10^5$ K) inside $r_{500}$. Similarly, we define the star fraction $f_\star \equiv M_\star(r<r_{500})/M_{500}$, where $M_\star(r<r_{500})$ is the stellar mass in $r_{500}$, and hence the baryon fraction $f_{\rm bary} = f_{\rm gas} + f_\star$.

The dark matter spin parameter, $\lambda_{\rm DM}$, measures the fraction of mechanical energy of the clusters due to rotation and is estimated using the relation by \cite{2001ApJ...555..240B}:
\begin{equation}
    \label{eq:spin-parameter}
    \lambda_{\rm DM} = \frac{j_{\rm DM}}{\sqrt{2} v_{\rm circ} r_{500}},
\end{equation}
where $j_{\rm DM}$ is the specific angular momentum (about the centre of potential) of the dark matter in $r_{500}$ and $v_{\rm circ}=\sqrt{GM_{500}/r_{500}}$ the circular velocity at $r_{500}$. We use $\lambda_{\rm DM}$ to classify slow and fast dark-matter rotators in the MACSIS sample. {In analogy to $\lambda_{\rm DM}$,} $\lambda_{\rm gas}$ instead uses the specific angular momentum of the hot gas $j_{\rm gas}$ to quantify the fraction of kinetic energy of the ICM associated with the bulk rotation. 

In Fig. \ref{fig:corner-plot-basic-properties}, we show these quantities for the MACSIS clusters at $z=0$ (pink) and $z=1$ (blue) in a corner plot, with the probability density functions (p.d.f.s) along the principal diagonal. For each p.d.f., a vertical dashed line indicates the median value, which we will use to split the $z=0$ sample in Section \ref{sec:results}. The $M_{500}$ p.d.f. on the top-left corner is the normalised halo mass function (HMF), whose shape is determined by the sample selection method of \cite{macsis_barnes_2017}. We will discuss the MACSIS HMF further in Section \ref{sec:results:z-dependence}, when selecting clusters to compare the $z=0$ and $z=1$ populations. Below the $M_{500}$ p.d.f., the $f_{\rm gas}$-$M_{500}$ and $f_{\rm bary}$-$M_{500}$ panels show the hot gas and baryon mass-scaling relations. These reproduce the results shown by \cite{macsis_barnes_2017}. The p.d.f. for $\lambda_{\rm DM}$ and $\lambda_{\rm gas}$ show a population distribution compatible with early results by \cite{2001ApJ...555..240B} and, in addition, suggest that $\lambda_{\rm gas}$ is correlated with $\lambda_{\rm DM}$, but both are only weakly correlated with the other basic cluster properties. Neither the $\lambda_{\rm gas}$ or $\lambda_{\rm DM}$ distributions change significantly with redshift.

\begin{figure}
    \centering
	\includegraphics[width=\textwidth]{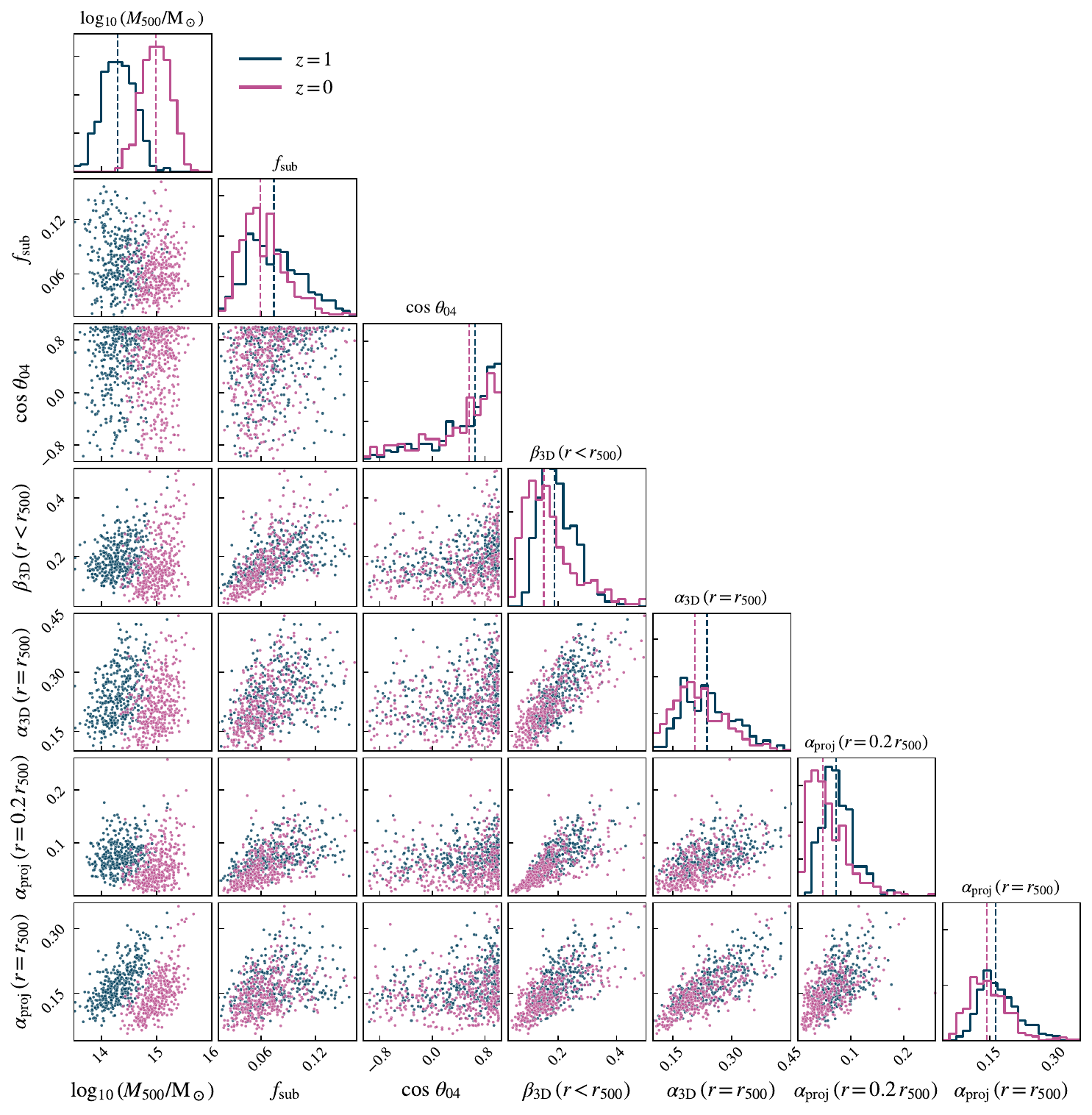}
    \caption{As in Fig. \ref{fig:corner-plot-basic-properties}, but focusing on the dynamical state metrics: the substructure mass fraction $f_{\rm sub}$, the angle between the angular momenta of the hot gas and the galaxies $\cos\,\theta_{04}$, the overall kinetic-to-thermal ratio $\beta_{\rm 3D}$ inside $r_{500}$, the 3D non-thermal pressure fraction at $\alpha_{\rm 3D}$ evaluated at $r_{500}$ and, finally, the projected non-thermal pressure fraction $\alpha_{\rm proj}$ evaluated at in the core ($0.2\,r_{500}$) and at at $r_{500}$.}
    \label{fig:corner-plot-dynamical-state}
\end{figure}

\subsection{Metrics for dynamical state and spin geometry}
\label{sec:alignment:spin}
To quantify how thermodynamically perturbed the ICM is, we use the kinetic-to-thermal (energy) ratio $\beta \equiv E_{\rm kin}/E_{\rm th}$ as a proxy. Here, $E_{\rm kin}=1/2 \sum m_i ({\bf v}_i - {\bf v_{\rm bulk}})^2$ is the total kinetic energy of the hot gas within $r_{500}$, computed from the mass $m_i$ and the velocity ${\bf v}_i$ of the particles after subtracting the bulk velocity ${\bf v_{\rm bulk}}$; assuming all gas is mono-atomic, the thermal energy $E_{\rm th}=3/2\, {\rm k_B}\sum T_i N_{{\rm H},i}$, where $T_i$ is the temperature of the gas particle, the number of hydrogen atoms is $N_{{\rm H},i}=m_i/(\mu\,m_{\rm P})$, $\mu\simeq 1.16$ is the mean molecular weight and $m_{\rm P}$ is the proton mass. Large values of the kinetic-to-thermal ratio ($\gtrsim 0.1$), typically measured in high-mass systems, are associated with a thermodynamically perturbed cluster atmosphere \citep[see e.g.][]{macsis_barnes_2017}. Throughout this work, $\beta_{\rm 3D}$ indicates the kinetic-to-thermal ratio computed for all hot gas particles in a 3D $r_{500}$ aperture.

From Fig. \ref{fig:corner-plot-dynamical-state}, we learn that 50\% of the MACSIS sample has a kinetic-to-thermal ratio lower than 0.15. However, we identify a handful of objects with $\beta_{\rm 3D}\approx0.5$. The hot gas in these systems have a particularly high kinetic energy (in the centre of potential, CoP, rest frame), which must be in the form of bulk rotation, internal bulk motions, and turbulence. The kinetic-to-thermal implicitly combines these two contributions. However, we can directly probe the kinetic energy in the rotational mode using $\lambda_{\rm gas}$, as defined above, and we use the non-thermal pressure, $P_{\mathrm{nth}}$, to quantify the kinetic energy in the form of turbulence and small-scale bulk motion \citep[e.g.][]{2018MNRAS.481L.120V}. After introducing the total (thermal plus non-thermal) pressure as $P_{\mathrm{tot}}$, we write the non-thermal support fraction as
\begin{equation}
    \alpha = \frac{P_{\mathrm{nth}}}{P_{\mathrm{tot}}} = \left[ 1 + \frac{3 k_{\rm{B}} T}{\mu m_{\rm{H}} \sigma_{\rm 3D}^2} \right]^{-1} = \frac{\beta_{\rm 3D}}{1 + \beta_{\rm 3D}}.
    \label{eq:alpha}
\end{equation}
In this expression, we use the 3D velocity dispersion, $\sigma_{\rm 3D}^2$, following \cite{2022arXiv221101239T}. In analogy to their approach, we adopt two methods, which we summarise below.
\begin{enumerate}
    \item {\bf Radial $\alpha_{\rm 3D}$ profiles}. We first compute the $j \in \{x, y, z\}$ components of the velocity dispersion along the three axes of the simulation box: 
    \begin{equation}
        \sigma_{j}^{2} = \frac{\sum_{i} W_{i} \left( v_{i,j} - \bar{v}_{j} \right)^{2}}{\sum_{i} W_{i}}.
        \label{eq:velocity-dispersion}
    \end{equation}
    Here, $\bar{v}$ is the bulk velocity of ensemble of particles $\{i\}$ within a spherical shell at radius $r$, and $W$ is a weighting function which in this case is set to the particle mass, $W_i = m_i$. Then, we find the total velocity dispersion by adding the components in quadrature as $\sigma_{\rm 3D}^2 = \sum_j \sigma_{j}^{2} \equiv \sigma_{x}^{2} + \sigma_{y}^{2} + \sigma_{z}^{2}$. This value is substituted in Eq.~\eqref{eq:alpha} to give the $\alpha_{\rm 3D}(r)$ profile. In this work, we use the value of $\alpha_{\rm 3D}(r = r_{500})$ as a metric for the turbulence derived from 3D profiles. We stress that $\alpha_{\rm 3D}$ computed in this manner is independent of the choice of LoS. Moreover, $\alpha_{\rm 3D}$ measures the \textit{local} value at $r_{500}$, unlike $\beta_{\rm 3D}$, which is an \textit{integrated} quantity.
    
    \item {\bf Projected $\alpha_{\rm 2D}$ profiles}. In our study, we employ an additional metric to estimate the non-thermal support fraction following the prescription of \cite{Roncarelli2018MeasuringObservations}. In Eq.~\eqref{eq:velocity-dispersion}, we use the X-ray emission measure as a weighing function, $W_{i}=\rho_i m_i$, and we construct one estimate of the total velocity dispersion for each axis of the simulation volume: $\sigma_{{\rm 3D}, j}^2 = 3\, \sigma_{j}^2$. These results lead to three projected profiles, $\alpha_{{\rm proj}, j}(r)$, estimated along perpendicular LoS's. As in the previous method, we aim to construct an direction-agnostic formulation of $\alpha$. We achieve this by averaging the projected $\alpha$ profiles along the three axes as $\alpha_{\rm proj}(r) = 1/3 \, \sum_j \alpha_{{\rm proj}, j}(r)$. Finally, we evaluate the averaged projected profile at $0.2\, r_{500}$ (in the core) and $r_{500}$, to obtain $\alpha_{\rm proj}(r = 0.2\, r_{500})$ and $\alpha_{\rm proj}(r = r_{500})$ respectively. The $\alpha_{\rm 2D}$ profiles are computed as the median of 12 azimuthal bins. This method was shown to be more effective in reducing local fluctuations then the spherical average method \citep[see e.g. Section 3.2 of][]{2022arXiv221101239T} used in the derivation of the $\alpha_{\rm 3D}$ profiles.
\end{enumerate} 
In Eq. \eqref{eq:alpha}, we emphasize that $\alpha$ can be written as a function of $\beta$. This connection exists because: (i) $({\bf v}_i - {\bf v_{\rm bulk}})^2$ in $E_{\rm kin}$ can be expressed in terms of the velocity dispersion $\sigma^2$ in $P_{\rm nth}$ and (ii) the (mass-weighted) temperature in $E_{\rm th}$ is the same used in the \textit{thermal} pressure fraction, which gives $P_{\rm tot}$. A detailed derivation is given in Section \ref{app:correlation:alpha-beta}.
To assess the dynamical state of a cluster, we also use the substructure fraction $f_{\rm sub}$, defined as the fraction of the mass in a FoF group bound to substructures inside $r_{500}$. For instance, \cite{2017MNRAS.465.3361H} use $f_{\rm sub}$ to classify relaxed clusters if $f_{\rm sub} < 0.1$. Crucially, substructures falling into the cluster's potential well are known to transfer angular momentum into the system and perturb the ICM, potentially enhancing the kSZ amplitude due to cluster rotation.

We show the {distribution} of these properties for the MACSIS sample in Fig. \ref{fig:corner-plot-dynamical-state}. We recover positive correlation between $\beta_{\rm 3D}$ and $f_{\rm sub}$ as expected and we also show a similar correlation between the three estimates for $\alpha$ at $z=0$ and $z=1$ with $M_{500}$, $\beta_{\rm 3D}$ and $f_{\rm sub}$. {None of the dynamical state metrics show very strong dependence on redshift.}

Next, we illustrate the framework for computing $\cos\theta_{04}$, {a metric for the alignment of the gas and galaxies spins,} and discuss its value in the context of the MACSIS clusters. The orientation of the angular momentum of the cluster components (gas, dark matter and stars) relative to each other is important when scaling, {reorienting} and stacking kSZ maps, for instance in the study by \cite{2019JCAP...06..001B}. In this work, we mimic their analysis method by computing the specific angular momentum (or simply \textit{spin}) of each cluster component as ${\bf j}_k = {\bf J}_k / M_k$ with $k\in\{{\rm gas,~DM,~stars}\}$, $M_k$ the component mass in $r_{500}$ and the angular momentum
\begin{equation}
    {\bf J}_k = \sum_{i:r<r_{500}} m_i ~ \left({\bf v_i} - {\bf v_{\rm bulk}}\right) \times \left({\bf r_i} - {\bf r_{\rm CoP}}\right) \Bigg|_{k\in\{{\rm gas,~DM,~stars}\}}.
\end{equation}
Note that ${\bf J}_k$ is computed about the centre of potential at position ${\bf r_{\rm CoP}}$ and in the cluster's rest frame, obtained by subtracting the bulk velocity ${\bf v_{\rm bulk}}$. Hereafter, we re-label the cluster components with the \texttt{ParticleType} notation used in \textsc{Gadget-3} as follows $k\in\{{\rm gas,~DM,~stars}\}\longrightarrow \{0, 1, 4\}$. We then compute the angle $\theta$ between the spin vectors of the components $(m,n) = (0,1), (0,4), (1,4)$ as
\begin{equation}
    \cos \theta_{mn} = \frac{{\bf j}_m \cdot {\bf j}_n}{|\, {\bf j}_m\, |\, |\, {\bf j}_n\, |},
    \label{eq:dotproduct}
\end{equation}
as also done by, e.g., \cite{2002ApJ...576...21V, 2010MNRAS.404.1137B} and \cite{2017MNRAS.466.1625Z}. From Eq.~\eqref{eq:dotproduct}, a value of $\cos \theta_{mn}\approx 1$ indicates that the spins are aligned, while anti-aligned spins return $\cos \theta_{mn}\approx -1$. 
For the calculation of the stellar spin, we only select the star particles in galaxies, following the definition in Section \ref{sec:datasets}.
While we only show data for $\cos \theta_{04}$ in Fig. \ref{fig:corner-plot-dynamical-state}, we report the other combinations $(m, n)$ combinations in Fig.~\ref{fig:corner-plot-all-properties}. 

We {find that, in all $\cos \theta_{mn}$ histograms,} most of the clusters have components with well-aligned spins, as expected for structures forming in a $\Lambda$CDM universe. We obtained sharply peaked distributions for $\cos \theta_{(01), (14)}\approx 1$, suggesting that the galaxies and the hot gas co-rotate with the DM halo, which accounts for the largest mass content (and angular momentum) in a cluster. The spins of the gas and the galaxies, however, are more poorly aligned with each other when compared to the (01) or the (14) pairs, shown by a broader tail in the distribution. The de-rotation method in \cite{2019JCAP...06..001B} uses the spin from the galaxies as proxy for the gas spin, however, our results prove that a MACSIS-like cluster population may have a large fraction of objects where this assumption is not valid and may affect the rkSZ amplitude in the stacked maps. We will corroborate this claim in Section \ref{sec:image-processing:stacking}. 

{We find no correlation between the alignment angles,
$\{ \cos \theta_{01}$, $\cos \theta_{04}$, $\cos \theta_{14} \}$, and the 
dynamical state indicators, 
$\{ \beta_{\rm 3D}$, $f_{\rm sub}$, $\alpha_{\rm 3D}$, $\alpha_{\rm proj}\}$, 
as illustrated by the $\cos \theta_{04}$ data in  Fig. \ref{fig:corner-plot-dynamical-state}}. The $\cos \theta_{mn}$ quantities are 
{also} not found to be correlated with the halo mass or the hot gas fraction, while they are positively correlated with each other. These results suggest that the tail of the $\cos \theta_{04}$ distribution is equally represented at all halo masses in the MACSIS sample. The effects of the misalignment of the galaxies and the hot gas spins are therefore expected to statistically affect the stacking of the rkSZ maps equally across the MACSIS mass range if the orientation of ${\bf j}_4$ is used as a proxy for that of ${\bf j}_0$. From Fig. \ref{fig:corner-plot-dynamical-state}, we find $\cos \theta_{04} < 0$ in about 20\% of the cluster sample. For these objects, the angle between the angular momenta of gas and galaxies is very large $>90^\circ$, with a few cases reaching $\approx 180^\circ$ (i.e. ${\bf j}_0$ and ${\bf j}_4$ are anti-aligned). Coherent rotation, expressed by small $\theta_{04}$ is a well-established result of tidal-torque theory and finding indications of counter-rotating gas and galaxy components may seem puzzling. We find that MACSIS clusters with $\cos \theta_{04} < 0$ also have low values of $\lambda_{\rm DM}$ and $\lambda_{\rm gas}$ (see Fig.~\ref{fig:corner-plot-all-properties} in Section \ref{app:correlation-coefficients}). Since $\lambda \propto |{\bf j}|$, as shown in Eq.~\eqref{eq:spin-parameter}, cluster with incoherent rotation tend to have a low angular momentum in $r_{500}$. This scenario can occur if the angular momenta of the individual gas particles (and galaxies) are not well-aligned aligned, leading to a small vector sum. We tested this hypothesis on 5 clusters with low $\cos \theta_{04}$ values at $z=0$ and verified that the distribution of the individual angular momenta about the CoP was overall isotropic for the gas, and even more so for the galaxies. Bulk rotation clearly has a small impact on the overall dynamics in clusters with this characteristic, and the orientation of ${\bf j}_0$ and ${\bf j}_4$ may simply be dictated by small excess contributions of, e.g., a substructure entering $r_{500}$ or exerting a gravitational tidal torque from nearby.

In Section \ref{app:correlation-coefficients}, we include the complete corner plot with both basic properties and dynamical state indicators; we also compute the Spearman correlation coefficients as a quantitative measure of the correlation between quantities.

\section{Image processing}
\label{sec:image-processing}

\subsection{De-rotation}
\label{sec:image-processing:derotation}
To recover the elusive rotational kSZ signal from noisy observations, or from substructure\hyp{}rich clusters with complex dynamics, we generate and stack projected maps of the rkSZ signal. Following the stacking method in \cite{2019JCAP...06..001B}, we first rotate the particle (or galaxies) positions and the velocity vectors such that ${\bf J}_{\rm k}$ aligns with the $z$-axis of the parent box, identified with the unit vector $\mathbf{\hat{z}}=(0, 0, 1)$. The rotation transformation is implemented using Rodrigues' rotation formula \citep[see derivations for an $SO(3)$ rotation group in e.g.][]{bauchau2003vectorial, dai2015euler}. We compute the rotation axis vector $\mathbf{q}$ by taking the outer product of the two vectors ${\bf J_{\rm k}}$ and $\mathbf{\hat{z}}$:
\begin{equation}
   \mathbf{q} \equiv \begin{pmatrix} q_x\\q_y\\q_z \end{pmatrix}=\frac{{\bf J_{\rm k}} \times \mathbf{\hat{z}}}{|{\bf J_{\rm k}} \times \mathbf{\hat{z}}|},
\end{equation}
and the angle between ${\bf J_{\rm k}}$ and $\mathbf{\hat{z}}$ given by the inner product $\cos \theta = {\bf J_{\rm k}} \cdot \mathbf{\hat{z}} / |{\bf J_{\rm k}}| = J_{{\rm k},z}/J_{\rm k}$. The skew-symmetric matrix is then defined as
\begin{equation}
\mathbf{Q}=
     \begin{bmatrix}
    0 & -q_z & q_y\\ q_z & 0 & -q_x\\ -q_y & q_x & 0,
    \end{bmatrix}
\end{equation}
and is used to compute the rotation matrix
\begin{equation}
\mathbf{R}=\mathbf{I}_3 + (\sin \theta) ~\mathbf{Q} + (1 + \cos \theta) ~\mathbf{Q}^2, 
\end{equation}
where $\mathbf{I}_3$ is the $3\times3$ identity matrix. We de-rotate the positions $\mathbf{r_i}$ and velocities $\mathbf{v_i}$ of the selected gas particles as follows
\begin{equation}
    {\bf r'}_i = \mathbf{R}^{-1}\, \left({\bf r}_i - {\bf r}_{\rm CoP}\right);\\
    {\bf v'}_i = \mathbf{R}^{-1}\, \left({\bf v}_i - {\bf v}_{\rm bulk}\right).
\end{equation}
By using the angular momentum of the gas, DM and galaxies, we construct three matrices $\mathbf{R}_k$ which align $\mathbf{J}_k$ with the $z$-axis of the box and produce different projections.

\begin{sidewaysfigure}
    \centering
	\includegraphics[width=0.95\textwidth,trim=0.5cm 0.5cm 0cm 0cm,clip]{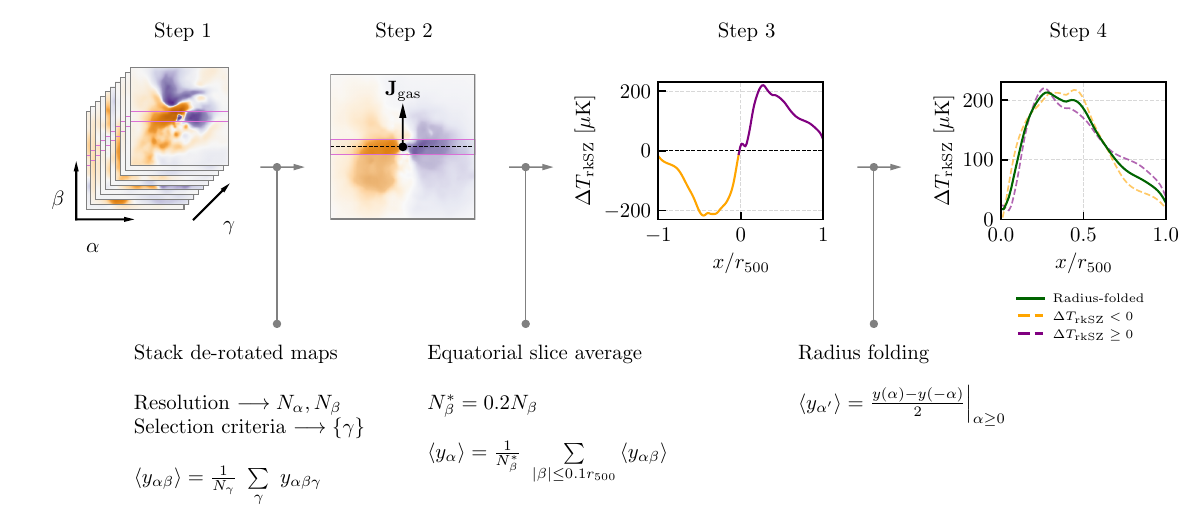}
    \caption{Workflow diagram for stacking rkSZ de-rotated and projected maps of MACSIS clusters, averaging the equatorial slice and folding the rotation profile to obtain the radial rkSZ profile. We define four steps and  The illustrations at the top show an example of this procedure by considering the rkSZ maps from the 10 most massive MACSIS clusters (step 1). In each map, the pixel coordinates are defined by the indices $(\alpha, \beta)$ and the index $\gamma$ runs over different maps. The 10 maps are averaged over $\gamma$ to produce the stacked map (step 2). Since these maps are constructed from an edge-on configuration, ${\bf J_{\rm gas}}$ points upwards in the plane of the map, as shown by the black arrow. From the stacked map, we select an equatorial slice (in red) of thickness $0.1\, r_{500}$, used to obtain the profile in step 3. Finally, this result is folded about the origin and averaged to produce the radial rkSZ profile in step 4. Throughout the diagram, we indicate positive Doppler boost in purple and negative in orange. In the right plot (step 4), the dashed lines are the positive and negative folded sections of the profile in step 3, and the solid line is their average.}
    \label{fig:stacking_workflow}
\end{sidewaysfigure}

\subsection{Projections}
\label{sec:image-processing:projections}
We then compute the $y_{\rm rkSZ}$ contribution for each particle using the integrand in Eq.~\eqref{eq:kSZ_definition} and sum the contributions along the LoS to produce projected maps. In this work, we consider seven projections: 
\begin{enumerate}
    \item In the first three projections, $\{x,y,z\}$, the clusters are not de-rotated and the LoS aligned to each of the axes of the parent box. In $\Lambda$CDM cosmology, the orientation of cluster spins follows a random distribution for a large sample of objects; stacking maps without de-rotating is expected to smooth out the coherent dipole signature from cluster rotation, producing a noisy signal, {consistent with no rotation}.
    \item \textit{gas-edge-on} projection, where ${\bf J}_{\rm gas}$ is aligned and the LoS is perpendicular to the angular momenta via the $\mathbf{R}_{\rm gas}$ rotation matrix. This configuration is expected to produce the largest possible rotational signal.
    \item \textit{gas-face-on} uses the same alignment as above, but with the LoS perpendicular to the plane of rotation. This result is obtained by projecting the particle data \textit{along} $\mathbf{J}_z$, rather than perpendicularly to it. In this case, the stacked maps are not expected to show any rkSZ temperature signal.\footnote{At second order in the rotational speed an $E$-mode polarization pattern appears in this case \citepalias{CM02}, however, we neglect this contribution.}
    \item In the \textit{galaxies-edge-on} projection, the clusters are de-rotated to align the galaxies angular momenta and then viewed edge-on via the $\mathbf{R_{gal}}$ rotation matrix. This projection is designed to yield the largest signal if the galaxies are used as proxy for the gas angular momenta see \citep[see][]{2019JCAP...06..001B}.
    \item In the \textit{dark-matter-edge-on} projection uses the same prescription as \textit{galaxies-edge-on}, but the alignment is based on ${\bf J_{\rm DM}}$ and the $\mathbf{R_{DM}}$ rotation matrix.
\end{enumerate}
After obtaining ${\bf r'_i}$ and ${\bf v'_i}$ from the rotation matrices, we compute the rkSZ projection maps using the discretised form of Eq.~\eqref{eq:kSZ_definition}.

\subsection{Stacking and averaging}
\label{sec:image-processing:stacking}

After {orienting} the cluster datasets, we combine the rkSZ maps to produce the radial rotation profiles in four steps, summarised in Fig. \ref{fig:stacking_workflow}. In this example, we selected gas-aligned edge-on rkSZ maps from the 10 most massive MACSIS clusters. This sample defines the set $\{\gamma \}$, where $\gamma$ is an index running over the selected maps. The number of maps in the selection is $N_\gamma = 10$ in this demonstration. For each map, we specify the Cartesian position of pixels by an index pair $(\alpha, \beta)$, where $\alpha \in [0, N_\alpha]$,  $\beta \in [0, N_\beta]$ and $N_\alpha \times N_\beta$ is the resolution of the map. We stress that, in this section, $\alpha$ and $\beta$ are indices running over the pixels of the maps, and do not refer to the cluster properties. Using this set-up, we stack the maps by averaging over $\gamma$, to obtain the average map $\langle y_{\alpha \beta}\rangle$, as shown in Fig.~\ref{fig:stacking_workflow}. We consider an equatorial slice of thickness $0.2 \times r_{500}$ with $N^*_\beta = 0.2 \times N_\beta$ pixels, spanning the horizontal axis of the map, then we compute the  mean of the pixels in each pixel column. This operation leads to step 3 and produces the \textit{equatorial} rotation profile $\langle y_{\alpha}\rangle$. To facilitate the model fitting strategy, we further fold the negative section of the rotation profile onto the positive section via a reflection about the origin, $\langle y(\alpha<0) \rangle \longrightarrow -\langle y(-\alpha) \rangle$, and average the two sections to produce the radial rkSZ profile $\langle y_{\alpha '} \rangle \equiv y_{\rm rkSZ}(R)$.

As a final step in our analysis pipeline, we estimate the statistical scatter in the radial rkSZ profiles using the bootstrapping technique \citep{efron1979bootstrap, efron1987better}. We randomly sample the maps in the simulation catalogue with replacement and then we stack these to produce a mean profile (as in Fig. \ref{fig:stacking_workflow}). This operation is repeated $10^4$ times, producing as many profiles for each sample selection. From these $10^4$ realisations, we then compute the median profile, and define the uncertainties the difference between the first and third quartiles. We use this method for presenting the results from the MACSIS sample in Section \ref{sec:results}.

\section{Results}
\label{sec:results}

\begin{sidewaystable}
    \renewcommand{\arraystretch}{0.9}
    \centering
    \caption{Summary of the selection criteria (column 1), the reference to the panel showing the radial rkSZ profile (column 2), the maximum measured rkSZ amplitude $A_{\rm max}$ and the peak-radius $r_{\rm max}$ for the gas-aligned profiles (columns 3 and 4), the galaxies-edge-on projection (columns 5 - 6) and the DM-edge-on projection (columns 8 and 9). In columns 7 and 10, we show the ratio between the maximum amplitude for the galaxies- and DM-aligned profiles with that of the gas-aligned profiles in the same sample of objects. These results are obtained at $z=0$ unless stated otherwise. To facilitate the visualisation of this table, we have highlighted the \textit{All clusters} row in grey; for the remaining rows, we alternate white and orange backgrounds to indicate the low- and high-quantity samples respectively. We will use the same colour coding in other tables throughout the paper.}
    \label{tab:max_measured}
    \small
    \rowcolors{9}{orange!20}{white}
    \begin{tabular}{lccccccccc}
    \toprule
     &
     \multicolumn{1}{c}{}    & 
     \multicolumn{2}{c}{\bf Gas-aligned} & 
     \multicolumn{3}{c}{\bf Galaxies-aligned} & 
     \multicolumn{3}{c}{\bf DM-aligned}\\ 
     
     \cmidrule(rl){3-4} \cmidrule(rl){5-7} \cmidrule(rl){8-10} \rule{0pt}{1ex}
     
      Selection criterion & Fig.  &  $A_{\rm max}$ & $r_{\rm max}$ &  $A_{\rm max}$ & $r_{\rm max}$ & Amplitude fraction & $A_{\rm max}$ & $r_{\rm max}$ & Amplitude fraction\\
                 & & [$\mu$K] & [$r_{500}$] & [$\mu$K] & [$r_{500}$] & $A_{\rm max}^{\rm (galaxies)} / A_{\rm max}^{\rm (gas)}$ &  [$\mu$K] & [$r_{500}$] & $A_{\rm max}^{\rm (DM)} / A_{\rm max}^{\rm (gas)}$ \\
    \midrule
    \rowcolor{gray!25}
    All clusters (377)                        &  \ref{fig:slices_projections}                                       &    80.2$\,\pm\,5.2$ & 0.20 & 32.1$\,\pm\,5.3$ & 0.20 & 0.40$\,\pm\,0.07$ & 50.1$\,\pm\,5.1$ & 0.20 & 0.63$\,\pm\,0.08$ \\
    $M_{500} < 9.7\times 10^{14}$ M$_\odot$   &  \cellcolor{white}                                                  &    32.5$\,\pm\,1.7$ & 0.23 & 12.0$\,\pm\,1.6$ & 0.28 & 0.37$\,\pm\,0.05$ & 20.7$\,\pm\,1.8$ & 0.24 & 0.64$\,\pm\,0.06$ \\
    $M_{500} > 9.7\times 10^{14}$ M$_\odot$   &  \cellcolor{white}\multirow{-2}{*}{\ref{fig:slices:properties}.A}   &    128.4$\,\pm\,10.8$ & 0.19 & 54.2$\,\pm\,10.7$ & 0.20 & 0.42$\,\pm\,0.09$ & 79.0$\,\pm\,9.8$ & 0.20 & 0.62$\,\pm\,0.09$ \\
    $f_{\rm gas}$ < 0.12                      &  \cellcolor{white}                                                  &    47.9$\,\pm\,3.4$ & 0.24 & 17.3$\,\pm\,3.2$ & 0.33 & 0.36$\,\pm\,0.07$ & 25.3$\,\pm\,2.5$ & 0.31 & 0.53$\,\pm\,0.06$ \\
    $f_{\rm gas}$ > 0.12                      &  \cellcolor{white}\multirow{-2}{*}{\ref{fig:slices:properties}.B}   &    108.7$\,\pm\,10.8$ & 0.18 & 49.2$\,\pm\,10.2$ & 0.18 & 0.45$\,\pm\,0.10$ & 78.5$\,\pm\,11.1$ & 0.13 & 0.72$\,\pm\,0.12$ \\
    $f_{\rm bary}$ < 0.12                      &  \cellcolor{white}                                                 &    55.1$\,\pm\,5.2$ & 0.24 & 22.5$\,\pm\,4.5$ & 0.27 & 0.41$\,\pm\,0.09$ & 29.9$\,\pm\,3.1$ & 0.28 & 0.54$\,\pm\,0.08$ \\
    $f_{\rm bary}$ > 0.12                      &  \cellcolor{white}\multirow{-2}{*}{\ref{fig:slices:properties}.C}  &    110.0$\,\pm\,11.1$ & 0.17 & 47.3$\,\pm\,9.9$ & 0.19 & 0.43$\,\pm\,0.10$ & 76.5$\,\pm\,10.7$ & 0.13 & 0.70$\,\pm\,0.12$ \\
    
    $M_{\star} < 9.7\times 10^{14}$ M$_\odot$   &  \cellcolor{white}                                                   &    31.7$\,\pm\,1.7$ & 0.22 & 11.7$\,\pm\,1.7$ & 0.26 & 0.37$\,\pm\,0.06$ & 21.8$\,\pm\,1.8$ & 0.23 & 0.69$\,\pm\,0.07$ \\
    $M_{\star} > 9.7\times 10^{14}$ M$_\odot$   &  \cellcolor{white}\multirow{-2}{*}{\ref{fig:slices:properties}.D}    &    128.8$\,\pm\,10.3$ & 0.20 & 53.7$\,\pm\,10.3$ & 0.20 & 0.42$\,\pm\,0.09$ & 79.1$\,\pm\,9.7$ & 0.20 & 0.61$\,\pm\,0.09$ \\
    
    $\lambda_{\rm DM}$ < 0.03                 &  \cellcolor{white}                                                     &    60.4$\,\pm\,3.9$ & 0.24 & 11.2$\,\pm\,3.2$ & 0.31 & 0.18$\,\pm\,0.05$ & 25.7$\,\pm\,5.4$ & 0.20 & 0.42$\,\pm\,0.09$ \\
    $\lambda_{\rm DM}$ > 0.03                 &  \cellcolor{white}\multirow{-2}{*}{\ref{fig:slices:properties}.E}      &    102.1$\,\pm\,12.4$ & 0.17 & 59.5$\,\pm\,8.9$ & 0.19 & 0.58$\,\pm\,0.11$ & 74.2$\,\pm\,7.7$ & 0.21 & 0.73$\,\pm\,0.12$ \\
    $\lambda_{\rm gas}$< 0.051                &  \cellcolor{white}                                                     &    64.0$\,\pm\,5.5$ & 0.26 & 18.5$\,\pm\,4.9$ & 0.26 & 0.29$\,\pm\,0.08$ & 30.4$\,\pm\,6.6$ & 0.19 & 0.48$\,\pm\,0.11$ \\
    $\lambda_{\rm gas}$ > 0.051               &  \cellcolor{white}\multirow{-2}{*}{\ref{fig:slices:properties}.F}      &    101.3$\,\pm\,11.8$ & 0.17 & 52.0$\,\pm\,8.4$ & 0.19 & 0.51$\,\pm\,0.10$ & 70.8$\,\pm\,7.2$ & 0.22 & 0.70$\,\pm\,0.11$ \\
    
    \midrule
    
    $\beta_{\rm 3D}$ < 0.15        &  \cellcolor{white}                                                   &    54.8$\,\pm\,4.1$ & 0.21 & 22.8$\,\pm\,4.0$ & 0.26 & 0.42$\,\pm\,0.08$ & 35.7$\,\pm\,3.6$ & 0.23 & 0.65$\,\pm\,0.08$ \\
    $\beta_{\rm 3D}$ > 0.15        &  \cellcolor{white}\multirow{-2}{*}{\ref{fig:slices:dynamical-state}.A}    &   105.5$\,\pm\,8.5$ & 0.24 & 43.9$\,\pm\,10.3$ & 0.19 & 0.42$\,\pm\,0.10$ & 65.9$\,\pm\,9.5$ & 0.19 & 0.62$\,\pm\,0.10$ \\
    $f_{\rm sub}$ < 0.18             &  \cellcolor{white}                                                   &    72.0$\,\pm\,5.4$ & 0.20 & 25.2$\,\pm\,5.0$ & 0.25 & 0.35$\,\pm\,0.07$ & 52.2$\,\pm\,6.7$ & 0.19 & 0.72$\,\pm\,0.11$ \\
    $f_{\rm sub}$ > 0.18             &  \cellcolor{white}\multirow{-2}{*}{\ref{fig:slices:dynamical-state}.B}    &   89.5$\,\pm\,8.1$ & 0.25 & 40.5$\,\pm\,9.0$ & 0.20 & 0.45$\,\pm\,0.11$ & 49.3$\,\pm\,6.6$ & 0.24 & 0.55$\,\pm\,0.09$ \\
    $\cos\, \theta_{04}$ < 0.56               &  \cellcolor{white}                                                       &    64.1$\,\pm\,5.1$ & 0.23 & 16.3$\,\pm\,6.0$ & 0.16 & 0.25$\,\pm\,0.10$ & 26.7$\,\pm\,5.3$ & 0.28 & 0.42$\,\pm\,0.09$ \\
    $\cos\, \theta_{04}$ > 0.56               &  \cellcolor{white}\multirow{-2}{*}{\ref{fig:slices:dynamical-state}.C}        &   96.9$\,\pm\,9.9$ & 0.19 & 79.4$\,\pm\,8.8$ & 0.19 & 0.82$\,\pm\,0.12$ & 77.2$\,\pm\,7.7$ & 0.19 & 0.80$\,\pm\,0.11$ \\
    
    $\alpha_{\rm 3D}$ < 0.21        &  \cellcolor{white}                                                           &    58.6$\,\pm\,4.0$ & 0.22 & 23.9$\,\pm\,4.0$ & 0.25 & 0.41$\,\pm\,0.07$ & 41.2$\,\pm\,4.6$ & 0.20 & 0.70$\,\pm\,0.09$ \\
    $\alpha_{\rm 3D}$ > 0.21        &  \cellcolor{white}\multirow{-2}{*}{\ref{fig:slices:dynamical-state}.D}       &    101.9$\,\pm\,11.5$ & 0.17 & 40.6$\,\pm\,9.7$ & 0.20 & 0.40$\,\pm\,0.11$ & 60.3$\,\pm\,10.9$ & 0.13 & 0.59$\,\pm\,0.13$ \\
    $\alpha_{\rm 2D}\,(r= 0.2\,r_{500})$ < 0.047         &  \cellcolor{white}                                      &    58.9$\,\pm\,4.1$ & 0.18 & 21.1$\,\pm\,3.2$ & 0.28 & 0.36$\,\pm\,0.06$ & 34.4$\,\pm\,3.4$ & 0.23 & 0.58$\,\pm\,0.07$ \\
    $\alpha_{\rm 2D}\,(r= 0.2\,r_{500})$ > 0.047         &  \cellcolor{white}\multirow{-2}{*}{\ref{fig:slices:dynamical-state}.E}      &    103.8$\,\pm\,8.5$ & 0.25 & 45.3$\,\pm\,10.0$ & 0.20 & 0.44$\,\pm\,0.10$ & 65.2$\,\pm\,9.4$ & 0.20 & 0.63$\,\pm\,0.10$ \\
    $\alpha_{\rm 2D}\,(r=r_{500})$ < 0.14        &  \cellcolor{white}                                                      &    44.8$\,\pm\,2.7$ & 0.20 & 20.0$\,\pm\,2.5$ & 0.25 & 0.45$\,\pm\,0.06$ & 30.9$\,\pm\,3.4$ & 0.20 & 0.69$\,\pm\,0.09$ \\
    $\alpha_{\rm 2D}\,(r=r_{500})$ > 0.14        &  \cellcolor{white}\multirow{-2}{*}{\ref{fig:slices:dynamical-state}.F}        &    116.1$\,\pm\,8.8$ & 0.24 & 46.4$\,\pm\,10.8$ & 0.19 & 0.40$\,\pm\,0.10$ & 69.5$\,\pm\,11.3$ & 0.13 & 0.60$\,\pm\,0.11$ \\
    
    \midrule
    $z=0$ (75 clusters)              &  \cellcolor{white}                                                    &    18.0$\,\pm\,1.4$ & 0.23 & 7.5$\,\pm\,1.4$ & 0.25 & 0.42$\,\pm\,0.09$ & 12.9$\,\pm\,1.3$ & 0.27 & 0.71$\,\pm\,0.09$ \\
    $z=1$ (70 clusters)               &  \cellcolor{white}\multirow{-2}{*}{\ref{fig:slices:redshift}}        &    25.7$\,\pm\,2.9$ & 0.31 & 11.6$\,\pm\,3.5$ & 0.34 & 0.45$\,\pm\,0.15$ & 18.4$\,\pm\,3.2$ & 0.24 & 0.72$\,\pm\,0.15$ \\
    \bottomrule
    \end{tabular}
\end{sidewaystable}

\subsection{Complete sample}
\label{sec:results:complete-sample}

\begin{figure}
    \centering
	\includegraphics[width=\textwidth]{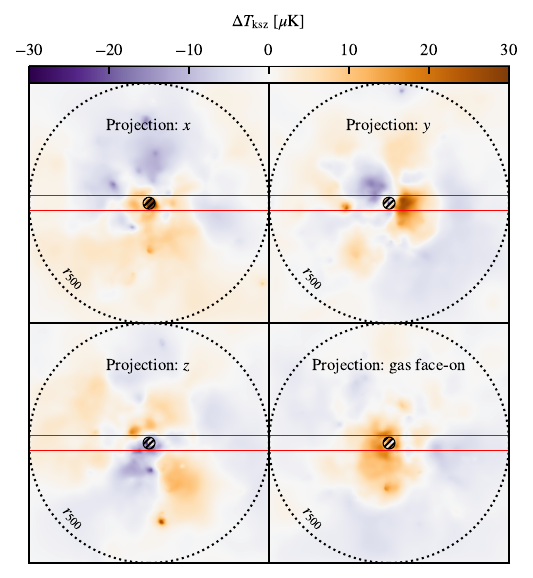}
    \caption{Maps showing $\Delta T_{\rm rkSZ}$ after stacking (and averaging) all MACSIS clusters considered in this work without any de-rotation (projections $x, y, z$ indicated at the top of each panel) and with ${\bf j}_0$ aligned and directed into the plane of the image (face-on projection). In all panels, we indicate $r_{500}$ as a black circle and with red rectangles the regions of the maps considered for computing the equatorial rotation profiles, spanning over $\pm\,r_{500}$ from the centre of the map in each direction.}
    \label{fig:rksz_gas_null}
\end{figure}

\begin{sidewaysfigure}
    \centering
	\includegraphics[width=\textwidth]{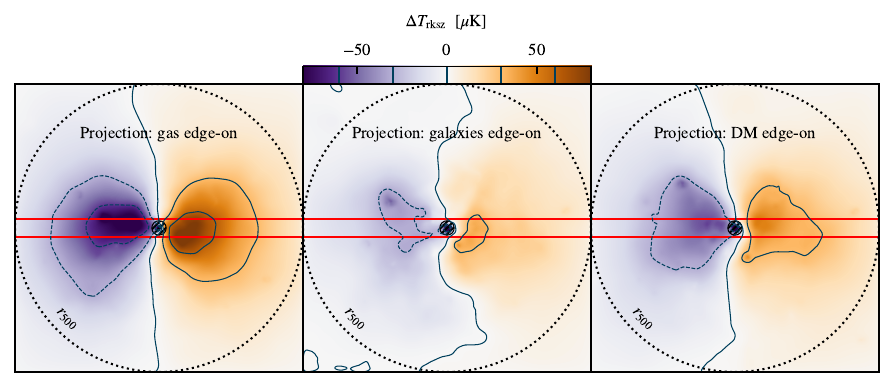}
    \caption{As in Fig. \ref{fig:rksz_gas_null}, but showing the stacked signal after de-rotating the clusters to an edge-on configuration. From left to right, the panels show the stacked signal when the alignment uses the angular momentum of the gas, the galaxies and the dark matter in $r_{500}$. The orange contours are evaluated for the same levels as in Fig. \ref{fig:rksz_model}. The largest rkSZ signal is obtained when orienting the clusters by aligning the gas spins.}
    \label{fig:rksz_gas_edge}
\end{sidewaysfigure}

\begin{figure}
    \centering
	\includegraphics[width=0.8\textwidth]{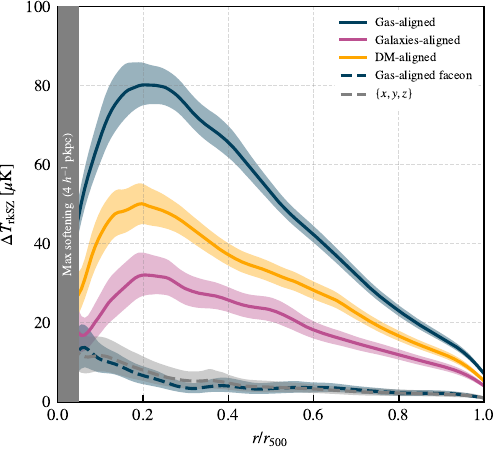}
    \caption{Rotational kSZ profiles obtained from the equatorial slices from the projections in Figs. \ref{fig:rksz_gas_null} (dashed lines) and \ref{fig:rksz_gas_edge} (solid lines) and folded about the centre of the map. The profiles are constructed from gas-aligned spins (blue), galaxies-aligned spins (orange) and dark matter-aligned spins (pink). We emphasize the gas-aligned face-on profile in blue, to be compared to its edge-on counterpart. The projections along the axes of the parent box are shown in grey. The grey vertical band represents the $r<0.05\, r_{500}$ region, which we exclude in our analysis. For each profile, the shaded bands represent the confidence intervals between the first and third quartiles, computed from $10^4$ realisations generated by bootstrapping the sample with repetition.}
    \label{fig:slices_projections}
\end{figure}

In Fig. \ref{fig:rksz_gas_null}, we show the maps for the mean rkSZ signal obtained from stacking the entire MACSIS sample (with the $M_{200}$ and $f_{\rm gas}$ selection of Section \ref{sec:datasets}) for the $\{x,y,z\}$ and the gas-face-on projections. Without de-rotation, or when the de-rotated angular momenta are oriented along the LoS (face-on), we expect the stacking method to suppress the rkSZ signal. The amplitude of the residual fluctuations in Fig.~\ref{fig:rksz_gas_null} is {$A_{\rm residual} \sim \langle \Delta T_{\rm rkSZ}^2 \rangle^{1/2} \approx 15~\mu$K}. Conversely, when the MACSIS clusters are de-rotated to an edge-on configuration, the rkSZ signal is stacked coherently and to produce amplitudes up to $A_{\rm max}\approx 80\,\mu$K, as shown in Fig. \ref{fig:rksz_gas_edge}. These maps were produced from an edge-on configuration aligning the angular momenta of the gas (left), galaxies (centre) and dark matter (left).

{Even without de-rotation, the maps in Fig. \ref{fig:rksz_gas_null} show features which could be mistaken for coherent motion. Their amplitude is often larger than that predicted by Poisson statistics, $A_{\rm residual, P} \approx 4~\mu$K}.\footnote{Assuming that the amplitude of residual fluctuation follows a Poisson distribution, then the variance should scale as $\propto 1/N$, where $N$ is the number of clusters in the sample. As more randomly-oriented clusters are stacked, we expect the amplitude of the fluctuations to decrease as $A_{\rm residual, P} \sim A_{\rm max} / \sqrt{N} \approx 4\, \mu$K, with $A_{\rm max}=80\,\mu$K and $N=377$.} {This result suggests that the amplitude of fluctuations are not generally Poisson-distributed. When probing the distribution of $A_{\rm residual}$, we found that the strongest signals in the maps of in Fig. \ref{fig:rksz_gas_null} originate from a subset of clusters where substructures produce intense rkSZ signals.} 
{We tested this hypothesis with two methods. (i) Firstly, we computed the \textit{median} of the $\{x,y,z\}$ and face-on projected signal pixel-wise. The median value is not affected by extreme data samples, unlike the average value. Using the median maps, we obtained fluctuations with typical amplitudes below $5~\mu$K, consistently with the Poisson estimate. (ii) In our second approach, we constructed a distribution of the peak amplitude, $\max(A_{\rm residual})=\max(|\Delta T_{\rm rkSZ}|)$, of the cluster maps and removed the clusters with high $\max(A_{\rm residual})$, above the 75$^{\rm th}$ percentile. The maps of the remaining clusters were stacked by averaging, and we recovered fluctuations of $5~\mu$K. Finally, we checked that the extreme objects that have been discarded in this test produce fluctuations $>15~\mu$K upon stacking. Such fluctuations are associated with substructures rich in dense gas and/or with extreme differential velocities along the LoS [see Eq. \eqref{eq:kSZ_definition}]. We leave an analysis of the impact of substructures on the rkSZ signal to future work.}

While the spatial distribution of the rkSZ amplitude is best probed using the maps in Figs. \ref{fig:rksz_gas_null} and \ref{fig:rksz_gas_edge}, the radial rkSZ profiles provide a quantitative comparison of the relative rkSZ amplitudes. In Fig.~\ref{fig:slices_projections}, we obtain the maximum amplitude (solid lines) when aligning the spin of the gas, $A_{\rm max}^{\rm (gas)}=80.2\,\mu$K (blue), while aligning to the spin of the galaxies produces an amplitude $A_{\rm max}^{\rm (galaxies)}=32.1\,\mu$K (pink), which is only $\simeq 40\%$ of the theoretical maximum signal. Aligning the spins of the dark matter halos produces an intermediate amplitude $A_{\rm max}^{\rm (DM)}=50.1\,\mu$K, {reproducing $\simeq 63\%$ of the maximal signal}. We summarise the measured rkSZ amplitudes $A_{\rm max}$ and the radius $r_{\rm max}$ where they peak in Table~\ref{tab:max_measured}. Although $A_{\rm max}$ varies significantly, in all cases we find $r_{\rm max}\simeq 0.2 \,r_{500}$, {which is in good agreement with \citetalias{CC02}.} The profile amplitudes of the edge-on configurations with all clusters can be clearly distinguished from the face-on and the $\{x,y,z\}$ set-ups, {which never exceed $\approx 15~\mu$K, even when objects with extreme fluctuations are included}. 

Given the results from stacking the MACSIS cluster sample \textit{in toto}, we now split the sample based on the the value of the cluster properties in Fig. \ref{fig:corner-plot-basic-properties}. For each property, we compute the median value and we define a high- and low-value sample. The rkSZ maps of clusters in are stacked separately for each of these two samples, allowing to probe difference in radial rkSZ profiles directly. These selection criteria are classified in three groups: the cluster properties at $z=0$ are presented in Section \ref{sec:results:cluster-properties}, the dynamical state metrics in Section \ref{sec:results:dynamical-state} and the redshift dependence in Section \ref{sec:results:z-dependence}. The results are also summarised in Table~\ref{tab:max_measured}. We conclude this discussion with a study of the rkSZ amplitude arising from differential motions in the ICM (Section \ref{sec:results:differential-motions}).

\subsection{Selection by cluster property}
\label{sec:results:cluster-properties}

\begin{figure}
	\includegraphics[width=\textwidth]{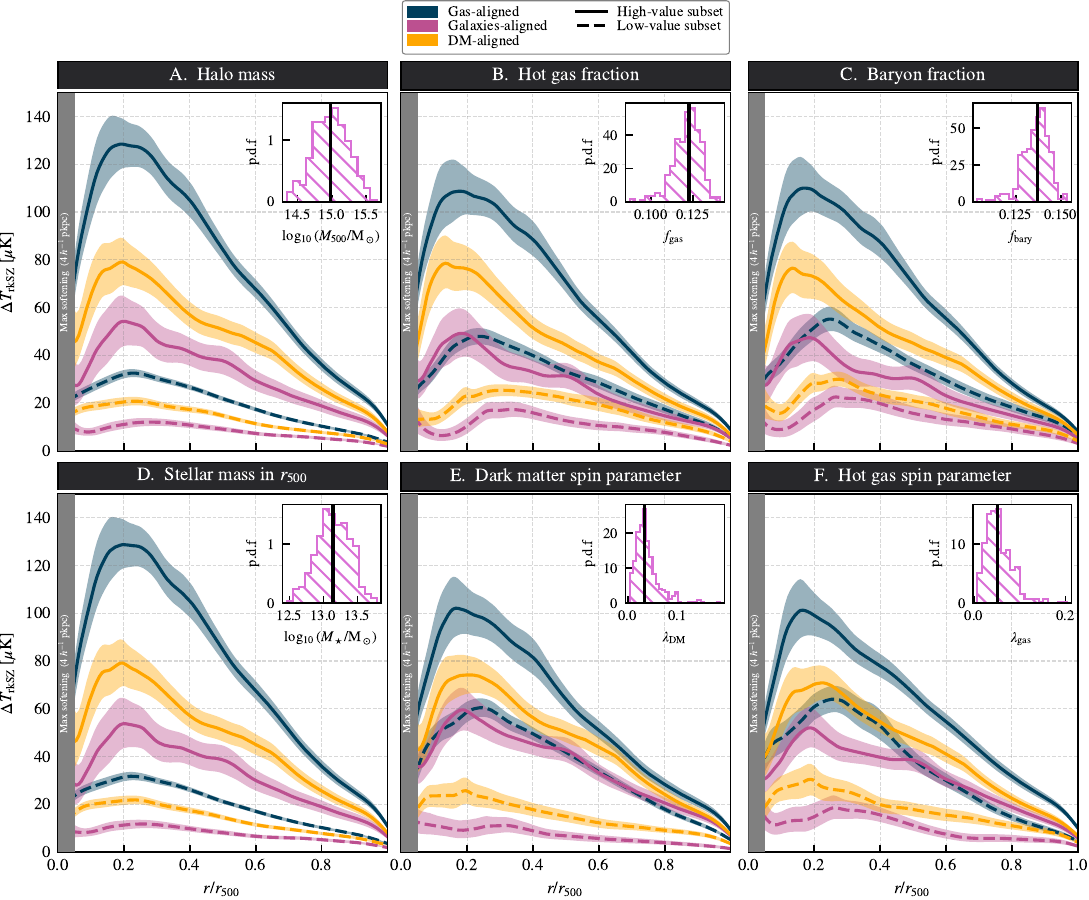}
    \caption{Radial rkSZ profiles in edge-on configuration for the MACSIS sample split according to the basic cluster properties of Section \ref{sec:alignment:properties}. The alignment methods are colour-coded as in Fig. \ref{fig:slices_projections}. The MACSIS sample is split into a high-value subset (solid lines), and a low-value subset (dashed lines). The rkSZ profiles are computed from stacking de-rotated rkSZ maps from clusters in each subset separately. The inset plot on the top-right of each panel shows the normalised p.d.f. of the quantity and the median, shown as a vertical solid line, defines the mass cut. At the top of each panel, we specify the property being examined, indexed from A to F.}
    \label{fig:slices:properties}
\end{figure}

Based on the median $M_{500}$ of the MACSIS clusters, $9.7 \times 10^{14}~{\rm M}_\odot$, we split the sample into a high- and low-mass sample, and we produce the radial rkSZ profiles in Fig. \ref{fig:slices:properties}.A. Here, the solid lines refer to the stacked (averaged) high-mass sample, while the dashed lines indicate the low-mass samples. The colours represent the gas, galaxies and dark matter spin alignment methods, similarly to Fig. \ref{fig:slices_projections}, and are used consistently throughout this work. We found that the high-mass MACSIS clusters produce a rotational signal $\approx 4$ times stronger than the low-mass ones. This ratio remains the same regardless of the spin alignment method. Cluster-sized dark matter halos form hierarchically by accreting smaller objects, which introduce angular momentum into the system until the turnaround point. High-mass halos virialise later than low-mass ones and they experience a longer angular momentum growth phase \citep[e.g.][]{2002MNRAS.332..325P}. Assuming that the gas distribution traces the dark-matter potential, \citetalias{CM02} predicted a direct correlation between halo mass and rkSZ signal strength and our results in Fig. \ref{fig:slices:properties} confirm this. Moreover, this increase exceeds the prediction of the scaling relation in Eq.~\eqref{eq:kSZ-scaling}. Taking the first and third quartiles of $M_{500}$ as the centroids of the mass bins, the self-similar relation predicts $y_{\rm kSZ}$ to be $\approx 1.8$ times higher in the high-mass sample than in the low-mass sample. Although Eq.~\eqref{eq:kSZ-scaling} underpredicts the kSZ signal from cluster rotation in MACSIS by a factor of 2, we have shown that the mass-dependent trend is consistent with self-similar expectations, noting that these should only be used as guidelines. Self-similar scaling does not account for the dynamical state of the cluster. In Section \ref{sec:results:dynamical-state}, we will show that unrelaxed clusters produce a stronger signal and, since these objects also tend to be the most massive (see Fig.~\ref{fig:corner-plot-dynamical-state}), the rkSZ amplitude is likely to exceed the self-similar prediction.

The galaxies-aligned and the DM-aligned profiles have amplitudes 60 and 35\% lower than the gas-aligned profiles. In fact, we expect the gas-aligned profiles to produce the strongest possible signal, since the rotation of the ICM is made coherent by construction before stacking. Remarkably, choosing the angular momentum of the galaxies as proxy for the rotation axis of the gas suppresses $\approx$60\% of the signal to 54 $\mu$K for the high-mass sample. 
The MACSIS clusters in the low-mass bin therefore predict an rkSZ amplitude which is $\approx 10$ times larger than the estimate by \citetalias{CC02}. Our results are compatible with the amplitude found by \cite{2017MNRAS.465.2584B} from the MUSIC clusters of comparable mass and also agree well with the analytic estimates of \citetalias{CM02} for recent mergers.
{We discuss possible causes of the difference with \citetalias{CC02} in Section \ref{sec:summary}}, but we anticipate that {one of the main effects is the larger spin of the MACSIS clusters.} 

We then split the MACSIS sample into subsets with high and low hot gas fractions and show the radial rkSZ profiles in Fig. \ref{fig:slices:properties}.B. The clusters with $f_{\rm gas}$ above the median value of 0.12 produce a rotational signal 2.5 to 3 times stronger than clusters with low gas fractions. Since $y_{\rm rkSZ} \propto n_e$ (see Eq.~\ref{eq:kSZ_definition}), clusters with large hot gas content are expected to produce large $\Delta T_{\rm rkSZ}$ contributions to the rotational profiles. Moreover, the $f_{\rm gas}$-$M_{500}$ scaling relation, shown in Fig. \ref{fig:corner-plot-basic-properties}, suggests that clusters with high $f_{\rm gas}$ are also massive. Therefore, we expect $A_{\rm max}$ to increase with $f_{\rm gas}$ also because $f_{\rm gas}$ increases with $M_{500}$. The baryon fraction $f_{\rm bary}$ also affects the rkSZ amplitude in a similar fashion.

The following two results in our analysis consider the spin parameter of the hot gas and the dark matter. The spin parameter, defined in Eq.~\eqref{eq:spin-parameter}, probes the fraction of the kinetic energy in a system which is associated to rotational motion, as opposed to unordered dynamics, or turbulence in the case of the hot gas. We begin by splitting the MACSIS sample using the median value of $\lambda_{\rm DM}$ and we present the radial rkSZ profiles in Fig. \ref{fig:slices:properties}. For MACSIS, we find that fast dark matter rotators 
{($\lambda_{\rm DM} > 0.03$)} 
produce a stronger rkSZ signal than clusters with low $\lambda_{\rm DM}$, as expected in a scenario where the gas traces the DM during the angular momentum growth phase. 

Conversely, in Fig. \ref{fig:slices:properties} we show a larger kSZ contribution from rotational motion in cluster with high $\lambda_{\rm gas}$, and a weaker signal in clusters with $\lambda_{\rm gas} < 0.051$. When a large fraction of the hot gas kinetic energy is associated with unordered motion (thermalised ICM), such as in the low $\lambda_{\rm gas}$ sample, the rkSZ amplitude is smaller than in the cluster sample with a larger fraction of energy associated to coherent rotation.

\subsection{Selection by dynamical state}
\label{sec:results:dynamical-state}

\begin{figure}
	\includegraphics[width=\textwidth]{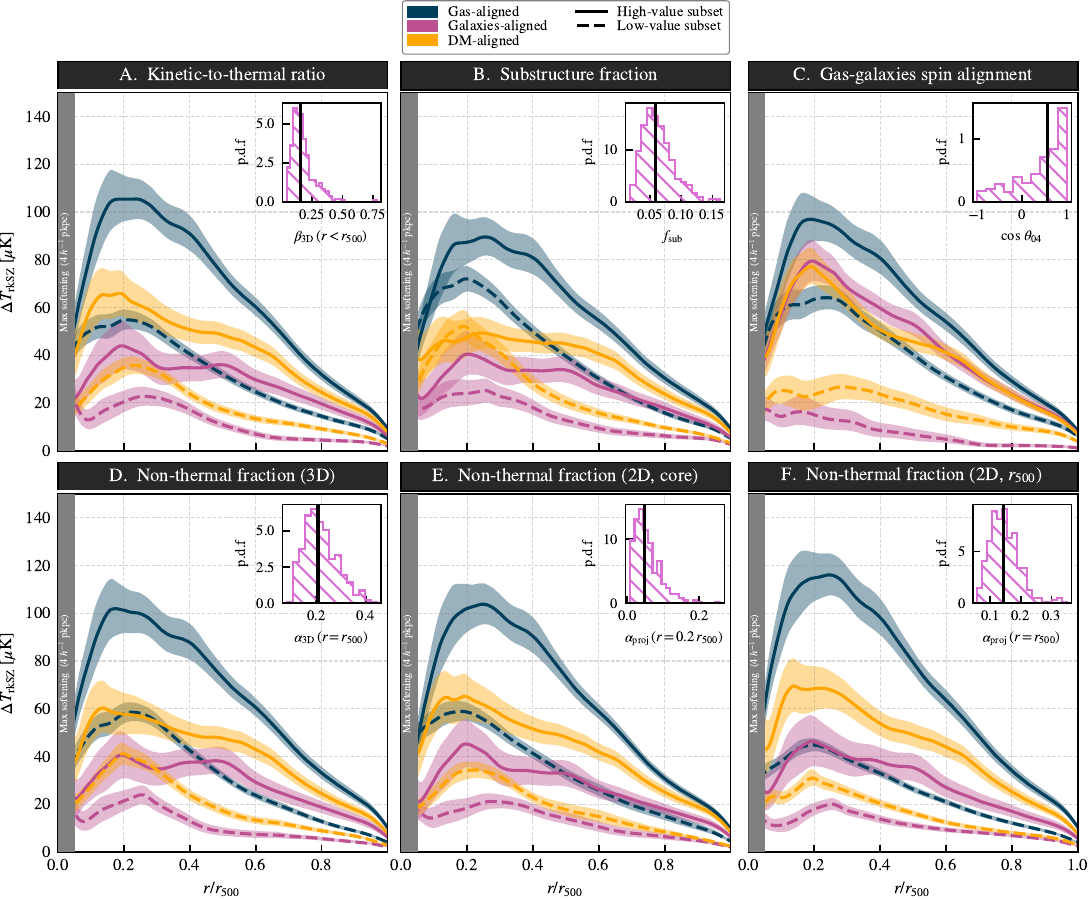}
    \caption{As in Fig. \ref{fig:slices:properties}, but splitting the MACSIS sample based on dynamical state descriptors (see Fig. \ref{fig:corner-plot-dynamical-state}).}
    \label{fig:slices:dynamical-state}
\end{figure}

Following the results on the hot gas and dark matter spin parameters, we investigate the role of the substructures in the rotational kSZ signal. As in the previous results, we show in Fig. \ref{fig:slices:dynamical-state} the radial rkSZ profiles for clusters with substructure fraction below and above the median. Clusters with high substructure fractions tend to produce, on average, an rkSZ signal larger than clusters with a small substructure population.

The kinetic-to-thermal ratio measures the relaxation state of the hot gas in clusters and here we use it to define a thermodynamically relaxed sample, with $\beta_{\rm 3D} < 0.15$, and a non-relaxed sample, $\beta_{\rm 3D} > 0.15$, as shown in Fig. \ref{fig:slices:dynamical-state}.A. We find that non-relaxed clusters produce an rkSZ signal twice as strong as the relaxed. Therefore, the rkSZ amplitude is enhanced in clusters with active mergers, where the transfer of angular momentum drives the formation of dipole-like feature in the $\Delta T_{\rm kSZ}$ map. {The rkSZ amplitude is thus a proxy for the dynamical state of the cluster, as anticipated previously.}

We also split the MACSIS sample based on the angle between the angular momenta of the hot gas and the galaxies in $r_{500}$. Here, the value of $\cos \theta_{04}$ is used to define a gas-galaxies \textit{aligned} samples and a \textit{misaligned} sample. Focusing on the aligned sample, composed of clusters with $\cos \theta_{04} > 0.56$, we report an 18\% reduction in amplitude when de-rotating the maps using the galaxies as proxy for the rotation axis instead of using the gas angular momenta. For the misaligned sample ($\cos \theta_{04} < 0.56$), on the other hand, the amplitude is suppressed by 75\%. {This clearly highlights the importance of understanding the degree of alignment of gas and galaxy spins.}

To quantify the role of substructures in the defining the direction of the gas spin, we computed the angle between the galaxies spin and the hot gas in $r_{500}$ with and without substructures. We find that the values of $\cos \theta_{04}$ for the two scenarios have a median difference of 0.4 \% and $\approx 4 \%$ at the 90$^{\rm th}$ percentile. Given these results, we predict that the hot gas substructures in $r_{500}$ do not affect the orientation of the total angular momentum significantly, for most MACSIS clusters. We therefore predict that the substructure contribution to the de-rotation procedure is small enough to leave the rkSZ profiles largely unaltered and that nearly all the angular momentum is associated with the ICM.

Finally, we quantify the effect of non-thermal pressure on the rkSZ profiles by splitting the sample based on values of $\alpha$ from 3D and projected profiles as discussed in previous sections. The bottom row in Fig. \ref{fig:slices:dynamical-state} shows that high non-thermal pressure, possibly associated with turbulence, enhances the rkSZ amplitude by $\approx 40 \%$, compared to a cluster sample with low values of $\alpha$. When comparing the rkSZ amplitudes between high- and low-$\alpha_{\rm proj}$ samples, we find that the $\alpha_{\rm proj}(r=r_{500})$ selection (Fig. \ref{fig:slices:dynamical-state}.F) gives a larger difference, 71.3~$\mu$K, than the  $\alpha_{\rm proj}(r=0.2\, r_{500})$ selection (Fig. \ref{fig:slices:dynamical-state}.E), 44.9 $\mu$K. These values are computed for the gas-aligned configuration, and the same relative differences are found for the galaxies-aligned and DM-aligned scenarios.

\subsection{Redshift dependence}
\label{sec:results:z-dependence}

From simple self-similar scalings, the rotational kSZ amplitude increases with halo mass and redshift (see Eq.~\ref{eq:kSZ-scaling}). However, typical halo masses are not independent of redshift: clusters at low redshift  are on average more massive, since these have undergone cosmological accretion for longer (in a $\Lambda$CDM universe, this result can be interpreted as the halo-mass function shifting towards higher masses as structures-form hierarchically). Therefore, if we were to track a fixed population of clusters through redshift, the difference in rkSZ amplitude that would be measured would not just be due to a redshift dependence {of the dynamical quantities}, but a \textit{combined redshift and halo-mass} dependence.

\begin{figure}
	\includegraphics[width=\textwidth]{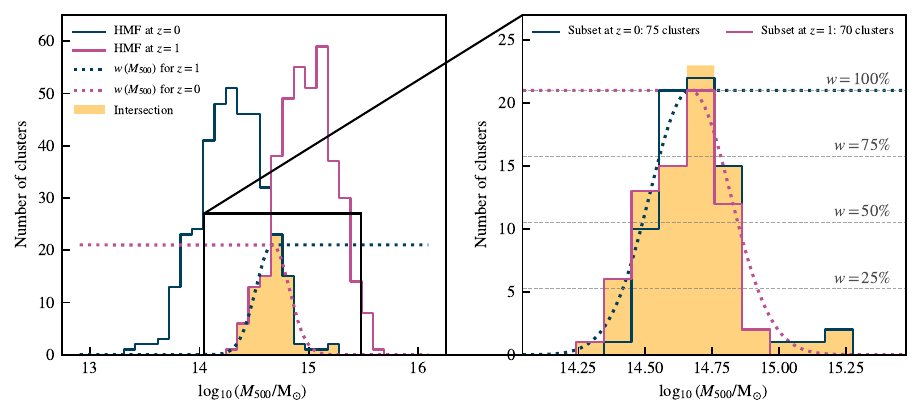}
    \caption{\textit{Left.} Halo mass functions for the complete MACSIS sample at $z=0$ (blue) and $z=1$ (pink). The weighting function $w$ is shown as dotted lines and its maximum value ($w=1$) is set to be the same as the peak of the Gaussian fit to the intersection region $N_\cap$ (yellow). \textit{Right.} Zoom-in view of the intersection region. Here, the halo mass functions are limited to objects that have been selected by the HMF-matching algorithm. We also show the quartile levels of the $w$ selection function as horizontal dashed lines. The median $M_{500}$ is $4.52 \times 10^{14}$ M$_\odot$ for the selected $z=0$ subset and $4.71 \times 10^{14}$ M$_\odot$ for the $z=1$ subset.}
    \label{fig:redshift_hmf_match}
\end{figure}

\begin{figure}
	\includegraphics[width=\textwidth]{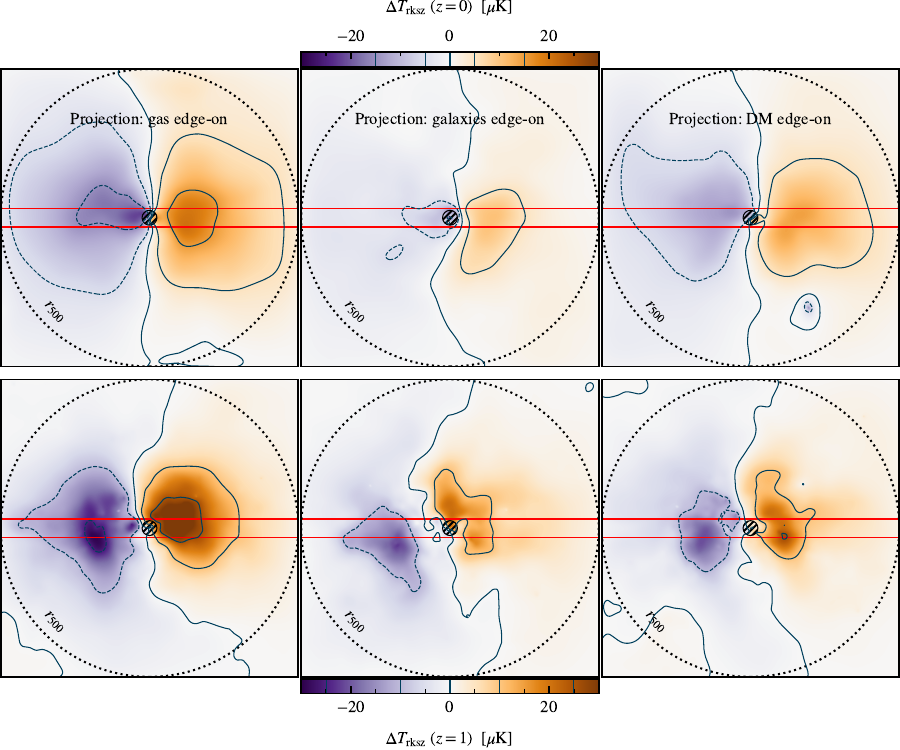}
    \caption{As in Fig. \ref{fig:rksz_gas_edge}, but comparing the edge-on configurations for the $z=0$ subset (75 clusters, top row) and the $z=1$ subset (70 clusters, bottom row). In all maps, the color bar limits are set to [-30, +30] $\mu$K and kept constant to allow a direct comparison of the rkSZ signal strength. We plot different contour levels to highlight the dipolar pattern at both redshifts: $\{0, \pm 5, \pm 15\}~\mu$K for $z=0$ and $\{0, \pm 10, \pm 25\}~\mu$K for $z=1$. The maximum rkSZ amplitude for the $z=1$ sample aligned with the gas spin ($\approx 30\, \mu$K) is larger than that of the $z=0$ sample ($\approx 20\, \mu$K).}
    \label{fig:maps:redshift}
\end{figure}

We describe a method for matching the halo-mass function (HMF) with the aim to sample clusters of similar $M_{500}$ at different redshifts and probe the rkSZ effect at fixed halo mass. In this study, we consider the MACSIS sample at two redshifts, $z=0$ and $z=1$, and their halo mass functions shown in Fig. \ref{fig:redshift_hmf_match}. The HMFs for the MACSIS sample are not monotonically decreasing with halo mass as in e.g. \cite{2008ApJ...688..709T}, because the object selection is only mass-limited for the 90 most massive halos \citep[see][for further details]{2017MNRAS.465.2584B}. As a result, the HMFs at the two redshifts overlap only for a limited range of masses ($10^{14.25}<M_{500}/{\rm M}_\odot < 10^{15.25}$) and we highlight this intersection in yellow in Fig. \ref{fig:redshift_hmf_match} in the left panel. This intersection defines the mass range of our HMF-matched subset and the object count in each overlapping bin determines how many objects in the original mass functions must be selected. Given the halo mass function at the two redshifts $N(M_{500}, z=0)$ and $N(M_{500}, z=1)$, we define the intersection as
\begin{equation}
    N_{\cap}=\min\left[N(M_{500}, z=0), N(M_{500}, z=1)\right].
\end{equation}
We then fit a Gaussian profile $w_{\rm G}$ to $N_{\cap}$, normalise $w_{\rm G}$ to range between $[0, 1]$, and constructed weight functions $w(M_{500},z)$ as follows. For the $z=1$ weight function (blue dotted line), we combine piece-wise the half-profile with positive gradient up to its maximum ($M_{500, \rm peak}$) with a constant value of one elsewhere:
\begin{equation}
 w(M_{500},z=1) = \begin{cases} 
      w_{\rm G} & M_{500}\leq M_{500, \rm peak} \\
      1 & M_{500} > M_{500, \rm peak}.
   \end{cases}
\end{equation}
This functional form ensures that the selection includes all the high-mass clusters at $z=1$, but suppresses the sampling at low masses, which are not covered by the HMF at $z=0$. The weight function at $z=0$ (magenta dotted line) follows a similar prescription, but the piece-wise components are swapped. In this case, we achieve a mass-limited sampling for the low-mass halos at $z=0$, but we suppress the sampling at the high-mass end, which is not covered by the $z=1$ HMF. The functional form of the $z=0$ weight function is therefore
\begin{equation}
w(M_{500},z=0) = \begin{cases} 
      1 & M_{500} \leq M_{500, \rm peak} \\
      w_{\rm G} & M_{500} > M_{500, \rm peak}.
   \end{cases}
\end{equation}
In Fig. \ref{fig:redshift_hmf_match}, the right panel shows a zoomed-in view of the intersection histogram $N_{\cap}$ and the weight functions for the two redshifts. In the same plot, we also show four levels of the weight functions, where the HMF is sampled at 25, 50, 75 and 100\%, with grey dashed lines to guide the eye.

The following step includes combining the HMF with the weight function to obtain the \textit{number of objects} to be selected in each mass bin. We define this subset-HMF $N_{\rm subset}(M_{500}, z=0) = N(M_{500}, z=0)\times w(M_{500},z=0)$  and similarly for $z=1$. In mass bins where $w=1$, then we select all the objects with those masses, but for $w<1$ we choose $N_{\rm subset}(M_{500})$ objects at random from the $N(M_{500})$ in the original HMF. When downsampling the HMF, we choose a random selection to avoid bias towards high or low masses, depending on the redshift considered.\footnote{We make the list of clusters for both redshift subsets publicly available online in our GitHub repository: \href{https://github.com/edoaltamura/macsis-cosmosim/tree/master/redshift_samples}{https://github.com/edoaltamura/macsis-cosmosim/tree/master/redshift\_samples} (see also the \textit{Data Availability} statement).}

The final selected subsets are shown in the right panel of Fig. \ref{fig:redshift_hmf_match} in the same colors as the HMF on the left. The $z=0$ subset contains 75 objects, located towards the low-end of the HMF at that redshift, while the $z=1$ subset contains 70 objects at the high-mass end of the HMF at that redshift. This method does not impose the subset size to match, and we therefore expect that the subset at the two redshifts may have a slightly different number of elements. The discrepancy is of the order of the Poisson noise, $\lfloor \sigma_{\rm P} \rfloor = 8$, and we find that imposing the same number of objects artificially does not change the final result for the rkSZ profiles. 

We note that a similar HMF-matching procedure was used by \cite{macsis_barnes_2017} and \cite{2020MNRAS.493.3274L} to combine the MACSIS sample at different redshifts consistently with the BAHAMAS data-set. Their method relies on a mass cut, which would be equivalent to $w(M_{500})$ being a step-function, where the jump-value is aligned with the minimum and maximum of the $N_{\cap}$ domain. This HMF-selection approach would overpopulate the high-redshift sample with low-mass objects, biasing the median $M_{500}$ towards lower masses, and overpopulate the low-redshift set with high-mass clusters, which would skew the p.d.f. in the opposite direction. While the differences between the HMF-matched and mass-cut methods are comparable to the Poisson noise due to the limited sample size of MACSIS, we emphasize that our HMF-matched approach is statistically robust and could deliver accurate mass-independent forecasts with much richer data sets from future large-volume hydrodynamic simulations.

This method introduced to match the HMF of the MACSIS sample is a novel procedure designed to produce a consistent mass coverage when comparing objects at different redshifts. The results are stable for HMF with a MACSIS-like shape. However, the functional form of $w(M, z)$ can be adapted to simulations with mass-limited HMFs, depending on the geometry of the overlapping region $N_{\cap}$.

\begin{figure}
    \centering
	\includegraphics[width=0.8\textwidth]{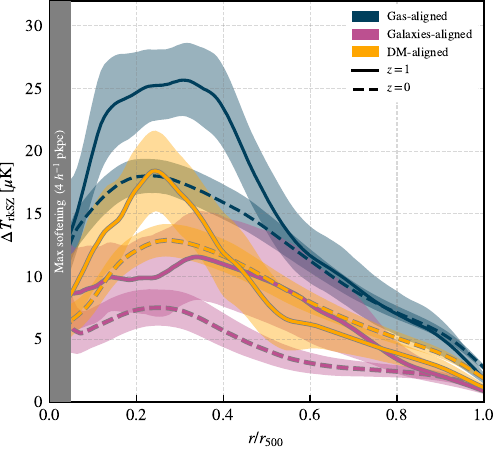}
    \caption{As in Fig. \ref{fig:slices:properties}, but comparing the HMF-matched sample at $z=0$ and $z=1$, derived from the rkSZ maps in Fig. \ref{fig:maps:redshift}. The galaxies- and DM-aligned profiles are shown with a grey edge for clarity.}
    \label{fig:slices:redshift}
\end{figure}

We show the rkSZ maps for the stacked cluster subset at $z=0$ and $z=1$ in Fig. \ref{fig:maps:redshift}; the corresponding equatorial profiles are reported in Fig. \ref{fig:slices:redshift}. The measured rkSZ amplitude at $z=1$ is found to be $\approx 1.4$ times larger than at $z=0$. In this instance, the self-similar scaling in Eq.~\eqref{eq:kSZ-scaling} overestimates this ratio, yielding $\left[E(z=1) / E(z=0)\right]^{5/3} = E(z=1)^{5/3} \approx 2.6$. Here, the mass dependence can be neglected because the HMF of the two subsets, and hence the median $M_{500}$, is matched by construction. By highlighting the redshift subsets in the corner plots in Figs. \ref{fig:corner-plot-basic-properties} and \ref{fig:corner-plot-dynamical-state}, we found that the $z=1$ have a larger median $\beta_{\rm 3D}$ value than the $z=0$ sample. In fact, most clusters at $z=1$ are captured during their accretion phase, when mergers cause them to be unrelaxed. This dynamical state leads to enhanced rkSZ signal, as we have shown in Section \ref{sec:results:dynamical-state} and panel A of Fig. \ref{fig:slices:dynamical-state}. The discrepancy in the amplitude between the MACSIS prediction and the self-similar scaling may be caused by the velocity scaling leading to Eq.~\eqref{eq:kSZ-scaling}, which expresses the circular velocity of virialised halos. In hydrodynamic simulations, this value is usually found to be higher than the tangential component of the velocities associated with the bulk rotation of the ICM \citep[see e.g.][]{2017MNRAS.465.2584B}, causing the self-similar scaling to have a stronger redshift dependence than we measured. We defer a detailed redshift study to future work.

While clusters similar to those in MACSIS at high redshift may contribute to the detected SZ signal more strongly, especially on small angular scales, their rkSZ amplitude is also greatly reduced when the de-rotation is based on the spin of the galaxies. Examining the galaxies-aligned profiles in Fig, \ref{fig:slices:redshift}, we measure amplitudes of $7.5~\mu$K for $z=0$, and $11.6~\mu$K for $z=1$. Both values are comparable to the rkSZ amplitude from the unordered substructure motions $\approx 5~\mu$K obtained from the same subsets ($\{x, y, z\}$-projections), and could potentially make the detection of the rkSZ signals from stellar proxies more challenging, as explained in Section \ref{sec:results:differential-motions}.

\subsection{Differential motions}
\label{sec:results:differential-motions}
The growth of galaxy clusters is driven by the infall of substructures, which contribute to creating a complex dynamic environment. When selecting a particular aperture to compute cluster properties, and particularly the angular momentum vector orientation, we actually obtain averaged quantities over the $r_{500}$ sphere. Clearly, the substructure clumps (individually) and their peculiar radial motion carry angular momentum which may be significantly different in magnitude and orientation than the aperture average. The \textit{differential motion} of ionised gas clouds inside the selected aperture can have an effect on how the angular momentum is estimated, and therefore on the de-rotation procedure too. Stacking de-rotated rkSZ maps heavily mitigates this effect, however, we have shown in Fig. \ref{fig:rksz_gas_null} that residual motions still appear in absence of de-rotation and even when the LoS is parallel to $\mathbf{J}$. The corresponding rkSZ profiles in Fig. \ref{fig:slices_projections} (grey and blue dashed lines) demonstrate that differential substructure motions can produce amplitudes as large as 20 $\mu$K near the centre. Observational studies relying on the angular momentum of galaxies to de-rotate the kSZ maps may often measure \textit{stacked} rkSZ amplitudes of $\approx 20~\mu$K (see profiles for galaxies-aligned, low-value subsets in Figs. \ref{fig:slices:properties} and \ref{fig:slices:dynamical-state}), which are comparable to the those produced by unordered differential motions. The question of whether the rkSZ features are due to ordered rotation or unordered substructure motions, even after stacking, makes the study of $\{x, y, z\}$-projection maps necessary and compelling.

\begin{figure}
	\includegraphics[width=\textwidth]{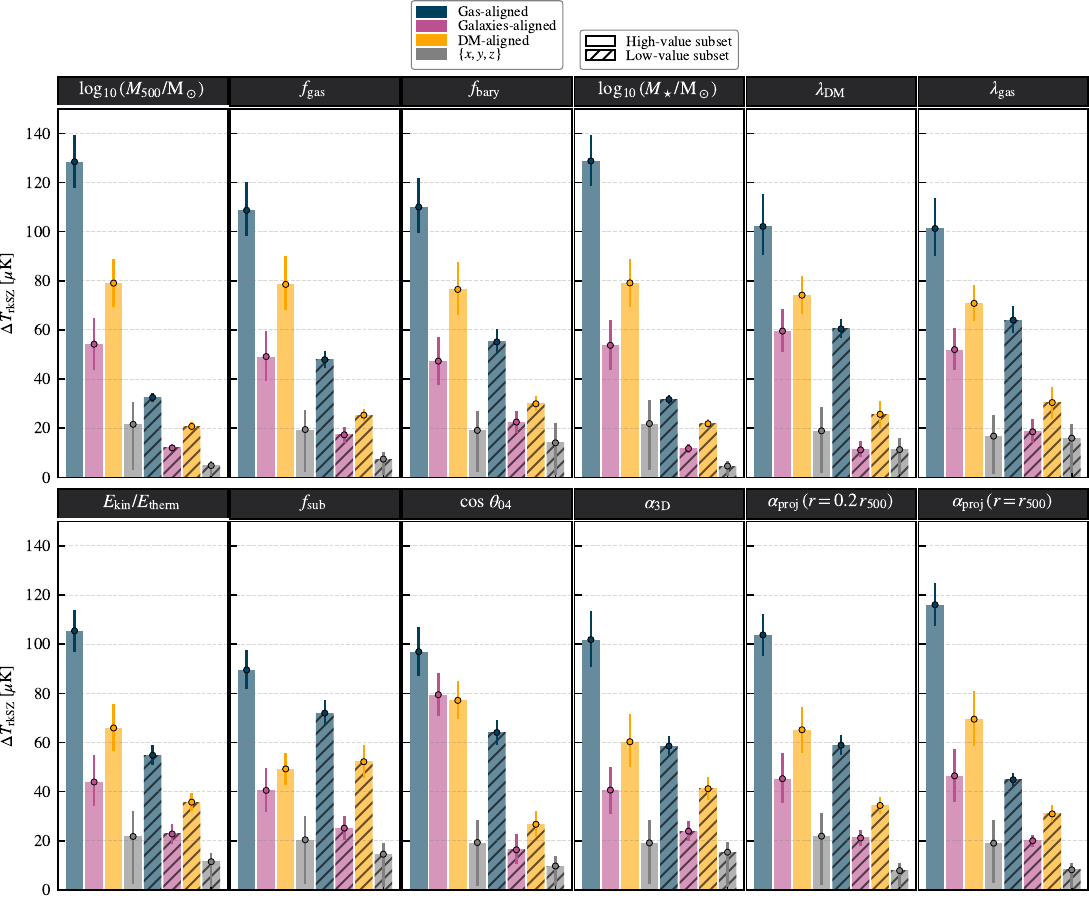}
    \caption{Summary of the rkSZ amplitude, $A_{\rm max}$, of the profiles in Figs. \ref{fig:slices:properties} (top row) and \ref{fig:slices:dynamical-state} (bottom row). The colours indicate the de-rotation configuration, as above; the high- and low-value subsets are represented by a solid colour and a hatch pattern respectively. In addition to the Gas, galaxies and DM alignment configurations, we show the amplitude of the non-de-rotated projections, quantifying the rkSZ signal due to differential motions for each population subset.}
    \label{fig:differential-motions}
\end{figure}

Using the median profiles explored in this section, we record the maximum amplitude or the rkSZ signal and we compare it with that of the $\{x, y, z\}$-projection profiles. In Fig. \ref{fig:differential-motions}, we show these results for the cluster subsets next to the corresponding three non-de-rotated projections averaged together. Identifying the rotational signature with a high confidence level demands that its amplitude is greater than the signal due to spurious features originating from unordered motion in the cluster's rest frame. An example of high-amplitude ratio is obtained when splitting the cluster population by $M_{500}$, in the top-left panel of Fig. \ref{fig:differential-motions}. The signal from differential motions of 188 stacked high-mass clusters\footnote{The MACSIS sample, reduced to 377 clusters after the $M_{200}$ and $f_{\rm gas}$ cuts, is split by the \textit{median} of a cluster property, producing two subsets of 189 and 188 clusters.} has an amplitude $A_{\rm max}^{\{x,y,z\}} \approx 20\, \mu$K, while $A_{\rm max}^{\rm (gas)}=128\, \mu$K is $\approx 6$ times larger. Conversely, the low-$\lambda_{\rm DM}$ and low-$\lambda_{\rm gas}$ samples show $A_{\rm max}^{\{x,y,z\}} \approx A_{\rm max}^{\rm (galaxies)}$, suggesting that the rkSZ signal after stacking these objects, obtained with a galaxy-based de-rotation criterion, will likely be due to unordered motion rather than coherent rotation. This result may impact the significance of rkSZ measurements, highlighting once again the importance of using a robust de-rotation method to maximise the overall amplitude upon stacking.

Improving the accuracy of the de-rotation, however, might not be sufficient to achieve a high $A_{\rm max} / A_{\rm max}^{\{x,y,z\}}$ ratio where the signal is intrinsically low, e.g. in low-mass clusters. The effect of differential motions could, however, be reduced in two ways: (i) increasing the number of stacked clusters would suppress the random features and (ii) varying the aperture used in the calculation of the bulk velocity (and angular momentum) would shift the rest frame of the cluster (and the de-rotation alignment) in phase space, suppressing some features from unordered motion and enhancing others, allowing to estimate the underlying coherent rotational signature more reliably. We delegate the detailed investigation of the impact of sample size and aperture selection to future work.

Based on this argument, we comment on the evidence for cluster rotation measured by \cite{2019JCAP...06..001B}. The sample used in their study includes low-redshift ($z \in [0.02-0.1]$) and  $M_{500} \in [10^{14} - 10^{15}]$ M$_\odot$ clusters, compatible with our MACSIS low-mass subset de-rotated using galaxy spins. Based on results in Table~\ref{tab:max_measured}, we predict $A_{\rm max}^{\rm (galaxies)} = (32.5 \pm 1.7)\, \mu$K, which is comparable to their measured amplitude. For this subset, we estimate a differential motion amplitude of $A_{\rm max}^{\{x,y,z\}} \approx 5\, \mu$K, which is $\approx 6$ times lower than the rkSZ amplitude. Our prediction supports their $2\sigma$ detection claim, but we recommend caution when using their result, given the limited sample sizes of 6 and 13 objects.

\section{Fitting a analytic model}
\label{sec:results:fitting}

We now provide a analytic model which can represent the rkSZ profiles of massive clusters. To obtain these results, we follow the procedure used by \cite{2017MNRAS.465.2584B}, which includes two steps: (i) fitting a \citeauthor{2006ApJ...640..691V} model to the 3D number density profile $n_e(r)$ and (ii) fitting an angular velocity profile $\omega(r)$ to the projected (2D) rkSZ maps, assuming the $n_e(r)$ found previously.

\begin{figure}
	\includegraphics[width=\textwidth]{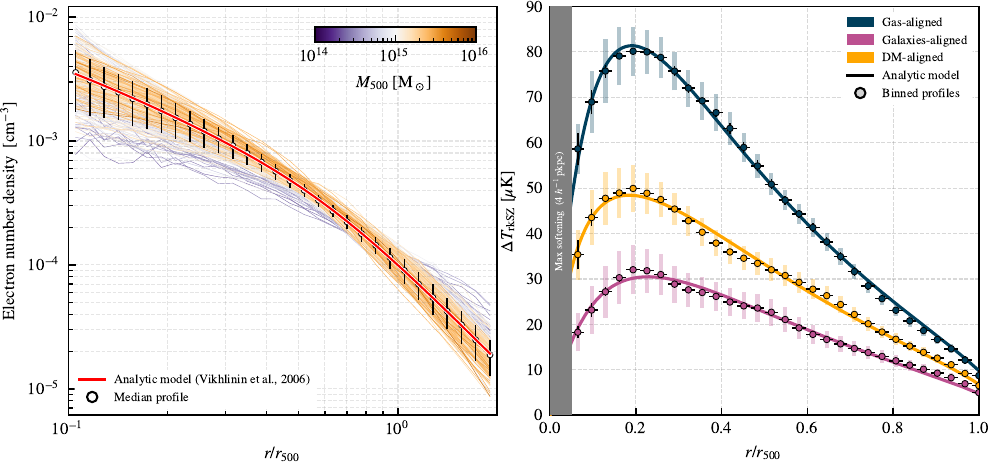}
    \caption{\textit{Left panel}. Density profiles (hot gas, $T>10^5$ K) for the MACSIS clusters colour-coded by $M_{500}$. The median profile is indicated by black markers and fitted by a \protect\citeauthor{2006ApJ...640..691V} model, shown in red. The black error bars indicate the first and third quartiles of the scaled density at each radial bin. \textit{Right panel.} The rkSZ profiles obtained after stacking all clusters (as in Fig. \ref{fig:slices_projections}), with colors indicating the de-rotation criterion as in previous figures. The profiles are down-sampled to 64 radial bins and fitted by the analytic rkSZ model in Eq.~\eqref{eq:kSZ-projection}. The black error bars indicate the 10$^{\rm th}$ and 90$^{\rm th}$ percentiles of the variation of the median profile in each bin, and the thicker coloured bars show the first and third quartiles of the profile from bootstrap resampling, as in Fig. \ref{fig:rksz_gas_edge}.}
    \label{fig:density-profiles}
\end{figure}

\begin{enumerate}
    \item \textbf{Number density profiles}. As a first step, we divide the hot ($T>10^5$ K) gas particles in 30 radial shells with log-spaced radii ranging from $0.05\, r_{500} -\, r_{500}$. The mean density $\rho(r)$ in each shell is therefore given by the total mass divided by the volume of the shell. From the density profile, $n_e(r) = \rho(r) / (\mu \, m_{\rm H})$, where the $\mu=1.14$ is the mean atomic weight and $m_{\rm H}$ is the mass of the hydrogen atom. In our procedure, we do not directly compute $n_e(r)$ and we scale $\rho(r)$ by the density profiles to the critical density of the Universe, and the radius by $r_{500}$. Using this method, we obtain the dimensionless density profile for each MACSIS cluster in the selection subsets.

    {To reduce the degeneracy of the \citeauthor{2006ApJ...640..691V} model, we propose constraining the range of the parameters and fixing the slope $\alpha$. For the MACSIS clusters, we found that imposing the bounds reported in Table~\ref{tab:mass-dependence-fit-params} greatly stabilise the model (see Fig. \ref{fig:density-profile-mass-bins}).}
    
    \begin{table}
    \centering
    \caption{{Constraints imposed on the parameters in the \protect\citeauthor{2006ApJ...640..691V} model (columns 2-3), and linear-fit parameters ($a, b$) describing the dependence of the $\mathcal{V}_6$ parameters on halo mass (columns 4-5). In all our fits to electron number density profiles, we fix $\alpha=1.5$. $^\star$For the $z=1$ sample, we raise the upper bound of the cool-radius to $r_c = 0.28\, r_{500}$ to match the larger and shallower density core.}}
    \label{tab:mass-dependence-fit-params}
    \begin{tabular}{lcccc}
    \toprule
    \multicolumn{1}{c}{\bf Parameters}    &   
    \multicolumn{2}{c}{\bf Constraints} &
    \multicolumn{2}{c}{\bf Mass dependence} \\
    \cmidrule(rl){2-3} \cmidrule(rl){4-5}
    \multicolumn{1}{c}{$\mathcal{V}_6$} & Min & Max & $a$ & $b$ \\
    \midrule
    $n_0 / (10^{-3}\, {\rm cm^{-3}}$) & 0.9 & 9.0 & -6.47 & 103 \\
    $r_c /r_{500}$ & 0.01 & 0.18$^\star$ & 0.139 & -1.99\\
    $r_s /r_{500}$ & 0.50 & 0.75 & -0.0847 & 1.89\\
    $\alpha$ \quad (fixed) & 1.5 & 1.5 & 0 & 1.5\\
    $\beta$ & 0.3 & 0.6 & 0.115 & -1.29\\
    $\varepsilon$ & 2 & 3 & 0.275 & -1.45\\
    \bottomrule
    \end{tabular}
    \end{table}   
    
    
    {Having fixed $\alpha=1.5$, we fit a 5-parameter ($\mathcal{V}_{\rm 5}$)} \citeauthor{2006ApJ...640..691V} model to the median profile via the L-BFGS-B optimisation method \citep{zhu1997algorithm} implemented in \textsc{Scipy} \citep{virtanen2020scipy}. An example of this procedure is shown on the left panel of Fig. \ref{fig:density-profiles}, where we included all MACSIS clusters. The 6 best-fit parameters $\mathcal{V}_6$ for the cluster sample in Fig. \ref{fig:density-profiles} and those using other cluster selection criteria are summarised in Table~\ref{tab:vikhlinin_z0}.
    
    \item \textbf{Angular velocity profiles}. Given the priors $\mathcal{V}_6$ setting the density profile, which we keep fixed, we now proceed in the evaluation of the projection integral in Eq.~\eqref{eq:kSZ-projection} to find the best-fit parameters $\mathcal{B}_{\rm g}=\{v_{\rm t0}, r_0, \eta\}$ for $\omega(r)$. We now fit this template to the median rkSZ profiles in Figs. \ref{fig:slices_projections} to \ref{fig:slices:redshift} to obtain a analytic model for each subset sample of clusters we introduced. Unlike for the number density profile, here we use the Sequential Least Squares Programming \citep[SLSQP,][]{kraft1988software, nocedal2006numerical} method to optimise the fitting functional form, since we found a faster convergence time compared to the L-BFGS-B method.
    We found that the $\omega(r)$ model in Eq.~\eqref{eq:omega-profile} allows for great flexibility to match the rkSZ profiles well at all radii, due to the unconstrained $\eta$ parameter, as we demonstrate in the right panel of Fig. \ref{fig:density-profiles}. As in the previous steps, we report the best-fit parameters $\mathcal{B}_{\rm g}$ for the generalised angular velocity profile in Table \ref{tab:baldi_z0}.
\end{enumerate}

\section{The rkSZ temperature power spectrum}
\label{sec:power-spectrum}
The results of the previous sections deliver a simple prescription for the electron density and angular velocity profiles of clusters at varying redshifts. {In principle,} these can be directly used to predict the expected temperature power spectrum contribution using a halo model approach, improving on the treatment given in \citetalias{CC02}. {We outline and discuss the framework to carry out such a calculation, leaving a full calculation to future work with large-volume hydrodynamic simulations.}

The main starting point for the computation of the rkSZ temperature power spectrum calculation is the halo model \citep[e.g.,][]{Sheth1999,Seljak:2000gq, Cooray2002}, which allows to define the comoving number density of halos as a function of mass and redshift, ${\rm d} N(z, M)/{\rm d}M{\rm d}V$. The treatment then is essentially like for the standard $y$-distortion power spectrum \citep{Refregier2000, Hill2013}, but with the Legendre transform of the cluster $y$-profiles being replaced by the corresponding transforms of the rkSZ profile. For the 1-halo contribution to the CMB temperature power spectrum, this then reads \citepalias{CC02}
\begin{align}
C_{\ell}^{\rm 1h} 
&=\int_{0}^{z_{\rm max}} {\rm d}z \frac{{\rm d}V}{{\rm d}z}\int_{M_{\rm min}}^{M_{\rm max}} {\rm d}M \frac{{\rm d}N}{{\rm d}M{\rm d}V} \,\frac{2}{3}\, \left|\, y^{\rm rkSZ}_{\ell}(M,z)\, \right| ^2.
\label{eq:yy_total}
\end{align}
The factor of $2/3=\langle\sin^2 i\rangle$ arises from the average over a random distribution of inclination angles, $i$. This expression directly uses the fact that the rkSZ signal causes a simple temperature perturbation, $\Delta T/T$, on the sky, see Eq.~\eqref{eq:kSZ_definition}.
The Legendre transform of the rkSZ signal profile can be expressed as
\begin{subequations}
\label{eq:yy_transfor}
\begin{align}
\left|\, y^{\rm rkSZ}_{\ell}(M,z)\, \right| ^2
&\simeq\frac{\pi}{2}\,
\left[
\int_0^{\theta_{200}}
\eta(\theta)\,J_1(\ell\,\theta)\,{\rm d}\theta
\right]^2
\\[1mm]
\eta(\theta)&=
\frac{2\sigma_T}{c}R(\theta)\,\int_{R(\theta)}^{r_{200}} \frac{n_e(r)\,r\,\omega(r){\rm d}r}{\sqrt{r^2-R(\theta)^2}}
\end{align}
\end{subequations}
where we recast the expressions of \citetalias{CC02} {to match our formalism}. Here, $J_1(x)$ is the Bessel function of the first kind. We also use the angular diameter distance, $d_A(z)$, to obtain the radius $R(\theta)=d_A \theta$ via the angular scale $\theta$. For readability, we suppress the explicit dependency of the variables on mass and redshift.

In \citetalias{CC02}, the electron density profile was modeled using the hydrodynamic equilibrium assumption in a Navarro-Frenk-White \citep[NFW,][]{1996ApJ...462..563N} dark matter profile. In addition, the angular velocity profile was assumed to be given by simple solid body rotation {(i.e., $\omega \sim$ constant, see also Section \ref{app:profile-fits})}. Both aspects can be improved upon using the results {presented in this work}.

{To model the rotational kSZ signature using the halo model, we require both $n_e(r)$ and $\omega(r)$ as functions of $M_{500}$ and redshift. Within the present simulations, it is difficult to assess the detailed redshift dependence of the profiles, and therefore we recommend using the scalings obtained at $z=0$ for all redshifts. {In the future, large-volume ($\, \gtrsim 1$ Gpc$^3$) simulations will provide more accurate estimates of the rkSZ amplitude over a range of redshifts.}}

Consulting Table~\ref{tab:baldi_z0}, for the circular velocity profile in the gas-aligned case relevant here, we find that the parameters $r_0$ and $\eta$ only depend weakly on the cluster mass. We thus recommend the average values $\bar{r}_0\simeq 0.16\,r_{500}$ and $\bar{\eta}\simeq 2.0$ for the gas-aligned case. For $v_{\rm t 0}$, a mass-dependence is present, however, given the limited mass resolution of the simulation, we recommend using $\bar{v}_{\rm t 0}\simeq 1.5\,  v_{\rm circ}$. 
These simple scaling relations provide a reasonable representation of the simulation results, and broadly should give a signal that is about one order of magnitude larger than in \citetalias{CC02}.

{
For the electron density profile, from Table~\ref{tab:vikhlinin_z0} we find indications for a dependence of the profile parameters on cluster mass. To verify this dependence, we split the MACSIS sample in 8 logarithimcally spaced mass bins and then computed the median density profiles of the clusters in each bin. However, the density model parameters are strongly degenerate, motivating us to adopt $\alpha=1.5$ in all our fits. We show the fit model for each mass bin in Fig.~\ref{fig:density-profile-mass-bins}).}

{
We find that, for the mass range $\log_{10}(M_{500} /{\rm M}_\odot) \in [14.4, 15.6]$, the density fit parameters ($\mathcal{V}_i \in \mathcal{V}_6$) can be generally described by a linear function with gradient $a$ and intercept $b$, expressed as
\begin{equation}
    \label{eq:Ne_approximation}
    \mathcal{V}_i(M_{500}) = a \times \log_{10}(M_{500}/{\rm M}_\odot)  + b.
\end{equation}
The linear parameters $a$ and $b$ are reported in Table~\ref{tab:mass-dependence-fit-params} and we show the individual scaling relations in Fig. \ref{fig:density-profile-mass-bins-parameters}.
}

Together with the standard mass and redshift dependencies of $M_{500}$, $r_{500}$, $v_{\rm circ}$ and $d_A$, these relations can be used to compute $y^{\rm rkSZ}_{\ell}$ from Eq.~\eqref{eq:yy_transfor}. From Eq.~\eqref{eq:yy_total}, one can then obtain the auto-power spectra caused by the rkSZ effect, e.g., by using {\tt CLASS-SZ} \citep{Bolliet:2017lha} to compute the average over the mass function. We note that the relations in Eq.~\eqref{eq:Ne_approximation} are derived from a cluster sample with a very narrow mass range and they may not be valid for values of $M_{500}$ outside the MACSIS range. We therefore recommend calibrating the $\mathcal{V}_6$ parameters for mass-limited cluster data sets, spanning over a wider mass range.

{A detailed exploration of the dependence of the rkSZ power spectrum signal on various parameters will be left to future work. However, we comment of a few important aspects. Firstly, one does expect contributions from the 2-halo term to become relevant at large angular scales \citepalias[see also][]{CC02}. To compute this signal, a model for the halo bias with respect to linear theory must be provided. This treatment also must account for the effect of spin alignments, which can affect the total amplitude of the 2-halo contribution in non-trivial ways, requiring additional investigation.}
{Secondly, here we have only provided a prescription for the 1-halo contribution from the large-scale rotation. As our analysis has revealed, significant kSZ contributions also arise from internal motions of substructures (see Section~\ref{sec:results:differential-motions}). The 1-halo kSZ contribution will therefore be enhanced by these effects, in particular at small scales corresponding to $\ell \gtrsim {\rm few}\times 10^3$.}
{Finally, the rkSZ is expected to add an irreducible noise floor to studies of the moving-lens effect \citep{Birkinshaw1983, 2019PhRvL.123f1301H} and cosmological vorticity modes \citep{2023arXiv230111344C}, possibly biasing the inference. Given that the estimates of \citetalias{CC02} seem to be on the low end, extra effort should be made to improve the modeling of the rkSZ contribution for future applications.} {We leave a full calculation and discussion of the temperature power spectrum to future work.}

\section{Discussion and Conclusions}
\label{sec:summary}

Cluster rotation can be detected via its dipole-like kSZ signature, provided that the orientation of the angular momentum orientation of the gas is well estimated, or reconstructed. Our study of the rkSZ effect with MACSIS clusters complements that of \cite{2017MNRAS.465.2584B} where a sample of six low-mass relaxed galaxy clusters was selected from the MUSIC simulations. To achieve these results, we followed the methodology in Fig. \ref{fig:stacking_workflow}. Then, in Section \ref{sec:results}, we study the rkSZ profiles over a mass range $M_{500} \sim 10^{15-16}$ M$_\odot$, wider than that in \cite{2017MNRAS.465.2584B}, and showed how the rkSZ signal strength varies when clusters are selected by their global properties or dynamical state descriptors.

Our rkSZ amplitudes are comparable to the study from the MUSIC simulations \citep[$\approx 30-50~\mu$K,][]{2017MNRAS.465.2584B}, stressing that this value should be compared the low-mass (Fig.~\ref{fig:slices:properties}.A) and the relaxed (Fig.~\ref{fig:slices:dynamical-state}.A) subsets. Moreover, our high-mass subset shows an rkSZ amplitude close to the regime of a maximally rotating cluster merger, used by \citetalias{CM02} to set an upper bound ($\approx 200~\mu$K) to the typical kSZ contribution from bulk rotation. Overall, this study shows agreement with the rkSZ signal strength at the relaxed, low-mass end of the cluster population \citep{2017MNRAS.465.2584B} and at the unrelaxed, high-mass end (\citetalias{CM02}). However, we note that none of these works predicts an rkSZ amplitude as small as \citetalias{CC02}. From MACSIS, we measured a rkSZ amplitude 10 times larger than the \citetalias{CC02} estimate of $\approx 3~\mu$K, computed for a cluster with virial mass $5\times 10^{14}$ M$_\odot$ at $z=0.5$. Assuming their cosmological parameters and our self-similar relations, we can re-scale their value to match the halo mass and redshift of our sample. This calculation indicates that rkSZ amplitude at $z=0$ would be lower by 40\% at their cluster mass; up-scaling this value self-similarly in mass to match the median value for MACSIS ($M_{200}=1.4\times 10^{15}$ M$_\odot$) would increase the amplitude by a factor of 2, giving an overall signal of $\Delta T_{\rm CC02}\approx 3.6~\mu$K. We will compare this value with $80.2~\mu$K obtained from the \textit{all-clusters} sample (see Table~\ref{tab:max_measured}). We identified three parameters which may have been underestimated by \citetalias{CC02}.

\begin{enumerate}
    \item The Universal \textit{baryon fraction} from their cosmology is 10\% lower than ours, causing the cluster gas mass to be underestimated by a factor of 0.9 \citep[see also][]{2000PhRvD..62j3506C}. Using the mass scaling in Eq.~\eqref{eq:kSZ-scaling}, the correction to the rkSZ amplitude is expected to be $\delta(f_{\rm b})\approx 1.1$.

    \item they used the value of $\approx 36~{\rm km\, s^{-1}}$ for the \textit{tangential velocity} at $\approx 0.2$ Mpc, while we found $\approx 90~{\rm km\, s^{-1}}$  at $r/r_{500} = 1/5$, i.e. the radius of the maximum amplitude, which could increase the rkSZ signal strength linearly by a factor of $\delta(v_{\rm tan})\approx 2.5$.

    \item Finally, they assume the mean \textit{spin parameter} for dark-matter halos $\bar{\lambda}_{\rm DM} \approx 0.04$, while we recommend using the hot gas spin parameter instead, which has a median value of $\lambda_{\rm gas} \approx 0.05$, leading to an additional correction of $\delta(\lambda)\approx 1.2$. 
\end{enumerate}

After combining these corrections, we find $\delta(f_{\rm b})\, \delta(v_{\rm tan})\, \delta(\lambda) \approx 3.3$, which is still not sufficient to explain the factor of 10 difference from our measurement. However, we have shown in Fig.~\ref{fig:slices:properties}.F that the rkSZ amplitude is dependent on $\lambda_{\rm gas}$, which can reach values of $\approx 0.15$. Clusters with high spin parameters are abundant: they could indeed contribute significantly to this estimate since they form an extended tail in the log-normal-like distribution at most halo masses \citep{2010MNRAS.404.1137B}. Now, assuming $\lambda_{\rm gas} \approx 0.15$, we find that \citetalias{CC02} may have underestimated the spin parameter by a factor of $\delta(\lambda)\approx 3.6$. When combined, these considerations lead to $\delta(f_{\rm b})\, \delta(v_{\rm tan})\, \delta(\lambda) \approx 10$, which could reconcile the prediction by \citetalias{CC02} with our study and \cite{2017MNRAS.465.2584B}.

We now highlight the following key findings from this work:
\begin{enumerate}
	\item \textbf{Mass dependence.} High-mass clusters produce a larger rkSZ amplitude than low-mass ones. This trend is consistent with self-similar scaling relations, and we find a mass dependence twice as strong as that suggested in Eq.~\eqref{eq:kSZ-scaling}. For a low-mass, relaxed sample of clusters, our rkSZ profile amplitudes are $\approx 30~\mu$K, consistent with \cite{2017MNRAS.465.2584B}.

	\item \textbf{Dynamical state.} Our metrics to assess the dynamical state of cluster atmospheres are correlated to the halo masses  (see Fig.~\ref{fig:slices:dynamical-state}) and indicate that unrelaxed clusters produce a rkSZ signal two times stronger than relaxed ones. While disentangling the dynamics of individual substructures in merging systems via the kSZ effect can be complex, stacking maps suppresses the effect of transient features and enhances the signal from bulk rotation. 

	\item \textbf{Spin alignment.} If stacking maps is a decisive step in retrieving the kSZ signal from cluster rotation, then de-rotating the maps and aligning the expected dipolar feature coherently are also crucial. This procedure relies on determining the spin orientation of the gas. We have shown that using the galaxy angular momentum orientation as a proxy for that of the hot gas, as implemented by \cite{2019JCAP...06..001B}, reduces the combined rkSZ signal by $\approx$ 60\% compared to when the de-rotation is based on the gas spin itself. This effect arises because galaxies and hot gas do not always co-rotate, as shown by a tail towards low values in the $\cos \theta_{04}$ distribution in Fig. \ref{fig:corner-plot-dynamical-state}. {Similarly, using the DM spin as a proxy for the gas spin would reduce the stacked rkSZ signal amplitude by about 40\%.}
	
	\item \textbf{Temperature power spectrum.} Adapting the formalism in \citetalias{CC02} for the temperature power spectrum calculation, we provide an improved method for estimating the one-halo term arising from cluster rotation. We remove the assumption of a generalised-NFW model and solid-body rotation and, instead, we input the rkSZ profiles from full-physics simulations. 
	%
	The description presented here can be directly used in, e.g., {\tt CLASS-SZ} \citep{Bolliet:2017lha} to compute the rkSZ temperature power spectrum contribution. While these are expected to be about one order of magnitude smaller than the usual kSZ effect \citepalias[see][]{CC02}, this could add a new cosmological noise-floor to studies of the moving lens effect \citep[e.g.,][]{2019PhRvL.123f1301H}. Future kSZ studies, e.g., with the Simons Observatory \citep{SOWP2019}, might also become sensitive to this additional kSZ component, with a possible contributions from internal substructure motion at small scales.
{Due to the difference in the amplitude of the rkSZ signal with respect to the model of \citetalias{CC02}, it will be important to consider these effects more carefully.}

\end{enumerate}
For microwave observations of the kSZ effect, an improved sensitivity may not guarantee a reliable reconstruction of the cluster rotation from the kSZ signal without a robust method of probing the orientation of the ICM spin. This step is critical for avoiding a potential $60$\% signal loss (see Section \ref{sec:results}) and observationally challenging \citep{2017MNRAS.465.2616M, 2019JCAP...06..001B}. {The \citet{2017MNRAS.465.2616M} sample consists of the most strongly rotating halos, as determined from galaxy LoS velocity data in SDSS-DR10. The resulting selection bias naturally favours strongly rotating halos, and could effectively lead to a signal loss smaller than we predicted without such selection. Provided that the spins of DM and galaxies are closely aligned, as we showed in Fig. \ref{fig:corner-plot-all-properties} for MACSIS, we can estimate the rkSZ signal suppression by using clusters with a strongly rotating DM halo (high $\lambda_{\rm DM}$) to represent those with a strongly rotating galaxy population. Indeed, the rkSZ amplitude of our high-$\lambda_{\rm DM}$ subset for the galaxies-aligned case is $59.5\, \mu$K, almost double that of the overall sample, $32.1\, \mu$K. On average, the signal suppression relative to $A_{\rm max}^{(\rm gas)}$ has a smaller impact on high-$\lambda_{\rm DM}$ clusters (42\%) compared to the overall sample (60\%, see Table~\ref{tab:max_measured}). Our study suggests that the object-selection strategy of \cite{2019JCAP...06..001B} can marginally boost the stacked rkSZ signal of galaxies-aligned clusters.}

Measuring the {LoS} velocity of the hot gas \textit{directly} may become feasible in the future with the advent of new high-resolution X-ray spectrometers \citep[see also][]{2013MNRAS.434.1565B}. The X-ray space observatory \textit{Athena}, developed by the European Space Agency \citep[see the white paper by][]{2013arXiv1306.2307N}, is planned to be launched in the late 2020s and will be equipped with the X-ray Integral Field Unit \citep[X-IFU,][]{2018SPIE10699E..1GB, 2018arXiv180706903G}. \textit{Athena}/X-IFU opens exciting prospects for a Doppler measurements of the {LoS} velocity field of the ICM, with a spectral resolution of $2.5-7$ eV in the soft X-ray band. These specifications are even superior to the capabilities of XRISM/Resolve micro-calorimeter array, managed by the Japan Aerospace Exploration Agency and capable of resolving Doppler speeds of $\simeq 300~{\rm km\, s^{-1}}$ \citep{2018arXiv180706903G, 2020SPIE11444E..22T, 2021JATIS...7c7001T} and, previously, of the Hitomi spectrometer which provided measurements of turbulent motions in the core of the Perseus cluster \citep{2016Natur.535..117H} by resolving speeds of $\simeq 100~{\rm km\, s^{-1}}$ at 6 keV \citep[][]{2018JATIS...4b1402T}.

More accurate predictions for the rkSZ amplitude will soon be made possible by future large-volume hydrodynamic simulations. In the very near future, an example of such runs is Virgo Consortium's flagship FLAMINGO project, which will contain a much richer cluster sample of $\sim 10^6$ objects modelled with full physics \citep{flamingo_schaye2023}. With clusters as massive as the MACSIS objects and a mass-limited HMF down to $M_{500}\simeq 10^{13}$ M$_\odot$, FLAMINGO will be able to reproduce cluster-count statistics and improve the estimates of our rkSZ profiles using particle-data snapshots and halo catalogues. In addition, the light-cone outputs from FLAMINGO will allow to construct Healpix \citep{2005ApJ...622..759G} all-sky maps over a selected redshift range. In a future work, we aim to use these data sets to forecast the rotational kSZ signal from clusters through power spectra and feature extraction methods \citep[see e.g.][]{1996MNRAS.279..545H, 2021MNRAS.507.4852Z}. {Together with an improved modelling of the relativistic SZ effect \citep[e.g.][]{2022MNRAS.517.5303L} this could refine the simulation-driven prescription of SZ clusters in cosmology.}

As an observational outlook, we will also illustrate how our framework could facilitate the study of CMB foregrounds and play a role in upcoming precision cosmology programs with stage-4 facilities, such as Simons Observatory \citep{SOWP2019} and SKA-2 (see \citealt{2016PhRvD..94d3522A} for a review, and \citealt{2016arXiv161002743A}). In particular, estimates from simulations could guide the development of four additional areas of research: matched filters for separating the cluster rotation from the moving-lens effect \citep{1986Natur.324..349G, 2007MNRAS.380.1023S, 2021PhRvD.103d3536H, 2021PhRvD.104h3529H}; the pairwise transverse velocity measurement with the Rees-Sciama effect \citep{2019ApJ...873L..23Y, 2019PhRvL.123f1301H}; the vector gravito-magnetic distortion, predicted to occur when rotating massive clusters induce space-time frame-dragging \citep{2021ApJ...911...44T, 2022MNRAS.510.3589B}; and the $\simeq 10\, \sigma$ measurement of cosmic filament rotation using the kSZ effect \citep{2023MNRAS.519.1171Z}.


\section*{Acknowledgements}

This work used the DiRAC@Durham facility managed by the Institute for Computational Cosmology on behalf of the STFC DiRAC HPC Facility (www.dirac.ac.uk). The equipment was funded by BEIS capital funding via STFC capital grants ST/K00042X/1, ST/P002293/1, ST/R002371/1 and ST/S002502/1, Durham University and STFC operations grant ST/R000832/1. DiRAC is part of the National e-Infrastructure. EA and IT acknowledge the STFC studentship grant ST/T506291/1. 
{JC was furthermore supported by the ERC Consolidator Grant {\it CMBSPEC} (No.~725456) and the Royal Society as a Royal Society University Research Fellow at the University of Manchester, UK (No.~URF/R/191023).}
The research in this paper made use of the following software packages and libraries: 
\textsc{Python} \citep{van1995python},
\textsc{Numpy} \citep{harris2020array},
\textsc{Scipy} \citep{virtanen2020scipy},
\textsc{Numba} \citep{lam2015numba},
\textsc{Matplotlib} \citep{hunter2007matplotlib, caswell2020matplotlib},
\textsc{SWIFTsimIO} \citep{Borrow2020} and
\textsc{Astropy} \citep{robitaille2013astropy, price2022astropy},
\textsc{Unyt} \citep{goldbaum2018unyt}. The colour scheme used throughout the document was generated using the open-source tool \texttt{ColorBrewer} \citep{harrower2003colorbrewer}.

\section*{Data Availability}
The MACSIS simulations were produced by \cite{macsis_barnes_2017}. Enquiries concerning the availability of raw snapshot data, \texttt{SUBFIND} halo catalogues and the version of the \texttt{Gadget-3} code used to run the simulations should be directed to the original authors. The code used for reading and querying the MACSIS data is publicly available on the first author's GitHub repository (\href{https://github.com/edoaltamura/macsis-cosmosim}{github.com/edoaltamura/macsis-cosmosim}) and we include the data products used to generate the figures presented throughout the document. The repository also contains the list of indices of the clusters in the $z=0$ and $z=1$ subsets of Section \ref{sec:results:z-dependence}, ordered by the  FoF of the MACSIS parent simulation. Additional information for reproducing our results is contained in \texttt{JSON} files; e.g. the parameters for the density profile model and the median values of $\{r_{500},\, r_{200},\, M_{500},\, M_{200},\, \lambda_{\rm gas},\, v_{\rm circ}\}$ for each selection subset. The bootstrap samples are not included, but can be reproduced using the code provided. Some intermediate data products, such as original rkSZ cluster maps, are too large to be hosted on GitHub and can be made available upon request to the corresponding authors.

\section{Correlations of cluster properties}
\label{app:correlation-coefficients}

\subsection{Analytic relation between \texorpdfstring{$\alpha_{\rm 3D}$}{alpha} and \texorpdfstring{$\beta_{\rm 3D}$}{beta}}
\label{app:correlation:alpha-beta}
To justify the correlation in Fig. \ref{fig:corner-plot-dynamical-state} between the non-thermal pressure fraction, $\alpha$, and the kinetic-to-thermal ratio, $\beta$, we prove the relation between these quantities introduced in Eq. \eqref{eq:alpha}.
We summarise the definitions of the kinetic and thermal energy:
\begin{subequations}
\begin{align}
    E_{\rm kin} &= \frac{1}{2} \sum_i m_i ({\bf v}_i - {\bf v_{\rm bulk}})^2
    \label{eq:energies-summary:kinetic}
    \\[1mm]
    E_{\rm th} &= \frac{3}{2}\,{\rm k_B}\sum_i \frac{T_i\, m_i}{\mu\,m_{\rm P}}.
    \label{eq:energies-summary:thermal}
\end{align}
\end{subequations}
We begin by studying the form of the equation for the kinetic energy. Since we compute the bulk velocity in the set of particles $\{i\}$, we have the expression
\begin{equation}
    \label{eq:bulk-motion-app}
    {\bf v_{\rm bulk}} = \frac{\sum_i m_i {\bf v}_i}{\sum_i m_i} = \frac{\sum_i m_i {\bf v}_i}{M} \equiv \langle {\bf v} \rangle = \sum_j \langle v_j \rangle,
\end{equation}
where $M\equiv \sum_i m_i$. By substituting the above in Eq. \eqref{eq:energies-summary:kinetic}, and expanding the calculation for each spatial component $j\in\{x, y, z\}$, one can obtain
\begin{subequations}
\begin{align}
    E_{{\rm kin}, j} &= \frac{1}{2} \sum_i m_i \left(  v_{i,j} - \langle v_j \rangle \right)^2
    \\[1mm]
    &= \frac{1}{2} \sum_i m_i \left(v_{i,j}^2 - 2\, v_{i,j} \langle v_j \rangle + \langle v_j \rangle ^ 2 \right)
    \\[1mm]
    &= \frac{1}{2} \biggl[ 
    \sum_i m_i v_{i,j}^2 - 2\, \langle v_j \rangle \underbrace{\sum_i m_i v_{i,j}}_{\substack{\text{$=M\,\langle v_j \rangle$} \\ \text{from Eq.~\eqref{eq:bulk-motion-app}}}}  +~ \langle v_j \rangle ^ 2 \underbrace{\sum_i m_i}_{\text{$=M$}} \biggr]
    \\[2mm]
    &= \frac{1}{2} \biggl[ \sum_i m_i v_{i,j}^2 - 2\, M \langle v_j \rangle^2 + M \langle v_j \rangle ^ 2 \biggr]
    \\[1mm]
    &= \frac{1}{2} \left( \sum_i m_i v_{i,j}^2 - M \langle v_j \rangle ^ 2 \right)
    \label{eq:kinetic-calculation-last}
    \\[1mm]
    &= \frac{1}{2} M\left(\langle v_{j}^2 \rangle - \langle v_j \rangle ^ 2 \right).
    \label{eq:kinetic-calculation-final}
\end{align}
\end{subequations}
The first term in Eq. \eqref{eq:kinetic-calculation-last} is just the mass-weighted square of the velocities in $\{i\}$ for the component $j$, which can be written as $\sum_i m_i v_{i,j}^2 = M \langle v_{j}^2 \rangle$, in analogy to Eq. \eqref{eq:bulk-motion-app}.
Eq. \eqref{eq:kinetic-calculation-final} contains the definition of the velocity dispersion $\sigma_j^2 \equiv \langle v_{j}^2 \rangle - \langle v_j \rangle ^ 2$. Finally, we can sum the kinetic energy for the three components of the velocity dispersion:
\begin{equation}
    \label{eq:kinetic-vel-dispersion}
    E_{\rm kin} = \frac{1}{2} M \sum_j \sigma_j^2 = \frac{1}{2} M (\sigma_x^2 + \sigma_y^2 + \sigma_z^2) \equiv \frac{1}{2} M \sigma^2.
\end{equation}

The next step involves rearranging Eq. \eqref{eq:energies-summary:thermal} and expressing the thermal energy in terms of the mass-weighted temperature for the ensemble of particles $\{i\}$
\begin{equation}
    \label{eq:mass-weighted-temperature}
    T_{\rm mw} = \frac{\sum_i m_i\, T_i}{\sum_i m_i} = \frac{\sum_i m_i\, T_i}{M}.
\end{equation}
This procedure is carried out as follows:
\begin{subequations}
\begin{align}
    E_{\rm th} &= \frac{3}{2}\,{\rm k_B}\sum_i \frac{T_i\, m_i}{\mu\,m_{\rm P}}
    \\[1mm]
    &= \frac{3}{2}\,\frac{\rm k_B}{\mu\,m_{\rm P}}\, \sum_i T_i\, m_i
    \\[1mm]
    &= \frac{3}{2}\,\frac{{\rm k_B}\, T_{\rm mw}}{\mu\,m_{\rm P}}\,M.
    \label{eq:thermal-energy-derivation}
\end{align}
\end{subequations}
Combining Eqs. \eqref{eq:kinetic-vel-dispersion} and \eqref{eq:thermal-energy-derivation} gives
\begin{equation}
    \beta = \frac{E_{\rm kin}}{E_{\rm th}} = \frac{\mu\,m_{\rm P}\, \sigma^2}{3\, {\rm k_B}\, T_{\rm mw}}.
\end{equation}
After defining the non-thermal pressure as 
\begin{equation}
    P_{\rm nth} = \frac{1}{3}\, \rho_{\rm gas} \sigma^2,
\end{equation}
where $\rho_{\rm gas}$ the local density of the hot gas \citep[e.g.][]{2014MNRAS.442..521S, 2022arXiv221101239T}, we can reproduce the formulation in Eq. \eqref{eq:alpha}
\begin{subequations}
\begin{align}
    \alpha &= \frac{P_{\rm nth}}{P_{\rm nth} + P_{\rm th}} 
    \\[1mm]
    &= \frac{1/3\, \rho \sigma^2}{1/3\, \rho \sigma^2 + \rho\, {\rm k_B}\, T_{\rm mw}/(\mu\,m_{\rm P})}
    \label{eq:alpha-derivation-gas-density}
    \\[2mm]
    &= \biggl[ 1 + \underbrace{\frac{3\, {\rm k_B}\, T_{\rm mw}}{\mu\,m_{\rm P}\, \sigma^2}}_{=1/\beta} \biggr]^{-1},
\end{align}
\end{subequations}
where $P_{\rm th}$ is the thermal pressure, derived from the ideal gas law. Rearranging the expression above, we finally obtain
\begin{equation}
    \alpha = \frac{\beta}{1 + \beta}.
\end{equation}
This relationship can explain the large correlation between the $\alpha$ quantities and $\beta_{\rm 3D}$ reported in Fig. \ref{fig:corner-plot-dynamical-state} and Section \ref{app:correlation-coefficients}. Furthermore, we note that $\beta_{\rm 3D}$ is a cluster-averaged property, since it is obtained for all hot gas particles in $r_{500}$, while $\alpha_{\rm 3D}(r)$ is a profile evaluated from particles in a thin spherical shell at radius $r$.

\subsection{Correlation coefficients}

\begin{figure}
    \centering
	\includegraphics[width=\textwidth]{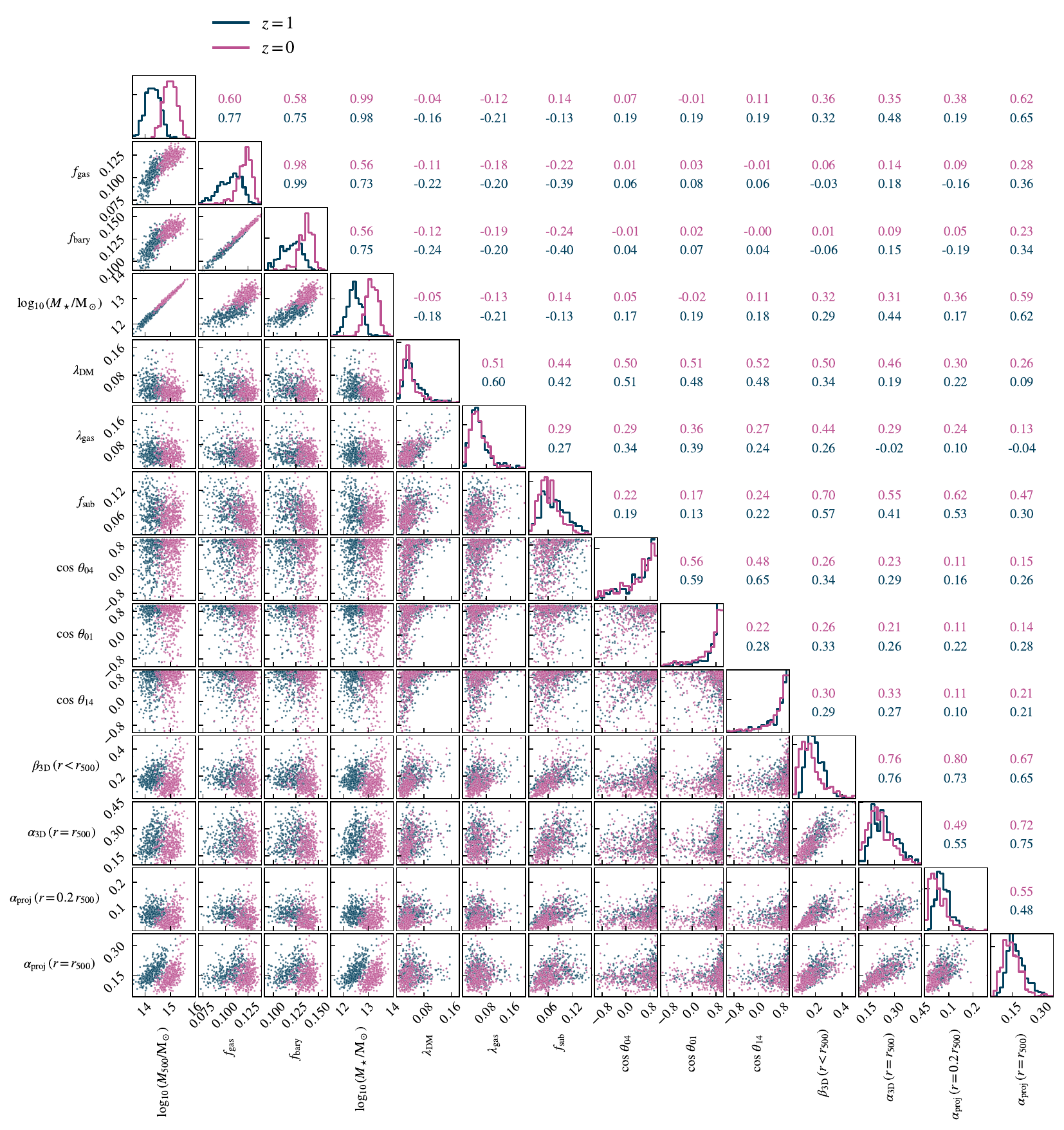}
    \caption{As in Fig. \ref{fig:corner-plot-basic-properties}. In the top right, we write the Spearman rank correlation coefficients for each data set pair, colour-coded by redshift.}
    \label{fig:corner-plot-all-properties}
\end{figure}

In addition to the corner plots in Figs. \ref{fig:corner-plot-basic-properties} and \ref{fig:corner-plot-dynamical-state}, we provide the complete set of correlation relations in Fig. \ref{fig:corner-plot-all-properties}. Here, we also quote the Spearman correlation coefficient, a parameter that quantifies the correlation between any two given sets of data, assuming that a monotonic relation is expected. The Spearman correlation coefficient of two variables $X$ and $Y$ is defined by
\begin{equation}
    r_{\rm S} = 1- {\frac {6 \sum d_i^2(X,Y)}{n(n^2 - 1)}},
\end{equation}
where $n$ is the number of data points, or realisations, and $d_i(X,Y)$ is the pairwise distances of the ranks of the variables $X_i$ and $Y_i$.

In Fig. \ref{fig:corner-plot-all-properties}, we also report the distributions for $\cos\, \theta_{01}$, corresponding to the angle between gas and dark matter components, and $\cos\, \theta_{14}$ for the angle between dark matter and stars. We can recover high values of $r_{\rm S} \simeq 0.6-1$ for known mass-scaling relations, such as gas fraction, baryon fraction and stellar mass in $r_{500}$. We find small, negative correlations for $\lambda_{\rm DM}$ and  $\lambda_{\rm gas}$ against $M_{500}$, in agreement with \cite{2010MNRAS.404.1137B}. The correlation coefficient between $\alpha_{\rm 3D}$ (at a particular radius) and $\beta_{\rm 3D}$ (integrated over $r_{500}$) quantities is $r_{\rm S} \simeq 0.6-0.7$, as expected from Section \ref{app:correlation:alpha-beta}. We also report significantly high correlations between $f_{\rm sub}$ and the quantities based on $\alpha_{\rm 3D}$ and $\beta_{\rm 3D}$, suggesting that MACSIS clusters with high substructure fractions tend to have a thermodynamically unrelaxed and turbulent atmosphere.

\section{Analytic fits to the rkSZ profiles}
\label{app:profile-fits}
\begin{figure}
    \centering
    \includegraphics[width=0.8\textwidth]{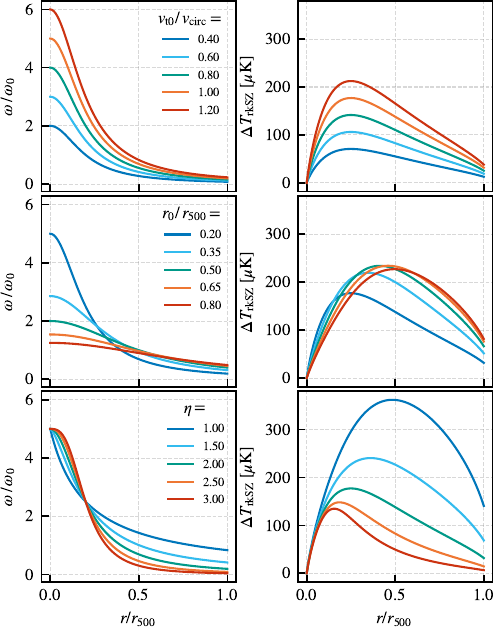}
    \caption{Effect of varying the parameter values in Eq.~\eqref{eq:omega-profile} on the angular velocity profile (left) and the rkSZ profile (right), assuming a \protect\cite{2006ApJ...640..691V} electron number density profile from the MACSIS 0 cluster at $z=0$. From top to bottom, we vary the amplitude $v_{\rm t0}$, the scale-radius $r_0$ and the slope $\eta$. The model with $\eta = 2$ corresponds to the \protect\cite{2017MNRAS.465.2584B} functional form.}
    \label{fig:generalised-baldi-model}
\end{figure}

To match the simulated rkSZ profiles with a analytic functional form, we adopt angular velocity profile which depends on three free parameters, as in Eq.~\eqref{eq:omega-profile}. We found that the $\omega(r)$ model in \cite{2017MNRAS.465.2584B} could not fit the slope of the MACSIS rkSZ profiles beyond $r_{\rm max}$. Therefore, we introduced the $\eta$ parameter, which controls the slope of the $\omega(r)$ profile with a pivot at $r_0$. In Fig.~\ref{fig:generalised-baldi-model}, we illustrate how $\omega(r)$ changes when varying each parameter individually, and the effect on the rkSZ profiles. For convenience, we normalise the $\omega(r)$ profile by the constant $\omega_0 \equiv v_{\rm circ}/ r_{500} = \sqrt{GM_{500}/r_{500}^3}$. For each column, we set $\mathcal{B}_{\rm g}=\{v_{\rm t0} = v_{\rm circ},\, r_0=r_{500}/5,\, \eta = 2 \}$ as default parameters and we only allow one parameter to vary. In the top panels, we show the variation of the tangential velocity scale, $v_{\rm t0}$, which controls the overall amplitude of the profiles. In the middle panels, we show that increasing the scale radius $r_0$ causes the $\omega(r)$ to become shallower, alters the maximum rkSZ amplitude $A_{\rm max}$ and changes the profile of the slope beyond $r_{\rm max}$. In the bottom panels, we show that the rkSZ amplitude decays to 0 $\mu$K faster with radius for higher values of $\eta$ ($\gtrsim 2$). By setting $\eta = 2$, we recover the model used in \cite{2017MNRAS.465.2584B}:
\begin{equation}
    \omega(r) = \frac{v_{\rm t0}}{r_0 \left[1 + (r/r_0)^2\right]}.
\end{equation}
For {the limiting case where} $\eta \longrightarrow 1$, the angular velocity profile becomes cusp-like at $r\simeq 0$; the rate of decay $d\omega/dr$ drops rapidly and $\omega(r)$ is still relatively large at $r \simeq r_{500}$. Finally, setting $\eta = 0$ yields $\omega(r) = v_{\rm t0} / (2\, r_0) \sim {\rm constant}$, similarly to the solid-body rotation model assumed by \citetalias{CC02} and \citetalias{CM02}.

\begin{figure}
    \centering
    \includegraphics[width=0.8\textwidth]{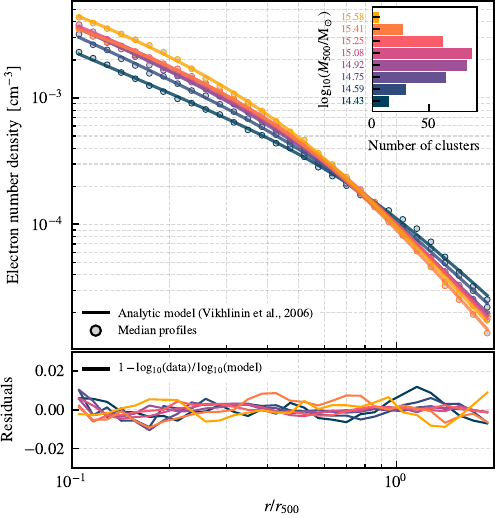}
    \caption{\textit{Top.} Median density profiles in 8 log-spaced mass bins and their analytic fits (solid lines) using a \protect\cite{2006ApJ...640..691V} model with 5 free parameters ($\alpha$ is fixed to 1.5). The inset at the top-right is the HMF of the MACSIS $z=0$ sample (as in Fig. \ref{fig:slices:properties}.A), shown using eight $M_{500}$ logarithmic bins. The central value of the bins is shown on the $y$-axis and the number of clusters in each bin on the $x$-axis. The bins are colour coded to match the fits in the main panel and the data for the median profiles. {\textit{Bottom.} Logarithmic percentile residuals taking the analytic model as baseline. The fits are well converged and the measured median profiles never deviate more than 1\% from the best fit model.}}
    \label{fig:density-profile-mass-bins}
\end{figure}

\begin{figure}
    \centering
    \includegraphics[width=0.75\textwidth]{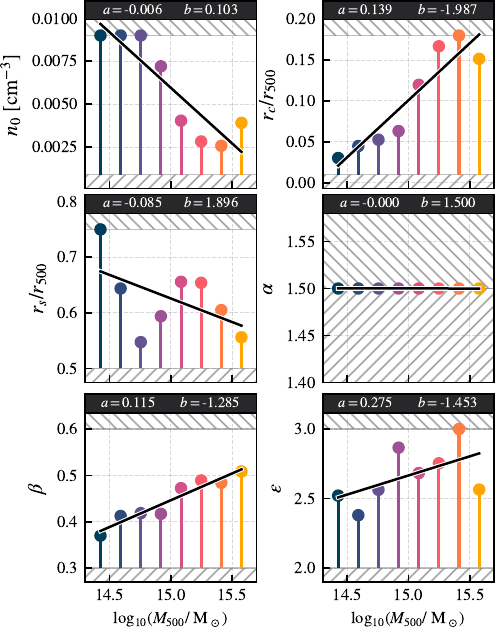}
    \caption{Summary of the $\mathcal{V}_6$ best-fit parameters for the electron density profiles shown in Fig. \ref{fig:density-profile-mass-bins} with increasing halo mass. The data points are coloured by mass bin as in Fig. \ref{fig:density-profile-mass-bins} and we indicate the upper and lower bounds imposed to the fit with hatched regions. For $\alpha$, the bounds edges coincide, allowing only for $\alpha=1.5$. For each parameter, we fit a linear model (black solid line) and we report the gradient $a$ and the intercept $b$ at the top of the corresponding panel.}
    \label{fig:density-profile-mass-bins-parameters}
\end{figure}

To fit the rkSZ profile template to the simulation data, we use the best-fit \cite{2006ApJ...640..691V} parameters, $\mathcal{V}_6$, as priors. We report their values in Table~\ref{tab:vikhlinin_z0} for each subset of the MACSIS sample. In addition to the selection methods used throughout Section \ref{sec:results}, we split the cluster population in 8 logarithmic mass bins and fit the median density profile for each bin, as shown in Fig.~\ref{fig:density-profile-mass-bins}. In the left panel, we show the median density profiles (circles) for each mass bin, fit by a  \cite{2006ApJ...640..691V} model. On the top-right, the HMF shows that most of the mass bins have a limited number of objects ($\sim 10$). To avoid over-fitting to transient features of the density profile, we reduced the dimensionality of the original \cite{2006ApJ...640..691V} parameter space to 5 free parameters, and we set $\alpha=1.5$. The values of the parameters $\mathcal{V}_6$ for each mass bin, shown on the right of Fig.~\ref{fig:density-profile-mass-bins-parameters}, provided the simple, linear mass-scaling relations reported in Eq.~\eqref{eq:Ne_approximation} with gradient and slope values listed in Table~\ref{tab:mass-dependence-fit-params}. In this work, these scaling relations are only used to provide guidelines for the temperature power spectrum calculation (see Section \ref{sec:power-spectrum}).

Given the $\mathcal{V}_6$ parameters in Table~\ref{tab:vikhlinin_z0}, we report the best-fit parameters $\mathcal{B}_{\rm g}$ for the analytic rkSZ profile in Table~\ref{tab:baldi_z0}, based on the method in Section \ref{sec:results:fitting}. {In addition to the value of the tangential scale velocity, we provide a parametrisation similar to the $\omega$ profile in Eq.~(10) of \citetalias{CC02}: $v_{\rm t0} = 3\, \xi_0 \lambda_{\rm gas} v_{\rm circ}$, where we express the spin parameter scaling explicitly and we introduce a positive coefficient $\xi_0$ as a free parameter.}
We commonly find $\xi_0\simeq 10$, which demonstrates that the simulation clusters rotate faster than assumed in \citetalias{CC02}. To give a reference, by comparing to $\omega_0=v_{\rm circ}/r_{500}$ (chosen in Fig.~\ref{fig:generalised-baldi-model}) for $\eta=2$ and $r_0=r_{500}/5$ we expect $\omega(r=0)/\omega_0= v_{\rm t0}/(r_0 \omega_0) = 5\, v_{\rm t0}/v_{\rm circ}$ as found in the upper left panel of Fig.~\ref{fig:generalised-baldi-model}. From the fits in Table~\ref{tab:baldi_z0}, we furthermore conclude that $3\, \xi_0 \lambda_{\rm gas}  \simeq 1.5$ for the gas-aligned case, which is about ten times larger than \citetalias{CC02}.


\begin{table}
    \centering
    \caption{Summary of the selection criteria (column 1) and the best fit parameters from the median electron number density profile using the \protect\cite{2006ApJ...640..691V} model (columns 2-6), with fixed $\alpha=1.5$. These results are obtained at $z=0$ unless stated otherwise.}
    \label{tab:vikhlinin_z0}
    \rowcolors{5}{orange!20}{white}
    \begin{tabular}{lccccc}
    \toprule
    \multicolumn{1}{c}{}    &   
    \multicolumn{5}{c}{\bf Vikhlinin parameters} \\ 
    \cline{2-6} \rule{0pt}{1ex}
    
    Selection criterion     &  $n_0$ &  $r_c$ &  $r_s$ &  $\beta$ &  $\varepsilon$ \\
                            &  [$10^{-3}~{\rm cm}^{-3}$] & [$r_{500}$] & [$r_{500}$] & [--] & [--] \\
    \midrule
    \rowcolor{gray!25}
All clusters                                       &    4.49 &   0.10 &      0.61 &     0.44 &   2.74\\
$M_{500} < 9.7\times 10^{14}$ M$_\odot$            &    9.00 &   0.05 &      0.60 &     0.42 &   2.63\\
$M_{500} > 9.7\times 10^{14}$ M$_\odot$            &    2.70 &   0.18 &      0.70 &     0.52 &   2.61\\
$f_{\rm gas}$ < 0.12                               &    2.69 &   0.15 &      0.75 &     0.48 &   2.49\\
$f_{\rm gas}$ > 0.12                               &    8.59 &   0.07 &      0.64 &     0.45 &   3.00\\
$f_{\rm bary}$ < 0.14                              &    4.50 &   0.09 &      0.62 &     0.41 &   2.68\\
$f_{\rm bary}$ > 0.14                              &    5.70 &   0.09 &      0.61 &     0.46 &   2.80\\
$M_{\star} < 1.4\times 10^{13}$ M$_\odot$          &    9.00 &   0.05 &      0.59 &     0.42 &   2.61\\
$M_{\star} > 1.4\times 10^{13}$ M$_\odot$          &    2.62 &   0.18 &      0.68 &     0.51 &   2.66\\
$\lambda_{\rm DM}$ < 0.03                          &    9.00 &   0.05 &      0.50 &     0.41 &   2.63\\
$\lambda_{\rm DM}$ > 0.03                          &    3.16 &   0.13 &      0.73 &     0.46 &   3.00\\
$\lambda_{\rm gas}$ < 0.05                         &    7.97 &   0.07 &      0.53 &     0.43 &   2.68\\
$\lambda_{\rm gas}$ > 0.05                         &    3.09 &   0.13 &      0.72 &     0.45 &   2.85\\
    \midrule
$\beta_{\rm 3D}$ < 0.15                            &    9.00 &   0.06 &      0.52 &     0.44 &   2.36\\
$\beta_{\rm 3D}$ > 0.15                            &    2.22 &   0.18 &      0.73 &     0.47 &   3.00\\
$f_{\rm sub}$ < 0.06                               &    9.00 &   0.06 &      0.52 &     0.45 &   2.39\\
$f_{\rm sub}$ > 0.06                               &    2.30 &   0.17 &      0.75 &     0.47 &   3.00\\
$\cos\, \theta_{04}$ < 0.56                        &    4.95 &   0.09 &      0.60 &     0.44 &   2.60\\
$\cos\, \theta_{04}$ > 0.56                        &    4.09 &   0.10 &      0.61 &     0.43 &   2.87\\
$\alpha_{\rm 3D}$ < 0.21                           &    4.80 &   0.10 &      0.59 &     0.46 &   2.45\\
$\alpha_{\rm 3D}$ > 0.21                           &    3.36 &   0.13 &      0.69 &     0.45 &   2.99\\
$\alpha_{\rm proj}\,(r= 0.2\,r_{500})$ < 0.05      &    9.00 &   0.06 &      0.54 &     0.46 &   2.34\\
$\alpha_{\rm proj}\,(r= 0.2\,r_{500})$ > 0.05      &    2.13 &   0.18 &      0.69 &     0.45 &   3.00\\
$\alpha_{\rm proj}\,(r=r_{500})$ < 0.14            &    9.00 &   0.05 &      0.56 &     0.42 &   2.53\\
$\alpha_{\rm proj}\,(r=r_{500})$ > 0.14            &    2.51 &   0.18 &      0.71 &     0.50 &   2.77\\
    \midrule
$z=0$ (75 clusters)                                &    9.00 &   0.05 &      0.61 &     0.41 &   2.42\\
$z=1$ (70 clusters)                                &    4.59 &   0.28 &      0.70 &     0.53 &   3.00\\
    \bottomrule
    \end{tabular}
\end{table}

\begingroup
\begin{sidewaystable}
    \renewcommand{\arraystretch}{0.9}
    \centering
    \caption{{Summary of the halo parameters and best fit parameters for the analytic rkSZ profiles. From left to right: selection criteria defining the cluster sub-sample (first column); the median value of the hot gas spin parameter, $\lambda_{\rm gas}$, and the circular velocity, $v_{\rm circ}$, at $r_{500}$ computed from the simulations for each sub-sample (\textit{subset medians}); the best fit parameters for the $\omega(r)$ profile from the \textit{gas}-edge-on projection, the \textit{galaxies}-edge-on projection and the \textit{DM}-edge-on projection. These results are obtained at $z=0$ unless stated otherwise.}}
    \label{tab:baldi_z0}
    \small
    \rowcolors{9}{orange!20}{white}
    \begin{tabular}{lccc|cccc|cccc|cccc}
    \toprule
    
    \multicolumn{1}{c}{}    &
    \multicolumn{2}{c}{\bf Subset medians}    & 
    \multicolumn{4}{c}{\bf Gas-aligned} & 
    \multicolumn{4}{c}{\bf Galaxies-aligned} & 
    \multicolumn{4}{c}{\bf DM-aligned}\\ 
    
    \cmidrule(rl){2-3} \cmidrule(rl){4-7} \cmidrule(rl){8-11} \cmidrule(rl){12-15} \rule{0pt}{1ex}    
    
     Selection criterion   & $\lambda_{\rm gas}$ & $v_{\rm circ}$ &  $v_{\rm t0}$ & $\xi_0$ & $r_0$ & $\eta$  & $v_{\rm t0}$ & $\xi_0$ &  $r_0$ & $\eta$ &  $v_{\rm t0}$ & $\xi_0$ & $r_0$ & $\eta$ \\ 
                           &  [-] & [km/s] & [$v_{\rm circ}$]   &  [-]  & [$r_{500}$]   & [-]  & [$v_{\rm circ}$] & [-]  & [$r_{500}$]   & [-]  & [$v_{\rm circ}$] & [-]  & [$r_{500}$]  & [-]  \\
    \midrule
    \rowcolor{gray!25}
All clusters                                  & 0.0505 &   1648 & 1.53 &    10.02 & 0.16 & 2.01 & 0.51 & 3.35 & 0.16 & 1.82 &  0.94 & 6.19 & 0.11 & 1.80 \\
$M_{500} < 9.7\times 10^{14}$ M$_\odot$       & 0.0542 &   1430 & 0.99 &    6.06  & 0.17 & 2.25 & 0.30 & 1.84 & 0.18 & 2.00 &  0.71 & 4.36 & 0.12 & 2.00 \\
$M_{500} > 9.7\times 10^{14}$ M$_\odot$       & 0.0462 &   1891 & 1.92 &    13.85 & 0.15 & 1.92 & 0.68 & 4.88 & 0.14 & 1.75 &  1.14 & 8.19 & 0.10 & 1.70 \\
$f_{\rm gas}$ < 0.12                          & 0.0532 &   1477 & 0.93 &    5.84  & 0.15 & 1.70 & 0.27 & 1.70 & 0.53 & 2.34 &  0.41 & 2.55 & 0.25 & 1.50 \\
$f_{\rm gas}$ > 0.12                          & 0.0462 &   1807 & 1.47 &    10.58 & 0.12 & 1.92 & 0.61 & 4.41 & 0.10 & 1.81 &  1.16 & 8.37 & 0.09 & 1.87 \\
$f_{\rm bary}$ < 0.14                         & 0.0547 &   1493 & 1.46 &    8.88  & 0.21 & 2.15 & 0.46 & 2.82 & 0.40 & 2.46 &  0.71 & 4.31 & 0.24 & 2.05 \\
$f_{\rm bary}$ > 0.14                         & 0.0458 &   1779 & 1.53 &    11.12 & 0.13 & 2.01 & 0.64 & 4.62 & 0.09 & 1.83 &  1.17 & 8.54 & 0.08 & 1.84 \\
$M_{\star} < 1.4\times 10^{13}$ M$_\odot$     & 0.0522 &   1430 & 0.55 &    3.51  & 0.09 & 1.69 & 0.16 & 1.03 & 0.07 & 1.48 &  0.43 & 2.76 & 0.05 & 1.60 \\
$M_{\star} > 1.4\times 10^{13}$ M$_\odot$     & 0.0474 &   1891 & 2.10 &    14.74 & 0.16 & 1.96 & 0.74 & 5.18 & 0.15 & 1.74 &  1.25 & 8.78 & 0.10 & 1.72 \\
$\lambda_{\rm DM}$ < 0.03                     & 0.0433 &   1649 & 0.69 &    5.31  & 0.11 & 1.61 & 0.16 & 1.24 & 0.05 & 1.55 &  0.36 & 2.79 & 0.05 & 1.56 \\
$\lambda_{\rm DM}$ > 0.03                     & 0.0619 &   1644 & 2.49 &    13.39 & 0.16 & 2.13 & 1.31 & 7.03 & 0.13 & 1.83 &  1.73 & 9.30 & 0.14 & 1.93 \\
$\lambda_{\rm gas}$ < 0.05                    & 0.0348 &   1702 & 0.70 &    6.67  & 0.23 & 2.23 & 0.18 & 1.68 & 0.29 & 2.32 &  0.39 & 3.76 & 0.05 & 1.58 \\
$\lambda_{\rm gas}$ > 0.05                    & 0.0725 &   1596 & 2.50 &    11.48 & 0.12 & 1.85 & 1.13 & 5.18 & 0.09 & 1.62 &  1.68 & 7.70 & 0.13 & 1.85 \\
\midrule
$\beta_{\rm 3D}$ < 0.15                       & 0.0417 &   1562 & 0.80 &    6.43  & 0.10 & 1.92 & 0.23 & 1.87 & 0.27 & 2.70 &  0.40 & 3.17 & 0.22 & 2.44 \\
$\beta_{\rm 3D}$ > 0.15                       & 0.0584 &   1761 & 2.66 &    15.18 & 0.21 & 2.20 & 0.91 & 5.21 & 0.08 & 1.53 &  1.78 & 10.16 & 0.08 & 1.68 \\
$f_{\rm sub}$ < 0.06                          & 0.0450 &   1638 & 0.91 &    6.70  & 0.12 & 2.08 & 0.28 & 2.05 & 0.17 & 2.27 &  0.59 & 4.36 & 0.18 & 2.55 \\
$f_{\rm sub}$ > 0.06                          & 0.0556 &   1703 & 2.60 &    15.56 & 0.24 & 2.41 & 1.03 & 6.16 & 0.18 & 1.94 &  1.47 & 8.79 & 0.08 & 1.63 \\
$\cos\, \theta_{04}$ < 0.56                   & 0.0439 &   1633 & 1.22 &    9.23  & 0.11 & 1.90 & 0.37 & 2.78 & 0.12 & 2.34 &  0.47 & 3.55 & 0.05 & 1.52 \\
$\cos\, \theta_{04}$ > 0.56                   & 0.0581 &   1681 & 2.04 &    11.70 & 0.18 & 2.09 & 1.62 & 9.27 & 0.18 & 2.08 &  1.63 & 9.35 & 0.14 & 1.97 \\
$\alpha_{\rm 3D}$ < 0.21                      & 0.0441 &   1559 & 0.99 &    6.06  & 0.17 & 2.25 & 0.30 & 1.85 & 0.18 & 2.01 &  0.71 & 4.36 & 0.11 & 2.00 \\
$\alpha_{\rm 3D}$ > 0.21                      & 0.0578 &   1759 & 1.92 &    13.85 & 0.15 & 1.92 & 0.68 & 4.88 & 0.14 & 1.75 &  1.14 & 8.19 & 0.10 & 1.70 \\
$\alpha_{\rm proj}\,(r= 0.2\,r_{500})$ < 0.05 & 0.0461 &   1536 & 0.93 &    5.84  & 0.15 & 1.70 & 0.27 & 1.69 & 0.53 & 2.34 &  0.41 & 2.55 & 0.25 & 1.51 \\
$\alpha_{\rm proj}\,(r= 0.2\,r_{500})$ > 0.05 & 0.0541 &   1778 & 1.47 &    10.58 & 0.12 & 1.92 & 0.61 & 4.41 & 0.10 & 1.82 &  1.16 & 8.36 & 0.09 & 1.87 \\
$\alpha_{\rm proj}\,(r=r_{500})$ < 0.14       & 0.0480 &   1484 & 1.46 &    8.88  & 0.21 & 2.15 & 0.46 & 2.82 & 0.40 & 2.46 &  0.71 & 4.31 & 0.24 & 2.10 \\
$\alpha_{\rm proj}\,(r=r_{500})$ > 0.14       & 0.0531 &   1855 & 1.53 &    11.12 & 0.13 & 2.00 & 0.64 & 4.62 & 0.09 & 1.83 &  1.17 & 8.54 & 0.08 & 1.84 \\
\midrule
$z=0$ (matched HMF, 75 clusters)              & 0.0560 &   1277 & 0.98 &    6.05  & 0.17 & 2.24 & 0.30 & 1.84 & 0.18 & 2.00 &  0.71 & 4.36 & 0.11 & 2.00 \\
$z=1$ (matched HMF, 70 clusters)              & 0.0430 &   1567 & 1.92 &    13.85 & 0.14 & 1.92 & 0.68 & 4.88 & 0.14 & 1.75 &  1.14 & 8.19 & 0.10 & 1.70 \\
    \bottomrule
    \end{tabular}
\end{sidewaystable}
\endgroup

%% file: Chapters/Chapter5.tex
\chapter{The entropy-core problem in EAGLE-like simulations models}
\label{chapter:5}


\section{Preface to the chapter}
\label{sec:ch5-preface}

Among the most fundamental properties of physical systems -- such as electric charge, mass and symmetries -- \textit{entropy} plays a special role. Conceptually coined by \cite{clausius_r_1865_1423700} and further probed by \cite{boltzmann1866mechanische} and \cite{gibbs1876equilibrium} through the lens of statistical physics, different formulations of entropy are nearly ubiquitous in today's sciences, spanning from error-correction strategies in quantum computing \citep[e.g.][]{grassl2022entropic} to stochastic modelling of economic \citep[see the reviews by][]{jakimowicz2020role, e15114909}.

Since entropy is associated with all irreversible processes in the Universe, most cosmological processes can be interpreted in terms of change in entropy as a function of other properties, e.g. temperature (as in the second law of thermodynamics), spatial scale and time. More generally, numerous studies \cite[e.g.][]{2004ApJ...616..643F, 2010ApJ...710.1825E} provided estimates for an \textit{entropy census} in the observable Universe,\footnote{Here, we refer to \textit{scheme 1} of \cite{2010ApJ...710.1825E}, which is computed for a comoving volume representing a causally-isolated particle horizon, equivalent to the size of the observable Universe.} finding that: 
\begin{itemize}
    \item the \textbf{total entropy} in the current Hubble horizon is $S_H\sim 10^{104}~k_{\rm B}$, where $k_{\rm B} = 1.380649\times 10^{-23}~{\rm J\, K^{-1}}$ is the Boltzmann constant;
    
    \item the largest entropy contribution in the Universe comes from super-massive \textbf{black holes} (SMBHs): $S_{\rm SMBHs} \sim (1 - 10^{-7})\, S_H \approx 99.99999\, \%~\textrm{of}~S_H$, followed by stellar-mass black holes, contributing $S_{\rm \star BHs} \sim 10^{97}~k_{\rm B}\approx 10^{-7}\, S_H$. Although black holes only account for a small \textit{mass} contribution in today's Universe (and in cosmological simulations), they are in fact remarkably entropy-dense. Moreover, their entropy is proportional to the square of the mass,\footnote{For a Schwarzschild black hole of mass $M^2_{\rm BH}$ described by the Bekenstein-Hawking theory \citep{PhysRevD.7.2333}, we have $S \propto M^2_{\rm BH}$.} resulting in galactic black holes holding much more entropy than the population of less massive stellar black holes.
    
    \item Next, \textbf{photons and relic neutrinos} add up to $S_{\gamma + \nu}\sim 10^{89}~k_{\rm B} \approx 10^{-15}\, S_H$, and finally
    
    \item the hot, ionised \textbf{intra-cluster medium} (ICM) and \textbf{inter-galactic medium} (IGM) account for $S_{\rm ICM + IGM}\sim 10^{89}~k_{\rm B} \approx 10^{-15}\, S_H$.
\end{itemize}

Although incredibly small compared to the total entropy in the observable Universe, the spatial distribution and evolution of $S_{\rm ICM + IGM}$ in virialised structures is crucial for the formation of galaxies, stars and all other baryonic processes included in sub-grid models.

This chapter and the next include hydrodynamic simulation studies of the
\begin{center}
    $\sim 0.0000000000001\, \%$ of the total entropy $S_H$
\end{center}
associated with the ICM and IGM, probing its connection with other processes, such as radiative cooling from metal, the quenching of star-formation, and, finally, feedback processes. 

Following up from \cite{2023MNRAS.520.3164A}, we then assess how the entropy distribution evolves with time in simulations of a group and a cluster (Chapter \ref{chapter:6}), with the aim to probe the \textit{entropy core problem} and describe the behaviour of the EAGLE model when forming cool-core (low-entropy) and non-cool-core (high-entropy) galaxy clusters. Chapter \ref{chapter:6} constitutes the basis for a publication (Altamura et al., in preparation), which will spin off from the doctoral research project.

\subsubsection*{A thermodynamic disclaimer}
Similarly to most flagship hydrodynamic simulations of large-scale structure formation, our set-up model the thermodynamic evolution of the gas treating it as an \textit{ideal gas}. This approximation is valid in galaxy and cluster scales because the gas density is low enough to make the atomic and molecular interactions negligible.

For a fluid with specific heat capacity at constant volume $c_V$, the thermodynamic entropy (also known as Clausius entropy) $S$ is given by
\begin{equation}
    S=c_V\, \log \left(k_{\rm B} T\, n^{1-\gamma}\right) = c_V\, \log \underbrace{\left(k_{\rm B} T\, n^{-2/3}\right)}_{\equiv K},
\end{equation}
where $\gamma = 5/3$ is the adiabatic index, also used in the SPH formulation, $T$ is the gas temperature, and $n$ the number density. In the expression above, $K=k_{\rm B} T\, n^{-2/3}$ is defined as the \textit{pseudo-entropy}, and can be mapped directly onto the (true) thermodynamic entropy via $c_V$.

In galaxy cluster astrophysics, $K$ is often adopted as the definition of thermodynamic entropy, although, in strict terms, it is classically defined as the pseudo-entropy \citep[e.g.][]{entropy_intro_bower_1997}.

\subsubsection*{Declaration of contributions}
The zoom-in simulation pipeline, including the initial conditions down to the final results, is attributed to the first author, who acknowledges technical support from the EAGLE-XL Collaboration and the DiRAC-COSMA administration team. Most of the text was produced by Edoardo Altamura, with contributions from Scott T. Kay in parts of the introduction (Section \ref{sec:entropy-profiles:intro}) and the discussion (Section \ref{sec:discussion}).

\newpage

\begin{spacing}{1.}

\topskip0pt
\vspace*{\fill}
\begin{center}

    {\Large{EAGLE-like simulation models do not solve the entropy core problem in groups and clusters of galaxies}}

    \vspace{1cm}
    
    \textbf{Edoardo Altamura}, Scott T Kay, Richard G Bower, Matthieu Schaller, Yannick M Bah{\'e}, Joop Schaye, Josh Borrow, Imogen Towler, Monthly Notices of the Royal Astronomical Society, Volume 520, Issue 2, April 2023, Pages 3164–3186, \href{https://doi.org/10.1093/mnras/stad342}{https://doi.org/10.1093/mnras/stad342}
    
    \vspace{1.3cm}
    
    \begin{minipage}{0.85\textwidth}
        \textsc{Abstract}
        \vspace{0.5cm}
        
        Recent high-resolution cosmological hydrodynamic simulations run with a variety of codes systematically predict large amounts of entropy in the intra-cluster medium at low redshift, leading to flat entropy profiles and a suppressed cool-core population. This prediction is at odds with X-ray observations of groups and clusters. We use a new implementation of the EAGLE galaxy formation model to investigate the sensitivity of the central entropy and the shape of the profiles to changes in the sub-grid model applied to a suite of zoom-in cosmological simulations of a group of mass $M_{500} = 8.8 \times 10^{12}~{\rm M}_\odot$  and a cluster of mass $2.9 \times 10^{14}~{\rm M}_\odot$. Using our reference model, calibrated to match the stellar mass function of field galaxies, we confirm that our simulated groups and clusters contain hot gas with too high entropy in their cores. Additional simulations run without artificial conduction, metal cooling or AGN feedback produce lower entropy levels but still fail to reproduce observed profiles. Conversely, the two objects run without supernova feedback show a significant entropy increase which can be attributed to excessive cooling and star formation. Varying the AGN heating temperature does not greatly affect the profile shape, but only the overall normalisation. Finally, we compared runs with four AGN heating schemes and obtained similar profiles, with the exception of bipolar AGN heating, which produces a higher and more uniform entropy distribution. Our study leaves open the question of whether the entropy core problem in simulations, and particularly the lack of power-law cool-core profiles, arise from incorrect physical assumptions, missing physical processes, or insufficient numerical resolution.

        \vspace{-5pt}
        \par\noindent\rule{\textwidth}{1pt}
        Published in \mnras~and referenced in this thesis as \citet{2023MNRAS.520.3164A}.     
        
    \end{minipage}
    
\end{center}
\vspace*{\fill}
\end{spacing}
\newpage

\section{Introduction}
\label{sec:entropy-profiles:intro}
Located at the centre of massive galaxies, Super-Massive Black Holes (SMBHs) are believed to be the main cause for the self-regulation of the baryon content and the thermodynamics of the local Inter-Galactic Medium \citep[IGM, e.g.][]{2013ARA&A..51..511K, 2014ARA&A..52..589H}. The SMBH can capture cold and dense material from its environment and can generate outflows of hot plasma into the IGM. These two processes, known as \textit{feeding} and \textit{feedback} respectively, can be detected in X-ray, optical and microwave/sub-mm bands and usually affect the IGM from scales of the order of 10 kpc to 1 Mpc \citep[e.g.][]{2012ARA&A..50..455F}. Active Galactic Nuclei (AGN) are excellent examples of high-redshift galaxies showing cooling flows, relativistic jets and shock-heated plasma, resulting from a complex interaction between SMBH feeding and feedback processes \citep[see e.g.][]{agn_review_eckert_2021}. These events have an impact on the thermodynamic state of the hot IGM in groups and clusters, altering quantities such as the gas density, temperature, metallicity, and also regulating the star formation (SF) and the growth of the SMBHs.

A useful quantity to probe the thermal state of the IGM is the thermodynamic entropy $K$, which is obtained by combining the temperature $T$ and electron number density $n_e$ of the ionised gas, as $ K=k_{\rm B}T/n_e^{2/3}$, where $k_{\rm B}$ is the Boltzmann constant \citep[see][for a derivation from classical thermodynamics with a monoatomic gas]{entropy_intro_bower_1997}. Entropy profiles, obtained from measured temperature and density profiles, are sensitive to the thermodynamic state of the group/cluster atmosphere, which is influenced by feedback, cooling flows, star formation and gravitational processes. They are, in other words, conceptually simple tools which can facilitate the study of complex and interdependent processes taking place in the IGM. Probing the distribution of entropy in groups and clusters can potentially shed new light on highly debated topics in galaxy formation, such as the role of the baryon cycle across cosmic time and the link between AGN activity and galaxy quenching \citep[see][ for a review and Kay \& Pratt, in preparation]{agn_review_eckert_2021}.

X-ray observations provide measurements of the temperature and density distribution of the hot gas. Using temperature and density profiles from ROSAT observations, \cite{1996ApJ...473..692D} showed that metallicity also has a significant impact on the entropy profiles. By applying similar analysis techniques, \cite{2009ApJS..182...12C} found that, on average, galaxy clusters have self-similar entropy profiles near the virial radius, while showing great variability in the inner regions. Using a sample of 31 nearby clusters (the REXCESS sample), \cite{entropy_profiles_pratt2010} showed that the central entropy $K_0$, the entropy excess above a power-law \citep{2005ApJ...630L..13D}, and the logarithmic slope of the profile at $0.075\, r_{500}$ follow bimodal distributions.\footnote{Following the spherical overdensity formalism, we define $r_{500}$ as the radius (from the gravitational potential minimum) at which the internal mean density exceeds the critical density by a factor of 500. Other properties such as the mass $M_{500}$ are computed from particles located within $r_{500}$ from the centre of the halo.} Clusters with a low value of $K_0\approx3~\mathrm{keV~cm^2}$ also show a steep slope, while objects with high $K_0\approx75~\mathrm{keV~cm^2}$ have shallower profiles in the core. Since the gas can reach low entropy by reducing its temperature and increasing its density, the clusters with power-law entropy profiles (and steep logarithmic slopes) are classified as cool-core (CC), while the ones with shallow profiles in the inner region and large deviation from a power-law are defined as non-cool-core (NCC).



Naturally, the gas found in cooling flows is denser and cooler than the IGM in quasi-hydrostatic equilibrium and must have low entropy. In the absence of non-gravitational processes, such as radiative cooling, star formation and feedback, we expect low-entropy gas to sink to the inner region and high-entropy gas to rise buoyantly and expand towards the outskirts of the systems. These conditions were recreated in the early \textit{non-radiative} hydrodynamics simulations of galaxy clusters and produced self-similar power-law entropy profiles, showing low entropy in the cluster cores and rising towards the virial radius \citep{2001ApJ...546...63T, vkb_2005}. {In particular, Lagrangian methods, such as smoothed-particle hydrodynamics (SPH), tended to produce power-law-like entropy profiles and small cores, while Eulerian (grid-based) codes produced solutions with larger entropy cores (see e.g. the \texttt{Bryan-SAMR} code in Fig. 18 of \citealt{1999ApJ...525..554F}, and Fig. 10 of  \citealt{2016MNRAS.459.2973S}).} A study by \cite{2009MNRAS.395..180M} attributed the lack of isentropic cores in non-radiative SPH simulations of clusters to the absence of gas mixing, which is a consequence of hydrodynamic instabilities \citep[see also][]{2007MNRAS.380..963A}. More recent formulations of SPH used in simulations include an artificial conduction term to reproduce the instabilities and capture shock-heating through artificial viscosity. The importance of thermal conduction and entropy mixing is confirmed in the nIFTy comparison project, where 12 codes were used to simulate a $10^{15}$ M$_\odot$ cluster, producing a core with constant entropy, high temperature and low density \citep{2016MNRAS.457.4063S}. 

In non-radiative simulations, entropy mixing can produce large isentropic cores by suppressing cooling flows, however, studies using radiative cooling and pre-heated gas found that cooling flows can also remove low-entropy gas from the centre of halos \citep[e.g.][]{2000MNRAS.317.1029P, 2001MNRAS.326.1228B, 2002MNRAS.336..527M, 2005MNRAS.361..233B}. They demonstrated that the densest gas in the centre of clusters can radiate away energy through free-free and line emission and attain very short cooling times. The formation of cooling flows can remove this {cold and} low-entropy material from the {hot,} X-ray emitting gas phase. {Higher-entropy gas is then allowed} to sink to the centre of the system, {establishing} a high entropy level. Pre-heating was introduced in these simulations to regulate the formation of strong cooling flows which could otherwise lead to unrealistically high stellar masses {and cold gas masses}. An important drawback of this method is the production of a large entropy excess in the cores of groups and clusters, which cannot be reconciled with the observations.

Motivated by observations, modern simulations of groups and clusters dropped the pre-heating approach and, instead, explicitly model SN-driven winds and AGN outflows (feedback) to regulate the cooling flows and produce realistic galaxy populations and cluster scaling relations \citep[e.g.][]{2010MNRAS.406..822M, 2014MNRAS.442.2304H, 2014MNRAS.445.1774P, 2015ApJ...813L..17R, 2017MNRAS.465.2936M, macsis_barnes_2017, ceagle.barnes.2017, 2018MNRAS.479.5385H, 2018MNRAS.481.1809B, 2020MNRAS.495.2930L, 2020MNRAS.498.3061R, 2021MNRAS.504.3922C}. As mentioned, the shape of the entropy profiles is an excellent metric for assessing the impact of stellar and AGN feedback and radiative cooling (free-free and metal lines) on the IGM. A number of these full-physics simulations show good agreement with X-ray observations. For instance, the BAHAMAS and MACSIS samples \citep{2017MNRAS.465.2936M, macsis_barnes_2017} produced entropy profiles at $z=0$ and $z=1$ in agreement with observations from \cite{entropy_profiles_pratt2010, 2015MNRAS.447.3044G} and \cite{2014ApJ...794...67M}. BAHAMAS and MACSIS used a purely kinetic SN feedback scheme, based on the prescription by \cite{2008MNRAS.387.1431D}, and a purely thermal AGN scheme following \cite{2009MNRAS.398...53B}. These two simulation schemes, with gas mass resolution $\sim 10^9$ M$_\odot$, consistently produce power-law entropy profile at low redshift. However, this behaviour drastically changes at higher resolution, with particle mass of $\sim 10^6$ M$_\odot$. The C-EAGLE simulations \citep{ceagle.barnes.2017, 2017MNRAS.470.4186B}, based on the EAGLE model of \cite{eagle.schaye.2015}, provide clear examples of the entropy-core problem. The clusters, comparable in mass to the REXCESS sample from \cite{entropy_profiles_pratt2010}, show median entropy profiles which are up to 4 times larger than REXCESS at $0.01\, r_{500}$. This discrepancy can also be identified in the temperature and density profiles, one being higher and the other lower than the observations (refer to the above definition of entropy). The ROMULUS-C simulation \citep{2019MNRAS.483.3336T} breaks this trend by producing power-law entropy profiles with small entropy cores at high resolution ($\sim 10^5$ M$_\odot$) and at low redshifts. They model {one single galaxy cluster} with metal cooling for low-temperature ($T<10^4$ K) gas {\citep[see Section 2.1 of][for further details]{2022MNRAS.515...22J}}. However, their run does not include radiative cooling from metals lines (i.e. gas with $T>10^4$ K), which is known to play a pivotal role in the shape of the thermodynamic profiles and the evolution of the IGM \citep[e.g.][]{2011MNRAS.412.1965M}. Despite potentially being able to form in any group or cluster, cooling flows can easily be disrupted by non-radiative (e.g. merging events and mixing) and non-gravitational processes (e.g. cooling, SN and AGN feedback), hence breaking the self-similarity of the entropy profiles \citep{vkb_2005, 2007MNRAS.376..497M, 2008MNRAS.391.1163P}.

The thermodynamic state of the hot IGM in the centres of groups and clusters is extremely sensitive to AGN feedback due to its proximity to the central SMBH. The energy output from feedback events can increase the temperature of the gas and decrease its density by generating outflows, with the overall effect of disrupting the CC state of the object {in most simulation set-ups. Instead, studies of the ROMULUS-C cluster system \citep{2021MNRAS.504.3922C} and the Rhapsody-G data \citep{2017MNRAS.470..166H} found that mergers can disrupt cool cores.} Simulations of isolated clusters, where the hot gas starts from a hydrostatic equilibrium state, show a widely variable entropy level in their inner region due to AGN activity \citep[e.g.][]{2019A&A...631A..60B, 2022MNRAS.tmp.1955N}. {Sub-grid prescriptions based on kinetic jets and cold accretion models have been shown to preserve the cool-core properties of idealised cluster simulations down to low redshift \citep{2015ApJ...811...73L, 2015ApJ...811..108P, 2016ApJ...829...90Y, 2023MNRAS.518.4622E}.} The simulations from \cite{2022MNRAS.tmp.1955N, 2022MNRAS.516.3750H} show how galaxy clusters in quasi-hydrostatic equilibrium can also preserve a cool core in the presence of anisotropic outflows and buoyantly rising bubbles produced by AGN feedback{, with the exception of the low-mass group ($M_{200} = 10^{13}$ M$_\odot$) in \citeauthor{2022MNRAS.tmp.1955N} (\citeyear{2022MNRAS.tmp.1955N}, Section 4.5 and Fig. 10), where the cool-core is disrupted and the central entropy rises above the 50 keV cm$^2$ level irreversibly. We note, however, that their models do not include cosmological accretion and merging events, which can destabilise pre-existing cool-cores \citep[e.g.][]{2021MNRAS.504.3922C}.} In turn, the bubbles can produce weak shocks which gently transport high-entropy gas outside the cluster core, as initially suggested by \cite{2001ApJ...554..261C} and \cite{2003MNRAS.344L..43F}. {Recent works modelling kinetic feedback jets with high resolution found that weak shocks can indeed heat the cluster atmosphere isotropically via sound-wave propagation \citep{2016ApJ...829...90Y, 2017ApJ...847..106L, 2019MNRAS.483.2465M, 2021MNRAS.506..488B, 2022MNRAS.516.3750H}.}

According to the estimates of \cite{entropy_profiles_pratt2010}, CC clusters account for about one third of the population, although it is worth highlighting that the bimodality in the central entropy and logarithmic slope could be a statistical fluctuation stemming from the limited sample size and complex selection criteria. Recent numerical simulations modelling hydrodynamics, AGN feedback and other sub-grid processes, on the other hand, produce very few to no CC clusters (C-EAGLE, \cite{ceagle.barnes.2017}; SIMBA, \cite{2019MNRAS.486.2827D}; TNG, \cite{2018MNRAS.473.4077P, 2018MNRAS.481.1809B}; FABLE, \cite{2018MNRAS.479.5385H}). The entropy in these simulated galaxy clusters does not decrease towards the centre, but it flattens to a constant level and gives rise to large entropy cores \citep[see also][for a review]{2021Univ....7..209O}. {The Rhapsody-G suite of cosmological zoom-simulations of clusters stand out as a sample which reproduce the observed CC-NCC dichotomy, although it should be noted their X-ray luminosity being too high \citep{2017MNRAS.470..166H}.} In this work, we investigate several processes which may give rise to the entropy cores in simulations, ultimately inhibiting the formation of CC clusters at low redshift. Our aim is to identify the causes of excess entropy in simulations to explain why galaxy formation models like EAGLE cannot reproduce the CC cluster population found by X-ray observations.

Since its initial formulation, the EAGLE model has played a pivotal role in interpreting observations and providing theoretical predictions. Yet, it fails to produce CC clusters at $z=0$ and power-law-like entropy profiles as measured by \cite{entropy_profiles_sun2009} and \cite{entropy_profiles_pratt2010}. The causes are unclear and may be related to the modelling of the hydrodynamics, the feedback implementation, the resolution or missing physical processes which are not included in the EAGLE model. In this work, we investigate the effects of thermal conduction, metal cooling, SN and AGN feedback to identify how they affect the entropy budget and inhibit the formation of cool cores in groups and clusters at EAGLE resolution and eight times lower mass resolution. We study the sensitivity of the group/cluster properties and the entropy profiles to changes in the model and we rule out possible causes for the lack of CC clusters in our simulation sample.

This work is organised as follows. In Section \ref{sec:simulation_methods} we introduce the sample of simulated objects, the initial conditions, as well as the simulation and analysis methods. There, we also describe the SWIFT-EAGLE model used in our investigation \citep[see also][]{bahe_2021_bh_repositioning, 2022MNRAS.tmp.1955N}, highlighting the key updates from the original EAGLE Ref simulations \citep{eagle.schaye.2015}; we then lay out the changes made to our reference model to explore different physical processes. Section \ref{sec:reference_model_results} highlights the entropy core problem by comparing the results from the simulated objects at $z=0$ against observed entropy profiles from groups and clusters of comparable mass. Section \ref{sec:results_model_variations} presents and discusses the main results for two objects in the simulated sample, run with different sub-grid models in order to investigate the effects on the system-wide properties and the 3D radial profiles. Finally, we discuss our results in Section \ref{sec:discussion} and present concluding remarks in Section \ref{sec:conclusion}.

Throughout this work, we assume the Planck 2018 cosmology, given by 
$\Omega_{\rm m}=0.3111$, 
$\Omega_{\rm b}=0.04897$, 
$\Omega_\Lambda=0.6889$, 
$h=0.6766$, 
$\sigma_8=0.8102$, 
$z_{\rm reion}=7.82$, 
$T_{\rm CMB}=2.7255$ K \citep{planck.2018.cosmology}.



\section{Simulations and numerical methods}
\label{sec:simulation_methods}
AGN feedback is known to affect low- and high-mass groups differently \citep[e.g.][]{2010MNRAS.406..822M, 2022MNRAS.tmp.1955N}. The energy output from the SMBH can lead to catastrophic outflows in the IGM of small groups, while large objects might just experience a re-distribution of the gas within their virial radius. In order to study the effects of AGN feedback in these regimes, we built an \textit{extended} sample of 27 objects ranging from group-sized to cluster-sized masses. In order to study the effect of sub-grid model changes in a more controlled and detailed setting, we further selected a \textit{reduced} sample of one low- and one high-mass object, referred to as \textit{group} and \textit{cluster} respectively in the following sections.

\subsection{Sample selection}
\label{sec:simulation_methods:sample_selection}

\begin{sidewaystable}
    \centering
    \caption{Mass resolution and softening lengths for different types of simulations. The mass of dark matter (DM) particles is expressed by $m_{\rm DM}$ and the initial mass of gas particles is $m_{\rm gas}$. $\epsilon_{\rm DM,c}$  and $\epsilon_{\rm gas,c}$ indicate the comoving Plummer-equivalent gravitational softening length for DM particles and the gas respectively, $\epsilon_{\rm DM,p}$ and $\epsilon_{\rm gas,p}$ indicate the physical maximum Plummer-equivalent gravitational softening length for DM particles and the gas respectively.}
    \begin{tabular}{llcccccc}
    \toprule
    Set-up      & Simulation type   & $m_{\rm DM}$           & $m_{\rm gas}$ & $\epsilon_{\rm DM,c}$ & $\epsilon_{\rm gas,c}$ & $\epsilon_{\rm DM,p}$ & $\epsilon_{\rm gas,p}$ \\ 
                &                   & (M$_\odot$)            & (M$_\odot$)   & (comoving kpc)        & (comoving kpc)         & (physical kpc)        & (physical kpc)         \\ \midrule
    Parent box  & DM-only           & $5.95 \times 10^{9}$   & --            & 26.6                  & --                     & 10.4                  & --                     \\
    Zoom (low-res)  & DM-only       & $9.32 \times 10^{7}$   & --            & 6.66                  & --                     & 3.80                  & --                     \\
    Zoom (high-res) & DM-only       & $1.17 \times 10^{7}$   & --            & 3.33                  & --                     & 1.90                  & --                     \\
    Zoom (low-res)  & DM+hydro  & $7.85 \times 10^{7}$   & $1.47 \times 10^{7}$ & 6.66           & 3.80                   & 2.96                  & 1.69                   \\
    Zoom (high-res) & DM+hydro  & $9.82 \times 10^{6}$   & $1.83 \times 10^{6}$ & 3.33           & 1.90                   & 1.48                  & 0.854                  \\
    \bottomrule
    \label{tab:resolutions}
    \end{tabular}
\end{sidewaystable}

The objects studied in this paper were selected from a $(300~\textrm{Mpc})^3$ volume (referred to as the \textit{parent} simulation), run using the \swift hydrodynamic code \citep{2016pasc.conf....2S, schaller_2018_swift} in gravity-only mode and $564^3$ DM particles with mass $m_{\rm DM} = 5.9 \times 10^{9}$ M$_\odot$ (see Table \ref{tab:resolutions}). The \velociraptor structure-finder code \citep{2019PASA...36...21E} was then used to identify DM halos in the snapshots and build a catalogue of  halos at $z=0$. We then selected the halos within the mass range $12.9 \leq \log_{10}(M_{500}/{\rm M}_\odot) \leq 14.5$ which formed in relatively isolated regions of the volume. The isolated objects are defined such that no field halos larger than 10\% of their mass can be found within a 10 $r_{500}$ radius from their centre of potential. We then split the $10^{12.9 - 14.5}~{\rm M}_\odot$ mass interval into three mass bins equally spaced in log-scale. From each bin, 9 isolated halos were randomly selected, giving an overall population of 27 objects, which we denote as the \textit{extended} halo catalogue. Spanning a wide range in masses, this sample is particularly well-suited for studying scaling relations and outlining the entropy-core problem by comparing simulated and observed entropy profile in groups and clusters (see section \ref{sec:reference_model_results:entropy_profiles}).

From the extended halo catalogue, we then focused on two objects, one of high mass and one of low mass, for studying the entropy distribution and other quantities with varying sub-grid models. The high-mass object, which we identify as \textit{cluster}, is the largest DM halo in the extended sample and has a $z=0$ mass of $M_{500} = 2.92 \times 10^{14}~{\rm M}_\odot$. The low-mass object, called \textit{group} in later sections, has $M_{500} = 8.83 \times 10^{12}~{\rm M}_\odot$ at $z=0$ and was randomly selected from the halos in the lowest mass bin. This particular subset of the extended catalogue forms the \textit{reduced} catalogue.

\begin{figure}
	\includegraphics[width=\columnwidth]{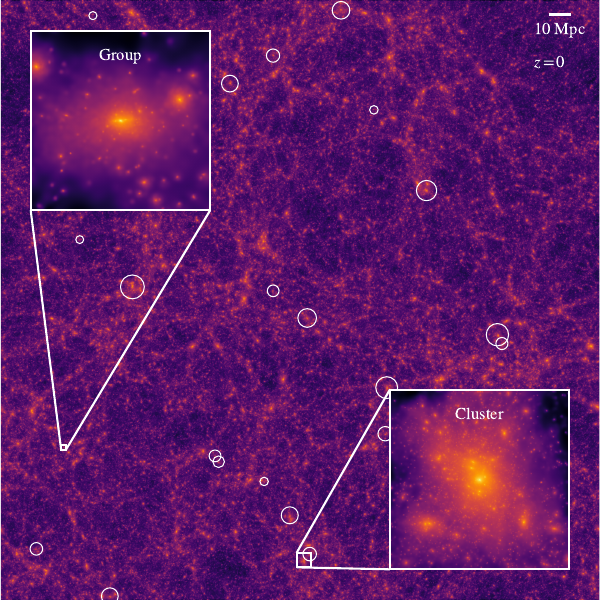}
    \caption{Projected DM mass map of the 300 Mpc volume at $z=0$, where the two insets show the objects in the reduced catalogue (group and cluster, as indicated) re-simulated in DM-only mode to $z=0$. The white circles mark the position of the remaining 25 objects in the extended catalogue; some are covered by the insets. The white circles represent the (projected) spherical 3D apertures, of radius $6\, r_{500}$, used to mask the high-resolution region in the zoom set-up. The size of the group and cluster maps is $12\, r_{500}$ along the diagonal. The spatial scale of the parent box is indicated in the top-right corner of the image.}
    \label{fig:dmo_box_with_halos}
\end{figure}

Fig. \ref{fig:dmo_box_with_halos} shows a projected DM mass image of the parent volume, with the location of the objects in the extended catalogue (white circles) and the re-simulated objects in the reduced catalogue (square insets, zooming onto the group and the cluster). 

\subsection{Zoom initial conditions set-up}
\label{sec:simulation_methods:initial_conditions}

The process of simulating the objects in the extended catalogue is based on the zoom simulation technique \citep{1993ApJ...412..455K, 1997MNRAS.286..865T} and was developed in three stages: producing the parent simulation,  constructing the initial conditions with variable particle resolution, and re\hyp{}simulating each halo at higher resolution, including hydrodynamics and sub\hyp{}grid physics.
The initial conditions for both the parent simulation and the zoom set\hyp{}ups were built from a set of DM particles arranged in a glass\hyp{}like structure. Each particle was then displaced and assigned a velocity vector determined from first (zooms) and second (parent) order Lagrangian perturbation theory to $z=127$. The displacement and velocity fields are computed using the method of \cite{2010MNRAS.403.1859J} implemented in \textsc{Panphasia} \citep{2013MNRAS.434.2094J}, an open\hyp{}source code which can generate realisations of primordial Gaussian random fields in a multi\hyp{}scale setting.\footnote{The phase descriptor containing the unique seed of the white noise field used for the parent simulation and the zoom simulations is [Panph1,L18,(74412,22732,260484),S3,CH1799108544,EAGLE\hyp{}XL\_L0300\_VOL1].}

The objects in the extended sample were then re-simulated individually using the zoom-in technique. For each selected halo in the parent box at $z=0$, a spherical high-resolution region was centred on the DM halo's potential minimum and extended to $6~r_{500}$. The initial conditions, generated with the \textsc{IC\_Gen} tool and \textsc{Panphasia} \citep{2013MNRAS.434.2094J}, were produced at two resolutions. The high resolution was chosen to match that of the EAGLE L100N1504 volume \citep{eagle.schaye.2015}. In the DM-only configuration at high resolution, we used particles with mass $m_{\rm DM}=1.17 \times 10^{7}~{\rm M}_\odot$, while full-physics simulations have $m_{\rm DM}=9.82 \times 10^{6}~{\rm M}_\odot$ and initial gas particle mass $m_{\rm gas}=1.83 \times 10^{6}~{\rm M}_\odot$. An additional set of simulations was produced at a mass resolution 8 times lower than EAGLE, corresponding to (initial) particle masses that are 8 times larger. Table \ref{tab:resolutions} summarises the masses of DM and gas particles at both resolutions, in the case of DM-only simulations and full-physics runs. We also report the comoving and physical Plummer\hyp{}equivalent gravitational softening length for DM and gas particles.

\subsection{The SWIFT-EAGLE subgrid model}
\label{sec:simulation_methods:ref_model}
In this study, we describe an updated version of the EAGLE sub-grid model of \cite{eagle.schaye.2015}, which we denote as SWIFT-EAGLE. This model, introduced in \cite{bahe_2021_bh_repositioning} and implemented in \swift, uses an equation of state based on a pressure floor to model the interstellar medium (\citealt{2008MNRAS.383.1210S}, using a polytrope index $\gamma=4/3$ as in Section 2.1 of \citealt{bahe_2021_bh_repositioning}), a {Kennicutt-Schmidt} star-formation law \citep{1959ApJ...129..243S, 1998ApJ...498..541K} and a metallicity-dependent star-formation number density threshold \citep{2004ApJ...609..667S, 2015MNRAS.450.1937C}. Stellar mass loss and chemical enrichment is implemented using the prescription of \cite{2009MNRAS.399..574W}. In this section, we outline the main changes that have been applied to the simulation technique since the original EAGLE simulations. Similarly to \cite{eagle.schaye.2015}, we define a fiducial sub-grid model, which we denote Ref, followed by additional ones, designed to investigate the effects of specific processes in the population of simulated clusters.

For this work, we use the SPHENIX Smoothed Particle Hydrodynamics (SPH) scheme described in \cite{sphenix_borrow2022} and configured with a quartic (M5) spline kernel. Unlike the SPH implementation ANARCHY, used in EAGLE, which uses the pressure\hyp{}entropy formalism \citep[see][for a description]{eagle.schaye.2015, 2015MNRAS.454.2277S}, SPHENIX uses the traditional density-energy method, which proved to give more accurate results for the particle-particle forces and energy conservation than pressure\hyp{}based schemes when coupled to astrophysical sub-grid models \citep{2021MNRAS.505.2316B}. In addition, SPHENIX makes use of an adaptive artificial viscosity (triggered in shocks using the Balsara switch, \citealt{1995JCoPh.121..357B} and \citealt{2010MNRAS.408..669C}) and artificial conduction, which is also present in ANARCHY, albeit with a different implementation. The artificial conduction is computed particle-wise and capped by the maximum artificial conduction coefficient $\alpha_{D, \rm max}\in [0, 1]$, which can be set between one, to fully enable conduction, and zero, to disable conduction completely \citep{2008JCoPh.22710040P}.

Our version of the EAGLE model includes the cooling tables presented by \cite{2020MNRAS.497.4857P}. This cooling model includes a modified version of the redshift\hyp{}dependent ($z\in [0, 9]$) UV/X-ray background from \cite{2020MNRAS.493.1614F}, self\hyp{}shielding for the cool interstellar medium (ISM) down to 10 K, dust, an interstellar radiation field and the effect of cosmic rays via the Kennicutt-Schmidt relation \citep{1998ApJ...498..541K}. When used with \swift, the radiative cooling for each particle is computed element-wise by interpolating the tables over redshift, temperature and hydrogen number density.

Starting from $z=19$, groups above a friends-of-friends (FoF) mass of $10^{10}$ M$_\odot$  are periodically identified and BHs with subgrid mass $M_{\rm sub} = 10^4$ M$_\odot$ are seeded within them, at the point where the local gas density is highest. Once seeded, the BHs can grow by either accreting gas or by merging with other BHs. The gas accretion rate can be estimated using the spherically symmetric Bondi-Hoyle-Lyttleton model \citep{1939PCPS...35..405H, 1944MNRAS.104..273B}, to which is applied an Eddington-limiter. Based on this prescription, the mass-accretion rate is given by
\begin{equation}
    \label{eq:bh_accretion}
    \dot{m}_{\rm BH} = \min \left[\alpha \cdot \frac{4 \pi G^2~m_{\rm BH}^2~\rho_{\rm gas}}{\left(c_{\rm s}^2 + v_{\rm gas}^2\right)^{3/2}},~\dot{m}_{\rm Edd} \right],
\end{equation}
where $m_{\rm BH}$ is the BH subgrid mass, $c_{\rm s}$ is the gas sound speed, $v_{\rm gas}$ the bulk velocity of the gas in the kernel relative to the BH, and $\rho_{\rm gas}$ the gas density evaluated at the BH position. $\dot{m}_{\rm Edd}$ is the \cite{1926ics..book.....E} rate, computed as
\begin{equation}
\label{eq:bh_eddington}
    \dot{m}_{\rm Edd} = \frac{4 \pi G m_{\rm P}}{\epsilon_{\rm r}~c~\sigma_{\rm T}}\cdot m_{\rm BH} \approx 2.218~{\rm M_\odot~yr^{-1}}\cdot\left( \frac{m_{\rm BH}}{10^8~{\rm M_\odot}} \right),
\end{equation}
for a radiative efficiency $\epsilon_{\rm r} = 0.1$ \citep{1973A&A....24..337S}, where $c$ is the speed of light, $m_{\rm P}$ the mass of the proton and $\sigma_{\rm T}$ the Thomson cross-section \citep[see Section 2.2.4 in][for further details]{bahe_2021_bh_repositioning}. In Eq. \ref{eq:bh_accretion}, the definition of the boost factor $\alpha$ follows the \cite{2009MNRAS.398...53B} model:
\begin{equation}
    \alpha = \max\left[ \left( \frac{n_{\rm H}}{n_{\rm H}^\star} \right)^\beta,~1 \right],
\end{equation}
where $n_{\rm H}$  is the gas number density and $n_{\rm H}^\star=0.1~\mathrm{cm}^{-3}$ is a reference density marking the threshold for the onset of the cold phase in the ISM \citep{2004ApJ...609..667S}. As in EAGLE \citep{eagle.schaye.2015, 2015MNRAS.450.1937C}, we set $\alpha=1$ and the free parameter $\beta=0$ in all models. Similarly to the AGN feedback in the EAGLE-like model in \cite{2022MNRAS.tmp.1955N}, we do not include the \cite{2015MNRAS.454.1038R} angular momentum limiter. We note that our runs have $\approx$18 and 146 times higher gas {particle} mass; $\approx$3 and 6 times larger gravitational softening {than the simulations in \cite{2022MNRAS.tmp.1955N}} (see Table \ref{tab:resolutions}).

Over one timestep of length $\Delta t$, the mass involved in BH accretion is $\Delta m=\dot{m}_{\rm BH}~\Delta t$. Of this amount, a fraction $\epsilon_{\rm r}$ is converted into energy, while a fraction $(1-\epsilon_{\rm r})$ is added to the subgrid mass of the BH. To conserve the mass-energy budget in the simulations, the SMBHs can accrete a mass $\Delta m$ by "nibbling" mass from their neighbours. In this scenario, the gas particles in the kernel of the SMBH lose a small amount mass, weighted by their density contribution, to the BH accretion process \citep[refer to Eq. 4 in][]{bahe_2021_bh_repositioning}. 

{Unlike the original EAGLE model, where AGN feedback is stochastic, the AGN feedback in our Ref model uses a deterministic approach with an energy reservoir \citep{2009MNRAS.398...53B}.} Thermal AGN feedback is implemented by raising the temperature of one gas particle by $\Delta T_{\rm AGN}$, which is a (free) parameter fixed to $10^{8.5}$ K for the reference model. For every timestep, the subgrid reservoir of each SMBH is incremented by an energy $\epsilon_{\rm f}\epsilon_{\rm r} \Delta m c^2$, where $\epsilon_{\rm f}$ is the gas coupling efficiency, expressing the fraction of mass-energy radiated via BH accretion which contributes to AGN feedback. All our simulations use $\epsilon_{\rm f}=0.1$. A feedback event occurs when the reservoir contains enough energy to heat the designated gas neighbour by a temperature $\Delta T_{\rm AGN}$ \citep{2009MNRAS.398...53B, bahe_2021_bh_repositioning}. In the SWIFT-EAGLE Ref model, the target gas particle involved in thermal AGN feedback is chosen to be the closest to the SMBH (i.e. minimum-distance scheme, see Section \ref{sec:simulation_methods:model_variations}), however, we also investigate the effect of different feedback distribution methods in Section \ref{sec:results_model_variations}. Our AGN feedback implementation is purely thermal in all our simulations. 

Throughout our simulations, SMBHs are repositioned at every timestep according to the \texttt{Default} method of \cite{bahe_2021_bh_repositioning}, which aims to compensate for the inconsistencies of (unresolved) dynamical friction of SMBHs by artificially displacing them towards the minimum of the local gravitational potential. The repositioning algorithm has the effect of improving mass-accretion profiles of the SMBHs through time, as well as a self-consistent energy output from AGN feedback, compared to runs without BH repositioning.

The chemical enrichment from stellar populations follows \cite{eagle.schaye.2015} and models the effect of stellar winds from AGB and massive stars, core-collapse supernovae (SNII) and type Ia supernovae (SNIa) based on stellar ages and metallicity \citep{2009MNRAS.399..574W}. In our model, massive stars of mass $M>8$ M$_\odot$ release $10^{51}$ erg at the end of their evolution. This energy is made available for thermal SN feedback and it is modulated using a function of density and metallicity \citep{2015MNRAS.450.1937C} with a SNII energy fraction in the interval $[0.5, 5]$, as in \cite{bahe_2021_bh_repositioning}. The SN feedback implementation is stochastic and purely thermal with a fixed SN heating temperature $\Delta T_{\rm SN}=10^{7.5}$ K \citep{2012MNRAS.426..140D}. For both the SN and AGN feedback, we choose the \textit{minimum-distance} scheme as the default particle-selection method (Section \ref{sec:simulation_methods:model_variations}).

\subsection{Variations on the reference model}
\label{sec:simulation_methods:model_variations}

\begin{figure}
    \centering
	\includegraphics[width=0.75\columnwidth]{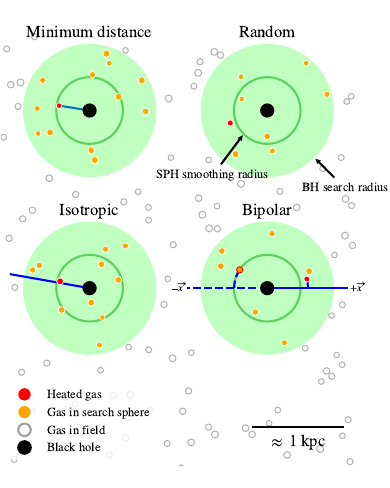}
    \caption{Schematic illustration of the rule used in the AGN feedback schemes for selecting the gas particle to heat (simplified to 2D). From the top-left to bottom-right, we illustrate the Minimum distance, Random, Isotropic and Bipolar schemes, as labelled in the figure. BHs are represented by black circles; their SPH smoothing radius is indicated by {empty} dark-green circles and the search radius by {filled} light-green circles{, around the smoothing length circle}. The gas particles are represented by empty grey circles if not found within the search radius of a BH, or filled orange circles otherwise. The gas particle selected by the heating mechanism is highlighted in red. For the Minimum distance method, we show the line connecting the BH with its nearest gas neighbour. For the Isotropic and Bipolar methods, we show the rays as solid blue lines cast from the BHs and the arc formed with the selected particle. For the Bipolar case, the orientation of the ray is randomly chosen between the $\pm \protect\overrightarrow{x}$ directions; for the positive case, we use solid lines for the ray and the arc, while we use dashed lines for the negative case. In the bottom right corner, we indicate the order of magnitude of the scale of the image.}
    \label{fig:heating_schemes}
\end{figure}

In addition to the minimum-distance method used in the Ref model summarised above, we implement three additional rules for selecting the neighbouring gas particle to be heated in SN and AGN feedback: Random, Isotropic and Bipolar. {To limit the number of scheme combinations used for SN and AGN feedback}, the energy-injection mechanism for SN feedback is the same as the one adopted for AGN feedback in all our models. {Any change in energy distribution scheme is applied to the SN and the AGN feedback in the same way.} We explain the particle selection rule using the diagram in Fig. \ref{fig:heating_schemes} \citep[see also][]{2022MNRAS.514..249C}. To produce this illustration, we generated a random distribution of gas particles (grey circles) and highlighted in orange those which are located within the search radius of the SMBH and are candidates in an AGN heating event. In all schemes, we define the SMBH \textit{search radius} to be approximately twice\footnote{For the quartic (M5) SPH kernel, we choose the search radius to be $2.018932\approx 2$ times the SPH smoothing length of the SMBH \citep[see][]{2010MNRAS.408..669C}.} its SPH smoothing length and we use it to evaluate the local gas properties. In Fig. \ref{fig:heating_schemes}, the search radius is represented by the dark-green circle around the BH. From the top-left to the bottom-right, we explain the particle selection methods as follows:
\begin{enumerate}
    \item \textit{Minimum distance}. The particle closest to the SMBH is selected and its temperature is raised by $\Delta T_{\rm AGN}$. We conventionally adopt this method in our reference model.
    
    \item \textit{Random}. One particle within the SMBH search radius is selected at random and its temperature is raised by $\Delta T_{\rm AGN}$. We note that this algorithm, also used in the original EAGLE model, results in mass-weighted sampling \cite{2022MNRAS.514..249C}.
    
    \item \textit{Isotropic}. A ray is cast from the BH in a random direction.\footnote{The probability density distribution of the polar angles $(\theta, \phi)$ of the ray (centred on the BH) are uniform in $\cos\theta\in \mathopen[-1, 1\mathclose[$ and $\phi\in \mathopen[0, 2\pi\mathclose]$. In the limit of a large number of neighbours within the BH search radius and a large number of AGN feedback events, the energy is then injected isotropically around the BH.} The gas particle that minimises the arc-length to this ray is selected and its temperature is raised by $\Delta T_{\rm AGN}$. To compute the arc-length between ray-particle pairs, we use the haversine formula reported in \cite{2022MNRAS.514..249C}. 
    
    \item \textit{Bipolar}. Same as Isotropic, but the ray is always cast along the $x$-axis of the simulation box, randomly flipping between the positive and negative directions. To decide on the sign of the $\protect\overrightarrow{x}$ direction, we draw a random float in the $\mathopen[0, 1\mathclose]$ interval and cast the ray in the $-\protect\overrightarrow{x}$ direction if the number is within $\mathopen[0, 0.5\mathclose]$ or choose $+\protect\overrightarrow{x}$ otherwise.
\end{enumerate}

 In this work, we also test the effects of the choice of $\Delta T_{\rm AGN}$ on the halo properties and the radial profiles. In addition to Ref, we introduce a model with stronger AGN feedback $\Delta T_{\rm AGN}=10^{9}$ K (AGNdT9) and one with weaker AGN feedback $\Delta T_{\rm AGN}=10^{8}$ K (AGNdT8). For models varying the heating scheme, namely Random, Isotropic and Bipolar, we adopt the fiducial $\Delta T_{\rm AGN}=10^{8.5}$ K. Crucially, different values of $\Delta T_{\rm AGN}$ do not alter the cumulative energy injected over the total simulation time. As a result, AGN feedback events are expected to occur more frequently in weaker-AGN models (e.g. AGNdT8), while stronger-AGN (e.g. AGNdT9) models lead to more "explosive" and sporadic gas-heating episodes. 

We note that the choice of a preferred AGN feedback scheme should not only be motivated by physical arguments. While the effective spatial resolution (Plummer-equivalent softening) of the simulations presented in this work is of order 1 kpc, the physical processes that govern the accretion of the SMBH in galaxies and, consequently, the exact dynamics of the relativistic jets and coupling with the local IGM occurs on sub-parsec scales, which are unresolved in our simulations. The aim of our research is to identify \textit{assumptions} on the energy-injection mechanism which can be implemented in a SMBH-based sub-grid model and lead to realistic halo properties and, crucially, are able to reproduce the observed entropy profiles.

In addition to the sub-grid models presented above, we also consider models without AGN or SN feedback, without artificial thermal conduction and without radiative cooling from metals. To switch off AGN feedback, we simply do not seed BHs at high redshift. SN feedback, instead, is switched off by imposing a zero heating temperature. Crucially, this method does not alter the metal enrichment of the gas, which is important for radiative cooling. In the SPH module, the artificial conduction is controlled by the $\alpha_{D, \mathrm{max}}$ parameter, which enters the equations of motion as a diffusive term \citep{sphenix_borrow2022}; in runs with no artificial conduction, we explicitly set $\alpha_{D, \mathrm{max}} = 0$, while we leave the default $\alpha_{D, \mathrm{max}} = 1$ for runs with artificial conduction.

In runs without metal cooling, we restrict the calculation of the cooling rates to hydrogen and helium and discard C, N, O, Ne, Mg, Si, S, Ca and Fe.

A summary of the runs used in this study is shown in Table \ref{tab:models}.

\subsection{Cluster properties}
\label{sec:analysis_methods}
The entropy is calculated by obtaining the number density of free electrons considering the chemical abundances from the chemical elements tracked by the sub-grid and hydrodynamics code. Assuming fully ionised gas, we define the total free-electron fraction for each hot ($T>10^5$ K) gas particle as
\begin{equation}
    X_{\rm e} \equiv \frac{n_{\rm e}}{n_{\rm H}} = \frac{m_{\rm H}}{f_{\rm H}}~\sum_\epsilon Z_\epsilon \frac{f_\epsilon}{m_\epsilon}
\end{equation}
and the ion fraction $X_i$
\begin{equation}
    X_{\rm i} \equiv \frac{n_{\rm i}}{n_{\rm H}} = \frac{m_{\rm H}}{f_{\rm H}}~\sum_\epsilon \frac{f_\epsilon}{m_\epsilon},
\end{equation}
which can be combined to compute the electron number density
\begin{equation}
    n_{\rm e} = \frac{X_{\rm e}}{X_{\rm e} + X_{\rm i}} \frac{\rho_g}{\mu~m_{\rm H}} = \rho_g ~ \sum_\epsilon Z_\epsilon \frac{f_\epsilon}{m_\epsilon},
\end{equation}
where $f_\epsilon$ is the gas particle mass fraction for chemical element $$\epsilon = \left\{{\rm H, He, C, N, O, Ne, Mg, Si, S, Ca, Fe}\right\},$$ $m_\epsilon$ the associated atomic mass and $Z_\epsilon$ the atomic number. $f_{\rm H}$ and $m_{\rm H}$ are the hydrogen mass fraction and atomic mass respectively. {Based on gas density-temperature phase-space plots for our groups and clusters, we found that the temperature cut at $T_{\rm cut}=10^5$ K is suitable for selecting the hot ionised gas and filtering out the colder, star-forming gas on the equation of state. The gas on the pressure floor and above the star formation density is neglected when calculating hot gas properties.} Finally, $\rho_g$ is the SPH density of the gas particle and $\mu$ its mean atomic weight, computed by summing the species contributions as follows:
\begin{equation}
    \mu = \left[ m_{\rm H} \sum_\epsilon \frac{f_\epsilon}{m_\epsilon} \cdot (Z_\epsilon + 1) \right]^{-1}.
\end{equation}
For our results, we present entropy, mass-weighted temperature and density profiles of the simulated groups and clusters. To compute the radial profiles, we consider 50 spherical shells centred on the centre of potential with a logarithmically increasing radius, spanning from $(0.01-2.5) \times r_{500}$. We then sum the contributions of the particles in each radial bin. 
For the $i^{\rm th}$ shell, we compute the density profile as
\begin{equation}
    \rho_{g,i} = \frac{\sum_j m_j}{V_i},
\end{equation}
where the sum is the total gas mass in shell $i$, divided by the volume $V_i$ of the shell. We also define the mass-weighted temperature $T_{{\rm MW}, i}$ as
\begin{equation}
    T_{{\rm MW}, i} = \frac{\sum_j m_j T_j}{\sum_j m_j}.
\end{equation}
The entropy profiles are then computed via the mass-weighted temperature profiles and the density profiles, as described in \cite{vikhlinin_2006_profile_slope}:
\begin{equation}
    K(r)=\frac{\mathrm{k_B}T(r)}{n_{\rm e}(r)^{2/3}}.
\end{equation}
The thermodynamic profiles are normalised to their self-similar values, appropriate for an atmosphere in hydrostatic equilibrium. We scale the density profiles to the critical density of the Universe
\begin{equation}
    \rho_{\rm crit}(z) = E^2(z) \frac{3 H_0^2}{8 \pi G},
\end{equation}
where $E^2(z)\equiv H^2(z) / H_0^2 =\Omega_{\rm m}(1+z)^3 + \Omega_\Lambda$. To select X-ray emitting gas, we only consider gas particles above a temperature of $10^5$ K. We normalise the temperature profiles to the characteristic temperature at $r_{500}$,
\begin{equation}
\label{eq:t500}
    k_\mathrm{B}T_{500}=\frac{G \bar{\mu} M_{500} m_{\rm H}}{2r_{500}},
\end{equation}
where $\bar{\mu} = 0.5954$ is the mean atomic weight for an ionized gas with primordial ($X = 0.76$, $Z = 0$) composition. Using the characteristic temperature, we also define the characteristic entropy:
\begin{equation}
    K_{500}=\frac{k_\mathrm{B}T_{500}}{\left(500 f_{\rm bary} \rho_{\rm crit}~/(\overline{\mu}_{\rm e} m_{\rm H})\right)^{2/3}},
\end{equation}
where $\overline{\mu}_{\rm e} = 1.14$ is the mean atomic weight per free electron and $f_{\rm bary}=0.157$ is the universal baryon fraction obtained by \cite{planck.2018.cosmology}.

\section{Reference model results}
\label{sec:reference_model_results}

As a first step, we introduce results for the extended catalogue at $z=0$, generated using the reference model only. We begin by illustrating the differences between the entropy profiles measured observationally and those from our simulations (Section \ref{sec:reference_model_results:entropy_profiles}), followed by a comparison between simulated and observed hot-gas fractions and star fractions (Section \ref{sec:reference_model_results:fractions}). The entropy profiles, gas and star fractions are derived from \textit{true} quantities, and do not attempt to include the hydrostatic mass bias.

\subsection{Entropy profiles}
\label{sec:reference_model_results:entropy_profiles}

\begin{figure}
	\includegraphics[width=\columnwidth]{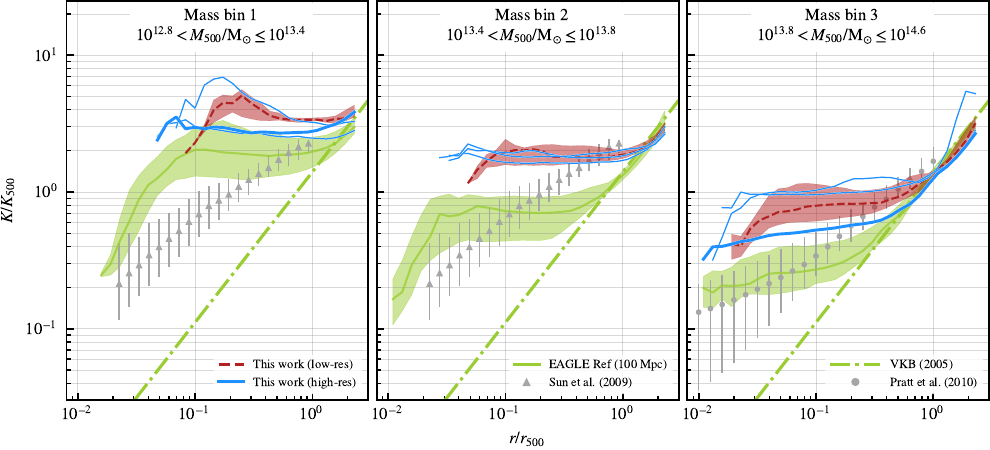}
    \caption{Comparison between the extended catalogue simulated with the Ref model (dark-red lines), the entropy profiles measured by \protect\cite{entropy_profiles_sun2009} and \protect\cite{entropy_profiles_pratt2010} using X-ray data (grey markers) and entropy profiles from the EAGLE 100 Mpc volume run with the Ref model \protect\citep[][green solid lines]{eagle.schaye.2015}. The objects in the extended sample are grouped by $M_{500}$ at $z=0$ in 3 bins, with mass ranges indicated by the label in each panel. For the objects in the extended sample, dashed red lines refer to the low-resolution simulations (combined to produce a median profile and the 16$^{\rm th}$ - 84$^{\rm th}$ percentile bands) and solid blue lines to high-resolution (plotted individually). The thicker blue lines highlight the high-res group and the high-res cluster (reduced sample) in the first and third mass bins respectively. The data from EAGLE and the observations are also reported with the percentile bands similarly to the extended sample. At small radii, the profiles are truncated where the number of particles in a 3D spherical shell falls below 50; this criterion is applied to the extended sample and the EAGLE objects. The entropy baseline profile from non-radiative simulations \protect\citep{vkb_2005} is displayed in all panels as a dot-dashed green line.}
    \label{fig:median_entropy_profiles_extended_sample}
\end{figure}

In Fig. \ref{fig:median_entropy_profiles_extended_sample}, we compare the entropy profiles of the simulated objects at $z=0$ with observational results from \cite{entropy_profiles_sun2009} and \cite{entropy_profiles_pratt2010} over a wide range in $M_{500}$, spanning between $8.83 \times 10^{12}$ M$_\odot$ and $2.92 \times 10^{14}$ M$_\odot$. We gathered the groups and clusters in the extended sample in three mass bins, containing 7, 9 and 11 objects run with low resolution and 3 objects each run with high resolution. For the extended sample at low resolution, we computed the median entropy profile and the associated 16$^{\rm th}$ and 84$^{\rm th}$ percentile bands in each mass bin (dark red dashed lines). The entropy profiles for the high-res simulations are shown individually in blue; the group and the cluster in the reduced sample are displayed as thicker lines in the first and third mass bin respectively. In the first two bins, we show the median entropy profile from the \cite{entropy_profiles_sun2009} sample (spectroscopic\footnote{Unlike the true quantities, computed from the unprocessed simulation data, \textit{spectroscopic} quantities are estimated assuming hydrostatic equilibrium and modelling X-ray emission. Spectroscopic masses estimated from simulations are often compared to masses measured from X-ray observations. We refer to \cite{2021MNRAS.506.2533B} and references therein for further details.} $M_{500} = 1.37\times10^{13}-1.37\times10^{14}$ M$_\odot$, $0.012 <z<0.12$); the error bars span between the 10$^{\rm th}$ and 90$^{\rm th}$ percentile level. For the third mass bin, we indicate the median entropy profiles from \cite{entropy_profiles_pratt2010}, computed using only the objects with estimated $M_{500}$ in the mass range of the simulations; the error bars span between the 16$^{\rm th}$ and 84$^{\rm th}$ percentile. {The \cite{entropy_profiles_pratt2010} data set shown throughout this work includes all REXCESS clusters in the mass range, regardless of their CC/NCC classification or morphology.} The left panel of Fig. \ref{fig:median_entropy_profiles_extended_sample} focuses on group-sized objects, comparable in mass with those in the sample of \cite{entropy_profiles_sun2009}. The right panel illustrates the simulated profiles of the clusters, together with the median profiles of the objects included in the study by \cite{entropy_profiles_pratt2010}. The median profile from the objects in the intermediate mass is only compared to the results from \cite{entropy_profiles_sun2009}, since their sample included more objects within this mass range than \cite{entropy_profiles_pratt2010}. We also indicate the entropy baseline predicted by \citeauthor{vkb_2005} (\citeyear{vkb_2005}, VKB) for non-radiative simulations. We also compare our entropy profiles of the extended sample with those from the EAGLE 100 Mpc simulation run with the Ref model \citep{eagle.schaye.2015, 2015MNRAS.450.1937C, 2016A&C....15...72M}. For each panel in Fig. \ref{fig:median_entropy_profiles_extended_sample}, the green lines indicate the median entropy profiles from the groups and clusters in the EAGLE volume, binned by $M_{500}$ similarly to the extended sample. The entropy profiles from the EAGLE volume were produced from the $z=0$ snapshot; the EAGLE data and those from the extended sample were reduced using the same analysis pipeline. In both data sets, we truncate the inner region of the entropy profiles at the radius where the number of particles in the shell falls below 50.

The radial profiles for the extended sample show excess entropy compared to observations across the three mass bins. The discrepancy is smallest around $r_{500}$ and increases towards the inner region of the groups and clusters. Besides an entropy excess in the core, the objects in the first and second mass bins show a flat entropy distribution, in contrast with the power-law-like profiles observed by \cite{entropy_profiles_sun2009}. For the group, we also highlight a local peak in the entropy at $\approx 0.2~r_{500}$ in the first mass bin, which is likely produced by feedback processes coupling to the diffuse gas within these groups of galaxies. The low-mass clusters in the third bin also show excess entropy in the core, compared to the \cite{entropy_profiles_pratt2010} sample, e.g. at $0.1~r_{500}$, the low-res simulations predict a dimensionless entropy of 0.8, while the observed median is 0.35. Similarly to the lower-mass bins, the shape of the simulated cluster profiles does not match the observations. Interestingly, the radial profiles from the groups and clusters in the EAGLE 100 Mpc volume have a lower entropy level than the extended sample, despite using the same feedback heating temperature. {The difference in central entropy may be due to the use of different hydrodynamic solvers (SPHENIX in our models and ANARCHY in EAGLE Ref), which can slightly alter the properties of the IGM \citep{2015MNRAS.454.2277S}, such as the entropy}. However, similarly to the extended sample, the EAGLE median profiles also show a flat-entropy core, which disagrees with the observational data. These results from Fig. \ref{fig:median_entropy_profiles_extended_sample} illustrate the effects of the entropy-core problem in simulations of groups and clusters of galaxies, highlighting the excess in entropy in the central IGM and an unusually flat entropy distribution across radii \citep[see also][for a review]{2021Univ....7..209O}. {We also found great similarity between the entropy profiles at $z=0$ and at $z=1$, suggesting that the entropy plateau was likely established early in the evolution of these groups and clusters.}

\subsection{Hot gas and star fractions}
\label{sec:reference_model_results:fractions}

\begin{figure}
	\includegraphics[width=\columnwidth]{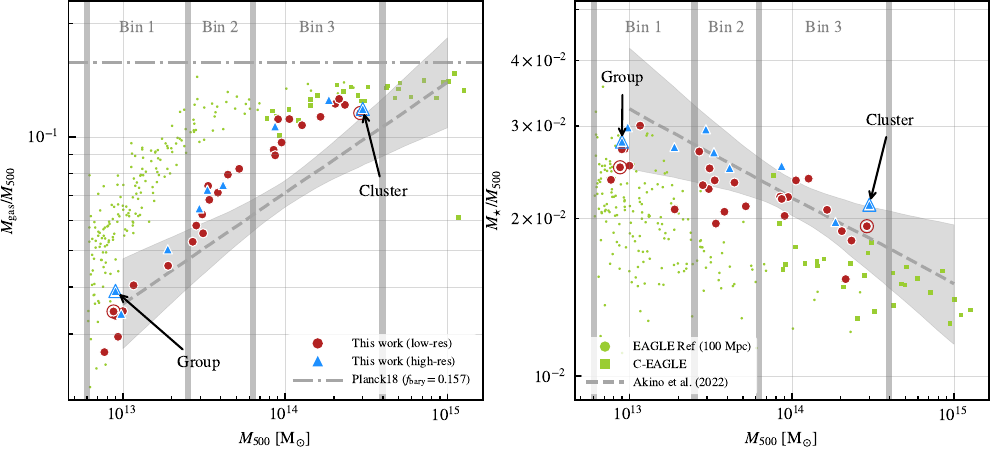}
    \caption{Scaling relations showing the gas fraction $M_{\rm gas}/M_{500}$ (left) and stellar mass fraction $M_{\star}/M_{500}$ (right) as a function of $M_{500}$ at $z=0$. In both panels, the dark-red circles and blue triangles represent the (extended) sample of objects presented in this work at low and high-resolution respectively. Double-edged markers highlight the two objects in the reduced sample (see annotations). In the left panel, the grey horizontal dash-dotted line indicates the universal baryon fraction $f_{\rm bary}=\Omega_{\rm b}/\Omega_{\rm m}=0.157$ derived from \protect\cite{planck.2018.cosmology}. The green markers are the objects in the EAGLE Ref 100 Mpc volume (dots) from \protect\cite{eagle.schaye.2015}, and the C-EAGLE objects (squares) from \protect\cite{ceagle.barnes.2017, 2017MNRAS.470.4186B}. In both panels, the dashed grey lines are the fitted scaling relations obtained by \protect\cite{xxl.baryons.akino2022} for the gas and stellar mass fractions, with the 16$^{\rm th}$ and 84$^{\rm th}$ percentile levels from the errors in the fit parameters, shown by the grey area. We emphasise that the grey area does not represent the intrinsic scatter, which instead extends for about 1 dex along the $x$-axis and 0.2 dex along the $y$-axis. In the right panel, the observations include the group and cluster sample from \protect\cite{xxl.baryons.akino2022}. The three mass bins specified in Fig. \ref{fig:median_entropy_profiles_extended_sample} are also indicated.}
    \label{fig:scaling_relations_extended_sample}
\end{figure}

For the extended sample, we also compute the hot gas fraction $f_{\rm gas} = M_{\rm gas}/M_{500}$, where $M_{\rm gas}$ defines the mass of the gas with temperature $T > 10^5$ K within $r_{500}$. Similarly, we define the star fraction as $f_\star = m_\star/M_{500}$, where $m_\star$ is the stellar mass within $r_{500}$. These results are used to construct the $f_{\rm gas}$-$M_{500}$ and $f_\star$-$M_{500}$ scaling relations shown in Fig. \ref{fig:scaling_relations_extended_sample}. 

The $f_{\rm gas}$-$M_{500}$ relation for the extended sample (dark-red circles for low-resolution and blue triangles for high-resolution) is compared to the results from the HSC-XXL weak lensing survey \cite[][136 objects with $0\leq z \leq 1$]{xxl.baryons.akino2022}. The weak lensing method of the HSC-XXL survey introduces a lower mass bias relative to X-ray measurements, motivating the use of true masses for our analysis \citep[e.g.][]{2012MNRAS.421.1073B}. Our sample of simulated objects shows good convergence at the two resolutions. When compared to the scaling relation from weak lensing, our sample shows larger $f_{\rm gas}$ for $M_{500}\sim 10^{14}$ M$_\odot$ clusters (we note, however, that the scaling relation from \cite{xxl.baryons.akino2022} was obtained by averaging a {binned} sample of 136 objects{; without binning, the individual data points from their sample} would otherwise show a much larger scatter at the $10^{14}$ M$_\odot$ level). From group-sized objects towards large clusters, the hot gas fraction approaches the universal baryon fraction measured by \cite{planck.2018.cosmology}, $f_{\rm bary}=\Omega_{\rm m}/\Omega_{\rm b}=0.157$. This is a well-known consequence of feedback processes being more efficient at expelling hot gas from low-mass groups than in large clusters, which can retain most of the baryons throughout their formation history \citep{2010MNRAS.406..822M}. 

In this comparison, we also include the groups and low-mass clusters from the EAGLE Ref 100 Mpc simulation \cite{eagle.schaye.2015} and the C-EAGLE clusters \citep{ceagle.barnes.2017}, both displayed as green markers. For groups ($10^{13}<M_{500}/{\rm M_\odot}<10^{14}$) the EAGLE Ref model produces systematically higher gas fractions than our reference model; this discrepancy is largest in the second mass bin, where our runs predict $f_{\rm gas}\approx0.07$, while EAGLE produces values around 0.12. The cause of this effect is unclear, as multiple changes have been applied to the simulation code (gravity and hydrodynamic solver) and the subgrid model \citep[see Appendix B of][for a comparison between the star-formation rate history in EAGLE and SWIFT-EAGLE]{bahe_2021_bh_repositioning}. Despite the gas fractions still being systematically higher than the observations, the $M_{500}$-$f_{\rm gas}$ relation produced with our objects is closer to the \cite{xxl.baryons.akino2022} data than EAGLE for groups. In the third mass bin, our results agree with those from the EAGLE and C-EAGLE samples, with the former showing a slightly lower $f_{\rm gas}$ than the latter. While the reference model used in this work uses the same AGN heating temperature as the EAGLE Ref model, the C-EAGLE sample was run with a higher value (equivalent to the AGNdT9 model described in \citealt{eagle.schaye.2015}) than EAGLE Ref, introduced to expel more hot gas from the clusters.
The high-resolution simulations, indicated with blue triangles in the $M_{500}$-$f_{\rm gas}$ relation, show good numerical convergence with the low-res objects. The group in the reduced sample aligns with the parametrised relation from the HSC-XXL survey, while the cluster is slightly more gas rich than the observed median relation ($f_{\rm gas}=0.1$ at $M_{500}=2.9\times10^{14}~{\rm M_\odot}$).

The star fractions are also compared to the observational results by \cite{xxl.baryons.akino2022}. In this case, the extended sample is in excellent agreement with the parametrised scaling relation from the HSC-XXL survey. The high-resolution simulations systematically yield a slightly higher stellar mass than their low-resolution version. 
%
In summary, the original EAGLE model produces a poor representation of gas and stellar properties, while showing a closer agreement between the entropy profiles and the observations. Our version of the same model yields instead more realistic gas and stellar properties, but much larger entropy cores. We stress that our SWIFT-EAGLE model has not been fully calibrated. {The stellar mass function, galaxy sizes and black hole-stellar mass relations were reasonably matched to observations, however this agreement was not tested for a simulated cosmological volumes as large as our parent simulation (300 Mpc).} Moreover, this model does not reproduce the original EAGLE galaxy properties presented in \cite{eagle.schaye.2015} and \cite{2015MNRAS.450.1937C}. We refer to \cite{2022JOSS....7.4240K} for an illustration of the sub-grid calibration methodology and to Borrow and EAGLE-XL Collaboration, in preparation, for a detailed discussion of the galaxy properties with the SWIFT-EAGLE model.

\begin{sidewaystable}
\centering
\caption{List of the models used for simulating the objects in the reduced catalogue. In this work, each model is identified with a label summarising its key features. These include the distribution scheme used in the AGN heating mode, the temperature increase $\Delta T_{\rm AGN}$ of the gas particles when heated by an AGN outburst, whether the SNe or the AGNs are enabled in \swift, whether the metals are included in the radiative cooling calculation, and the $\alpha_{D,\mathrm{max}}$ parameter, which is set to 1 if the SPH scheme uses artificial thermal conduction and to 0 otherwise.}
\begin{tabular}{llccccc}
\toprule
Model label            & AGN distribution mode & $\log_{10} (\Delta T_{\rm AGN}/\mathrm{K})$ & SN active & AGN active & Metal cooling enabled & $\alpha_{D,\mathrm{max}}$ \\ \midrule
Ref                      & Minimum distance & 8.5     & Yes         & Yes          & Yes                     & 1          \\
No-conduction            & Minimum distance & 8.5     & Yes         & Yes          & Yes                     & 0          \\
No-metals                & Minimum distance & 8.5     & Yes         & Yes          & No                      & 1          \\
No-SN                    & Minimum distance & 8.5     & No          & Yes          & Yes                     & 1          \\
No-AGN                   & Minimum distance & 8.5     & Yes         & No           & Yes                     & 1          \\
AGNdT8                   & Minimum distance & 8.0     & Yes         & Yes          & Yes                     & 1          \\
AGNdT9                   & Minimum distance & 9.0     & Yes         & Yes          & Yes                     & 1          \\
Random                   & Random           & 8.5     & Yes         & Yes          & Yes                     & 1          \\
Isotropic                & Isotropic        & 8.5     & Yes         & Yes          & Yes                     & 1          \\
Bipolar                  & Bipolar          & 8.5     & Yes         & Yes          & Yes                     & 1          \\
\bottomrule
\label{tab:models}
\end{tabular}
\end{sidewaystable}

\section{Sub-grid model variations}
\label{sec:results_model_variations}

\begin{figure}
	\includegraphics[width=\columnwidth]{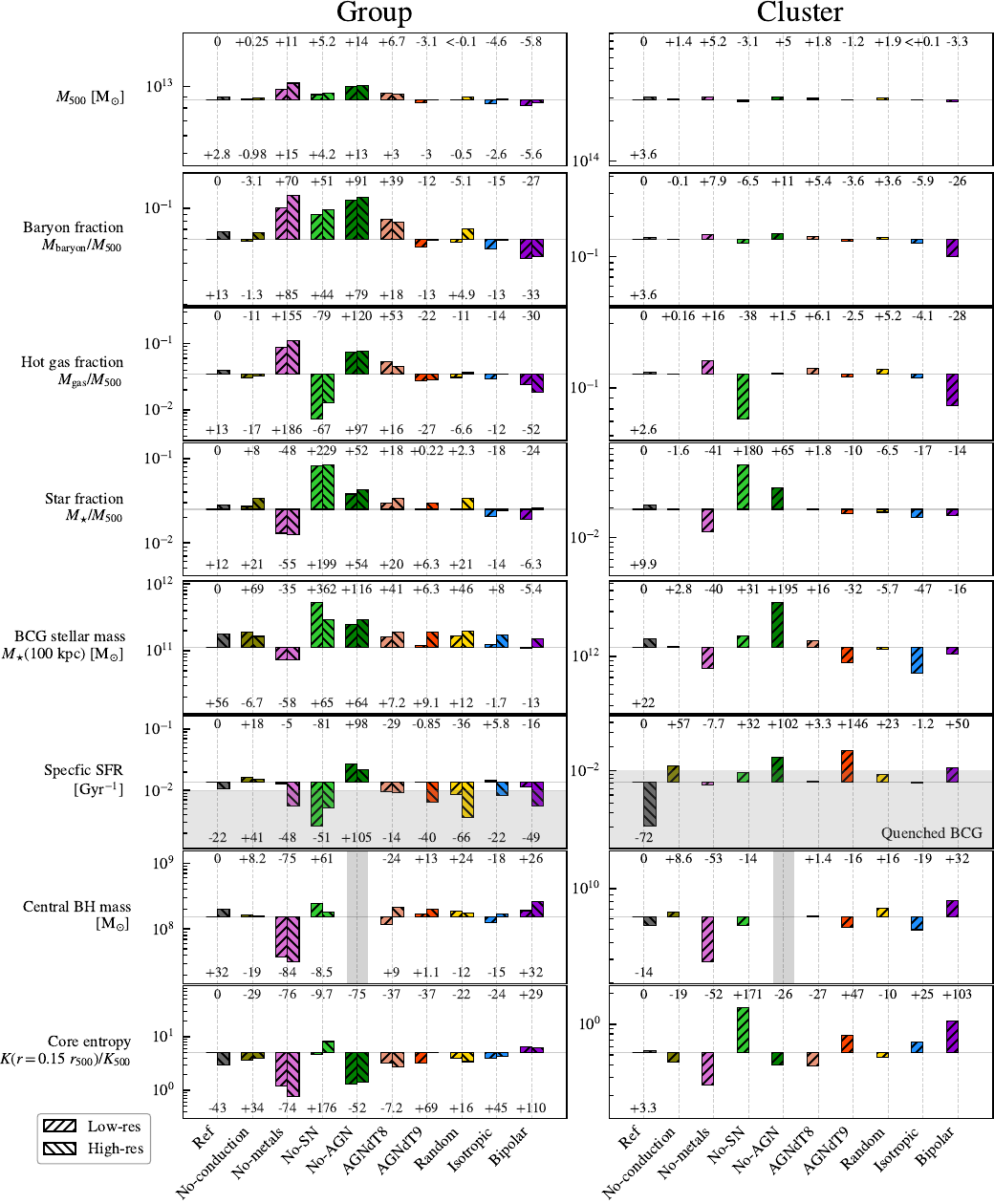}
    \caption{Summary of the $z=0$ properties of the group (left panels) and the cluster (right panels) in the reduced catalogue. For each panel, the quantity of interest is represented on the vertical axis, while the horizontal axis spans the models listed in Table \ref{tab:models}. For both objects, the results at low resolution (1/8$^{\rm th}$ EAGLE) and high resolution (EAGLE) are represented by hatched bars with forward and backward slashes respectively. The layout highlights the relative differences compared to the Ref model at low resolution, with the relative change, in per cent, for the low and high resolution annotated at the top and the bottom of each panel respectively. For the No-AGN model, the central BH mass is not defined, as indicated by the excluded grey area.}
    \label{fig:z0properties}
\end{figure}

Given these results, we now investigate the sensitivity of the hot gas distribution to changes in the sub-grid model. The sub-grid parameters which are changed between different models are summarised in Table \ref{tab:models}. In Fig. \ref{fig:z0properties}, we show the variation of $M_{500}$, $f_{\rm gas}$ and $f_\star$, the stellar mass of the BCG within a 100 kpc spherical aperture $M_\star(\rm 100 kpc)$, the specific SFR\footnote{We average the SFR over 1 Gyr using the birth scale factor of stars in a 100 kpc spherical aperture.} of the BCG within 100 kpc and the mass of the central BH, $M_{\rm BH}$, all computed at $z=0$ for the reduced sample, i.e. 1 group and 1 cluster. In addition, we also report the entropy in the core, obtained from interpolating the radial profiles and normalised to $K_{500}$. The sub-grid models are distributed along the horizontal axis and listed at the bottom of the figure. Each row of plots focuses on one of the quantities mentioned above; the panels on the left show the properties of the group, whilst those on the right refer to the cluster. In order to highlight the relative differences in the quantities with changing sub-grid model and mass resolution, we fixed the low-resolution Ref model at the centre of the vertical plot range, where we show a grey solid line to guide the eye. The variation in the quantities for each model and resolution can be seen by inspecting the height of the bars from the horizontal grey line. The bars are grouped in pairs, with the one on the left-hand side representing the low resolution run and the one on the right-hand side the high-resolution run. At the top (bottom) of each panel, we report the percentile change of the quantities for the low (high) resolution simulations, relative to the low (high)-res Ref runs. The No-AGN simulations do not seed BHs in the centres of galaxies, and therefore the central BH mass is not defined for this model, as represented by the crossed out region in the bottom two panels.

We complement our discussion by presenting the 3-dimensional mass-weighted radial profiles of the density, mass-weighted temperature and entropy profiles for the hot ICM. These are included in Figs. \ref{fig:profiles_noconduction} - \ref{fig:profiles_schemes}. For each figure, the top three panels show the entropy, temperature and density profiles (from left to right) of the group, while the bottom panels show those for the cluster, in the same order. The sub-grid variations use the same colour scheme as in Fig. \ref{fig:z0properties}. The profiles for low- and high-resolution runs are illustrated using solid and dashed lines respectively. In all panels, the horizontal axis indicates the radius scaled by $r_{500}$ and the vertical axes show the dimensionless entropy $K/K_{500}$, the scaled temperature $T/T_{500}$, and the normalised density $\rho/\rho_{\rm crit} (r/r_{500})^2$. In all profiles, we use the same axis ranges in order to facilitate a visual comparison between the different figures.

In the rest of this section, we describe the overall variation in the halo mass across different models and resolutions, before detailing the results from models with varying AGN heating temperature (Ref, AGNdT8 and AGNdT9), AGN energy-distribution scheme (Ref, Random and Isotropic), the effects of directional heating (Isotropic vs Bipolar), artificial conduction (Ref vs No-conduction), the contribution from individual feedback processes (Ref, No-AGN and No-SN) and the effect of metal cooling (Ref vs No-metals).

For most models, we do not find significant differences in $M_{500}$, with the exception of the group run with AGNdT8, No-AGN, No-SN and No-metals due to their larger baryon fractions. The variation in sub-grid model is, in fact, expected to have minimal influence on the halo mass at $z=0$ (although effects are still significant for precision cluster cosmology, e.g. \citealt{2021MNRAS.505..593D}), since the gravitational potential is dominated by dark matter, particularly at $r_{500}$. This claim finds confirmation in Fig. \ref{fig:z0properties} for both the group and the cluster, with the latter exhibiting even smaller differences in $M_{500}$ due to its deeper potential well. We note that the same effect also applies to the quantities that are directly linked to hydrodynamics, such as the hot gas mass and hot gas fraction. Most models produce objects with well-converged $M_{500}$ at both resolutions. We find this not to be true for the No-metals model, which generated a 4\% discrepancy in the group's mass at different resolutions. We discuss the effect of changes in the models on the other properties in the sections below.

\subsection{Effect of artificial conduction}
\label{sec:results_model_variations:conduction}

Switching off the artificial conduction has small effects on the hot gas fractions. The change in star fraction compared to Ref is $\sim$1\% for the cluster, however, we report a 8\% increase in the group at low resolution, and a 21\% increase at high resolution. Overall, the baryon fractions of both objects run with Ref and No-conduction are very similar.

When switching off artificial conduction, we report larger differences in the BCG stellar mass, which is 69\% higher in the group at low resolution. The same object run without conduction at high resolution, however, forms a BCG 7\% less massive than with Ref. The mass of the central BH is not strongly impacted by the change in $\alpha_{D, {\rm max}}$ in either of the objects in the reduced sample. The same is true for the sSFR in the group's BCG, while the cluster's BCG shows a 18\% increase in sSFR, which makes it non-quenched at $z=0$.

\begin{figure}
	\includegraphics[width=\columnwidth]{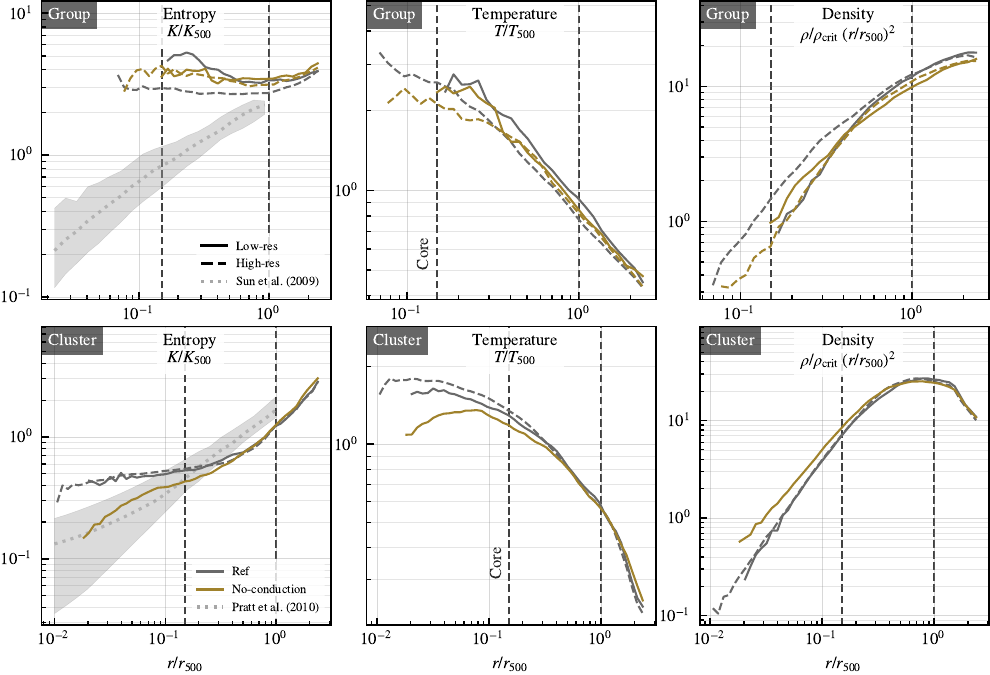}
    \caption{Radial profiles at $z=0$ for models with (Ref) and without (No-conduction) artificial conduction. \textit{Top row:} results from the group at both resolutions; \textit{bottom row:} results from the cluster at low resolution (and high resolution for the reference model). The panels on the left show the entropy profiles, normalised to $K_{500}$, the self-similar entropy scale. The central panels show the temperature profiles, scaled by $T_{500}$, and the right panels show the hot gas density profile in dimensionless units. We use 25 log-spaced bins from $0.01~r_{500}$ out to $2.5~r_{500}$. For the inner part, the profiles are truncated where the number of hot gas particles in the spherical shell drops below 50. The vertical dashed lines indicate $0.15~r_{500}$ (which we use to define the core as in \citealt{entropy_profiles_pratt2010}) and $r_{500}$.}
    \label{fig:profiles_noconduction}
\end{figure}

As shown in Fig. \ref{fig:profiles_noconduction}, artificial conduction does not impact the shape of the group's entropy profile, which remains flat at both resolutions. For the cluster, switching off artificial conduction lowers the entropy inside the core radius compared to the reference model. Artificial conduction causes more gas to mix and prevents low-entropy gas from sinking towards the centre of the halo. In the No-conduction model, instead, low-entropy gas does not mix with high-entropy gas and can collapse into denser structures, as can be seen in the cluster's density profile.

The No-conduction model also produces cooler cores in both the group and the cluster than in the Ref model. This result is compatible with the $\alpha_{D, {\rm max}}=0$ model allowing the inner IGM to collapse further and to form denser cores. By increasing the density, the gas loses more energy due to radiative cooling, which can be observed in Fig. \ref{fig:profiles_noconduction}.

{\cite{2015ApJ...813L..17R} also produced simulations with and without artificial conduction to probe its effect on the entropy profiles of their clusters.} The cluster run without artificial conduction does not form a cool core by $z=0$, unlike what \cite{2015ApJ...813L..17R} found for their "cluster-2" object with similar mass. Their simulations, however, are run at about 10 times lower resolution than EAGLE and use the SPH scheme by \cite{2016MNRAS.455.2110B}. These differences could have an impact on the mixing of high- and low-entropy gas. We emphasise that these SPH schemes require artificial conduction to correctly reproduce hydrodynamic instabilities and the No-conduction model gives an unphysical representation of the simulated objects \citep[e.g.][]{sphenix_borrow2022}. This comparison is nevertheless useful to show that artificial conduction cannot dictate the formation of a power-law-like entropy profile.

\subsection{Effect of metal cooling}

{When metal line cooling is included in the simulation model, the cooling rate of the gas increases, and so does the amount of material that cools down and exits the low-density phase, characteristic of the intra-cluster medium. Moreover, this process favours the formation of a multi-phase IGM, because it removes some of the gas from the hot phase ($T > 10^7$ K, near the virial temperature), and allows that gas to cool down to the warm and cool gas phases ($T\sim 10^5$ K and below, \citealt{2007ApJS..168..213G, 2013MNRAS.434.1043O}). The regulation of the amount of gas in the cold phase available to form stars, and potentially the cold gas fuelling SMBHs, is therefore tightly linked to the cooling of the metals.}

Without metal cooling, heavy elements radiate energy away on longer cooling timescales and fewer gas particles become cool and dense enough to form stars \citep[e.g.][]{2011MNRAS.412.1965M}. Correspondingly, Fig. \ref{fig:z0properties} confirms this claim by showing a lower $r_{500}$ star fraction (and BCG mass), as well as more hot gas in the No-metals runs. No-metals also produces a larger baryon fraction, which can be interpreted as the net effect of the reduced amount of stars and the consequent abundance of non-star-forming gas \citep{2021Univ....7..209O}. The baryon fraction is therefore a reliable metric for assessing the impact of feedback \citep[e.g.][]{2010MNRAS.406..822M}, which appears to be weaker in the No-metals runs. This effect could be due to the suppressed star formation at high redshift, which leads to weaker stellar winds and, more importantly, less SN feedback. Compared to Ref, the central BH also grows to a smaller mass and the overall AGN feedback is consequently weakened. These effects have a stronger impact on the group than on the cluster, since the former has a shallower gravitational potential and therefore is more susceptible to the amount of energy injected into the IGM.

\begin{figure}
	\includegraphics[width=\columnwidth]{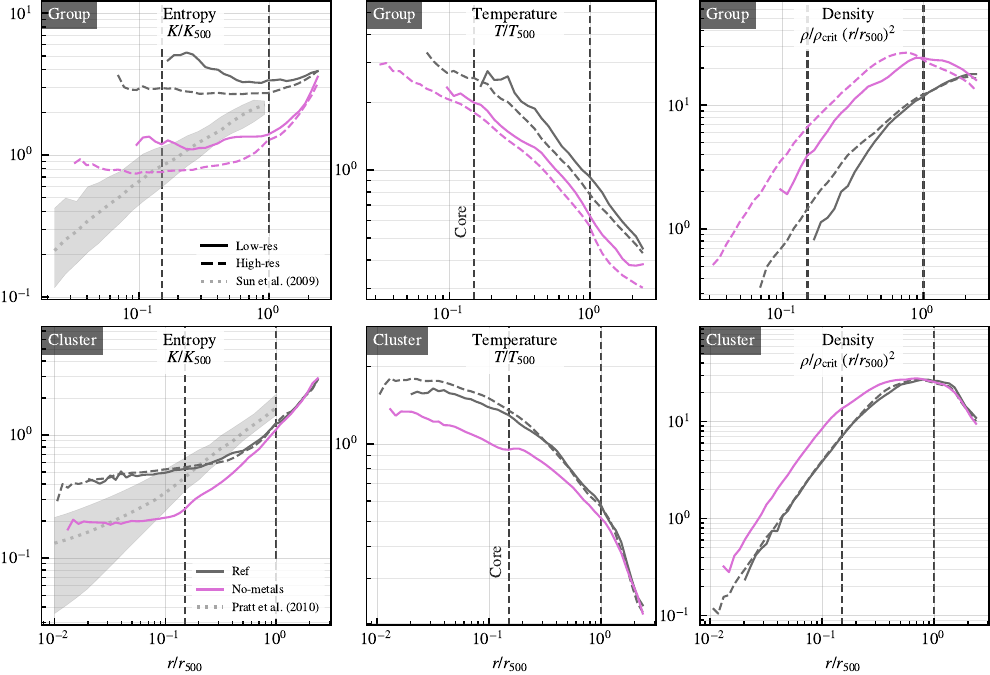}
    \caption{As in Fig. \ref{fig:profiles_noconduction}, but comparing models with and without metal cooling.}
    \label{fig:profiles_nometals}
\end{figure}

The absence of metal-line cooling also has important consequences for the profiles, as we show in Fig. \ref{fig:profiles_nometals}. The reduced effect of thermal feedback and the suppressed star formation allow the formation of cooler and denser cores, which in turn lead to lower entropy. While the entropy level in the core is reduced, the central plateau persists and prevents the formation of a power-law entropy profile in the central region. Simulated at a mass resolution of $2.1 \times 10^5$ M$_\odot$ (i.e. $\sim 10$ times higher resolution than our high-res set-up) without metal cooling, the ROMULUS-C cluster does form a cool core and a plateau-less entropy profile \citep{2019MNRAS.483.3336T}, {and maintains it until $z=0.3$, when the onset of a merger forms an entropy core \citep{2021MNRAS.504.3922C}}. Our sub-grid set-up differs from that of ROMULUS-C in several ways, however, it seems that, without metal cooling, the cluster run with the EAGLE model still cannot form a stable cool core. {Our results also agree with those of \cite{2011MNRAS.417.1853D}, where a suite Adaptive Mesh Refinement (AMR) simulations of a galaxy cluster (with similar mass to ours) showed a smaller entropy core without metal cooling (\texttt{AGNHEATrun}) and a larger core at the $K\approx 200$ keV cm$^2$ level when metal cooling was included (\texttt{ZAGNHEATrun})}. As in the runs without artificial conduction, the No-metals model is unphysical and cannot produce the correct global properties of clusters, groups, and especially galaxies. This comparison shows that, despite the moderate reduction in the central entropy level, switching off metal cooling simply does not lead to a cool-core scenario in our simulated objects. 

\subsection{Models without feedback}

\begin{figure}
	\includegraphics[width=\columnwidth]{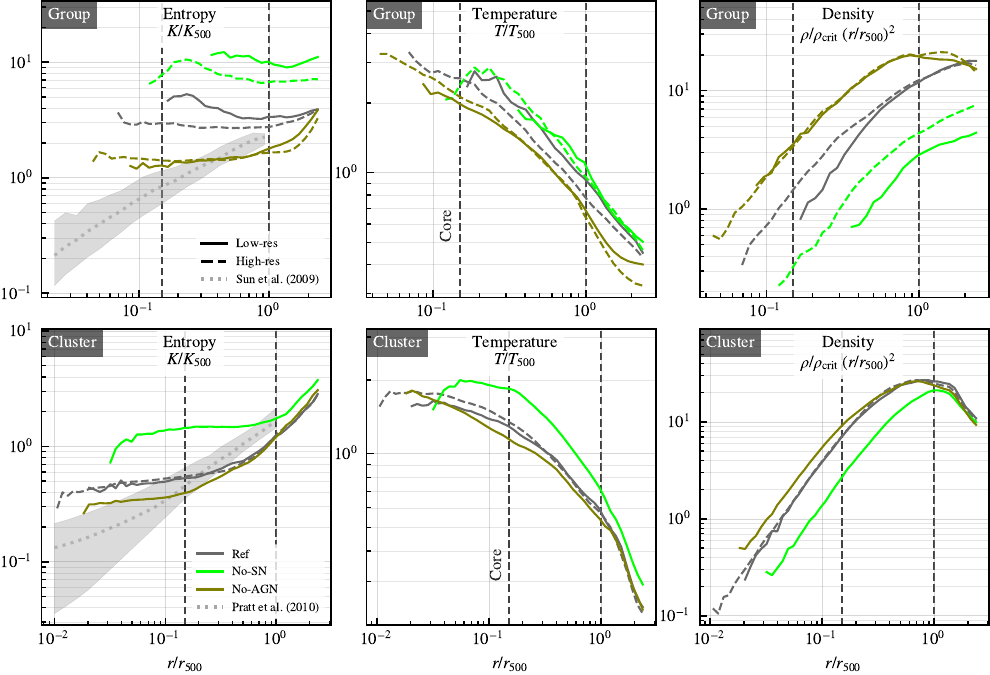}
    \caption{As Fig. \ref{fig:profiles_noconduction}, but including models with AGN or SN feedback switched off.}
    \label{fig:profiles_nofeedback}
\end{figure}

We now compare the Ref model, which includes both SN and AGN feedback processes, with runs where they are switched off one at a time. We present these results by considering the properties in Fig. \ref{fig:z0properties} and the profiles in Fig. \ref{fig:profiles_nofeedback}. In both the No-SN and No-AGN group runs, we measure a 50-90\% larger baryon fractions than in Ref, which are qualitatively consistent with what is expected from weaker feedback. The cluster's total baryon budget is less susceptible to these changes and only shows per-cent level differences at low resolution \citep[see also][]{2017MNRAS.465...32B}. The absence of one feedback channel, however, has major consequences on the baryon cycle, i.e. on the hot gas and star fraction, as we discuss below.

All our No-SN runs show two peculiarities: a 80-70\% lower hot gas fraction, and a doubled star fraction. We begin by discussing the first effect. The missing hot gas could not have \textit{only} been expelled by AGN feedback, {since the central SMBH has grown to a similar mass and could not have produced AGN feedback outflows strong enough to justify such a large depletion of hot gas. Instead}, as the higher baryon {(and star)} fraction suggests, {the missing hot gas} must have formed a larger amount of stars than Ref. As for the second effect, SN feedback regulates star formation by preventing gas from cooling down and collapsing. When no SNe are present, the collapse of the IGM on small scales is unregulated and the objects experience enhanced star formation, which leads to large stellar masses at $z=0$ and up to three times more massive BCGs. The absence of SNe also affects the central BH. In the group, the BH grows 10-60\% more massive than in Ref. 
We did not find evidence of this mechanism in the cluster, whose BH grows slightly less in the No-SN run than in Ref. 

The entropy profiles for the No-SN runs, shown in  Fig. \ref{fig:profiles_nofeedback}, are higher than in the Ref runs. While the shape of the profiles remains flat as in Ref, the overall normalisation is affected, resulting in upward-shifted profiles. A change in entropy normalisation is, in fact, expected to take place when a large amount of cool and low-entropy gas becomes star-forming and is removed from the core \citep[see Section 4.1 of][]{2002ApJ...576..601V}, as our density profiles suggest. The remaining hot and non-star-forming gas also has a higher mass-weighted temperature. Fig. \ref{fig:z0properties} also shows that SN feedback is tightly connected with the BCG star formation history and can affect its quenching time. The specific SFR (sSFR) is used to define whether a BCG is quenched at $z=0$; using the threshold adopted by \citeauthor{2012MNRAS.424..232W} (\citeyear{2012MNRAS.424..232W}, Section 3.1), we define a BCG with sSFR $<10^{-2}$ Gyr$^{-1}$ quenched, and non-quenched otherwise. The BCG in the No-SN group shows a lower sSFR than Ref, suggesting that the intense star-burst epoch that converted large amounts of gas into stars could have taken place earlier than in Ref, resulting in a quenched BCG populated by old stars.
Motivated by the $\sim$ 1 dex increase in entropy and the $\sim$ 1 dex decrease in density profiles compared to Ref, we checked that the SWIFT code, coupled with the EAGLE model, models the star formation accurately even for such an extreme set-up. We re-ran the group at EAGLE resolution {and the cluster at 8 times lower resolution} with a reduced SN heating temperature of $10^{7}$ K to study a scenario where $\Delta T_{\rm SN} < 10^{7.5}~{\rm K}$ and cooling losses are more prevalent than in Ref. As expected, we obtained profiles with intermediate values between those in Ref and No-SN, suggesting that the star-formation processes were captured as expected. {We show the profiles for these additional runs, labelled as SNdT7, in Appendix \ref{appx:sn-heating-temperature}.}

In the No-SN cluster run, however, the central BH was found outside the BCG, likely due to the repositioning algorithm. In the rare cases where the SMBH, and not the host galaxy, dominates the local gravitational potential, the repositioning stops being effective and the SMBH is no longer shifted towards the centre of the galaxy \citep[see][for a discussion of this behaviour]{bahe_2021_bh_repositioning}. This effect is nevertheless unimportant in our analysis. In fact, the BH in the No-SN cluster appears to have reached the expected mass at the end of its rapid growth phase \citep{2018MNRAS.481.3118M}, and delivered most of its feedback \textit{before} the onset of the undesired repositioning. Since BH accretion rates become rapidly suppressed at low redshift, we estimate that the BH mass could only have been minimally affected by the off-centering and therefore does not impact our results and conclusions.

In the simulations where BHs are removed and no AGN feedback is present, the hot gas fraction of the group doubles compared to Ref (the cluster is largely unaffected), and the star fraction/BCG mass are also greater by 50-100\%. {We note that the absence of SMBHs also avoids some gas being accreted and removed from the local environment. Nevertheless, the accreted gas would be negligible compared to the total gas content and, therefore, it does not affect our analysis.} Unlike the No-SN model, a smaller amount of gas cools and becomes star-forming because the SN feedback is now able to regulate this process. On the other hand, less hot gas was expelled via thermal feedback due to the missing AGN channel (the baryon fraction is higher than in Ref), allowing more gas to remain within $r_{500}$ and, while being regulated by SN feedback, form stars and add to the stellar budget of the system, in agreement with \cite{2012MNRAS.422.2816B}. Due to SN feedback, the No-AGN runs do not show the signature of a runaway star formation as in the No-SN model, and produce cores with lower entropy (see Fig. \ref{fig:profiles_nofeedback}). The absence of AGN feedback, once again, does not lead to power-law-like entropy profiles; the entropy plateaus still persist in all No-AGN runs, although their entropy level (i.e. normalisation) is lower than in Ref. The role of AGN feedback in stopping star formation and quenching the BCG is commonplace in current studies of galaxy formation \citep[see e.g.][]{2013MNRAS.428.2885D} and the No-AGN runs illustrate a scenario where, instead, this process is absent, i.e. unphysical. In Fig. \ref{fig:z0properties}, we show that the No-AGN runs consistently produce non-quenched BCGs, where significant star formation is still ongoing at $z=0$.

Switching off one feedback channel completely (SN or AGN in turn) does not transform the core of our simulated objects from non-cool to cool. While the presence of feedback is causally linked to a change in entropy, the shape of the profile, as well as the entropy plateaus, do not appear to be set by the thermal feedback in the model we consider. We stress that the reduced sample may only contain objects whose accretion history does not admit the formation of a cool core, however, we have shown here evidence that feedback cannot be solely responsible for the formation of a power-law-like entropy profile in our simulations. 

\subsection{AGN heating temperature}
\begin{figure}
	\includegraphics[width=\columnwidth]{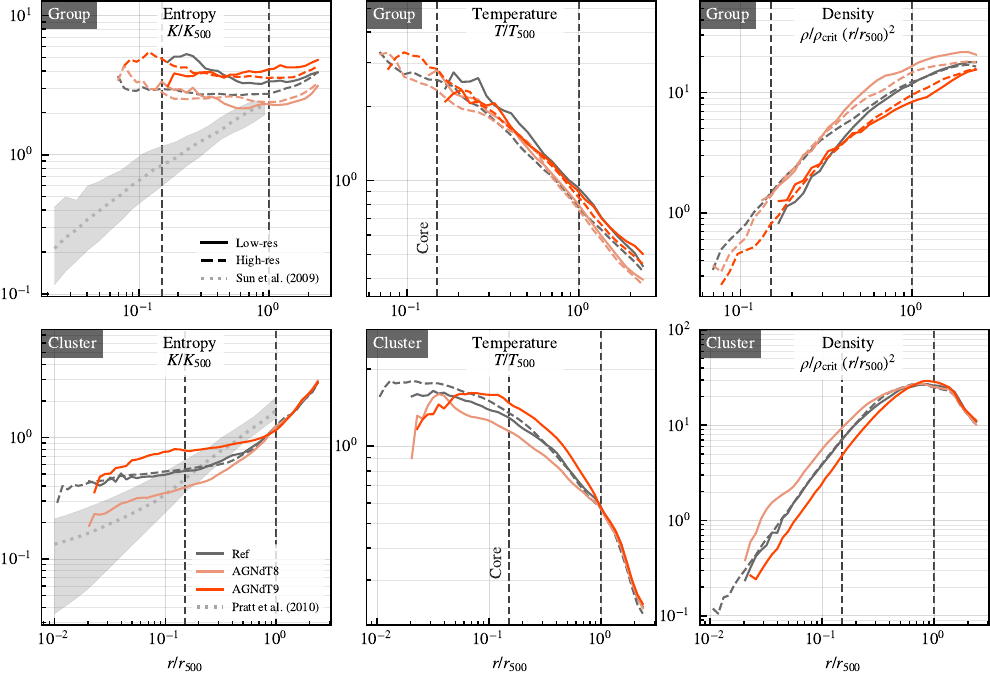}
    \caption{As in Fig. \ref{fig:profiles_noconduction}, but comparing models with varying AGN heating temperature.}
    \label{fig:profiles_agndt}
\end{figure}

To study the effects of the AGN heating temperature on the cluster properties, we compare runs performed using EAGLE's fiducial value of $10^{8.5}$ K, with a lower AGN heating temperature, $10^{8}$ K in AGNdT8, and a higher one, $10^{9}$ K in AGNdT9. As shown in Fig. \ref{fig:z0properties}, we find that AGNdT8 produces objects with larger baryon fraction than Ref, as expected from a more continuous energy injection consisting of more numerous but gentler heating events. A lower hot gas fraction is measured in the AGNdT9 runs, which use fewer and more energetic events. Similar results were found for the (lower-resolution) Cosmo-OWLS models \citep{2014MNRAS.441.1270L}. Moreover, varying the AGN heating temperature results in the baryon fraction changing in the same way for both the group and the cluster. Our calculations show that higher AGN heating temperature leads to more baryons being expelled from the system. The hot gas fraction follows the same behaviour as the baryon fraction in both objects. The star fractions and the BCG masses indicate that the star formation is well regulated in both AGNdT8 and AGNdT9, with values broadly consistent with Ref.

 We find that none of the models considered in this section produce a quenched group at low resolution. The high-resolution simulations show a consistently lower sSFR in the group and produce a quenched BCG in the group run with AGNdT8. For the group at high resolution, AGNdT8 yield an sSFR 29\% lower than the Ref version and the AGNdT9 an 1\% lower sSFR. For the cluster, we register a 3\% higher sSFR for the AGNdT8 and a 146\% higher sSFR in the AGNdT9, in contrast with what was found by \cite{2010MNRAS.401.1670F}. Previous simulation studies have confirmed AGN activity is the main cause of the quenching of the BCG at low redshifts \citep{2010MNRAS.406..822M, 2012MNRAS.420.2859M}. While this effect can be measured for an ensemble of objects, individual ones may exhibit discrepant behaviour, such as in our case.

The mass of the BH at the centre of the BCG increases by $\approx$20\%, with the high-resolution group hosting a 30-40\% more massive BH than the low-resolution counterparts. We note that the BH mass is largely determined by its accretion history. This variation can be attributed to different accretion events occurring at different times in the formation history of the host galaxy.

As anticipated in section \ref{sec:reference_model_results:entropy_profiles}, the group shows entropy profiles that are relatively flat regardless of $\Delta T_{\rm AGN}$ (Fig. \ref{fig:profiles_agndt}), but with a mild change in normalisation, especially for AGNdT8 (a factor of $\approx 2$ lower than Ref). This effect can be explained by the lower amount of energy injected into the IGM per AGN feedback event (note that the \textit{cumulative} energy injected by AGNs throughout cosmic time remains similar). Crucially, we demonstrate that the Ref and AGNdT9 profiles are very similar and that the minimum-distance AGN model does not produce a significant entropy excess when increasing $\Delta T_{\rm AGN}$ by 0.5 dex. More appreciable differences in the entropy profiles can be seen in the cluster. All models produce consistent profiles beyond $r_{500}$, however, AGNdT8 generates lower entropy in the core of this object compared to Ref (by a factor of $\approx 2$). The AGNdT9 model, on the other hand, predicts an $\approx$ 80\% higher entropy level than Ref in the core. At small radii, the AGNdT9 profile for the cluster turns downwards, resembling the shape of a power law. 


The dependence of the hot gas fraction on the AGN heating temperature can also be inferred from the density profiles in Fig. \ref{fig:profiles_agndt}. In the two right panels, the AGNdT8 (orange) yields a density profile that is always higher than in Ref, while the AGNdT9 profile is the lowest. This result confirms that AGNs are responsible for the removal of hot gas from the centre of the objects. The group is affected by AGN feedback out to $2r_{500}$, visible as an offset between the AGNdT8, Ref and AGNdT9 models. The cluster, on the other hand, has very similar profiles around $r_{500}$ and only shows differences in the inner region due to the feedback outflows. The cooling flow in the cluster run with AGNdT8 contributes to the lowering of the entropy level in the core, by allowing high-density, low-temperature gas to sink towards the centre of the halo. 

By comparing runs with different AGN heating temperatures, we conclude that the entropy core cannot be removed by simply varying $\Delta T_{\rm AGN}$. The AGN heating temperature impacts the overall normalisation of the entropy profiles in the core, but does not modify their shape from flat to a power-law. While AGNdT8 seems to produce a steeper slope of the cluster entropy profile which approaches the slope measured by \cite{entropy_profiles_pratt2010}, we emphasise that lowering $\Delta T_{\rm AGN}$ too much reduces the amount of gas expelled from the system and produces objects with unrealistically high gas (and baryon) fractions (see the discussion in \citealt{eagle.schaye.2015}, and the motivation for using AGNdT9 in the C-EAGLE simulations in \citealt{ceagle.barnes.2017}).

\subsection{AGN {and SN} feedback distribution schemes}
\begin{figure}
	\includegraphics[width=\columnwidth]{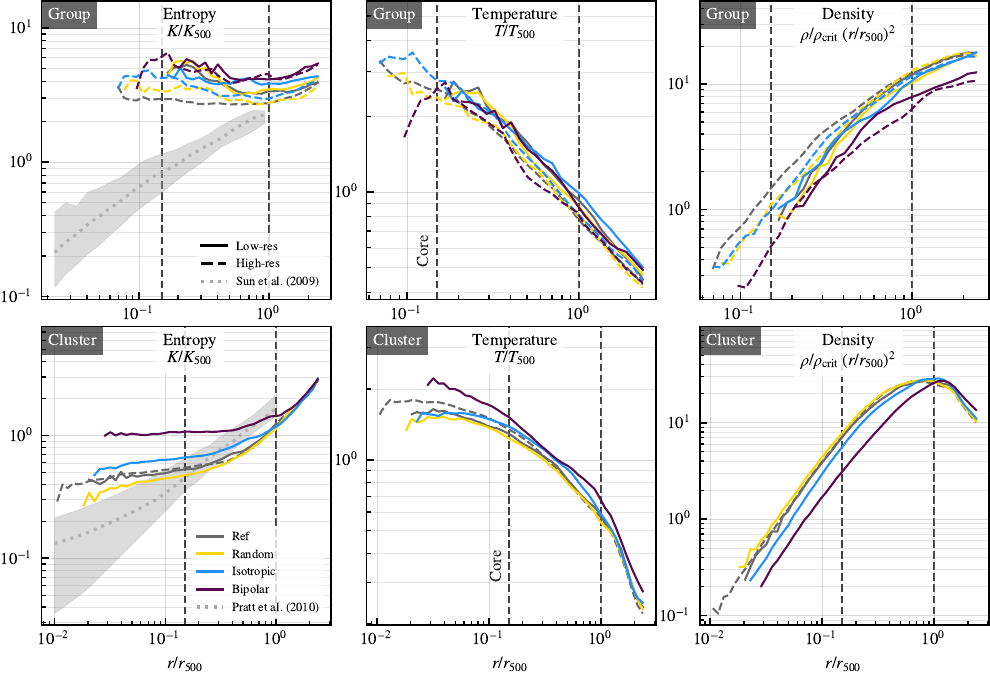}
    \caption{As in Fig. \ref{fig:profiles_noconduction}, but with varying AGN feedback distribution schemes (minimum distance, random, isotropic and bipolar).}
    \label{fig:profiles_schemes}
\end{figure}

Finally, we compare runs with different AGN energy distribution schemes: Minimum-Distance (i.e. Ref), Isotropic, Random (used in the EAGLE Ref model) and Bipolar.
We first discuss the properties of the group and cluster with reference to Fig. \ref{fig:z0properties} focusing on the grey, yellow and blue bars, which correspond to the first three models mentioned above. Isotropic and Random, configured with fixed $\Delta T_{\rm AGN}=10^{8.5}$ K, produce objects with baryon fractions comparable to Ref and show minimal differences in the hot gas and star fractions ($\sim$ 10\%) at both resolutions. Similarly, the choice of heating scheme does not seem to impact other properties significantly. An exception involves the Isotropic cluster run, which produced a BCG 47\% less massive than Ref. Despite the slightly lower baryon fraction, this difference is unlikely to be the result of slightly more efficient AGN feedback. {The central SMBH is less massive than Ref and, consequently, it could not have produced stronger outflows. However,} SN feedback could potentially be responsible for this effect. {The thermal energy injected by SNe into the IGM is distributed to the gas on small scales. This process has two effects: raising the temperature of the gas locally and producing pressure gradients which raise the velocity dispersion. Therefore, when SN feedback is more effective, we expect that star formation is delayed, leading to a less massive BCG.}

At constant  $\Delta T_{\rm AGN}=10^{8.5}$ K, the entropy profiles shown in Fig. \ref{fig:profiles_schemes} for the group and the cluster are insensitive to the choice of the feedback distribution scheme. This similarity is evident between the Ref and the Random model, which only differ in the inner region around $0.02\, r_{500}$. This result differs from what found by \cite{2022MNRAS.514..249C}. In their study, the Random (mass-weighted) distribution scheme produced less efficient feedback than the Minimum-distance scheme and led to significant differences in the properties of the IGM. The entropy profile produced by the Isotropic model is instead consistently higher within $r_{500}$. This effect is best observed in the cluster, although it also affects the group to a smaller extent. Crucially, the overall shape of the inner entropy profiles remains flat regardless of the the energy distribution scheme.


The results shown in Fig. \ref{fig:profiles_schemes} show that the choice of AGN distribution scheme does not have a large impact on the shape or the normalisation of the profiles for both objects at both resolutions. We note that the rule for selecting the target gas particles to be heated is a purely numerical choice, without any direct physical implication. Albeit still debated, the actual heating mechanism coupling the BHs and the gas may occur on much smaller scales{, yet unresolved. Other studies have implemented feedback mechanisms which couple the energy output to the IGM via jets on \textit{resolved} scales \citep{2016ApJ...829...90Y, 2017ApJ...847..106L, 2017MNRAS.472.4707B, 2017MNRAS.470.4530W, 2019MNRAS.483.2465M, 2021MNRAS.506..488B}.}


To expand our investigation, we also explore the effect of choosing a particular fixed direction (the x-axis of the parent box) to select the particle to heat, as opposed to heating particles in random directions at every feedback event. This scheme is implemented in the Bipolar model, and can be regarded as a limiting case of a collimated outflow. From Fig. \ref{fig:z0properties}, we immediately notice that the Bipolar model has the overall effect of reducing the baryon fraction of both objects, compared to the Isotropic case. Since the star fractions are similar, the variation in total baryon fraction is dominated by the differences in hot gas fraction. The group run with bipolar feedback contains $\sim$10\% less baryons than Isotropic, and this change is greater in the cluster, which shows a 12\% reduction in baryon fraction when switching from the Isotropic to the Bipolar model.

We find that changing the directionality of the AGN feedback also has an impact on the profiles, as shown in Fig. \ref{fig:profiles_schemes}. The group is, once again, less affected, but the Bipolar cluster run produced a larger isentropic core and a flatter entropy profile compared to the other schemes. We find that a collimated AGN feedback scheme does not allow the formation of a cool core, but rather has the opposite effect of producing an even flatter entropy profile. However, our implementation of the Bipolar model is remarkably simple and different results may be produced by a more sophisticated method based on semi-analytic models \citep[e.g.][]{2022MNRAS.516.3750H}.

\section{Discussion}
\label{sec:discussion}

Large entropy cores, defined by flat entropy profiles with extended plateaus, are commonplace in recent cluster simulations run at high resolution (e.g. C-EAGLE, \citealt{ceagle.barnes.2017}, SIMBA, \citealt{2019MNRAS.486.2827D}, TNG, \citealt{2018MNRAS.473.4077P, 2018MNRAS.481.1809B}, FABLE, \citealt{2018MNRAS.479.5385H}) . Power-law entropy profiles, typical in cool-core systems, are absent in all these examples, except for the higher-resolution Romulus-C cluster which, however, excludes the effects of metal cooling. Interestingly, these cores/plateaus appear to form irrespective of the hydrodynamics methods and feedback schemes. The study in this paper shows that this effect also persists when varying relevant parameters (e.g. feedback temperature, particle heating method) within a fixed model. In this section, we discuss alternative scenarios which could possibly explain the absence of cool-core clusters in contemporary simulations.

\begin{enumerate}
    \item \textbf{Object selection}. Cool cores in groups and clusters are rare and we have not performed \textit{ad hoc} selections of such objects in our sample of 27 objects. In particular, a power-law-like, cool-core system will only form when the cluster is moving into its relaxed phase and is no longer experiencing merger activity. We expect such objects to be less common on cluster scales given their complex dynamical youth. While we cannot rule out this explanation, we note that the general profile shape of X-ray-selected groups and clusters, such as those studied by \cite{entropy_profiles_sun2009} and \cite{entropy_profiles_pratt2010}, is considerably different from our simulated systems. However, X-ray luminosity selection can implicitly exclude objects with under-dense cores, particularly in lower-mass groups, where the high entropy is a likely result of AGN feedback, as concluded by \cite{2022A&A...663A...2C}. Furthermore, profiles from larger samples e.g. in the TNG simulation \citep{2018MNRAS.481.1809B}, are also at odds with observations, so we do not regard object selection as a likely explanation.
    
    \item \textbf{Hydrostatic mass bias}. We have not included the effects of hydrostatic mass bias explicitly in our analysis. The expected positive bias (i.e. a lower hydrostatic mass than the true mass) would not affect the shape of the profile; instead, it would shift the radial scale ($r_{500}$) and entropy scale ($K_{500}$) to smaller values for the simulations, leading to a higher normalisation in the scaled profile at fixed scaled radius. This would result in even poorer agreement with observations in the central region, as we found in the group and cluster run with the Ref model. Furthermore, we have verified that typical bias levels for our simulated objects are 10-20\% at $r_{500}$, in line with previous work. Therefore, hydrostatic bias does not strongly affect the entropy profiles in our analysis.
    
    \item \textbf{Projection effects}. Obtaining the 3D entropy profile from X-ray data is a complex procedure. A significant source of uncertainty comes from the deprojection of data (X-ray emission-measure/surface brightness and temperature profiles), based on the assumption of spherical symmetry. Projection effects are unlikely to dramatically change the shape of the entropy profile. To produce a cool-core system, we would require a radially-dependent bias that results in a decrease in the central temperature and/or increase in the central density. On the other hand, incorrectly accounting for gas at larger projected distances from the centre would likely have the opposite effect since the gas there is at higher entropy. Nevertheless, we have verified this by modelling projected data and using this to reconstruct the 3D entropy profile. 
    
    \item \textbf{X-ray weighting}. A related issue to the previous point is that X-ray observables are not mass-weighted. Firstly, since the gas density is derived from the emission measure (which depends on the square of the density), it is susceptible to clumping effects \citep[see e.g.][]{2022arXiv221101239T}. However, recent analyses take this effect into account by calculating the azimuthal median profile \citep[e.g. in X-COP,][]{2017AN....338..293E} in order to mitigate substructure effects. Secondly, the X-ray spectroscopic temperature is also well known to be biased relative to the mass-weighted value, due to the assumption of isothermality when fitting the spectrum. Spectroscopic temperatures are lower than mass-weighted temperatures \citep[e.g.][]{ceagle.barnes.2017, 2021MNRAS.506.2533B}, so the effect of this bias would reduce the entropy. {We have conducted simple tests to check the effect of X-ray weighting on both the group and the cluster run with the Ref model. We compared the mass-weighted thermodynamic profiles with the spectroscopic-like profiles, weighted by the square of the hot-gas density, and the profiles weighted by the X-ray luminosity computed using the same cooling tables used in the sub-grid gas cooling \citep{2020MNRAS.497.4857P}. We did not detect any changes in the shape of the scaled profiles and discrepancies relative to Ref were $\sim 10\%$.} Nevertheless, the X-ray weighting does not necessarily reduce the size of the entropy cores, as shown by the \textit{spectroscopic} profiles of C-EAGLE and IllustrisTNG \citep{ceagle.barnes.2017, 2018MNRAS.481.1809B}.
    
    \item \textbf{Jet feedback}. Observational evidence for AGN feedback in clusters comes from radio observations of {relativistic} jets. {Their effect of} displacing the thermal X-ray-emitting gas with bubbles of super-heated relativistic plasma is not modelled in our simulations. Such {\it jet feedback} is widely thought to be the solution to the cluster cooling flow problem, quenching star formation in the BCG, but the exact mechanisms by which the jets couple to the gas are poorly understood. Directional feedback may reduce the high entropy levels if these are mainly produced by quasi-spherical shock-heating in the current models (inspection of our simulation outputs show that such events do occur). However, it is difficult to imagine such directional heating being sustained on cosmological timescales, given that there is evidence for jet precession and that we expect precipitation (and star formation) along axes perpendicular to the jet direction \citep{2021A&A...646A..38T}. \\
    {Sub-grid models tracking the spin (magnitude and orientation) of the SMBH explicitly and dividing AGN feedback into quasar/radio modes have also shown that SMBH spins re-orient as mergers, gas accretion and SN feedback take place \citep[e.g.][]{2014MNRAS.440.2333D, 2019A&A...631A..60B}. Recent implementations of jet feedback, based on the \cite{1977MNRAS.179..433B} self-similar jet model, were successfully tuned to match the properties of the circum-galactic medium of observed in Seyfert galaxies with active SMBHs \citep{2021MNRAS.504.3619T}. Idealised simulations of galaxy clusters showed that the SMBH spins can re-orient more frequently in high-mass systems than in low-mass ones, and produce pressurised lobes which can uplift the cold, low-entropy gas and remove it from the core \citep{2022MNRAS.516.3750H}.} \\
    {The SIMBA simulations attempted to include the effects of jet feedback in a cosmological setting by using bipolar kinetic jets \citep{2019MNRAS.486.2827D}. Their implementation temporarily decouples the jetted gas particles from the hydrodynamic scheme, avoiding a directly interaction on small scales. This strategy led to realistic hot-gas and star fractions, but still produced large entropy cores \citep{2020MNRAS.498.3061R}. Finally, simulations of kinetic jets with \textit{chaotic cold gas accretion} \citep{2013MNRAS.432.3401G} in cluster simulations have been able to maintain cool-cores on long timescales, albeit only using idealised initial conditions \cite[e.g.][]{2015ApJ...811...73L, 2015ApJ...811..108P, 2016ApJ...829...90Y, 2023MNRAS.518.4622E}.} 
    
    \item \textbf{AGN heating cycle}. The shape and level of the inner region of the entropy profiles is largely dictated by the AGN activity. \cite{2022MNRAS.tmp.1955N} and \cite{2022MNRAS.516.3750H} have identified periodic behaviour in the core entropy, which rises as the AGN injects energy into the surrounding medium and decreases as the radiative cooling then takes over and prompts the formation of a cool core. Our simulations do not show such periodicity. Therefore, the high entropy and lack of cool cores at $z=0$ could not have been accidentally captured during a particularly active phase of a possible AGN heating cycle. A detailed time-evolution study of the entropy distribution will appear in a follow-up paper.

\end{enumerate}

We appear to be left with the scenario in which entropy cores are the outcome of feedback prescriptions currently being used to regulate star formation and black hole growth in the current generation of galaxy formation simulations. Interestingly, the problem is less acute at lower resolution (e.g. in \citealt{2015ApJ...813L..17R}, Dianoga Ref, \citealt{2020A&A...642A..37B}, BAHAMAS, \citealt{2017MNRAS.465.2936M} and MACSIS, \citealt{macsis_barnes_2017}) where good agreement with observed entropy profiles is found. Two possibilities for this difference are as follows. Firstly, it may be that more low entropy gas is being heated and ejected from proto-group and proto-cluster regions at early times in the higher-resolution simulations. This would lead to the larger characteristic entropy scale of the remaining gas, largely unaffected by the feedback. However, as discussed by \cite{ceagle.barnes.2017}, such a scenario is hard to justify when the low-redshift gas fractions are larger than observed. A second possibility is that the feedback is coupling more strongly to the gas (through shocks) as a result of improved resolution. This leads to the conclusion that we may be missing (or not resolving) important physics, allowing the energy to be deposited in a gentler way in sound waves (as suggested by X-ray observations of Perseus) or other, non-thermal mechanisms such as cosmic-ray heating and transport processes that would require us to model magnetic fields.

{A study based on the Dianoga simulations \citep{2021MNRAS.507.5703P} suggested that not only can the shock-heating mechanism resulting from AGN feedback thermalise the gas, but it can also transfer energy into populations of relativistic cosmic rays via diffusive shock acceleration (\citealt{2007ApJ...669..729K}, and the review by \citealt{2007NuPhS.165..122B}) and therefore alter the thermodynamic properties of the IGM. Furthermore, magnetic fields themselves can suppress the thermal conduction via processes of magnetic draping \citep{2006MNRAS.373...73L}, which is important in shock fronts generated by e.g. AGN outflows or mergers \citep{2022MNRAS.512.2157C}. Where thermal conduction is reduced, so is the mixing between high- and low-entropy gas; this scenario could indeed help preserving the low-entropy gas phase and cool-cores, as suggested by our No-conduction simulations (Section \ref{sec:results_model_variations:conduction}). On the other hand, magnetic fields could also favour the emergence of processes which produce the opposite effect. For instance, if AGN-produced magnetised bubbles are preserved, they can drive small-scale, often resolution-dependent, hydrodynamic instabilities. These motions can lead to a powerful turbulent cascade, which ultimately thermalises the local IGM and can compromise the integrity of a cool core \citep{2005MNRAS.363..891F, 2019ApJ...886...78B}.} Current models are also missing a cold (i.e., $T < 10^4$ K) interstellar gas phase that could also play an important role in how the central galaxy grows and how the feedback interacts with the gas, {similarly to the chaotic cold accretion mode in \cite{2023MNRAS.518.4622E}.}

\section{Conclusions}
\label{sec:conclusion}
Using a new implementation of the EAGLE galaxy formation model, we produced zoom-in simulations of 27 objects selected from a 300 Mpc parent volume. The re-simulated groups and clusters of galaxies showed reasonable agreement with the star fraction-halo mass relation from \cite{xxl.baryons.akino2022}, despite overshooting the hot gas fractions at halo masses $\sim 10^{14}$\,M$_\odot$. In spite of this shortcoming, our simulations provide hot gas and star fractions closer to the observed values than the EAGLE model in \cite{eagle.schaye.2015}. On the other hand, our simulations predict larger entropy cores than EAGLE, which still cannot be reconciled with the observations from \cite{entropy_profiles_sun2009} and \cite{entropy_profiles_pratt2010}. To investigate the sensitivity of the entropy profiles to sub-grid model changes, we selected one group and one cluster from the object sample and produced a suite of simulations with different sub-grid schemes and resolutions. We summarise the main results as follows:
\begin{enumerate}
    \item Artificial conduction does not have a large impact on the global properties of the objects; the thermodynamic profiles for the group are also generally unaltered. The cluster profiles, however, show the formation at $z=0$ of a colder, denser and anisentropic core when artificial conduction is switched off, which can be attributed to a smaller degree of mixing between gas in the high- and low-entropy phase. While yielding more power-law-like entropy profiles for clusters, the hydrodynamic scheme without artificial conduction cannot reliably reproduce instabilities and smooth particle behaviour at interfaces between fluids \citep{2008JCoPh.22710040P}.
    
    \item The simulations without metal cooling produced objects with gas fractions nearly twice as large and 50\% lower star fractions than the reference model. In the group, the lower-mass BCG does not differ significantly in its specific SFR, while the central BH is more than 70\% lighter in the runs without metal cooling. Both the group and the cluster have lower entropy overall and the cluster forms a smaller entropy core, with a power-law-like profile outside the core, in analogy to the results by \cite{2019MNRAS.483.3336T}, whose simulations also did not include metal cooling. The absence of metal cooling only affects the inner region of the cluster, but has a large impact on the group's environment out to $\approx2.5~r_{500}$.
    
    \item The runs without AGN also produced a large isentropic core, although the entropy level in the inner regions is lower than for the reference model. Turning off SN feedback, instead, produced a vast increase in entropy throughout the group and cluster atmospheres, associated with hotter and less dense cores. The absence of thermal SN feedback has a large impact on the self-regulation of the central BH, boosts the star formation in the group and cluster core.
    
    \item When increasing the AGN heating temperature from $10^8$ to $10^9$ K in steps of 0.5 dex, we observed an entropy increase at all radii in the group, which develops an overall less dense atmosphere. The cluster, on the other hand, only developed a larger entropy core, without impacting the hot gas outside $r_{500}$. High values of $\Delta T_{\rm AGN}$ lead to less frequent, but more energetic feedback events, which remove the low-entropy gas from the core.
    
    \item The minimum-distance, random and isotropic AGN feedback distribution schemes produced similar thermodynamic profiles for the group, and minimal differences can be seen in the cluster core. We conclude that, for the two objects considered in this comparative study, the entropy distribution is generally insensitive to the choice of distribution scheme for thermal AGN feedback. Our simple implementation of a fixed-direction bipolar scheme produces a marginally less dense, higher entropy IGM in the group. Larger effects are observed in the cluster core, where the entropy profile flattens to a constant level out to $0.5~r_{500}$, correlated with a higher core temperature and lower density. The bipolar feedback scheme is more effective in expelling hot gas from the system, however, it does not preserve the low-entropy gas and produces larger entropy cores.
    
\end{enumerate}
We find a close similarity between entropy profiles at $z=0$ and at $z=1$, suggesting that the formation of large, isentropic cores may occur at higher redshift, during the phase of maximum AGN activity. {The AGN favours the rise if entropy in the cores the group and the cluster. Additionally, a further investigation has shown that the shape of the entropy profiles changes significantly during and after merger episodes. This scenario is similar to that of the ROMULUS-C system \citep{2021MNRAS.504.3922C}. Indeed, the progenitor halo of our cluster appears to show a cool core at very high redshift ($z\approx 4$), but a core/plateau forms during the complex formation process that involves both mergers and AGN activity. The merger between our cluster and its neighbour produced large shock waves, which could play an important role in boosting the entropy amplification in the core.} In a follow-up paper, we provide a detailed overview of how the thermodynamic profiles and gas properties of the group and cluster evolve over time. The AGN model used in our simulations uses a simple prescription, which cannot simultaneously reproduce the gas properties and entropy profiles of groups and clusters. More sophisticated feedback mechanisms will be implemented and studied in future work.

\section*{Acknowledgements}
{The authors thank the anonymous referee for providing comments which improved the quality, clarity and depth of our work.} This work used the DiRAC Durham facility managed by the Institute for Computational Cosmology on behalf of the STFC DiRAC HPC Facility (www.dirac.ac.uk). The equipment was funded by BEIS capital funding via STFC capital grants ST/K00042X/1, ST/P002293/1, ST/R002371/1 and ST/S002502/1, Durham University and STFC operations grant ST/R000832/1. DiRAC is part of the National e-Infrastructure. EA acknowledges the STFC studentship grant ST/T506291/1. YMB acknowledges support from NWO under Veni grant number 639.041.751. {The work received support under the Project HPC-EUROPA3 (INFRAIA-2016-1-730897), with the support of the European Commission Research Innovation Action under the H2020 Programme; in particular, EA gratefully acknowledges the support of the Sterrewacht at Leiden University and the computer resources and technical support provided by SURFsara, the Dutch national high-performance computing facility.} The research in this paper made use of the \swift open-source simulation code (\href{https://swiftsim.com/}{swiftsim.com}, \citealt{schaller_2018_swift}) version 0.9.0. In addition, this work made use of the following software packages and libraries: 
\textsc{Python} \citep{van1995python},
\textsc{Numpy} \citep{harris2020array},
\textsc{Scipy} \citep{virtanen2020scipy},
\textsc{Numba} \citep{lam2015numba},
\textsc{Matplotlib} \citep{hunter2007matplotlib, caswell2020matplotlib},
\textsc{SWIFTsimIO} \citep{Borrow2020} and
\textsc{Astropy} \citep{robitaille2013astropy, price2022astropy}.

\section*{Data Availability}

The \swift and \velociraptor structure-finding codes are public and open-source, and can be downloaded from GitHub or GitLab. The initial conditions for the halos, the snapshots and the halo catalogues can be made available upon reasonable request to the corresponding author. The code used in the analysis is publicly available on the corresponding author's GitHub repository (\href{https://github.com/edoaltamura/entropy-core-problem}{github.com/edoaltamura/entropy-core-problem}) and we include the data used to generate the figures presented throughout the document.

\section{Reduced SN heating temperature}
\label{appx:sn-heating-temperature}

\begin{figure}
	\includegraphics[width=\columnwidth]{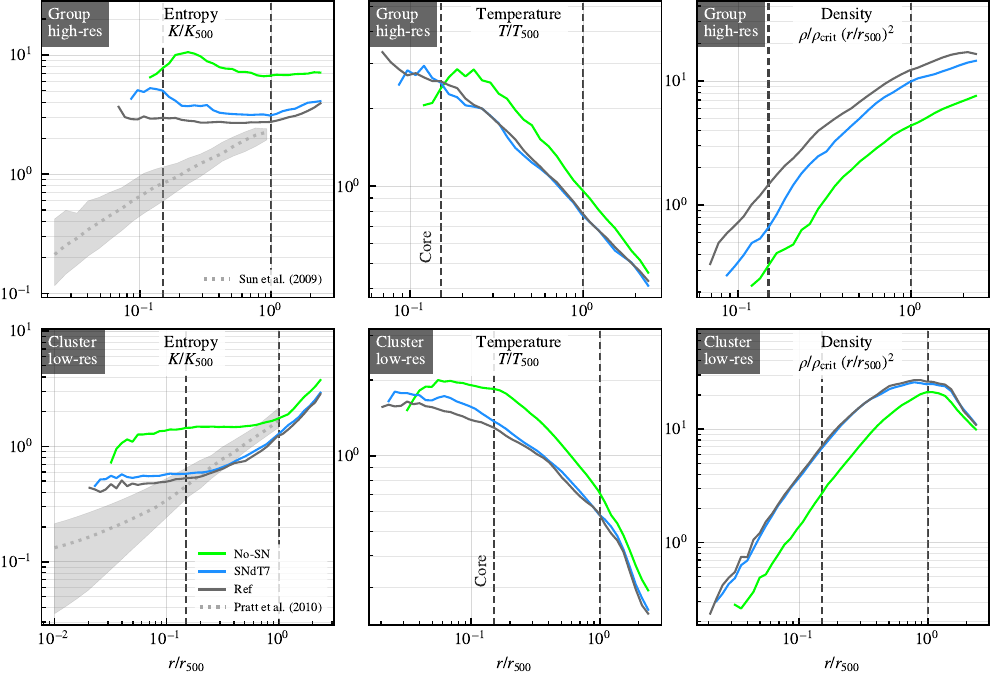}
    \caption{As in Fig.~\ref{fig:profiles_nofeedback}, but including runs with reduced heating temperature in SN feedback (blue). Here, we only show the profiles for the group at high-res (top) and the cluster at low-res (bottom).}
    \label{fig:sndt_variations}
\end{figure}

{To further examine the onset of cooling losses leading to anomalous star-formation in models with no SN feedback, we have produced additional simulations with reduced SN heating temperature, $\Delta T_{\rm SN} = 10^{7}~{\rm K}$. In Fig. \ref{fig:sndt_variations}, we compare this model, labelled as SNdT7, with Ref ($\Delta T_{\rm SN} = 10^{7.5}~{\rm K}$) and No-SN ($\Delta T_{\rm SN} = 0~{\rm K}$) for the group at high resolution (same as EAGLE) and the cluster at low resolution (8 times lower than EAGLE) at $z=0$. We found that the group was more susceptible to a decrease in $\Delta T_{\rm SN}$: the density of the gas drops at all radii, especially in the core. This effect results in higher entropy, despite the mass-weighted temperature profile being consistent with Ref. On the other hand, the SNdT7 cluster shows profiles similar to Ref. The temperature profile being slightly higher in the core means that the entropy is higher, but the magnitude of this effect is far smaller than disabling SN feedback completely (No-SN).\\
For the group at high-res, the Ref model produces a star fraction (in $r_{500}$) $f_\star = 0.028$, the SNdT7 variation returns $f_\star = 0.046$ and the No-SN solution gives $f_\star = 0.084$. Similarly, the low-res cluster gives $f_\star = 0.019$ for Ref, $f_\star = 0.023$ for SNdT7 and $f_\star = 0.054$ for No-SN. In both objects, we measure an increasing star fraction associated with a decreasing hot gas fraction. These results are consistent with the predictions of \cite{2012MNRAS.426..140D} and other simulations of galaxy-sized halos which found a larger stellar mass in weak SN feedback regimes \citep[e.g.][]{2014MNRAS.444.2837R}. Here, we show that cooling losses are more prevalent in the group than in the cluster and the overall effect is to raise the entropy profile.}

%% file: Chapters/Chapter6.tex
\chapter{Probes for entropy generation over cosmic time}
\label{chapter:6}

\section{Introduction}
\label{sec:ch6-intro}

In Chapter \ref{chapter:5}, we have shown that variations of the EAGLE galaxy formation model predicts amounts of high-entropy gas in the cores of groups and galaxies of clusters up to one order of magnitude higher than estimated observationally. Given these results, we have set the goal to investigate the formation of such entropy cores over cosmic time to identify the leading processes driving the generation of entropy. This study includes preliminary work which we plan to continue in the upcoming months after submitting of this document. We consider the hydrodynamic simulations of the group and the cluster of \cite{2023MNRAS.520.3164A} at EAGLE resolution and 8 times lower. Additionally, we introduce new simulation run using \swift and using the Ref model: the group at 8 times higher resolution than EAGLE (high-res). The parameters corresponding to the three resolutions were also discussed in Tab.~\ref{tab:resolutions-particle-load}.

While we previously restricted the analysis to the $z=0$ simulation snapshots, we now also include 2620 \swift outputs and \velociraptor halo catalogues in the range $z\in[0, 6]$, sampled at regular 10 Myr time-intervals. This output list only refers to the group at low- and mid-res and the cluster at low-res. Instead, we only saved 64 outputs for the group at high-res and the cluster at mid-res due to their larger file size. For these objects, the output redshift values are spaced in regular scale-factor intervals,\footnote{This output output redshift list is based on that of the EAGLE Ref 100 Mpc run \citep{eagle.schaye.2015}.} with an additional 2 snapshots at regular 10 Myr intervals immediately before $z=\{7\times 10^{-4} \approx 0,\, 0.21,\, 0.96,\, 2.06,\, 3.09,\, 4.72\}$ and 2 snapshots immediately after. In summary, our time sampling consists of 36 key snapshots of which 6 are anchor-points for groups of 5 snapshots sampled 10 Myr apart. The additional snapshots will be used to estimate the variation of the properties and profiles over time-scales of 10 Myr, matching the time-sampling used in the other runs. We do not investigate such variation in this document, but we may refer to it upcoming drafts of Altamura (in preparation), as required.

\begin{sidewaystable}
    \centering
    \caption{Mass resolution and softening lengths for different types of simulations. The simulations introduced in \protect\cite{2023MNRAS.520.3164A} are highlighted in yellow, and the high-resolution group introduced in this chapter is highlighted in orange. The mass of DM particles is expressed by $m_{\rm DM}$ and the initial mass of gas particles is $m_{\rm gas}$. $\epsilon_{\rm DM,c}$  and $\epsilon_{\rm gas,c}$ indicate the comoving Plummer-equivalent gravitational softening length for DM particles and the gas respectively, $\epsilon_{\rm DM,p}$ and $\epsilon_{\rm gas,p}$ indicate the physical maximum Plummer-equivalent gravitational softening length for DM particles and the gas respectively. In works using moving mesh codes, such as \protect\cite{2021MNRAS.506..488B}, $r_{\rm cell}$ indicates the cell radius. $^\star$Indicates full-physics simulations of isolated halos, with the highest particle-mass resolution in the inner 100 kpc and quadratically degrading resolution out to $r_{500}$, except for \protect\cite{2016MNRAS.461.1548B}, where the particle mass is fixed inside $r_{500}$. The DM halo is modelled with an external potential following an NFW profile. $^\dagger$The EAGLE-like model differs from \protect\cite{2023MNRAS.520.3164A} in the black-hole accretion and the star-formation scheme. $^\ddagger$The EAGLE-like model implements the changes of \protect\cite{2022MNRAS.tmp.1955N}, and models AGN feedback using self-consistent jets.}
    \label{tab:resolutions-comparison-other-works}
    \begin{tabular}{lllcccccc}
    \toprule
    Set-up                          & Model                 & References                    & $m_{\rm DM}$           & $m_{\rm gas}$         & $\epsilon_{\rm DM,c}$ & $\epsilon_{\rm gas,c}$ & $\epsilon_{\rm DM,p}$ & $\epsilon_{\rm gas,p}$ \\ 
                                    &                       &                               & (M$_\odot$)            & (M$_\odot$)           & (ckpc)         & (ckpc)         & (pkpc)        & (pkpc)       \\
    \midrule
    \rowcolor{yellow!25}
    Group and cluster (low-res)     & SWIFT-EAGLE           & \protect\cite{2023MNRAS.520.3164A}    & $7.85 \times 10^{7}$   & $1.47 \times 10^{7}$  & 6.66           & 3.80           & 2.96          & 1.69         \\
    \rowcolor{yellow!25}
    Group and cluster (mid-res)     & SWIFT-EAGLE           & \protect\cite{2023MNRAS.520.3164A}    & $9.82 \times 10^{6}$   & $1.83 \times 10^{6}$  & 3.33           & 1.90           & 1.48          & 0.854        \\
    \rowcolor{orange!25}
    Group (high-res)                & SWIFT-EAGLE           & This work                     & $1.23 \times 10^{6}$   & $2.29 \times 10^{5}$  & 1.67           & 0.95           & 1.74          & 0.427        \\
    \rowcolor{orange!25}
    Group and cluster (low-res)     & Non-radiative         & This work                     & $7.85 \times 10^{7}$   & $1.47 \times 10^{7}$  & 6.66           & 3.80           & 2.96          & 1.69         \\
    \rowcolor{orange!25}
    Group and cluster (mid-res)     & Non-radiative         & This work                     & $9.82 \times 10^{6}$   & $1.83 \times 10^{6}$  & 3.33           & 1.90           & 1.48          & 0.854        \\
    \rowcolor{orange!25}
    Group (high-res)                & Non-radiative         & This work                     & $1.23 \times 10^{6}$   & $2.29 \times 10^{5}$  & 1.67           & 0.95           & 1.74          & 0.427        \\
    ROMULUS-C                       & ROMULUS               &  \protect\cite{2019MNRAS.483.3336T}   & $3.40 \times 10^{5}$   & $2.10 \times 10^{5}$  & --             & --             & --            & 0.250        \\
    EAGLE 100 Mpc (Ref)             & EAGLE Ref             &  \protect\cite{2015MNRAS.446..521S}   & $9.60 \times 10^{6}$   & $1.80 \times 10^{6}$  & --             & --             & --            & 0.700        \\
    EAGLE 25 Mpc (high-res)         & EAGLE Ref             &  \protect\cite{2015MNRAS.446..521S}   & $1.21 \times 10^{6}$   & $2.26 \times 10^{5}$  & --             & 1.33           & --            & 0.350        \\
    C-EAGLE/Hydrangea               & EAGLE-AGNdT9          &  \protect\cite{ceagle.barnes.2017}    & $9.60 \times 10^{6}$   & $1.80 \times 10^{6}$  & --             & --             & --            & 0.700        \\
    TNG50                           & TNG                   &  \protect\cite{2019MNRAS.490.3234N}   & $4.50 \times 10^{5}$   & $8.50 \times 10^{4}$  & 0.288          & 0.074          & --            & --           \\
    Two groups and cluster$^\star$  & SWIFT-EAGLE$^\dagger$ &  \protect\cite{2022MNRAS.tmp.1955N}   & --                     & $(1-42) \times 10^{5}$& --             & --             & --            & 0.30         \\
    Two groups and cluster$^\star$  & SWIFT-EAGLE$^\ddagger$&  \protect\cite{2022MNRAS.516.3750H}   & --                     & $(1-42) \times 10^{5}$& --             & --             & --            & 0.30         \\
    Groups \textit{c10kHR}$^\star$  & Kinetic feedback      &  \protect\cite{2016MNRAS.461.1548B}   & --                     & $2.13 \times 10^{6}$  & --             & --             & --            & 1.40         \\
    Cluster                         & Fiducial jet-AGN      &  \protect\cite{2021MNRAS.506..488B}   & $6.86 \times 10^{7}$   & $1.94 \times 10^{7}$  & --             & --             & $2.5\, r_{\rm cell}$  &  4.00\\
    \bottomrule
    \end{tabular}
\end{sidewaystable}

\begin{figure}
\centering
    \begin{subfigure}{\textwidth}
        \centering
        \includegraphics[width=0.85\linewidth]{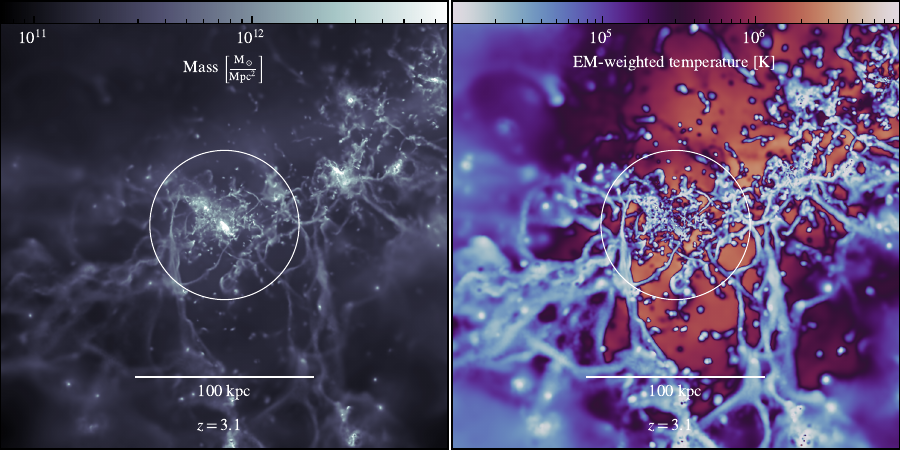}
    \end{subfigure}
    \begin{subfigure}{\textwidth}
        \centering
        \includegraphics[width=0.85\linewidth]{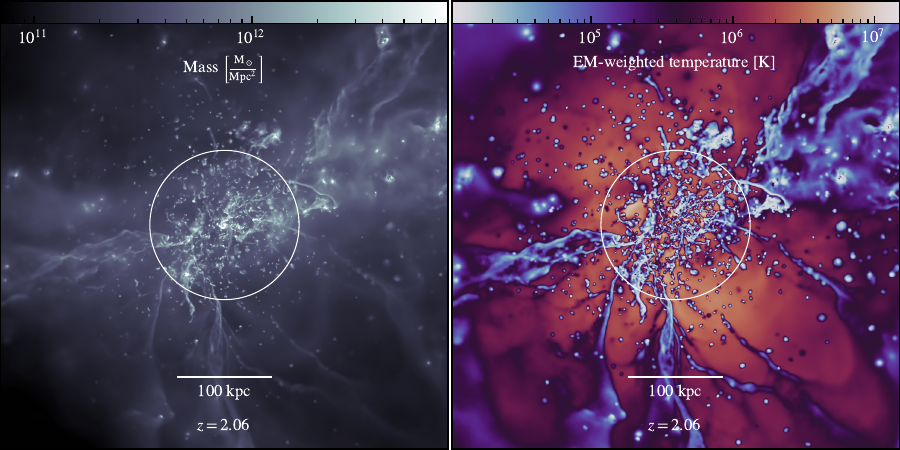}
    \end{subfigure}
    \begin{subfigure}{\textwidth}
        \centering
        \includegraphics[width=0.85\linewidth]{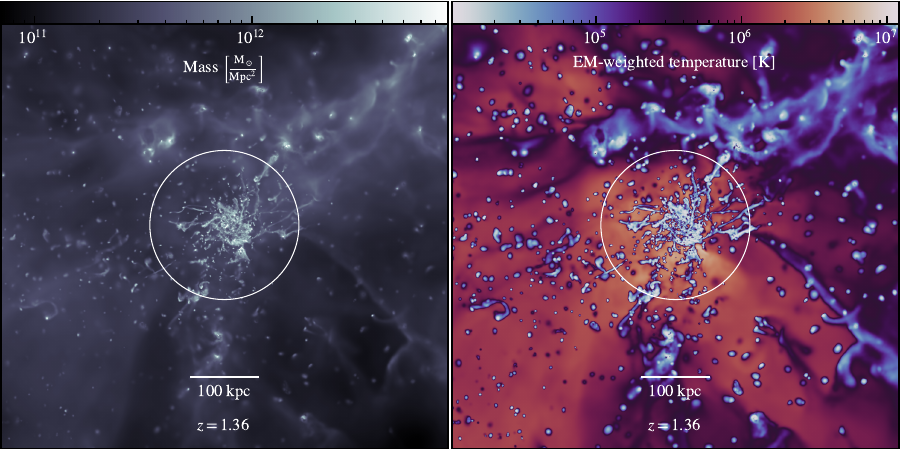}
    \end{subfigure}
\caption{Projection maps of the gas mass (left) and emission measure (EM)-weighted temperature (right) for the high-resolution proto-group. We present three moments in the early evolution of this system. \textit{Top:} during the initial rapid growth ($z=3.1$) several cold filaments are falling towards the centre of the halo. \textit{Mid:} immediately after the early accretion ($z=2.06$) the mass is increased to $M_{500}\approx 10^{12}$ M$_\odot$ and the gas is gravitationally heated to temperatures $\approx 10^{6.5}$ K. \textit{Bottom:} at $z=1.36$, the proto-group is about to experience a second accretion phase, driven by the cold gas filament on the top-right. In each map, we indicate $r_{500}$ with a circle and the physical 100 kpc scale (bottom). The spatial extent of the maps is $\pm 3\, r_{500}$ in both $(x,y)$ dimensions, and the depth of the projection is $\pm 3\, r_{500}$ around the centre of potential along the $z$-axis. We only select warm-hot gas with a temperature $T>10^4$ K.}
\label{fig:highres-group-evolution}
\end{figure}

In Fig.~\ref{fig:highres-group-evolution}, we present projection gas maps of the high-res group during three moments of its early evolution. On the left, we show the mass distribution of the warm-hot gas, with temperature above $10^4$ K, and on the right the gas temperature weighted by the emission-measure (EM) expressed as the square of the gas density ($\rho_{\rm g}$): EM $\propto \rho_{\rm g}^2$. At $z=3.1$ (top row), a large number of cold filaments falling towards the centre of the halo suggests a phase of intense growth, followed by the gravitational heating of the inter-galactic medium (IGM) at $z=2.06$ (mid row). Finally, we show the group a few 10s of Myr before starting a second accretion phase (bottom row), driven by the infall of the cold filament on the top-right. We show the projected gas contained within a cube of side $6\, r_{500}$ centred in the centre of potential. The spatial extent of the map, $\pm 3\, r_{500}$ is slightly larger than the largest radius used in the thermodynamic profiles ($2.5\, r_{500}$).

By introducing the high-res group, we aim to investigate the effect of emerging small-scale hydrodynamic processes on group and cluster atmospheres, bridging the gap between the scientific output of EAGLE-type simulations (with cosmological accretion) and some of the upcoming projects lead by members of the Virgo Consortium, striving to push the mass resolution down to $10^{4}$ M$_\odot$ (COLIBRE Collaboration, in preparation). In Tab.~\ref{tab:resolutions-comparison-other-works}, we compare the resolution parameters and sub-grid models of full-physics hydrodynamic simulations similar to those we introduced in Chapter \ref{chapter:5} (yellow rows) and the high-res group introduced in this chapter (orange row). The particle mass resolution of the latter is comparable to that of the ROMULUS-C cluster \citep{2019MNRAS.483.3336T} and approximately equivalent to the EAGLE 25 Mpc (high-res) volume, run with the EAGLE-Ref model \citep{2015MNRAS.446..521S}.

\section{The time-evolution of cluster metrics}
\label{ch6:metrics-evolution}

\begin{figure}
    \centering
    \includegraphics[width=\textwidth]{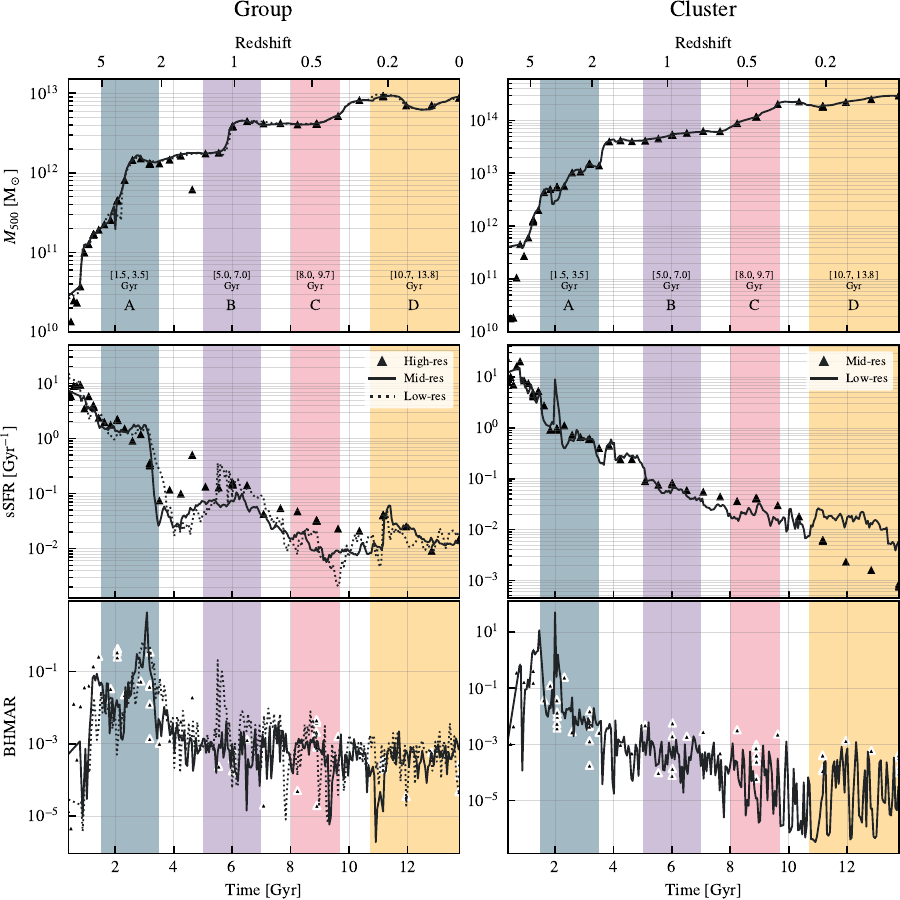}
    \caption{Evolution of the mass ($M_{500}$, top row), the specific star-formation rate (sSFR, mid row), and the black hole mass accretion rate (BHMAR, bottom row) with cosmic time (bottom scale) and redshift (top scale). \textit{Left}: properties of the group at low-res (dotted line), mid-res (solid line) and high-res (triangles). \textit{Right}: properties of the cluster at low-res (solid line) and mid-res (triangles). The four phases are indicated with coloured bands, with the time intervals reported on the top row. The intervals are fixed for both objects.}
    \label{fig:properties-evolution}
\end{figure}

In Fig.~\ref{fig:properties-evolution}, we show the evolution of the halo mass ($M_{500}$), the specific star-formation rate (sSFR) in Gyr$^{-1}$ and the black hole mass accretion rate (BHMAR) in units of the Eddington accretion rate $\dot{m}_{\rm Edd}$ computed for the central SMBH only. These quantities are defined as follows. $M_{500}$ is the total mass within $r_{500}$, as in previous chapters. Second, the sSFR is computed using star particles in a 100 kpc spherical aperture and considered over one time-step (i.e. instantaneous sSFR). Third, the Eddington accretion rate  \citep{1926ics..book.....E} used to scale the BHMAR is defined for a spherically symmetric matter distribution as
\begin{equation}
    \dot{m}_{\rm Edd} = \frac{4 \pi G m_{\rm P}}{\epsilon_{\rm r}~c~\sigma_{\rm T}}\cdot m_{\rm BH} \approx 2.218~{\rm M_\odot~yr^{-1}}\cdot\left( \frac{m_{\rm BH}}{10^8~{\rm M_\odot}} \right),
\end{equation}
as in Eq.~\eqref{eq:bh_eddington}.

In analogy to \cite{2019MNRAS.483.3336T}, we identified four evolutionary phases for the two objects considered. We specify these moments and outline the key properties of the group and cluster systems below. 
\begin{itemize}
    \item \textbf{Phase A} spans from 1.5 to 3.5 Gyr ($z\approx 2-5$). During this time, the \textit{proto-group} is a galaxy-size object with a median mass $M_{500}=1.31 \times 10^{12}$ M$_\odot$ in rapid growth. The BCG experiences intense star-formation activity until the end of this phase, when a sharp decline is correlated with the central black hole increase its mass accretion rate above $0.1\, \dot{m}_{\rm Edd}$. The \textit{proto-cluster} also shows a rapidly growing mass with a median $M_{500}=8.94 \times 10^{12}$ M$_\odot$. During this time, the proto-cluster has a similar mass as the fully-formed group at $z=0$. The sSFR of the BCG is in decline as the central black hole grows in mass and shows super-Eddington mass accretion rates up to $\approx 10\, \dot{m}_{\rm Edd}$. 
    
    \item \textbf{Phase B} spans from 5 to 7 Gyr ($z\approx 1$). The \textit{proto-group} experiences a second growth phase from $M_{500}= 2\times 10^{12}$ to $5\times 10^{12}$ M$_\odot$, together with an increase in sSFR. During this time, the mass accretion rate of the SMBH decreases down to $\sim 10^{-3}\, \dot{m}_{\rm Edd}$. The \textit{proto-cluster}, instead, does not experience a significant increase in mass, while its BCG becomes progressively less active (decreasing sSFR). The BHMAR also decreases to $\sim 10^{-3}\, \dot{m}_{\rm Edd}$, similar to the level of the proto-group.
    
    \item \textbf{Phase C} spans from 8.0 to 9.7 Gyr ($z\approx 0.5$). The \textit{group} is concluding a phase with little-to-no accretion ($M_{500} \approx$ constant between $0.5 < z < 1$). During this phase, the sSFR of the group's BCG decreases below the quenching threshold of $10^{-2}$ Gyr$^{-1}$. The BHMAR is approximately the same as in phase B. For the \textit{cluster}, we report a period of mass accretion, with $M_{500}$ increasing from $\approx 8\times 10^{13}$ M$_\odot$ to $\approx 2\times 10^{14}$ M$_\odot$. The sSFR of the BCG also decreases gently down to $\approx 2 \times 10^{-2}$ Gyr$^{-1}$ and the BHMAR also decreases below $\sim 10^{-3}\, \dot{m}_{\rm Edd}$.
    
    \item \textbf{Phase D} spans from 10.7 to 13.8 Gyr, equivalent to $0 \leq z < 0.3$. The \textit{group} shows a decrease in $M_{500}$ followed by a subsequent accretion due the infall of a substructure. The timing of this event correlates with a rise in sSFR of the BCG, which becomes active again ($\approx 4 \times 10^{-2}$ Gyr$^{-1}$). At the start of this phase, the BHMAR reaches its all-time low of $\sim 10^{-6}\, \dot{m}_{\rm Edd}$ (mid-res) before levelling just below $\approx 10^{-3}\, \dot{m}_{\rm Edd}$. The mass evolution of the \textit{cluster} also shows an accretion event at low redshift. The BCG becomes quenched by $z=0$ at both resolutions, although the quenching threshold is crossed earlier in the mid-res run than in the low-res one. The BHMAR decreases further towards $\sim 10^{-5}\, \dot{m}_{\rm Edd}$, which is $\sim 100$ times lower than the $z=0$ BHMAR of the group.
\end{itemize}

Overall, the metrics for the group and the cluster at different resolutions are consistent throughout the simulation, with three exceptions. Firstly, the sSFR of the BCG in the group at high-res is systematically higher than the other two resolutions between 5-10 Gyr (phases B and C). During this time, the BHMAR remains approximately constant $\approx 10^{-3}\, \dot{m}_{\rm Edd}$, suggesting that the higher sSFR in the high-res group may not have been favoured by a reduced black hole activity, and instead may arise from the ability to resolve hydrodynamics on smaller scales, associated with cold, star-forming gas. Secondly, the sSFR of the cluster during phase D is lower in the mid-res run than in the low-res one. While the BCG is quenched at $z=0$ in both cases (sSFR $\leq10^{-2}$ Gyr$^{-1}$), the mid-res cluster has a BCG 3 times less active than at low-res. Although the time-sampling of the BHMAR in the mid-res cluster is more sparse than for the low-res simulations, the black hole seems to remain on average more active in the mid-res simulation than at low-res. This scenario could be compatible with the residual AGN feedback quenching the BCG further. Finally, we report a discrepant value of $M_{500}$ in the group at high-res between phase A and phase B. This jump towards lower masses is an artifact due to the mis-identification of the central halo by the \velociraptor halo finder. We aim to correct for this effect in the future. We emphasise that this artifact does not affect any other result, since it is not included in the time intervals of interest.

\section{The time-evolution of entropy profiles}
\label{ch6:entropy-profiles-evolution}

\begin{figure}
    \centering
    \includegraphics[width=\textwidth]{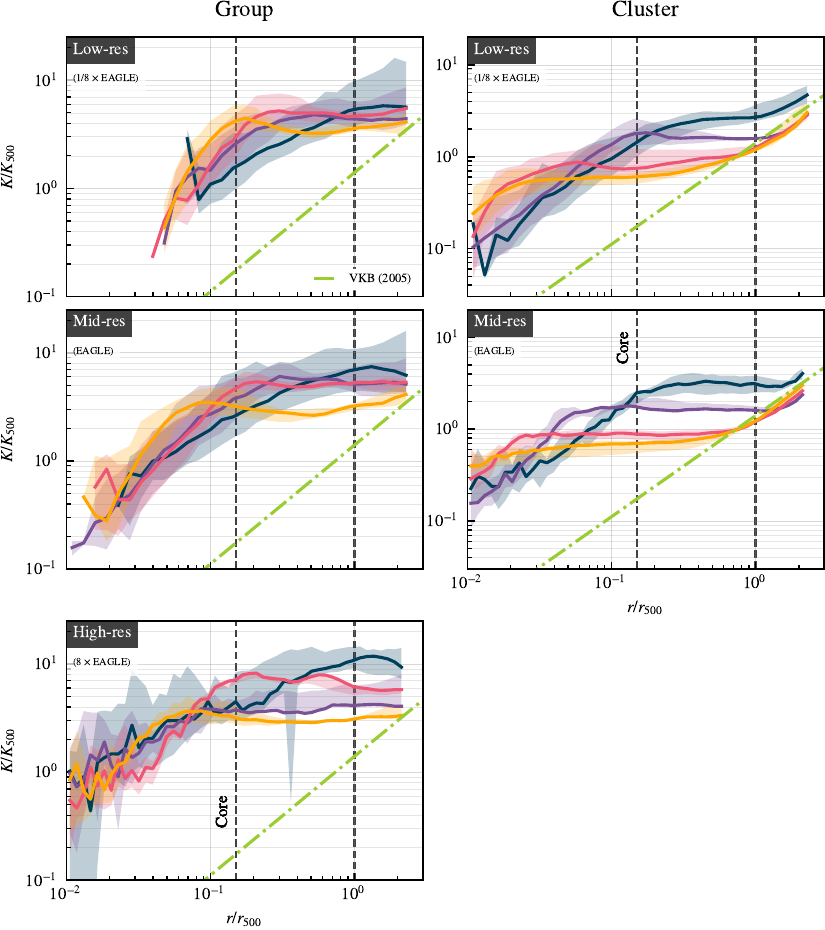}
    \caption{Scaled entropy profiles for the group (left) and the cluster (right) at different resolutions: low-res (top row), mid-res (mid row) and high-res (group only, bottom row). In each panel, the four profiles are the median of the snapshots within each phase and the shaded regions indicate the first and third quartile levels. The colour coding for the time-bins is the same as in Fig.~\ref{fig:properties-evolution}. The green dash-dotted line is the self-similar non-radiative entropy baseline \citep[VKB,][]{vkb_2005} and the vertical dashed lines are guidelines for the core radius ($0.15\, r_{500}$, as indicated) and $r_{500}$.}
    \label{fig:entropy-profile-scaled}
\end{figure}

Next, we select the entropy profiles of the objects during each of the four evolutionary phases defined in Section \ref{ch6:metrics-evolution}. For each phase, we compute the median profile, together with the first and third quartiles, and we present these results for the group and cluster in Fig.~\ref{fig:entropy-profile-scaled}. As in \cite{2023MNRAS.520.3164A}, the entropy profile $K(r)$ is obtained from the mass-weighted temperature $T(r)$ and the electron number density $n_{\rm e}(r)$ profiles, as
\begin{equation}
    K(r)=\frac{\mathrm{k_B}T(r)}{n_{\rm e}(r)^{2/3}}.
\end{equation}
We calculate $n_{\rm e}(r)$ from the gas density profile $\rho_{\rm g}(r)$, weighting the density of each gas particle by the chemical element contributions:
\begin{equation}
   n_{\rm e} = \rho_g ~ \sum_\epsilon Z_\epsilon \frac{f_\epsilon}{m_\epsilon},
\end{equation}
where the symbols are defined in Section \ref{sec:analysis_methods}. In our analysis, we only include the fully-ionised gas in the X-ray emitting phase, defined by the temperature cut $T>10^5$ K.

In Fig.~\ref{fig:entropy-profile-scaled}, we report the group entropy profiles on the left and the cluster on the right. The first row includes the low-res simulations, the second row at mid-res and the third row the high-res group. Here, we use the same colour coding as in Fig.~\ref{fig:properties-evolution}: the blue profile corresponds to the highest redshift ($z\approx 3$), purple to $z\approx 1$, pink to $z\approx 0.5$ and, finally, yellow to low redshift $z\in [0, 0.3]$. We will first describe the evolution of the entropy profile of the group and then we will discuss that of the cluster.

During phase A, the proto-group shows a power-law-like entropy profile between $r_{500}$ and the core radius ($0.15\, r_{500}$). The entropy at $r_{500}$, $K(r=r_{500})$, is higher than the self-similar gravitational baseline shown in green  \citep[VKB,][]{vkb_2005}, suggesting a departure from self-similarity which was also found in other simulations of group-sized objects \citep[e.g.][]{2014ApJ...783L..10G}. In phases B and C, the entropy near the core-radius at low- and mid-res increases, while it remains approximately constant at $r_{500}$. Finally, by $z=0$ the entropy plateau in the profile extends to smaller radii, while the slope of the profile inside $0.15\, r_{500}$ remains steep. The evolution of the profiles through the four phases is similar at all resolutions, however, we find that the plateau established at late times reaches smaller radii in the group at high-res and mid-res than at low-res.

Focusing on the cluster, we find that the high\hyp{}redshift median entropy profile is relatively steep, resembling that of the proto\hyp{}group in phase A. However, when comparing the entropy of the proto\hyp{}cluster in phase A to the entropy of the fully\hyp{}formed group in phase D, which both have a similar mass (though at different times), we note that the group has an extended plateau and a slightly larger entropy around the core radius, while the entropy profile of the proto\hyp{}cluster is monotonically decreasing towards smaller radii, on average. Between phase A and B, the entropy profile of the proto\hyp{}cluster decreases due to a shift in the $K_{500}$ normalisation deriving from the increase in $M_{500}$. As a consequence, the profile at $r_{500}$ approaches the gravitational self-similar baseline, as expected for cluster-sized objects with a gas distribution dominated by gravitational heating in a DM\hyp{}dominated potential \citep{vkb_2005}. While the scaled entropy at $r_{500}$ decreases, the entropy at the core radius increases, forming an extended plateau between the radii $(0.15-1)\times r_{500}$. In phase B, the isentropic distribution extends to smaller radii in the cluster at mid\hyp{}res than at low-res, similarly to what found in the group. Between phases B and C, the entropy level of the plateau at $0.15\, r_{500}$ decreases by a factor of 2, setting on levels very close to $K_{500}$. Only by $z=0$ do we measure a modest increase in the gradient, which nevertheless remains much shallower than the high\hyp{}redshift profile during phase A.

\section{Conclusions}
Our preliminary results do not indicate a clear correlation between merger events and the flattening of the entropy profiles. In fact, the entropy baseline forms from phase A to B both in the proto-group, which experiences high-accretion, and in the proto-cluster, which instead does not. Between phase B and C, we measure the same variation in the shape of the entropy profile, with the group captured during a quiescent phase and the cluster undergoing accretion.

The merger event associated with the \textit{proto-group} in phase B, however, coincides with a rise in sSFR \citep[e.g.][]{2008MNRAS.385L..38M, 2018MNRAS.476.2591V}. An enhanced star formation activity can then correlate with cold low-entropy gas being converted into stars, and thus decreasing the density and enhancing the entropy in the central region \citep[see also the No-SN model of][]{2023MNRAS.520.3164A}. The rise in the entropy profile at $0.15\, r_{500}$ for the group appears to be compatible with this hypothesis, although further analysis is required to probe the causal connection between enhanced sSFR and $K(r=0.15\, r_{500})/K_{500}$. An additional contribution could be associated with the effects of AGN feedback from of the central SMBH during phase A, which could remove the low-entropy gas from the central region at early times.

Throughout the evolution of the \textit{cluster}, merger events appear to have a smaller impact on the sSFR of the BCG. This effect manifests itself as a slower rate of decrease (see phases B and C) of sSFR rather than an enhancement as phase B of the group. For this reason, we do not find evidence of correlations between star formation and entropy generation in the cluster core.

%% file: Chapters/Chapter7.tex
\chapter{Conclusions and future work}
\label{chapter:7}

\section{Summary}
\label{sec:final-summary}

In cosmology and extra-galactic astrophysics, numerical simulations have become irreplaceable tools that observers and theorists use to solve complex problems and lead the path to new breakthroughs. As approximate models of the Universe, however, simulations present advantages and disadvantages. On one hand, they empower scientists to experiment hypothetical scenarios in a controlled virtual environment and predict astrophysical phenomena difficult to probe with observations. On the other hand, insights from simulations are prone to being extrapolated, leading to skepticism and misinterpretation. In Chapter \ref{chapter:1}, we begin this thesis by framing the role of simulations in science through the lens of philosophy. We highlight that simulations, regarded as epistemic tools, can bridge the \textit{knowledge by acquaintance} (empirical) to the \textit{knowledge by description} (fundamental, universal). We noted the validation of the outcomes is a pivotal step in this epistemic framework, and can dictate the success of a simulation, as well as the balance between trustworthiness and skepticism.

Then, in Chapter \ref{chapter:2} we outlined the processes leading to the formation of structures in the Universe. Starting from seed primordial perturbations, we describe how these evolve in the linear regime at early times and then in the non-linear regime. Crucially, numerical simulations are used to probe the collapse of structures in the non-linear regime at late times. Moreover, modern simulations combine the calculations of non-linear structure formation with hydrodynamics and baryonic physics, establishing direct comparisons between the synthetic and observational data.

As a next step, in Chapter \ref{chapter:3} we illustrate the technical details of numerical simulations. In particular, we demonstrate the production pipeline of zoom-in simulations, which are used to model individual objects in detail by increasing the particle mass resolution in a region of interest (where the object is found), while degrading the resolution of the surrounding space. Given our interest in modelling groups and clusters of galaxies, we produced a catalogue of 27 objects, sampled from a 300 Mpc cosmological volume, and re-simulated them using the zoom-in technique. Our pipeline consists of 14 main steps, some of which required the development of novel methods to mitigate known shortcomings. To mitigate the contamination from low-resolution dark matter particles invading the high-resolution region, we developed the \textit{topological closure} technique. This method combines a set of three operations to compute a spatial mask which includes particles in the initial conditions which may be located in topologically disconnected sub-domains. Finally, we assessed the weak-scaling performance of the \swift code in solving a purely hydrodynamic problem of a Kelvin-Helmholtz (KH) instability in 3D using the COSMA-7/8 HPC systems. For the test with the largest number of particles, $N_{\rm p}\approx5 \times 10^{11}$, we measures a increase in run-time of $\approx 5$\% relative to one node, which makes \swift one of the hydrodynamic codes with the highest parallel efficiency.

In Chapter \ref{chapter:4}, we present our prediction of the kinetic SZ effect arising from the bulk rotation of the ICM in simulated galaxy clusters. This effect, referred to as the rotational kinetic SZ (rkSZ), was originally predicted by \cite{Chluba2001Diploma, CC02} and \cite{CM02} and modelled using low-mass relaxed clusters from the MUSIC simulations \citep{2017MNRAS.465.2584B}. Our study provides the first prediction of the rkSZ amplitude for high-mass clusters from the MACSIS simulations \citep{macsis_barnes_2017}. We study the dependence of the rkSZ signal amplitude on global cluster properties and dynamical state metrics, finding a maximum signal of $\gtrsim$100 $\mu$K, i.e. approximately 30 times stronger than predicted by \cite{CC02} using self-similar analytic models. This work has been submitted to \mnras~ and is currently under review.

Chapter \ref{chapter:5} probes a tension between the distribution of pseudo\hyp{}entropy measured from observations and predicted by simulations of groups and clusters of galaxies. Firstly, we highlight that most recent hydrodynamic simulations systematically over\hyp{}predict the entropy profiles by up to one order of magnitude, leading to profiles which are shallower and higher than the power\hyp{}law like entropy profiles which have been observed. We discuss the dependence of the entropy distribution on different hydrodynamic and sub\hyp{}grid parameters, namely artificial conduction, radiative cooling from metals, SN and AGN feedback and energy distribution schemes. Not only did this comparative study isolate the fundamental cause of the high entropy levels in the group and cluster cores, but it also provides a library of sub\hyp{}grid variations of the EAGLE model which could become pivotal for future work. The results from this chapter were published on \mnras~\citep{2023MNRAS.520.3164A}.

Following from Chapter \ref{chapter:5}, in Chapter \ref{chapter:6} we briefly illustrate the evolution of global properties of the group and cluster of galaxies, together with the entropy profiles as a function of cosmic time. After identifying four key phases in the evolution of these systems, we report power-law-like entropy profiles at high redshift for both objects. At late times, however, an entropy plateau develops and alters the shape of the profile. This effect occurs at low resolution and mid (EAGLE) resolution. We introduce an additional run of the group at 8 times the resolution of EAGLE, and report a similar evolution of the entropy distribution. This work forms the basis of a paper in preparation.

\section{Future work}
\label{sec:future}

\subsection{Load balancing in HPC architectures}
\label{sec:future:zoom-load-balance}
Because of their multi-resolution layout, zoom-in simulations can be challenging to produce at scale. For instance, Virgo Consortium's recent flagship SIBELIUS-DARK project \citep{2022MNRAS.512.5823M}\footnote{The SIBELIUS-DARK simulation consisted in a zoom-in region with 131 billion collision-less DM particles, evolved using the \swift code running on 160 compute nodes of the COSMA-7 HPC system \citep{2022MNRAS.512.5823M}.} required significant modifications to the \swift code in order for the key simulation to run to completion. To extend the calculation across many compute nodes, the simulation volume is divided into \textit{cells} using the Parallel Graph Partitioning and Fill-reducing Matrix Ordering (\texttt{ParMETIS}) library and then the individual tasks were managed by the \texttt{MPI} library. If the simulation volume is partitioned with a regular cubic grid, however, the high-resolution region containing most of the particles will likely be completely enclosed within a single cell (or a few at most), causing most of the computational workload to be allocated to the \texttt{MPI} \textit{rank} assigned to that particular group of cells. The other \texttt{MPI} ranks, instead, would only receive tasks involving the less-demanding time-stepping of boundary DM particles. In this scenario, the \textit{imbalance} arising from the excessive load assigned to the ``high-resolution`` ranks compared to the ``boundary`` ranks increases the total run-time, since the boundary ranks are forced to wait for the high-resolution (slower) ones to complete their tasks.

In response to the increasing computational requirements for higher-resolution zoom simulations of galaxy clusters, a team of \swift developers led by William Roper (University of Sussex, UK) is currently perfecting new domain decomposition strategies to mitigate load imbalances. This method introduces a finer grid within the high-resolution region and, based on the decomposition of the finer mesh, distributes the \textit{sub-tasks} for high-resolution particles across the available \texttt{MPI} ranks. Preliminary tests of the refined mesh on FLARES dark-matter-only cosmological zoom-in set-ups \citep{2021MNRAS.500.2127L} indicate an speed-up factor of 2-3 over the \texttt{master} branch of \swift. Further tests on set-ups where the high-resolution volume is much smaller than that of the parent volume (e.g. C-EAGLE, \citealt{ceagle.barnes.2017} or MACSIS, \citealt{macsis_barnes_2017}) will be conducted in the near future to probe the numerical stability of the zoom-enhanced \swift code and accelerate the scientific progress of galaxy cluster simulations towards higher resolutions and a more comprehensive array of physical processes with projects such as BEEHIVE.

\subsection{Environmental sustainability}
\label{sec:future:environmental-sustainability}

Several research studies have attempted to quantify the carbon footprint of astronomy research and identified potential solutions to reduce their environmental impact. In the context of astronomy and astrophysics, \cite{2020NatAs...4..843S} and \cite{2020NatAs...4..823B} attributed a large fraction of carbon emissions to business travel, while \cite{2022NatAs...6..503K} highlighted the contribution from building and managing research facilities. Focusing on the Netherlands, whose policy and governance have set the pace for the European Union's commitment to cut carbon emission, \cite{2021NatAs...5.1195V} found that astronomy research is already benefiting from renewable energy sources, however, more support from political institutions will be necessary to take the next steps towards the net-zero target \cite[see also][]{2021NatAs...5..864W, 2021NatAs...5..857B}.

High-performance computing (HPC) systems are large consumers of energy. As scientists require progressively larger systems to perform increasingly complex calculations, HPC usage on the rise, while the HPC power consumption raises concerns about the environmental impact through carbon dioxide (CO$_2$) emissions. Carbon emissions in HPC can be attributed to various factors, including the power consumption of the computing hardware, cooling infrastructure, and data center facilities. As HPC systems continue to scale up in size and complexity, it is essential to address the sustainability challenges associated with carbon emissions to ensure the long-term viability of this crucial technology.

In a recent study, \cite{2020NatAs...4..819P} stressed that the CPU usage, approximately proportional to the energy consumed, highly depends on the software stack used to run simulations. Using the Astrophysical Multipurpose Software
Environment \citep[AMUSE,][]{2018A&A...616A..85P}, it has emerged that interpreted languages, such as \texttt{Python} are far slower and more energy-consuming than compiled languages, such as \texttt{C} and \texttt{C++}. Given these results, \swift should be considered a ``virtuous`` code. However, the benchmarks from the AMUSE software suite were only run on a workstation and not deployed on a large-scale HPC system, as would be the case in massively-parellel cosmological simulations. Moreover, the results from the interpreted languages do not examine the impact of run-time optimisations, which in \texttt{Python} could be enabled via the \texttt{Numba} and \texttt{Cython} libraries.

\begin{figure}
    \centering
    \includegraphics[width=\textwidth]{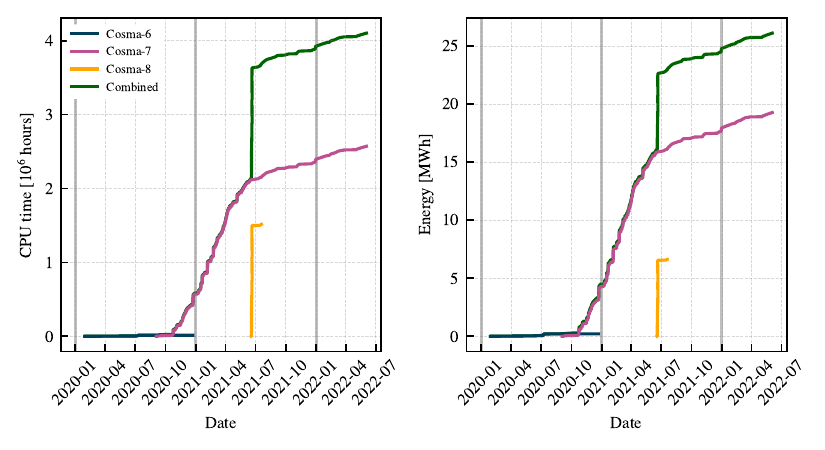}
    \caption{\textit{Left:} cumulative CPU hours used on the COSMA machines between January 2020 and July 2022. The COSMA-6 usage is indicated in blue, COSMA-7 in pink and COSMA-8 in yellow. We also show the total cumulative usage in gree. \textit{Right:} the energy (in MWh) used by the CPUs, assuming they operate at base frequency with all the cores active, computed from the thermal design power provided by the manufacturers (see Tab.~\ref{tab:cosma-cpus}). The vertical grey lines indicate the start of each calendar year.}
    \label{fig:cosma-usage}
\end{figure}

Here, we outline a possible extension of previous studies assessing the environmental impact of HPC in astronomy, starting by an assessment of the computational resources used to deliver the research presented in this document. In the left panel of Fig.~\ref{fig:cosma-usage}, we show the cumulative CPU hours on the COSMA-6/7/8 HPC systems used for our research between January 2020 and July 2022. We define the CPU hours used by a task as $t_{\rm CPU} = t_{\rm run}\,  N_{\rm cores}$, where $t_{\rm run}$ is the run-time and $N_{\rm cores}$ is the number of physical CPU cores used for the calculation. During the time considered, the COSMA-6 system was used at the start of the research program for preliminary analysis of existing simulations; later in 2020, the COSMA-7 usage corresponds to the production of the zoom-in simulations in \cite{2023MNRAS.520.3164A} and the efforts for calibrating the sub-grid model as part of the EAGLE-XL zoom-assisted tuning. The COSMA-8 usage is limited to a few days and is associated with the \swift performance profiling in Section~\ref{sec:computational-performance}, when the entire HPC system was made available for testing during the early commissioning phase.

\begin{table}
    \centering
    \caption{Summary of the CPU architectures for the COSMA HPC systems used to run the simulations in this work. From left to right: the name of the HPC system; the number ($N_{\rm CPU}$) of physical chips mounted on the motherboard of the compute nodes; the manufacturer and make of the CPU, associated with a link to the complete data sheet; the number ($N_{\rm cores}$) of physical cores in each CPU; the thermal design power ($P_{\rm tdp}$), defined as the average power dissipated by the processor when operating at base frequency with all the cores active; the the thermal design power per physical CPU core ($P_{\rm tdp}/N_{\rm cores}$).}
    \label{tab:cosma-cpus}
    \begin{tabular}{lclccc}
    \toprule
        HPC system  & $N_{\rm CPU}$     & CPU make & $N_{\rm cores}$      & $P_{\rm tdp}$ & $P_{\rm tdp}/N_{\rm cores}$    \\
                    &        &          &          & (W) & (W)          \\
    \midrule
        COSMA-6 & 2 & Intel Xeon E5-2670    (\href{https://ark.intel.com/content/www/us/en/ark/products/64595/intel-xeon-processor-e52670-20m-cache-2-60-ghz-8-00-gts-intel-qpi.html}{info})    & 8 & 115 &  14.38    \\
        COSMA-7 & 2 & Intel Xeon Gold 5120  (\href{https://www.intel.com/content/www/us/en/products/sku/120474/intel-xeon-gold-5120-processor-19-25m-cache-2-20-ghz/specifications.html}{info}) & 14 & 105 &  7.50    \\
        COSMA-8 & 2 & AMD EPYC 7H12         (\href{https://www.amd.com/en/products/cpu/amd-epyc-7h12}{info})   & 64   &  280       &  4.38 \\
    \bottomrule
    \end{tabular}
\end{table}

The COSMA compute nodes are configured to run at the base frequency and, therefore, the power used by the CPUs can be quantified by the the thermal design power ($P_{\rm tdp}$), measured by the manufacturers as the power dissipated when all CPU cores are active. In Tab.~\ref{tab:cosma-cpus}, we list the CPU makes installed in the COSMA machines. All nodes are equipped with a dual-CPU motherboard. While COSMA-6 and COSMA-7 use Intel Xeon CPUs, AMD chips were used for COSMA-8. Our summary of $P_{\rm tdp}$ per CPU and $P_{\rm tdp}$ per core indicates that, while the AMD chips use more power, they enclose 8 times more cores than the Intel chips in COSMA-6, resulting in a reduction of 70\% in the power per core. As a next step, we use $P_{\rm tdp}$ to compute the cumulative energy corresponding to the cumulative CPU hours, weighting the CPU usage for every machine by the $P_{\rm tdp}$ specifications of the corresponding chip make. The cumulative\footnote{In this section, we denote the cumulative quantities with a asterisk ($^*$).} energy is described by
\begin{equation}
    E^{*}= t^{*}_{\rm run}\,  N_{\rm cores}\, \frac{P_{\rm tdp}}{N_{\rm cores}} \equiv t_{\rm run}\, P_{\rm tdp},
\end{equation}
where $t^{*}_{\rm run} \equiv \int t_{\rm run} d\tau$ is the cumulative run-time integrated over the calendar dates $\tau$. The cumulative energy is shown in the right panel of Fig.~\ref{fig:cosma-usage}, where we report a total consumption of over 25 MWh. Given that the sharp rise in COSMA-8 usage was concentrated in one day of run-time, our results set a lower bound of $\approx 6$ MWh on the daily energy usage of the CPUs in COSMA-8 during its commissioning phase.

\begin{figure}
    \centering
    \includegraphics[width=0.75\textwidth]{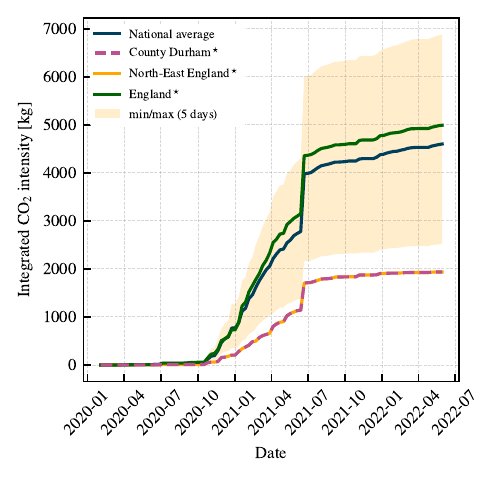}
    \caption{Estimates of the carbon emissions from the combined COSMA HPC usage in Fig.~\ref{fig:cosma-usage} assuming the national UK average CO$_2$ intensity (blue), the national UK minimum and maximum intensity in a 5-day time window (orange band), the County Durham (pink), the North-East England region (orange) and the England average (green). $^\star$The \textit{regional} carbon intensity (County Durham, North-East England and England) is estimated based on net power flows across regional boundaries processed with a set of machine learning models \citep{staffell2017measuring}.}
    \label{fig:emissions}
\end{figure}

Finally, we use the Carbon Intensity API\footnote{The Carbon Intensity API is accessible at \href{https://carbonintensity.org.uk/}{https://carbonintensity.org.uk/} and can be used with a \texttt{Python} interface to query a remote database.} to obtain estimates for the carbon intensity produced in the UK at a national and regional level. The carbon intensity $I=dm_{\rm CO_2}/dE$ expresses the mass of CO$_2$ ($m_{\rm CO_2}$) associated with the production of one kWh of energy. Considering the mean weekly carbon intensity $\bar{I}^{\rm [1~week]}$ between the date range specified above, we estimate the mass of CO$_2$ as
\begin{equation}
    M_{\rm CO_2} = \bar{I}^{\rm [1~week]} \times E,
\end{equation}
noting that the $\bar{I}^{\rm [1~week]}$ depends on time, with an update frequency of 1 week, and geographical area.
Finally, the integrated (or cumulative) CO$_2$ intensity is given by integrating $M_{\rm CO_2}$ over time, so that
\begin{equation}
    M^*_{\rm CO_2}(\tau) = \int_{\tau_0}^{\tau} \bar{I}^{\rm [1~week]}(\tau')\, E(\tau') ~ d\tau',
\end{equation}
where the start date is $\tau_0=$ January 2020 and the end date is in the range $\tau\in[$January 2020, July 2022$]$.

We present the integrated carbon emission in Fig.~\ref{fig:emissions}, where we consider $\bar{I}^{\rm [1~week]}$ (or simply $\bar{I}$) computed across the UK energy network (blue), limited to England (green), limited to the North-East of England (orange) and restricted to Country Durham (purple dashed). In addition, we show the carbon emission computed from the national average assuming the energy was utilised when $\bar{I}$ was at its minimum value in a 5-day time window and at its maximum (orange band). As the COSMA systems are hosted by Durham University, their energy usage is associated to the estimate for the County Durham carbon intensity. Under these conditions, the research activities discussed in this thesis are responsible for the production of at least 2 tons of CO$_2$, approximately equivalent to that associated with one round trip between London Heathrow and New York JFK for one passenger in Economy class on board of a Boeing-777 aircraft. If we assume the national average, the total mass of CO$_2$ increases to 4.5 tons. In fact, a large fraction of energy produced in the North-East comes from wind farms. As a renewable energy source, wind power has low values of $\bar{I}$. In the South of England, the power network is not dominated by renewable energy, leading to a larger $\bar{I}$ which increases the overall estimate of the carbon emission.

We stress that our estimates are based on $P_{\rm tdp}$, and therefore only consider the CPU usage alone. However, the COSMA HPC systems include numerous additional devices, such as storage units, cooling, and interconnect switches, which are always used when performing tasks and should be accounted in the overall calculation. Our analysis provides a proof-of-concept framework for estimating the carbon emissions associated with computational cosmology research. We encourage developing this approach further to optimise the usage of computational resources and monitor the effect of actions taken by scientists and policy-makers towards future sustainability goals. 

\subsection{Cosmological forecasts for the rkSZ effect}
\label{sec:future:rkSZ}
Future observations of CMB foregrounds through the SZ effects will continue to progress in the era of precision cosmology. In this context, studies of high-order $y$-type CMB spectral distortions, such as the relativistic and rotational kinetic SZ effects are likely to play a pivotal role in upcoming survey designs and observational programs. In Chapter \ref{chapter:4}, we presented the most up-to-date study of the rotational kSZ (rkSZ) effect using over 300 simulated clusters from the MACSIS simulations \citep{2023arXiv230207936A}. Then, we discussed how our study could serve as a \textit{proof-of-concept} for delivering even more precise predictions of the rkSZ amplitude using upcoming large-volume simulations.

Currently being developed by a large team in the Virgo Consortium, the FLAMINGO project \citep{flamingo_schaye2023} will provide a unique data set for study high-order SZ contributions thanks to a vast statistical sample of groups and clusters of galaxies. In particular, we envisage the following four potential areas of development:
\begin{itemize}
    \item \textit{Dependence on cluster properties}. In Section \ref{sec:cluster-properties} we split the MACSIS sample into a high- and low-value subset of fundamental cluster properties (Fig.~\ref{fig:corner-plot-basic-properties}) and metrics of dynamical state (Fig.~\ref{fig:corner-plot-dynamical-state}). With large-volume simulations, it will become possible to split the sample into many more smaller bins, enabling a detailed study of the dependence of the rkSZ amplitude as a function of cluster properties, as well as the variance and the correlations.

    \item \textit{Temperature power spectrum}. In \cite{2023arXiv230207936A}, we outline a method to compute the temperature power spectrum associated with the rkSZ signal. Since our method relies on scaling relations of the rkSZ amplitude with cluster properties, the analysis proposed in the point above could allow a precise determination of the rkSZ power spectrum from the 1-halo term. Moreover, the 2-halo term could be calculated from the the FLAMINGO lightcone outputs, allowing a powerful prediction of this effect for observations with upcoming CMB stage-4 experiments.   
    
    \item \textit{Extended mass range and sample completeness}. The rkSZ effect is most significant in high-mass clusters \citep[e.g.][]{2002A&A...396..419C}, such as those in the MACSIS sample \citep{macsis_barnes_2017}. For this reason, in Chapter \ref{chapter:4} we decided to focus our analysis on the MACSIS simulations. However, MACSIS-like objects are much rarer in the Universe than lower-mass clusters and groups of galaxies ($M_{500}\sim 10^{13}$ M$_\odot$). Although we predict that low-mass objects contribute less to the SZ signal \citep{2023arXiv230207936A}, their greater abundance suggests a compelling case for extending our study to the low-$M_{500}$ cluster population. The BAHAMAS simulations \citep{2017MNRAS.465.2936M} is a suitable candidate for extending our study to lower masses. Given the status of readiness of the FLAMINGO project, we recommend using the FLAMINGO large-volume simulations, which provide an updated sub-grid model and a complete halo-mass function throughout the domain.

    \item \textit{Extreme objects}. We reported that clusters constituting the high-spin parameter tail of the log-normal distribution (Fig.~\ref{fig:corner-plot-basic-properties}) could enhance the rkSZ signal significantly more than previously estimated. For this reason, clusters with $\lambda\approx0.1$ could be subject of an interesting case study, although they are extremely rare. The large cosmological volume of the FLAMINGO simulations is expected to contain several such \textit{fast rotators}, enabling a detailed study of this peculiar population.
    
\end{itemize}

\subsection{Entropy amplification through shocks}
\label{sec:future:shocks}

When shock waves propagate through gas, the local thermodynamic properties are altered. The Rankine–Hugoniot jump conditions predict that the entropy of the gas increases irreversibly as it moves through a shock front. Measuring the thermodynamic properties across before (subscript 2) and after a shock-front crossing (subscript 1), the specific entropy $s$ varies as
\begin{equation}
    s_1 - s_2 = c_{\rm V}\, \log\left( \frac{P_2}{P_1} \right) - c_{\rm P}\, \log\left( \frac{\rho_2}{\rho_1} \right),
\end{equation}
where $c_{\rm V}$ is the specific heat capacity at constant volume (isochoric) and $c_{\rm P}$ at constant pressure (isobaric), with the adiabatic index $\gamma = c_{\rm P} / c_{\rm V}$ \citep{LANDAU1959310}. Therefore, we expect shocks propagating through the cluster atmosphere to directly alter the entropy distribution and interact with a potentially pre-existing cool core (CC).

The effect of shocks in the CC/NCC cluster populations was studied by the Dianoga project team \citep{2021MNRAS.507.5703P}. However, we showed evidence that the EAGLE model produced high-entropy levels more systematically \citep[see also][]{ceagle.barnes.2017}, suppressing the CC population entirely \citep{2023MNRAS.520.3164A}. Using the time\hyp{}differential of the velocity divergences ($S=\partial_t \nabla \cdot \mathbf{v}$) provided by the \textsc{Sphenix} SPH scheme \citep{sphenix_borrow2022}, we plan to investigate the \textit{correlation} and the possible \textit{causal connection} between high-entropy gas and shock fronts. If such connection exists, it would strongly motivate further study of the evolution of the shock\hyp{}heated particles following cosmological accretion events (mergers) and feedback\hyp{}induced shocks. Zoom-in simulations such as those we produced and described in Chapters \ref{chapter:5} and \ref{chapter:6} will contribute in answering the question of which process(es) are the major drivers of the entropy amplifications in the EAGLE-like and other galaxy formation sub-grid models.

\section{Scientific impact}
\label{sec:impact}

The research presented in this document probes two fundamental physical quantities: \textit{angular momentum} and \textit{entropy}. 

Angular momentum is generated during the collapse of large-scale structures from a net-zero initial condition, and is then transferred to smaller objects, such as galaxy clusters. This step is crucial to trigger a cascade of angular momentum transfer to even smaller scale, such as thin-disc galaxies, star-formation regions, down to planetary systems. Focusing on one step particular step of this angular momentum cascade, we provide the scientific community with a comprehensive framework for probing the angular momentum at a \textit{cluster} level and study its distribution in simulations of massive clusters. Given the nature of the rotational contribution to the SZ signal, our scientific contribution can find application in (i) cluster astrophysics, e.g. with studies of the gas turbulence and (ii) precision cosmology, with the subtraction of CMB foregrounds (Simons Observatory and CMB Stage-4 experiments).

In complex multi-component systems such as groups and clusters of galaxies, the distribution of entropy can provide clues about numerous processes as a single thermodynamic variable. The entropy-core problem could arise as the consequence of missing physics or the absence of processes which cannot be resolved at current resolutions. In addition to searching for the fundamental cause of high entropy levels in groups and clusters of galaxies, our research encourages future studies and establishes a target for assessing of the entropy distribution in next-generation high-resolution simulations. 

In conclusion, the field of hydrodynamic simulations of galaxy formation is experiencing a rapid growth on two fronts: (i) devising new solution to overcome computational challenges and (ii) testing scientific hypotheses and exposing the shortcomings of our current understanding of the Universe. In addition to fundamental physical laws implemented \textit{ab initio} (e.g. gravity and hydrodynamics), the role of observational findings in the calibration of sub-grid models is necessary to steer the development of simulations in a direction that maximises their predictive potential. Given that the design of simulation set-ups is influenced by observational results leads to the definition of numerical simulations as \textit{digital twins} of the observed Universe. The concept of digital twins has recently grown in popularity amongst industry partners and corporate strategy firms \citep[e.g.][]{nguyen2022toward, kenett2022digital, sadri2023integration}. When planning a sub-grid calibration strategy or testing the validity of a theoretical result, approaching cosmological simulations as ``digital twins`` could facilitate or promote two aspects of the future developments: (i) formulating scientific questions and compelling science cases, and (ii) introducing innovative methods with the potential to shift the way upcoming simulations will be produced and analysed.   

%% file: Appendices/AppendixA.tex
\chapter{Parameter files}
\label{appendix:1}

\begin{lstlisting}[language=bash,caption={Parameter file supplied to \textsc{IC\_Gen} to produce the initial conditions for the parent simulation \texttt{EAGLE-XL\_L0300N0564}.},label={app1:parent-sim-ics}]
[dc-alta2@login7a [cosma7] EAGLE-XL_L0300N0564]$ cat EAGLE-XL_L0300N0564.inp

1003007067 Version tag - do not edit this line.
------------------------------------------------------------------------------
# Only the values of the parameters should be altered. Changing the number of
# lines in this file (including comments like these) or swapping lines may
# lead to errors on execution, or worse nonsensical output.
# Note however lines beginning with a '%' are ignored. ARJ 16/08/2016
------------------------------------------------------------------------------
---------**********  Put text description of ICs below in quotes ********* ---
'EAGLE-XL L0300N0564 box'
------------------------------------------------------------------------------
# The three parameters below define the type of output
------------------------------------------------------------------------------
2                    ! Output format: (2 - hdf5) (1 - Gadget) (0 - ARJ BINARY)
2                    ! Indexing: (0 - 4 byte),(1 - 8 byte),(2 - ph), (3 -CSI)
21                   ! nbit - used for P-H indices
1                    ! Endianess:(0 - Big)(1 - Little)(2 - Native)
------------------------------------------------------------------------------
# Gadget file parameters (ignored for ARJ BIN FORMAT-but required to be here)
# The values chosen for the coordinate shift become the new origin in the
# output file.
# The create SPH and species division factors should be given in the form
# of log10(mparticle division/mass of all particles) - i.e. negative reals.
------------------------------------------------------------------------------
4                              ! Precision of output file (4-single) (8-double)
------------------------------------------------------------------------------
0.0 0.0 0.0                    ! Coordinate shift in Gadget Units (0,0,0)=none
0                              ! App sph truncation about origin?  0 == No
1.25                           ! Radius of spherical truncation
0                              ! Add SPH particles?  0 - No  1 - Yes
0.15742424178301725            ! Omega_b/Omega_0
-10.2                          ! Create SPH particles for dm below this mass
1.0                            ! Split factor:  0 - place SPH on dark matter
1                              ! Number of species (excluding gas == sp-0)
-9.5                           ! Division sp-1 and sp-2 (if present)
-9.0                           ! Division sp-2 and sp-3 (if present)
-8.0                           ! Division sp-3 and sp-4 (if present)
-6.0                           ! Division sp-4 and sp-5 (if present)
--short_param.inp-------------------------------------------------------------
# Resimulation parameters below:
 -----------------------------------------------------------------------------
1                              ! 2lpt ICS:  0-No 1-Yes (Cosm)  2-Yes (Resim)
0                              ! Write out Lagrangian positions 1-Yes,0-No
0                              ! For Resim ics only:  1: output linear density; 0 nothing
------------------------------------------------------------------------------
-1                             ! 0 - Import file : -1 -  cosm glass file:  > 0 the 1-d cubic grid size
'../../glass/Eagle_glass_file_47'  ! (Imported) Part load basename
12                                 ! Part load Rep factor
'./data/EAGLE-XL_L0300N0564_ICs'   !output file base name
------------------------------------------------------------------------------
'../../EAGLE_XL_powspec_18-07-2019.txt'         ! Input Power spectrum to use
------------------------------------------------------------------------------
# ********* Details of the parent simulation *****************
202.98           lbox              ! Parent box size in Mpc/h
0.3111, 0.6889   omega0,lambda0    ! Cosmological parameters of parent sim
0.8102           sigma8_p          ! sigma_8(z=0) of parent simulation
0.6766           Hubble_parameter  ! Redshift zero Hubble constant in units 100 km/s/Mpc
127.0            zshift            ! Redshift of the required displacement field
------------------------------------------------------------------------------
# ********* Phases used: place the panphasian descriptor on the line below ***
 [Panph1,L18,(74412,22732,260484),S3,CH1799108544,EAGLE-XL_L0300_VOL1]
0                             ! Number of constraint files to import
#-------------- List constraint phase descriptors and (newline) file path, and levels to use
-----------------------------------------------------------------------------
1             igrid           ! Fourier grid number
0.5           xcentre         ! x - Centre and size in units of lbox
0.5           ycentre         ! y -
0.5           zcentre         ! z -
1.0           side0           ! side-length
179406144     nparent         ! Number of particles in parent simulation
1536          ndim_fft        ! Dimension of fourier mesh used
0             cic_correct     ! Correct for CIC assign  1 - Yes
1             iquad           ! Quadratic interp - 1 -Yes(Cancels cic).
0             idep            ! Number of parent fourier grid
------------------------------------------------------------------------------
0                             ! Multigrid initial conditions?    0 : No   1: Yes
------------------------------------------------------------------------------
           1  basis function choice: 1 - Use SET8 basis function :0- use SET1
          12  Openning inverse angle (integer) used for multigrid ics only.
          1 1 1 1 1 1 1 1 1 1 1 1 1 1 1 1 Include contribution from levels - 1=YES
------------------------------------------------------------------------------
 #*********** Location of high resolution region *****************
 29.223        xcentre  (Mpc/h)
 18.618        ycentre  (Mpc/h)
 50.823        zcentre  (Mpc/h)
 17.6          side length (Mpc/h)
 68.56d6       Effective resolution     ! Number of particles in high res
 ------------------------------------------------------------------------------
\end{lstlisting}
\newpage

\begin{lstlisting}[language=bash,caption={Parameter file supplied to \textsc{IC\_Gen} to produce the initial conditions for the low-resolution zoom simulation of the cluster \texttt{L0300N0564\_VR18\_-8res}.},label={app1:vr18-ics}]
[dc-alta2@login7a [cosma7] L0300N0564_VR18_-8res]$ cat params.inp

1003007067 Version tag - do not edit this line.
------------------------------------------------------------------------------
# Only the values of the parameters should be altered. Changing the number of
# lines in this file (including comments like these) or swapping lines may
# lead to errors on execution, or worse nonsensical output.
# Note however lines beginning with a '%' are ignored. ARJ 16/08/2016
------------------------------------------------------------------------------
---------**********  Put text description of ICs below in quotes ********* ---
'L0300N0564_VR18_-8res'
------------------------------------------------------------------------------
# The three parameters below define the type of output
------------------------------------------------------------------------------
2                    ! Output format: (2 - hdf5) (1 - Gadget) (0 - ARJ BINARY)
1                    ! Indexing: (0 - 4 byte),(1 - 8 byte),(2 - ph), (3 -CSI)
21                   ! nbit - used for P-H indices
1                    ! Endianess:(0 - Big)(1 - Little)(2 - Native)
------------------------------------------------------------------------------
# Gadget file parameters (ignored for ARJ BIN FORMAT-but required to be here)
# The values chosen for the coordinate shift become the new origin in the
# output file.
# The create SPH and species division factors should be given in the form
# of log10(mparticle division/mass of all particles) - i.e. negative reals.
------------------------------------------------------------------------------
4                              ! Precision of output file (4-single) (8-double)
------------------------------------------------------------------------------
0.0 0.0 0.0                    ! Coordinate shift in Gadget Units (0,0,0)=none
0                              ! App sph truncation about origin?  0 == No
0.0                            ! Radius of spherical truncation
0                              ! Add SPH particles?  0 - No  1 - Yes
0.0                            ! Omega_b/Omega_0
0.0                            ! Create SPH particles for dm below this mass
1.0                            ! Split factor:  0 - place SPH on dark matter
2                              ! Number of species (excluding gas == sp-0)
-10.05                         ! Division sp-1 and sp-2 (if present)
0.00                           ! Division sp-2 and sp-3 (if present)
0.0                            ! Division sp-3 and sp-4 (if present)
0.0                            ! Division sp-4 and sp-5 (if present)
--short_param.inp------------------------------------------------------------
# Resimulation parameters below:
----------------------------------------------------------------------------
0                              ! 2lpt ICS:  0-No 1-Yes (Cosm)  2-Yes (Resim)
0                              ! Write out Lagrangian positions 1-Yes,0-No
0                              ! For Resim ics only:  1: output linear density; 0 nothing
----------------------------------------------------------------------------
0                             ! 0 - Import file : -1 -  cosm glass file:  > 0 the 1-d cubic grid size
'./ic_gen_submit_files/L0300N0564_VR18_-8res/particle_load/fbinary/PL'  ! (Imported) Part load basename
1                             ! Part load Rep factor
'./ic_gen_submit_files/L0300N0564_VR18_-8res/ICs/L0300N0564_VR18_-8res'  !output file base name
----------------------------------------------------------------------------
'EAGLE_XL_powspec_18-07-2019.txt'         ! Input Power spectrum to use
----------------------------------------------------------------------------
# ********* Details of the parent simulation *****************
202.98000000           lbox             ! Parent box size in Mpc/h
0.31110000  0.68890000  omega0,lambda0  ! Cosmological parameters of parent sim
0.81020000            sigma8_p          ! sigma_8(z=0) of parent simulation
0.67660000         Hubble_parameter     ! Redshift zero Hubble constant in units 100 kms/Mpc
127.00000000           zshift           ! Redshift of the required displacement field
----------------------------------------------------------------------------
 # ********* Phases used: place the panphasian descriptor on the line below ***
[Panph1,L18,(74412,22732,260484),S3,CH1799108544,EAGLE-XL_L0300_VOL1]
0
#-------------- List constraint phase descriptors and (newline) file path, and levels to use
%dummy
%dummy
%dummy
%dummy
%dummy
%dummy
----------------------------------------------------------------------------
1             igrid           ! Fourier grid number
0.5           xcentre         ! x - Centre and size in units of lbox
0.5           ycentre         ! y -
0.5           zcentre         ! z -
1.0           side0           ! side-length
11451483064   nparent         ! Number of particles in parent simulation
1536          ndim_fft        ! Dimension of fourier mesh used
0             cic_correct     ! Correct for CIC assign  1 - Yes
1             iquad           ! Quadratic interp - 1 -Yes(Cancels cic).
0             idep            ! Number of parent fourier grid
----------------------------------------------------------------------------
1                             ! Multigrid initial conditions?    0 : No   1: Yes
----------------------------------------------------------------------------
           1  basis function choice: 1 - Use SET8 basis function :0- use SET1
          12  Openning inverse angle (integer) used for multigrid ics only.
          1 1 1 1 1 1 1 1 1 1 1 1 1 1 1 1 Include contribution from levels - 1=YES
----------------------------------------------------------------------------
 # *********** Location of high resolution region *****************
98.48054373     xcentre  (Mpc/h)
16.06847645     ycentre  (Mpc/h)
58.56566873     zcentre  (Mpc/h)
54.36964286     side length (Mpc/h)
220075365       Effective resolution     ! Number of particles in high res region (=603^3)
----------------------------------------------------------------------------

\end{lstlisting}
\newpage

\begin{lstlisting}[language=bash,caption={Parameter file supplied to \textsc{make\_mask.py} to produce the high-mass resolution for the cluster \texttt{L0300N0564\_VR18\_-8res}, in preparation for the particle load.},label={app1:vr18-mask}]
# This lists all the available options, not to be used as a param file in its native form.
# All params, e.g. the coordinates, are expected to be in the native units of the snapshot.
# NOTE: If you are using an automatic group selection, make sure you are using a Velociraptor
# output catalogue (.properties). SUBFIND output catalogues are deprecated as of 17-09-2020.
# If using automatic selection through the Velociraptor catalogue a sorting key must be selected
# and the GroupNumber is the index of the group in the list sorted by decreasing mass (from largest to
# smallest).

# SET-UP MASK #
fname:             L0300N0564_VR18     # The save name for the mask file
snap_file:         /cosma/home/dp004/dc-alta2/data7/xl-zooms/parent_volume/L0300N0564/snapshots/EAGLE-XL_L0300N0564_DMONLY_0036.hdf5  # The location of the snapshot we are creating the mask from
bits:              21                  # The number of bits used in the particle IDs for the Peano-Hilbert indexing (EAGLE runs use 14)
shape:             sphere              # Shape of the region to reproduce. Available are: 'sphere', 'cubiod', 'slab'
data_type:         swift               # Can be 'gadget' or 'swift' (default 'swift')
divide_ids_by_two: False               # True if you need to divide PIDs by two to get back to ICS (needed for eagle)
min_num_per_cell:  5                   # Minimum particles per cell (default 3). Cells with less than `min_num_per_cell` particles are ignored
mpc_cell_size:     1.1                 # Cell size in Mpc / h (default 3.)
select_from_vr:    1                   # If set to 1, it enables automatic groups search from the Velociraptor catalogue. Set to 0 for manual selection

# TOPOLOGICAL CLOSURE #
topology_fill_holes:       1           # Toggle algorithm for filling holes. Set value to 1 to enable, 0 to disable (default 1)
topology_dilation_niter:   1           # Number of iterations of the algorithm for extrusion. Set value to 0 to disable (default 0)
topology_closing_niter:    1           # Number of iterations of the algorithm for rounding edges. Set value to 0 to disable (default 0)

# AUTOMATIC GROUP SELECTION #
vr_file:             /cosma/home/dp004/dc-alta2/data7/xl-zooms/parent_volume/L0300N0564/snapshots/stf_swiftdm_3dfof_subhalo_0036/stf_swiftdm_3dfof_subhalo_0036.VELOCIraptor.properties.0 # The location of the Velociraptor group catalogue (to find coordinates of given GN)
sort_m200crit:       0                 # If 1, sorts groups in the VR catalogue by M_200crit. Overrides `sort_m500crit` if both specified
sort_m500crit:       1                 # If 1, sorts groups in the VR catalogue by M_500crit
GN:                  18                # The Group-Number of the halo in the structure-finding catalogue (requires `vr_file`). GN relative to the sorting rule
#highres_radius_r200: 5.                 # How many times r200 of the passed group do you want to re-simulate. Overrides `highres_radius_r500` if both selected
highres_radius_r500: 6.                # How many times r500 of the passed group do you want to re-simulate

# OUTPUT DIRECTORY #
output_dir:           /cosma/home/dp004/dc-alta2/phd_thesis/ # Directory where the mask image and hdf5 output are saved

\end{lstlisting}
\newpage

\begin{lstlisting}[language=bash,caption={Parameter file supplied to \swift~to produce the high-resolution simulation for the cluster \texttt{L0300N0564\_VR18\_+1res}. This file also contains full information on the sub-grid physics used in our SWIFT-EAGLE Ref model.},label={app1:vr18-subgrid}]
[dc-alta2@login8b VR18_+1res_Ref]$ cat params.yml
# Define some meta-data about the simulation
MetaData:
  run_name:   L0300N0564_VR18_+1res_MinimumDistance_fixedAGNdT8.5_dec2021

# Define the system of units to use internally.
InternalUnitSystem:
  UnitMass_in_cgs:     1.98841e43    # 10^10 M_sun in grams
  UnitLength_in_cgs:   3.08567758e24 # Mpc in centimeters
  UnitVelocity_in_cgs: 1e5           # km/s in centimeters per second
  UnitCurrent_in_cgs:  1             # Amperes
  UnitTemp_in_cgs:     1             # Kelvin

# Cosmological parameters
Cosmology:
  h:              0.67660        # Reduced Hubble constant
  a_begin:        0.00781250     # Initial scale-factor of the simulation
  a_end:          1.00000000     # Final scale factor of the simulation
  Omega_cdm:      0.26213000     # Matter density parameter
  Omega_lambda:   0.68890000     # Dark-energy density parameter
  Omega_b:        0.04897000     # Baryon density parameter

# Parameters governing the time integration
TimeIntegration:
  dt_min:     1e-10 # The minimal time-step size of the simulation (in internal units).
  dt_max:     1e-2  # The maximal time-step size of the simulation (in internal units).

# Parameters governing the snapshots
Snapshots:
  basename:            snap
  delta_time:          1.02
  scale_factor_first:  0.02
  output_list_on:      1
  output_list:         ./snap_redshifts.txt
  compression:         0
  subdir:              snapshots

# Parameters governing the conserved quantities statistics
Statistics:
  delta_time:           1.02
  scale_factor_first:   0.05

Gravity:
  eta:                              0.025     # Constant dimensionless multiplier for time integration.
  MAC:                              adaptive  # Use the geometric opening angle condition
  theta_cr:                         0.7       # Opening angle (Multipole acceptance criterion)
  use_tree_below_softening:         1
  allow_truncation_in_MAC:          1
  epsilon_fmm:                      0.001
  mesh_side_length:                 512
  comoving_DM_softening:            0.00332742 # Comoving softening for DM (3.32 ckpc)
  comoving_baryon_softening:        0.00190215 # Comoving softening for baryons (1.79 ckpc)
  max_physical_DM_softening:        0.00147885 # Physical softening for DM (1.30 pkpc)
  max_physical_baryon_softening:    0.00084540 # Physical softening for baryons (0.70 pkpc)
  softening_ratio_background:       0.020

# Parameters for the hydrodynamics scheme
SPH:
  resolution_eta:                    1.2348     # Target smoothing length in units of the mean inter-particle separation (1.2348 == 48Ngbs with the cubic spline kernel).
  h_min_ratio:                       0.01       # Minimal smoothing length in units of softening.
  h_max:                             0.5        # Maximal smoothing length in co-moving internal units.
  CFL_condition:                     0.2        # Courant-Friedrich-Levy condition for time integration.
  minimal_temperature:               100.0      # (internal units)
  initial_temperature:               268.7      # (internal units)
  particle_splitting:                1          # Particle splitting is ON
  particle_splitting_mass_threshold: 0.00073348 # (internal units, i.e. 7e6 Msun ~ 4x initial gas particle mass)

# Parameters of the stars neighbour search
Stars:
  resolution_eta:        1.1642   # Target smoothing length in units of the mean inter-particle separation
  h_tolerance:           7e-3
  luminosity_filename:   ../photometry/

# Parameters for the Friends-Of-Friends algorithm
FOF:
  basename:                        fof_output  # Filename for the FOF outputs.
  min_group_size:                  256         # The minimum no. of particles required for a group.
  linking_length_ratio:            0.2         # Linking length in units of the main inter-particle separation.
  seed_black_holes_enabled:        1           # Enable seeding of black holes in FoF groups
  black_hole_seed_halo_mass_Msun:  1.0e10      # Minimal halo mass in which to seed a black hole (in solar masses).
  scale_factor_first:              0.05        # Scale-factor of first FoF black hole seeding calls.
  delta_time:                      1.00751     # Scale-factor ratio between consecutive FoF black hole seeding calls.

Scheduler:
  max_top_level_cells:   32
  cell_split_size:       200

Restarts:
  enable:             1
  save:               1          # Keep copies
  onexit:             1
  subdir:             restart    # Sub-directory of the directory where SWIFT is run
  basename:           swift
  delta_hours:        5.0
  stop_steps:         100
  max_run_time:       71.5       # In hours
  resubmit_on_exit:   1
  resubmit_command:   ./auto_resubmit

# Parameters related to the initial conditions
InitialConditions:
  file_name: /cosma7/data/dp004/dc-alta2/xl-zooms/ics/ic_gen/run/ic_gen_submit_files/1LPT_8bitID/L0300N0564_VR18_+1res/L0300N0564_VR18_+1res.hdf5
  periodic:   1
  cleanup_h_factors: 1               # Remove the h-factors inherited from Gadget
  cleanup_velocity_factors: 1        # Remove the sqrt(a) factor in the velocities inherited from Gadget
  generate_gas_in_ics: 1             # Generate gas particles from the DM-only ICs
  cleanup_smoothing_lengths: 1       # Since we generate gas, make use of the (expensive) cleaning-up procedure.
  remap_ids: 1                       # Re-map the IDs to [1, N] to avoid collision problems when splitting

# Parameters of the line-of-sight outputs
LineOfSight:
  basename:            eagle_los
  num_along_x:         0
  num_along_y:         0
  num_along_z:         100
  scale_factor_first:  0.1
  delta_time:          1.1

# Impose primoridal metallicity
EAGLEChemistry:
  init_abundance_metal:     0.
  init_abundance_Hydrogen:  0.752
  init_abundance_Helium:    0.248
  init_abundance_Carbon:    0.0
  init_abundance_Nitrogen:  0.0
  init_abundance_Oxygen:    0.0
  init_abundance_Neon:      0.0
  init_abundance_Magnesium: 0.0
  init_abundance_Silicon:   0.0
  init_abundance_Iron:      0.0

# EAGLE cooling parameters
EAGLECooling:
  dir_name:                ../coolingtables/
  H_reion_z:               7.5                 # Planck 2018
  H_reion_eV_p_H:          2.0
  He_reion_z_centre:       3.5
  He_reion_z_sigma:        0.5
  He_reion_eV_p_H:         2.0

# COLIBRE cooling parameters
COLIBRECooling:
  dir_name:                ../UV_dust1_CR1_G1_shield1.hdf5 # Location of the cooling tables
  H_reion_z:               7.5               # Redshift of Hydrogen re-ionization (Planck 2018)
  H_reion_eV_p_H:          2.0
  He_reion_z_centre:       3.5               # Redshift of the centre of the Helium re-ionization Gaussian
  He_reion_z_sigma:        0.5               # Spread in redshift of the  Helium re-ionization Gaussian
  He_reion_eV_p_H:         2.0               # Energy inject by Helium re-ionization in electron-volt per Hydrogen atom
  delta_logTEOS_subgrid_properties: 0.3      # delta log T above the EOS below which the subgrid properties use Teq assumption
  rapid_cooling_threshold:          0.333333 # Switch to rapid cooling regime for dt / t_cool above this threshold.

# EAGLE star formation parameters
EAGLEStarFormation:
  SF_threshold:                      Subgrid      # Zdep (Schaye 2004) or Subgrid
  SF_model:                          PressureLaw  # PressureLaw (Schaye et al. 2008) or SchmidtLaw
  KS_normalisation:                  1.515e-4     # The normalization of the Kennicutt-Schmidt law in Msun / kpc^2 / yr.  KS_exponent:                       1.4          # The exponent of the Kennicutt-Schmidt law.
  min_over_density:                  100.0        # The over-density above which star-formation is allowed.
  KS_high_density_threshold_H_p_cm3: 1e3          # Hydrogen number density above which the Kennicut-Schmidt law changes slope in Hydrogen atoms per cm^3.
  KS_high_density_exponent:          2.0          # Slope of the Kennicut-Schmidt law above the high-density threshold.
  EOS_entropy_margin_dex:            0.3          # When using Z-based SF threshold, logarithm base 10 of the maximal entropy above the EOS at which stars can form.
  threshold_norm_H_p_cm3:            0.1          # When using Z-based SF threshold, normalisation of the metal-dependant density threshold for star formation in Hydrogen atoms per cm^3.
  threshold_Z0:                      0.002        # When using Z-based SF threshold, reference metallicity (metal mass fraction) for the metal-dependant threshold for star formation.
  threshold_slope:                   -0.64        # When using Z-based SF threshold, slope of the metal-dependant star formation threshold
  threshold_max_density_H_p_cm3:     10.0         # When using Z-based SF threshold, maximal density of the metal-dependant density threshold for star formation in Hydrogen atoms per cm^3.
  threshold_temperature1_K:          1000         # When using subgrid-based SF threshold, subgrid temperature below which gas is star-forming.
  threshold_temperature2_K:          31622        # When using subgrid-based SF threshold, subgrid temperature below which gas is star-forming if also above the density limit.
  threshold_number_density_H_p_cm3:  10           # When using subgrid-based SF threshold, subgrid number density above which gas is star-forming if also below the second temperature limit.

# Parameters for the EAGLE "equation of state"
EAGLEEntropyFloor:
  Jeans_density_threshold_H_p_cm3: 1e-4      # Physical density above which the EAGLE Jeans limiter entropy floor kicks in expressed in Hydrogen atoms per cm^3.
  Jeans_over_density_threshold:    10.       # Overdensity above which the EAGLE Jeans limiter entropy floor can kick in.
  Jeans_temperature_norm_K:        800       # Temperature of the EAGLE Jeans limiter entropy floor at the density threshold expressed in Kelvin.
  Jeans_gamma_effective:           1.3333333 # Slope the of the EAGLE Jeans limiter entropy floor
  Cool_density_threshold_H_p_cm3: 1e-5       # Physical density above which the EAGLE Cool limiter entropy floor kicks in expressed in Hydrogen atoms per cm^3.
  Cool_over_density_threshold:    10.        # Overdensity above which the EAGLE Cool limiter entropy floor can kick in.  Cool_temperature_norm_K:        10.        # Temperature of the EAGLE Cool limiter entropy floor at the density threshold expressed in Kelvin. (NOTE: This is below the min T and hence this floor does nothing)
  Cool_gamma_effective:           1.         # Slope the of the EAGLE Cool limiter entropy floor

# EAGLE feedback model
EAGLEFeedback:
  use_SNII_feedback:                    1               # Global switch for SNII thermal (stochastic) feedback.
  use_SNIa_feedback:                    1               # Global switch for SNIa thermal (continuous) feedback.
  use_AGB_enrichment:                   1               # Global switch for enrichement from AGB stars.
  use_SNII_enrichment:                  1               # Global switch for enrichement from SNII stars.
  use_SNIa_enrichment:                  1               # Global switch for enrichement from SNIa stars.
  filename:                             ../yieldtables/ # Path to the directory containing the EAGLE yield tables.
  IMF_min_mass_Msun:                    0.1             # Minimal stellar mass considered for the Chabrier IMF in solar masses.
  IMF_max_mass_Msun:                  100.0             # Maximal stellar mass considered for the Chabrier IMF in solar masses.
  SNII_energy_fraction_function: EAGLE
  SNII_min_mass_Msun:                   8.0             # Minimal mass considered for SNII stars in solar masses.
  SNII_max_mass_Msun:                 100.0             # Maximal mass considered for SNII stars in solar masses.
  SNII_feedback_model:                  MinimumDistance       # Feedback modes: Random, Isotropic, MinimumDistance, MinimumDensity
  SNII_sampled_delay:                   1               # Sample the SNII lifetimes to do feedback.
  SNII_delta_T_K:                       3.16228e7       # Change in temperature to apply to the gas particle in a SNII thermal feedback event in Kelvin.
  SNII_energy_erg:                      1.0e51          # Energy of one SNII explosion in ergs.
  SNII_energy_fraction_min:             0.5             # Minimal fraction of energy applied in a SNII feedback event.
  SNII_energy_fraction_max:             5.0             # Maximal fraction of energy applied in a SNII feedback event.
  SNII_energy_fraction_Z_0:             0.0012663729    # Pivot point for the metallicity dependance of the SNII energy fraction (metal mass fraction).
  SNII_energy_fraction_n_0_H_p_cm3:     1.4588          # Pivot point for the birth density dependance of the SNII energy fraction in cm^-3.
  SNII_energy_fraction_n_Z:             0.8686          # Power-law for the metallicity dependance of the SNII energy fraction.
  SNII_energy_fraction_n_n:             0.8686          # Power-law for the birth density dependance of the SNII energy fraction.
  SNII_energy_fraction_use_birth_density: 0             # Are we using the density at birth to compute f_E or at feedback time?
  SNII_energy_fraction_use_birth_metallicity: 0         # Are we using the metallicity at birth to compuote f_E or at feedback time?
  SNIa_DTD:                             Exponential     # Functional form of the SNIa delay time distribution.
  SNIa_DTD_delay_Gyr:                   0.04            # Stellar age after which SNIa start in Gyr (40 Myr corresponds to stars ~ 8 Msun).
  SNIa_DTD_exp_timescale_Gyr:           2.0             # Time-scale of the exponential decay of the SNIa rates in Gyr.
  SNIa_DTD_exp_norm_p_Msun:             0.002           # Normalisation of the SNIa rates in inverse solar masses.
  SNIa_energy_erg:                     1.0e51           # Energy of one SNIa explosion in ergs.
  AGB_ejecta_velocity_km_p_s:          10.0             # Velocity of the AGB ejectas in km/s.
  stellar_evolution_age_cut_Gyr:        0.1             # Stellar age in Gyr above which the enrichment is down-sampled.  stellar_evolution_sampling_rate:       10             # Number of time-steps in-between two enrichment events for a star above the age threshold.
  SNII_yield_factor_Hydrogen:           1.0             # (Optional) Correction factor to apply to the Hydrogen yield from the SNII channel.
  SNII_yield_factor_Helium:             1.0             # (Optional) Correction factor to apply to the Helium yield from the SNII channel.
  SNII_yield_factor_Carbon:             0.5             # (Optional) Correction factor to apply to the Carbon yield from the SNII channel.
  SNII_yield_factor_Nitrogen:           1.0             # (Optional) Correction factor to apply to the Nitrogen yield from the SNII channel.
  SNII_yield_factor_Oxygen:             1.0             # (Optional) Correction factor to apply to the Oxygen yield from the SNII channel.
  SNII_yield_factor_Neon:               1.0             # (Optional) Correction factor to apply to the Neon yield from the SNII channel.
  SNII_yield_factor_Magnesium:          2.0             # (Optional) Correction factor to apply to the Magnesium yield from the SNII channel.
  SNII_yield_factor_Silicon:            1.0             # (Optional) Correction factor to apply to the Silicon yield from the SNII channel.
  SNII_yield_factor_Iron:               0.5             # (Optional) Correction factor to apply to the Iron yield from the SNII channel.

# EAGLE AGN model
EAGLEAGN:
  subgrid_seed_mass_Msun:             1.0e4           # Black hole subgrid mass at creation time in solar masses.
  use_multi_phase_bondi:              0               # Compute Bondi rates per neighbour particle?
  use_subgrid_bondi:                  0               # Compute Bondi rates using the subgrid extrapolation of the gas properties around the BH?
  with_angmom_limiter:                0               # Are we applying the Rosas-Guevara et al. (2015) viscous time-scale reduction term?
  viscous_alpha:                      1e6             # Normalisation constant of the viscous time-scale in the accretion reduction term
  with_boost_factor:                  0               # Are we using the model from Booth & Schaye (2009)?
  boost_alpha:                        1.              # Lowest value for the accretion effeciency for the Booth & Schaye 2009 accretion model.
  boost_beta:                         2.              # Slope of the power law for the Booth & Schaye 2009 model, set beta to zero for constant alpha models.
  boost_n_h_star_H_p_cm3:             0.1             # Normalization of the power law for the Booth & Schaye 2009 model in cgs (cm^-3).
  with_fixed_T_near_EoS:              0               # Are we using a fixed temperature to compute the sound-speed of gas on the entropy floor in the Bondy-Hoyle accretion term?
  fixed_T_above_EoS_dex:              0.3             # Distance above the entropy floor for which we use a fixed sound-speed
  fixed_T_near_EoS_K:                 8000            # Fixed temperature assumed to compute the sound-speed of gas on the entropy floor in the Bondy-Hoyle accretion term
  radiative_efficiency:               0.1             # Fraction of the accreted mass that gets radiated.
  use_nibbling:                       1               # Continuously transfer small amounts of mass from all gas neighbours to a black hole [1] or stochastically swallow whole gas particles [0]?
  min_gas_mass_for_nibbling:          9e5             # Minimum mass for a gas particle to be nibbled from [M_Sun]. Only used if use_nibbling is 1.
  max_eddington_fraction:             1.              # Maximal allowed accretion rate in units of the Eddington rate.
  eddington_fraction_for_recording:   0.1             # Record the last time BHs reached an Eddington ratio above this threshold.
  coupling_efficiency:                0.1             # Fraction of the radiated energy that couples to the gas in feedback events.
  AGN_feedback_model:                 MinimumDistance # Feedback modes: Random, Isotropic, MinimumDistance, MinimumDensity
  AGN_use_deterministic_feedback:     1               # Deterministic (reservoir) [1] or stochastic [0] AGN feedback?
  use_variable_delta_T:               0               # Switch to enable adaptive calculation of AGN dT [1], rather than using a constant value [0].
  AGN_with_locally_adaptive_delta_T:  0               # [UNUSED] Switch to enable additional dependence of AGN dT on local gas density and temperature (only used if use_variable_delta_T is 1).
  AGN_delta_T_mass_norm:              1e8             # [UNUSED] Normalisation temperature of AGN dT scaling with BH subgrid mass [K] (only used if use_variable_delta_T is 1).
  AGN_delta_T_mass_reference:         1e8             # [UNUSED] BH subgrid mass at which the normalisation temperature set above applies [M_Sun] (only used if use_variable_delta_T is 1).
  AGN_delta_T_mass_exponent:          0.666667        # [UNUSED] Power-law index of AGN dT scaling with BH subgrid mass (only used if use_variable_delta_T is 1).
  AGN_delta_T_crit_factor:            1.0             # [UNUSED] Multiple of critical dT for numerical efficiency (Dalla Vecchia & Schaye 2012) to use as dT floor (only used if use_variable_delta_T and AGN_with_locally_adaptive_delta_T are both 1).
  AGN_delta_T_background_factor:      0.0             # [UNUSED] Multiple of local gas temperature to use as dT floor (only used if use_variable_delta_T and AGN_with_locally_adaptive_delta_T are both 1).
  AGN_delta_T_min:                    1e7             # [UNUSED] Minimum allowed value of AGN dT [K] (only used if use_variable_delta_T is 1).
  AGN_delta_T_max:                    3e9             # [UNUSED] Maximum allowed value of AGN dT [K] (only used if use_variable_delta_T is 1).
  AGN_delta_T_K:                      3.16228e8       # Change in temperature to apply to the gas particle in an AGN feedback event [K] (used if use_variable_delta_T is 0 or AGN_use_nheat_with_fixed_dT is 1 AND to initialise the BHs).
  AGN_use_nheat_with_fixed_dT:        1               # Switch to use the constant AGN dT, rather than the adaptive one, for calculating the energy reservoir threshold.
  AGN_use_adaptive_energy_reservoir_threshold: 0      # Switch to calculate an adaptive AGN energy reservoir threshold.
  AGN_num_ngb_to_heat:                1.              # Target number of gas neighbours to heat in an AGN feedback event (only used if AGN_use_adaptive_energy_reservoir_threshold is 0).
  max_reposition_mass:                1e20            # Maximal BH mass considered for BH repositioning in solar masses (large number implies we always reposition).
  max_reposition_distance_ratio:      3.0             # Maximal distance a BH can be repositioned, in units of the softening length.
  with_reposition_velocity_threshold: 0               # Should we only reposition to particles that move slowly w.r.t. the black hole?
  max_reposition_velocity_ratio:      0.5             # Maximal velocity offset of a particle to reposition a BH to, in units of the ambient sound speed of the BH. Only meaningful if with_reposition_velocity_threshold is 1.
  min_reposition_velocity_threshold: -1.0             # Minimal value of the velocity threshold for repositioning [km/s], set to < 0 for no effect. Only meaningful if with_reposition_velocity_threshold is 1.
  set_reposition_speed:               0               # Should we reposition black holes with (at most) a prescribed speed towards the potential minimum?
  threshold_major_merger:             0.333           # Mass ratio threshold to consider a BH merger as 'major'
  threshold_minor_merger:             0.1             # Mass ratio threshold to consider a BH merger as 'minor'
  merger_threshold_type:              2               # Type of velocity threshold for BH mergers (0: v_circ at kernel edge, 1: v_esc at actual distance, with softening, 2: v_esc at actual distance, no softening).
  merger_max_distance_ratio:          3.0             # Maximal distance over which two BHs can merge, in units of the softening length.
  \end{lstlisting}